\documentclass[a4paper,12pt]{article}

\usepackage{soul}
\usepackage[usenames,dvipsnames]{color}
\usepackage{jheppub}
\usepackage{amsmath, amssymb, slashed, epsf, color, graphicx, latexsym}
\usepackage{epsfig}
\newcommand{\n}{\nu}
\newcommand{\m}{\mu}
\newcommand{\s}{\sigma}
\newcommand{\nn}{\nonumber}
\usepackage{color}

\begin{document}
\preprint{TIFR/TH/16-41}
\title{Currents and Radiation from the large $D$
Black Hole  Membrane}
\author[a]{Sayantani Bhattacharyya,} 
\author[a]{Anup Kumar Mandal,}
\author[b] {Mangesh Mandlik,}
\author[c]{Umang Mehta,}
\author[b]{Shiraz Minwalla,}
\author[c]{Utkarsh Sharma}
\author[b]{and Somyadip Thakur}

\affiliation[a]{Indian Institute of Technology Kanpur, Kanpur, India-208016}
\affiliation[b]{Tata Institute of Fundamental Research, Mumbai, India-400005}
\affiliation[c]{Indian Institute of Technology Bombay, Mumbai, India-400076}
\emailAdd{sayanta@iitk.ac.in}
\emailAdd{anupkm@iitk.ac.in}
\emailAdd{mangesh@theory.tifr.res.in}
\emailAdd{umangmehta@iitb.ac.in}
\emailAdd{minwalla@theory.tifr.res.in}
\emailAdd{utkarsh.sharma@iitb.ac.in}
\emailAdd{somyadip@theory.tifr.res.in}

\abstract{It has recently been demonstrated that black hole dynamics 
in a large number of dimensions $D$ reduces to the dynamics of a codimension
one membrane propagating in flat space. In this paper we define a stress 
tensor and charge current on this membrane and explicitly determine these 
currents at low orders in the expansion in $\frac{1}{D}$.
We demonstrate that dynamical membrane equations of motion derived in 
earlier work are simply conservation equations for our stress tensor and 
charge current. Through the paper we focus on solutions of the membrane 
equations which vary on a time scale of order unity. 
Even though the charge current and stress tensor are not parametrically small 
in such solutions,  we show that the radiation sourced by the corresponding 
membrane currents is generically of order $\frac{1}{D^D}$.  
In this regime it follows that the `near horizon' membrane degrees of freedom are decoupled 
from asymptotic flat space at every perturbative order in the $\frac{1}{D}$ 
expansion. We also define an entropy current on the membrane and use the 
Hawking area theorem to demonstrate that the divergence of the entropy 
current is point wise non negative. We view this result as a local form of 
the second law of thermodynamics for membrane motion.}

\maketitle

\section{Introduction}

\subsection{Review of Black hole - Membrane duality}

The classical dynamics of black holes in asymptotically Minkowski spacetimes has recently been shown to simplify in a large number of dimensions $D$. Consider a violent dynamical process such as a collision between two black holes. The dynamics of this situation is complicated when the black holes first 
`collide' . After a time of order $1/D$ after the `merger' however, 
it turns out that 
the spacetime metric settles down into a configuration whose near horizon geometry is a union of overlapping patches, each of size $1/D$. The geometry of each patch closely resembles that of a Schwarzschild or Reissner Nordstrom black hole. The effective radius, boost velocity and charge of these patches varies 
on the event horizon over time and length scales of order unity. The subsequent evolution of the 
spacetime is governed by an effective dynamical system whose variables are the effective shape of the event horizon (one function) together with its local boost velocity field ($D-2$ functions) and charge density field (one function), a total of $D$ functions of $D-1$ variables. The dynamical evolution of these variables is governed by a set of local membrane equations of motion. The underlying  Einstein-Maxwell equations that govern the dynamics of this system 
uniquely determine the membrane equations in a power series expansion in $1/D$. At leading order in $1/D$ the membrane
equations of motion take the form
\begin{equation}\label{memeom}
\begin{split}
\hat\nabla\cdot u &=0\\
 p^\nu_\mu \left(u\cdot\hat\nabla\right) u_\nu&=
p^\nu_\mu\left( \frac{ \hat\nabla^2u_\nu-(1-Q^2) \hat\nabla_\nu K+ K ~(u^\alpha K_{\alpha\nu} ) } {K(1+Q^2)}  \right),
\\
u^\nu\hat \nabla_\nu  \left( KQ \right)  &=\hat\nabla^2Q -{K}Q ~(u^\alpha K_{\alpha\beta} u^\beta),
\end{split}
\end{equation}\vspace{.5cm}

\eqref{memeom}\footnote{
The equations \eqref{memeom} were first obtained in 
the papers \cite{Bhattacharyya:2015dva,Bhattacharyya:2015fdk} building on the 
earlier work \cite{Emparan:2013moa,Emparan:2013xia,Emparan:2013oza,Emparan:2014cia,Emparan:2014jca,
Emparan:2014aba,Emparan:2015rva}. See also \cite{Emparan:2015hwa,Suzuki:2015iha,Tanabe:2015isb, Tanabe:2016opw} for the independent derivation of membrane equations in for the special case 
of stationary solutions. \eqref{memeom} had been generalized in \cite{Dandekar:2016fvw} to include 
first correction in $1/D$ for the special case of uncharged black hole membranes.
\cite{Emparan:2015gva,Suzuki:2015axa,Tanabe:2015hda,Emparan:2016sjk,
Tanabe:2016pjr} have also independently derived the equations of membrane 
dynamics in the so called `black brane' limit. At least for the case of uncharged black 
holes, the equations of \cite{Emparan:2015gva,Suzuki:2015axa,Tanabe:2015hda,Emparan:2016sjk,
Tanabe:2016pjr} were demonstrated in \cite{Dandekar:2016jrp} to be a special case 
(a special scaling limit) of the equation \eqref{memeom}. See 
\cite{Sadhu:2016ynd,Herzog:2016hob,Rozali:2016yhw,Chen:2015fuf,Giribet:2013wia,
Prester:2013gxa,Chen:2016fuy} for recent related work. }
\footnote{The notation used in this equation goes as follows. 
Here we view the membrane as embedded in flat Minkowski space. Small Greek indices denotes the intrinsic coordinates along the membrane worldvolume.  $\hat\nabla_\mu$ denotes the covariant derivative with respect to the intrinsic metric of the membrane, $g^{(ind,f)}_{\mu\nu}$. All raising and lowering of indices are also done using this intrinsic metric. 
$K_{\mu\nu}$ is the 
extrinsic curvature of the membrane , $K=K^\mu_\mu$ is 
the trace of the extrinsic curvature, $p_{\mu\nu}$ is the projector 
orthogonal to the velocity field 
$${p}_{\mu\nu}=g^{(ind,f)}_{\mu\nu}+u_\mu u_\nu, $$
$u_\mu$ is the velocity.} 
 are a set of $D$ equations for as many variables. It follows that  \eqref{memeom} defines a well posed initial value problem for membrane dynamics. 

We have presented the membrane equations \eqref{memeom} at leading order in the expansion in 
$\frac{1}{D}$; as a consequence all terms in each of the equations \eqref{memeom} are of the same 
order in $D$, where orders of $D$ are counted according to the rules spelt out in 
\cite{Bhattacharyya:2015fdk}. According to the rules of that paper in particular,
all divergences and Laplacians are of order $D$, while contractions of indices 
of the form $A_M B^M$ are of order unity. As an example of an application of 
this rule, $\nabla^2 u^M$ and $K= \nabla_A n^A$ are both taken to be 
of order ${\cal O}(D)$ while $(u^AK_{AB}u^B)$ is assigned order ${\cal O}(1)$ This rule applies irrespective of whether we are dealing with space-time indices or worldvolume indices.  See \cite{Bhattacharyya:2015fdk} for an explanation of the rational behind this rule.

Using the rule spelt out in the previous paragraph, it 
follows that the LHS of the first equation in \eqref{memeom} is of order $D$. 
Every term in the third equation in \eqref{memeom} is also of order $D$. However each term 
in the second equation of \eqref{memeom} is of order unity.

The membrane whose dynamics is described by \eqref{memeom} may be thought of in the 
following picturesque terms. The membrane consists of a bunch of `particles' of density
$u^0= \gamma$ whose velocity is given by $\frac{u^i}{u^0}$. $u^M$ is the `density current' 
of these `particles' and the first equation in \eqref{memeom} is a statement of the 
conservation of this density current. With this interpretation, the conservation of this 
density current is simply the statement that our fictional particles flow from one 
point to another but are never created or destroyed \footnote{As we will see below, the `particles' in 
question will turn out to be the basic carriers of entropy of the membrane, and the 
`particle density current' mentioned here is closely related to the membrane's entropy current. 
The conservation of entropy density holds only at first order; we will show below that the divergence 
of the entropy current is generically nonzero (but positive) at second order in the expansion in 
$1/D$. This means that the fictional `particles' mentioned in the text above are created in dynamical 
flows at second and higher order in $1/D$.} 
The second equation in \eqref{memeom} may be regarded as a statement of Newton's laws for 
the constituent particles of the membrane. This equation asserts that the acceleration of any
given membrane particle is governed by `forces' (the RHS of the second equation in \eqref{memeom}) which 
depend on the trajectories of neighbouring particles. \footnote{We have a $D-2$ parameter set 
of particles which execute a $D-2$ parameter set of particle flows. The $D-1$ dimensional 
membrane world volume is simply the congruence of these flow lines. Note that the extrinsic curvature of the membrane at any given point is completely determined by the shape of particle flow lines in the neighbourhood of that point.} The terms on the RHS of the second of \eqref{memeom} are reminiscent of the 
force terms that act on a regular fluid. The first term on the RHS of \eqref{memeom} captures the 
force of shear viscosity while the second term is analogous to a pressure force, with the role 
of the pressure played by $K$ the trace of the extrinsic curvature of the membrane. This term drives flows that reduces gradients of $K$ and works to iron out 
wrinkles in the membrane world volume that might otherwise have developed 
over the course of a dynamical flow. In some sense this term is responsible for
stitching the independent particle world lines (or, more visually, world 
threads) into a smooth membrane surface.  

The last equation in \eqref{memeom} asserts that our particles carry a separate independent `charge' 
- with density 
proportional to ${K} Q$. This charge is carried along by our particles as they move. In addition it `diffuses' between particles in the manner specified by the RHS of the third equation in \eqref{memeom}.
This charge density is, of course, closely related to the electromagnetic charge current of the membrane, 
a statement we will make precise in this paper. 

Let us re-emphasize the main point. If we wait for a time large compared to $1/D$ after a 
cataclysmal event, the equations that govern black hole dynamics reduce to the equations
that govern the motion of a relativistic membrane that propagates in 
flat space. At first nontrivial order, the membrane may usefully be thought of as generated 
by the flow lines of a  collection of `particles' which interact with each other locally as they flow. 
The membrane equations \eqref{memeom} - which define a good initial value 
problem for the membrane shape and velocity field - are simply a rewriting 
of Einstein's equations for black hole dynamics at leading order in $1/D$ and in 
the appropriate regime.

\subsection{Membrane coupling to radiation: qualitative discussion}

In this paper we refer to all degrees of freedom that vary on time and length scales of order 
unity (rather than, say, $1/D$) as slow. The collective coordinate membrane motions described above 
are one set of slow degrees of freedom in black hole spacetimes. A second simpler set of
slow degrees of freedom are gravitons and photons that live far away from the black hole and 
have wavelengths of order unity or larger. It is natural to wonder how these two distinct classes of 
slow modes interact with each other. In this paper we present a detailed analysis of the coupling 
of these two classes of slow modes. We demonstrate, in particular, that the coupling between 
membrane modes and light  gravitons is of order $\frac{1}{D^\frac{D}{2}}$, 
and so is nonperturbatively small in the $1/D$ expansion.  

As we explain in section \ref{bgm} below 
the smallness of this coupling at large $D$ may be understood as follows. 
The slow modes that describe the collective coordinate motions of membranes are 
localized to a region very near the the black hole horizon by a large potential barrier. The barrier 
is kinematical in origin and schematically takes the form of a repulsive potential 
$V(r)= \frac{D^2}{4 r^2}$ in an effective one dimensional Schrodinger problem.  
In order to escape as radiation, a membrane mode which lives at 
the edge of the black hole of radius $r_0$ has to to tunnel through this barrier all the way out 
to $r \approx \frac{D}{2\omega}$ before it can start to propagate.  The amplitude for this tunneling 
process is suppressed by the area under the potential curve, and is of order  
$e^{-\frac{D}{2} \ln \frac{D}{2 \omega r_0}} \sim \left( \frac{2 \omega r_o}{D} \right)^\frac{D}{2}$. When $r_0 \omega$ is of order unity, 
this amplitude is nonperturbatively small in the 
$1/D$ expansion. It follows that membrane motions on time scale of order 
$1/r_0$ do not source radiation at any finite order in the $1/D$ expansion.

The discussion of the previous paragraph is reminiscent of 
Maldacena's argument for the decoupling of the near horizon geometry of 
a D3 brane from the external bulk in the context of the AdS/CFT 
correspondence \cite{Maldacena:1997re}. Indeed at energies of order unity, the 
limit $D \to \infty$ is effectively a decoupling limit for the near horizon 
region of the Schwarzschild and Reissner Nordstrom black holes, analogous in 
many respects to the Maldacena decoupling limit in which energies are 
held fixed as $\alpha'$ is taken to zero.

We would like to emphasize that the decoupling between membrane degrees of freedom and asymptotic 
infinity is accurate only for the classical theory of gravity and appears to fail quantum mechanically, 
even semiclassically. The reason for this is simply that near horizon modes with 
$\omega \sim \frac{D}{r_0}$ do not decouple from infinity. 
As we will review below, however, the Hawking temperature of a black hole of radius $r_0$ 
scales like $\frac{D}{r_0}$ at large $D$. It follows that the Hawking radiation emitted by 
a black hole at large $D$ does not decouple from infinity.  This observation suggests that it is 
misguided to hope that there exists a quantum microscopic theory of the large $D$ membrane described in this paper. Such a theory 
- which might have been hoped to stand in the same relation to the membrane equations \eqref{memeom} as 
${\cal N}=4$ Yang Mills theory does to the hydrodynamics of \cite{Bhattacharyya:2008mz,Bhattacharyya:2008jc} 
- appears never to decouple from asymptotic infinity. In other words the analysis of this 
paper should be viewed purely in terms of the classical equations of gravity and not as the first step 
in a programme to quantize gravity at large $D$.

\subsection{Membrane coupling to radiation: quantitative discussion}

Although membrane degrees of freedom couple very weakly to external 
gravitons and photons at large $D$, they do couple to these modes 
at any finite $D$ no matter how large. In other words membrane motions source gravitational and 
electromagnetic radiation. One of the principle accomplishments of this paper 
is the derivation of a formula for the radiation sourced by any given 
membrane motion. 

In order to obtain this formula we first note that the explicit $\frac{1}{D}$ expansion
of spacetime solutions dual to membrane motions (see  
\cite{Bhattacharyya:2015dva,Bhattacharyya:2015fdk, Dandekar:2016fvw})  
is valid only at points whose distance from the event horizon, $S$, obeys the 
inequality $S \ll r_0$ ( here $r_0$ is the local black hole radius).
\footnote{More precisely $r_0=\frac{D}{ K}$ where ${K}$ is the trace of
the extrinsic curvature of the membrane surface. We use the notation of \cite{Bhattacharyya:2015fdk}
through this paper. Recall that ${K}$ is of order $D$ so $r_0$ is of order unity. 
} 
When, on the other hand, $S \gg \frac{r_0}{D}$ the solution reduces to 
a small fluctuation about flat space. In this region the solution is  
well approximated by a solution of the 
Einstein Maxwell equations linearized about flat space. Notice that the 
domains of validity of these two approximations overlap: the 
$1/D$ expansion of \cite{Bhattacharyya:2015dva,Bhattacharyya:2015fdk,Dandekar:2016fvw}
and linearization are both valid approximations in the overlap regime\footnote{We explicitly verify below that the metric and gauge field presented in \cite{Bhattacharyya:2015fdk}
is a solution of the linearized Einstein Maxwell equations in this regime.}
\begin{equation}\label{overlap}
\frac{r_0}{D} \ll S \ll r_0.
\end{equation}

In the previous subsection we have explained that the  radiation field first 
begins to propagate at distances $S$ of order $\frac{D}{\omega}$ away from the 
membrane. These distances lie well outside the regime of the 
$1/D$ expansion of \cite{Bhattacharyya:2015dva,Bhattacharyya:2015fdk, 
Dandekar:2016fvw}. However 
the radiation fields are extremely small, and so are well described by the linearized 
Einstein Maxwell equations. In order to obtain the radiation field due to a given membrane motion, 
all we need to do is to identify the effective linearized solution that 
the spacetimes of \cite{Bhattacharyya:2015dva,Bhattacharyya:2015fdk, Dandekar:2016fvw} reduce to in the overlap region \eqref{overlap} and then continue this linearized 
solution to infinity. 

The implementation of this programme is, however, 
complicated by an important detail. In order to explain this point we first 
pause to provide a qualitative description of space of linearized solutions 
to the Einstein Maxwell equations away from the membrane , i.e. at 
distances $S \gg \frac{r_0}{D}$ 
to the exterior of the membrane. 
The linearized solutions in this region  turn out to be a superposition of two classes of modes; modes whose integrated flux 
decays towards infinity (we call these the decaying modes)  and modes  
whose integrated flux grow towards infinity (we call these the growing modes).
These can be understood as the decaying and growing modes
of the effective Schrodinger problem under the potential barrier
$V(r)= \frac{D^2}{4r^2}$ mentioned in the previous subsection. 
As we show in section \ref{bgm} below, decaying modes of the effective 
Schrodinger problem start out at order unity very near the membrane and 
decay rapidly upon progressing outwards. 
On the other hand growing modes start out at order $1/D^D$ near the membrane but 
grow equally rapidly away from the membrane. The growing modes catch up 
in magnitude with the decaying mode at a distances of order 
$\frac{D}{2 \omega} $ away from the 
membrane. This is also precisely the point beyond which both the modes emerge out from under 
the effective potential barrier. 
At larger distances the modes cease to grow or decay but oscillate, propagating in form of radiation 
fields. The integrated flux of both modes stays constant as $r$ is further increased. 

As mentioned above, the  $\frac{1}{D}$ expansion of \cite{Bhattacharyya:2015dva,Bhattacharyya:2015fdk, Dandekar:2016fvw}
is valid simultaneously with the linearized approximation only in the region \eqref{overlap}.
The decaying solution is sizeable in this region and is perfectly captured by the $\frac{1}{D}$
expansion. On the other hand the growing mode is of order $\frac{1}{D^D}$ in 
this region. It is thus  nonperturbatively 
small and so is completely invisible to the $\frac{1}{D}$ expansion of \cite{Bhattacharyya:2015dva,Bhattacharyya:2015fdk}. 
In other words the solutions of \cite{Bhattacharyya:2015dva,Bhattacharyya:2015fdk}
capture only half of the information of the linearized solution in the overlap region 
\eqref{overlap}. In order to complete our specification of the 
linearized solution and to extend it into the radiation region we need more information.
The extra data comes from the physical expectation that 
radiation from the membrane motion is necessarily outgoing 
at infinity. The absence of ingoing radiation at infinity provides the second piece 
of data needed to continue the linearized solutions to large $S$. 

We now explain how the membrane solutions may actually be continued to infinity 
in a practically useful manner. In this paper we demonstrate that the decaying part of a linearized 
solution of the Einstein- Maxwell equations uniquely defines a stress tensor and 
a charge current on the membrane at large $D$. 
\footnote{The existence of such a map is plausible from a counting perspective; 
both sides of the map depend on a single piece of data on a slice (think the membrane) 
of spacetime.}. The sources thus defined may be thought of as giving rise
to (the decaying part of) the linearized solution we started with. 
More precisely the 
convolution of a Greens function against this source produces a response 
whose decaying part agrees with the solution we started out with. 
\footnote{This convolution procedure also produces a growing mode.  The 
detailed magnitude of that growing mode - which is always of order $1/D^D$ 
- depends on the Greens function we use. } The absence of ingoing radiation 
at infinity dictates that we use the retarded Greens function. This 
convolution produces the correct solution in and outside the 
overlap region \eqref{overlap}. In the overlap region the convolution 
produces the nonperturbatively small growing part of the solution in addition 
to the decaying piece obtained from the solutions of 
\cite{Bhattacharyya:2015dva,Bhattacharyya:2015fdk}. In the region 
$r \gg \frac{D}{\omega}$ the convolution produces the  radiation field 
that we wished to calculate.
 
In sections \ref{mstcc} and \ref{mclsdc} below - the technical heart of this paper - we explain 
in detail how the map between decaying solutions of the Einstein-Maxwell system
and a stress tensor and charge current on the membrane is constructed. Though
the derivation takes a lot of work the final prescription is very simple. 
The charge current $J_B$ is given by $$J_B=J^{(out)}_B - J^{(in)}_B.$$ 
Here 
\begin{equation}\label{jbout}
J^{(out)}_B = n^A F^{out}_{AB},
\end{equation}
where $F^{out}_{AB}$ is the field strength of the decaying part of 
external solution that was given to us, evaluated on the membrane, 
 and $n^A$ is the 
outward pointing unit normal to the membrane. Note that $n^B J^{(out)}_B=0$. 
It follows that this current may also be viewed as the current $J^{(out)}_\mu$ 
that lives on the world volume of the membrane. \footnote{See section 
\eqref{ccom} for the precise relationship between  $J^{(out)}_B$ and 
$J^{(out)}_\mu$.} In a similar manner the current $J^{(in)}_B$ turns out to obey 
$n^B J^{(in)}_B=0$ and 
can also be thought of as a current $J^{(in)}_\mu$ that lives on the membrane world volume. 
It turns out
\begin{equation}\label{innercurrent}
J^{(in)}_\mu= - \frac{\delta} {\delta A_\mu} S_{ctrA},
\end{equation}
where, to first order in the expansion in $1/D$, 
\begin{equation}\label{lcam}
S_{ctrA}= - \frac{1}{4}\int \frac{F_{\mu\nu} F^{\mu\nu}}{\sqrt{\mathcal R}},
\end{equation}
where ${\mathcal R}$ is the Ricci scalar on the world volume of the membrane
and $F_{\mu\nu}$ is the field strength of the linearized external solution 
restricted to the membrane.
\footnote{At leading order in the large $D$ expansion the Gauss Codazzi 
equations may be used to show that $\sqrt{ {\mathcal R}}={K}$. }

In a similar manner the stress tensor $T_{AB}$ on the membrane is given by 
\begin{equation}\label{stmdef}
T_{AB}= T_{AB}^{(out)} -T_{AB}^{(in)}.
\end{equation}
Here
\begin{equation}\label{tout}
8 \pi T_{AB}^{(out)}={\cal K}^{(out)}_{AB}-{\cal K}^{(out)} {\mathfrak p}^{(out)}_{AB} ~,
\end{equation}
is the Brown York stress tensor of the external solution evaluated on the 
membrane surface. Here ${\cal K}^{(out)}_{AB}$ and ${\mathfrak p}^{(out)}_{AB}$ are the extrinsic curvature and the projector on the membrane 
world volume viewed as a submanifold of the bulk whose metric is that 
of Minkowski space perturbed by the decaying external solution. As above, 
$T_{AB}^{(out)}$ and $T_{AB}^{(in)}$ are both tangential to the membrane world volume 
and so can equally well be regarded as stress tensors, $T_{\mu\nu}^{(out)}$ and $T_{\mu\nu}^{(in)}$
that live on the membrane world volume. \footnote{Once again see section \ref{ccom} for the precise 
relationship between the spacetime and world volume stress tensors.} 
It turns out that 
\begin{equation}\label{tin}
\sqrt{-g^{(ind)}} T_{\mu\nu}^{(in)}= -\frac{\delta}{\delta g_{(ind)}^{\mu\nu}} S_{(in)},
\end{equation}
where 
\begin{equation}\label{sin}
S_{(in)}=-\frac{1}{8\pi} \int \sqrt{-g^{(ind)}}\left[ \sqrt{ {\mathcal R}}+\frac{1}{2}\left(\frac{{\cal R}_{\mu\nu}{\cal R}^{\mu\nu}}{{\cal R}_{(in)}^\frac{3}{2}}\right) + {\cal O}\left(\frac{1}{D}\right)\right],
\end{equation}
${\cal R}$, ${\cal R}_{\mu\nu}$ and $g^{(ind)}_{\mu\nu}$ are respectively the intrinsic Ricci scalar , intrinsic Ricci tensor and the intrinsic metric of the membrane.\\
The stress tensors \eqref{tin} and \eqref{tout} are both evaluated on the 
membrane world volume using the prescribed external solution.  
Recall the external solution is  flat space plus the decaying linearized 
solution of Einstein's equations, which we assume is given to us. 
In the particular case of interest to this paper, this decaying linearized 
solution is given by matching with the metric presented in 
\cite{Bhattacharyya:2015dva,Bhattacharyya:2015fdk}. 

\subsection{Explicit formula for the Membrane Stress Tensor and Charge Current}

It is not difficult to implement the procedure described in the previous subsection 
on the solutions of \cite{Bhattacharyya:2015dva,Bhattacharyya:2015fdk} and so obtain 
a formula for the membrane stress tensor and charge current. We find 

\begin{equation}\label{stcc}
\begin{split}
 T_{\mu\nu}= &~\left(\frac{1}{8\pi}\right)\bigg[ \left( \frac{K}{2} \right) (1+Q^2) u_\mu u_\nu + \left( \frac{1-Q^2}{2}  \right)K_{\mu\nu} - \left(   \frac{\hat\nabla_\mu u_\nu +\hat \nabla_\nu u_\mu}{2} \right) \\
&-\left(\frac{{K}Q^2}{2D} + \frac{2Q\hat\nabla^2Q}{K} + Q^2 u^\alpha u^\beta K_{\alpha\beta}\right)u_\mu u_\nu-\left( u_\mu{\cal V}_\nu + u_\nu{\cal V}_\mu\right) \\
&~-\bigg[\left(\frac{1+Q^2}{2}\right)\left(u^\alpha u^\beta K_{\alpha\beta}\right)+\left(\frac{1-Q^2}{2}\right)\left(\frac{{K}}{D}\right)\bigg]g^{(ind,f)}_{\mu\nu}\bigg]\\
&~ + {\cal O}\left(\frac{1}{D}\right),\\
\\
J^\mu
=&~\left(\frac{Q}{2\sqrt{2\pi}}\right)\left[{K} u^\mu -\left(\frac{p^{\nu\mu}\hat\nabla_\nu Q}{Q}\right) - (u\cdot\hat\nabla)u^\mu- \left(\frac{\hat\nabla^2 u^\mu}{K}\right) +K^{\alpha\mu} u_\alpha\right]\\
&~+ {\cal Q}~ u^\mu + {\cal O}\left(\frac{1}{D}\right),\\
\end{split}
\end{equation}
where
\begin{equation}\label{calvint}
\begin{split}
{\cal V}_\mu =&~Q~\hat\nabla_\mu Q +Q^2(u^\alpha K_{\alpha\mu} ) +\left(\frac{2Q^4-Q^2-1}{2}\right)\left(\frac{\hat\nabla_\mu{K}}{K}\right)\\
&-\left(\frac{Q^2 + 2Q^4}{2}\right)(u\cdot\hat\nabla)u_\mu+ \left(\frac{1+Q^2}{K}\right)\hat\nabla^2 u_\mu,\\
{\cal Q} =&~  \left(\frac{Q}{2\sqrt{2\pi}}\right)\left[ \frac{\hat\nabla^2{K}}{{K}^2}-\frac{2K}{D} -\frac{(u\cdot\hat\nabla) K}{K} -\left(\frac{2\hat\nabla^2 Q +K(u\cdot\hat \nabla) Q}{Q~K}\right) +\left(u^\alpha u^\beta K_{\alpha\beta}\right)\right].
\end{split}
\end{equation}
Here $g^{(ind,f)}_{\mu\nu}$ denotes the induced metric on the membrane as embedded in flat space and $\hat\nabla_\mu$ denotes the covariant derivative with respect to  $g^{(ind,f)}_{\mu\nu}$. Extrinsic curvature of the membrane is denoted by $K_{\mu\nu}$ and $K$ is the trace of the extrinsic curvature.
%
%
%

According to the rules for $D$ counting explained earlier in this introduction, the 
first term on the RHS for the expressions for stress tensor and charge currents presented in \eqref{stcc}
are each of order $D$. All other terms in both expressions are of order unity. We emphasize, in 
particular, that the membrane stress tensor and charge current are not parametrically small 
in the large $D$ limit. The radiation sourced by these currents is nonetheless 
nonperturbatively small in the appropriate regimes, for the kinematical reasons 
- the heavily damped grey body factor - described earlier in this introduction. 

Several terms in the stress tensor and charge current above have familiar
hydrodynamical interpretations. In particular, relativistic fluids 
propagating on fixed background manifolds always have a contribution to 
their stress tensor proportional to $- \eta \sigma_{MN}$ where $\sigma_{MN}$ 
is the symmetrized derivative of the velocity field projected orthogonal 
to the velocity and $\eta$ is called the shear viscosity of the fluid. 
An inspection of the first line of \eqref{stcc} reveals that our membrane 
stress tensor also has such a contribution with effective value of 
$\eta= \frac{1}{16 \pi}$. Below we will see that the entropy density of 
the membrane is given, to leading order, by $\frac{1}{4}$. It follows that 
the ratio of shear viscosity to entropy density for our membrane 
equals $\frac{1}{4\pi}$, in agreement with \cite{Kovtun:2004de}.

Keeping only the leading terms (i.e the terms that scale like $D$) in \eqref{stcc} we find the 
much simplified expressions
\begin{equation}\label{stccs}
\begin{split}
T_{\mu\nu}=& \left( \frac{K}{16\pi} \right) (1+Q^2) u_\mu u_\nu, \\
J_\mu=& \frac{1}{2\sqrt{2\pi}} \left(QKu_{\mu} \right).
\end{split}
\end{equation}
Note that the leading order stress tensor and charge current is simply that of a collection 
of pressure free `dust' particles. Note, in particular, that the leading order stress tensor 
lacks a surface tension term (a term proportional to $\Pi_{\mu\nu}$). In this 
respect the stress tensor of the large $D$ black hole membrane differs significantly from
more familiar membranes like soap bubbles or $D2$ branes. 

\subsection{Equations of motion from conservation}

As the fractional loss of energy to radiation is non perturbative in the 
large $D$ 
limit, it follows that membrane energy, momentum and charge are conserved at 
each order in the $\frac{1}{D}$ expansion. In fact a stronger 
result must hold; in order for the formula for gravitational 
and electromagnetic radiation from the membrane to be gauge invariant, 
the membrane stress tensor and charge current must be conserved 
currents. Indeed the conservation of the membrane stress tensor 
and charge current turn out to be an alternate - and conceptually very 
satisfying - way of restating the membrane equations of motion 
\eqref{memeom}. The fact that the membrane equations \eqref{memeom} 
are simply statements of conservation of an appropriate membrane 
stress tensor and charge current emphasizes that our membrane equations 
are hydrodynamical in nature. 

We have explained above that the expressions for the stress tensor and charge current 
\eqref{stcc} each have one term of order $D$ and several terms of order unity. The 
reader may at first suppose that only the leading order terms (those of order $D$) are 
needed to obtain the leading order membrane equations of motion via conservation. This 
is indeed the case for the first equation \eqref{memeom}. The divergence of the 
leading order stress tensor a term of order $D^2$. This term is proportional to 
$K u^\mu \nabla.u$. It follows that the term in $\partial_\nu T^{\nu\mu}$ proportional 
to $u^\mu$ indeed receives its leading contribution from the order $D$ part of the 
stress tensor; the condition that this term vanish is simply the first equation of 
\eqref{memeom}

Let us turn our attention, however, to the projection of  $\partial_\nu T^{\nu\mu}$ orthogonal to $u_\mu$. 
According to the rules of large $D$ counting summarized earlier in this introduction, this projected expression is of order $D$ rather than of order $D^2$. At leading order (order $D$) this 
expression receives contributions both from the order $D$ as well as the order unity contributions to the 
stress tensor (recall that the divergence of a tensor or vector of order unity 
is generically of order $D$). The order $D$ piece of $T^{\mu\nu}$, \eqref{stccs}, 
yields the LHS of the second equation in \eqref{memeom}; the 
RHS of that equation is obtained from the 
divergence of the order unity parts of the stress tensor \eqref{stcc}. A similar 
statement is true of the relationship between the  conservation of the charge current 
and the third equation in \eqref{memeom}.  

In summary, in order to obtain the first equation in \eqref{memeom} we needed to know only 
the leading order stress tensor \eqref{stccs}. In order to obtain the second and third 
equations of \eqref{memeom}, however, we need to know the subleading terms in \eqref{stcc} as well.

\subsection{Entropy Current}

We have, so far, focused our attention on the conserved currents that 
live on the membrane. A key fact about black holes, however, is that that 
they carry entropy in addition to charge and energy. While charge and 
energy obey the first law of thermodynamics, and so are conserved, entropy 
obeys the second law and so is a non decreasing function of time. 

The entropy carried by a black hole is mirrored in the fact that the 
membrane carries an entropy current. In this paper we define this current 
and demonstrate that it obeys a local version of the second law of
thermodynamics, i.e.  
$$\hat\nabla_\mu J^\mu_S \geq 0.$$ 
Our construction of the membrane entropy current proceeds in a manner 
analogous to the construction of \cite{Bhattacharyya:2008xc}. The current is constructed by 
pulling the area form on the event horizon back onto the membrane. 
A local form of the  Hawking area increase theorem then 
ensures that the divergence of this entropy current is point wise non negative 
for every membrane motion. At first 
leading and subleading order in the $1/D$ expansion we find the extremely 
simple result 
\begin{equation}\label{entcur1}
J_{S}^M= \frac{u^M}{4},
\end{equation}
(see \eqref{entcur} for the correction to this equation at second 
subleading order 
in the special case of uncharged fluids).
By explicit use of equation 1.5 of \cite{Dandekar:2016fvw}
at leading nontrivial order in $\frac{1}{D}$ we find   
\begin{equation}\label{diventc}
\partial_M J_{S}= \frac{1}{8 {K} } \sigma_{AB} \sigma^{AB},
\end{equation} 
where 
\begin{equation}\label{sigma}
\sigma_{AB}=\left( \partial_M u_N +\partial_N u_M \right) P_A^M P_B^N.
\end{equation}
Note in particular that entropy production vanishes at leading order 
if and only if the fluid velocity flow is shear free. As the flow is 
always also divergence free, it follows that every time independent
(i.e. stationary) velocity vector field is proportional to a killing 
vector on the membrane world volume \cite{Caldarelli:2008mv}. This observation may be used as 
the first step in a systematic classification of stationary solutions 
of the membrane equations, a topic we hope to return to in the near future. 

\subsection{Radiation from small fluctuations}

In the Appendix \ref{bra} to this paper we develop the general theory of 
radiation for the Maxwell and Einstein equations \eqref{eomfa} coupled to 
sources after linearization. In that appendix we work in a particular 
Lorentz frame, expand all modes in spherical harmonics and present very 
explicit radiation formulae. As an application of these formulae, 
in the main text we evaluate the radiation that results from a general 
linearized fluctuation about a spherical membrane. It follows from 
the formulae of that section that energy lost to radiation per unit time is  
smaller by a factor of $1/D^D$ when compared to the membrane energy stored 
in the fluctuation, providing a clear demonstration of the smallness of radiation.

\subsection{Organization of this paper} 

This very long paper is organized as follows. In section \ref{bgm} 
we review the properties of retarded Greens functions in arbitrary dimensions 
with a special emphasis on the large $D$ limit. In section \ref{ccom} 
we review the structure of currents and stress tensors localized on 
a codimension one membrane. Sections \ref{mstcc} and \ref{mclsdc} 
are the technical heart of this paper. In these sections we construct 
a membrane charge current and stress tensor dual to any decaying linearized 
solution of the Einstein Maxwell equations in the exterior neighbourhood 
of the membrane world volume. In section \ref{ccst} we apply the general 
formalism of the previous two sections to the special case of the 
membrane spacetimes of \cite{Bhattacharyya:2015fdk}, and find the stress tensor and charge 
current that lives on the membrane dual to large $D$ black holes at leading
order in $\frac{1}{D}$. In section \ref{mec} we define an entropy current 
on the membrane and demonstrate that its divergence is point wise non negative.
In section \ref{bra} we proceed to review and develop the general theory of 
linearized radiation from localized sources for 
 the Einstein Maxwell equations in an arbitrary number of dimensions. 
 We then proceed, in section \ref{rfl}, to use these formulae 
to determine the radiation sourced by small fluctuations about the 
spherical membrane solution. Finally in section \ref{disc} we present 
a discussion of our results. Our paper also includes several appendices 
in which we present details of algebraically intensive computations.

\section{Review of background material: Greens functions 
in general dimensions} \label{bgm}

In this section we review elementary background material on Greens functions
in arbitrary dimensions, with a focus on the large $D$ limit. In the 
rest of this paper we will use the results of this subsection for qualitative
as well as quantitative purposes. The key qualitative results from this 
subsection that will be of importance to us below are
\begin{itemize} 
\item In the large $D$ limit distinct Greens functions (e.g.  retarded and 
Feynman Greens functions) differ from each other only at order 
$1/D^{\frac{D}{2}}$ at spatial distances and time frequencies of order unity  
(see subsection \ref{ldgf} below).
\item The fractional energy loss per unit time 
into gravitational radiation, from a stress tensor that varies over distance 
and time scales of order unity, is of order $1/D^D$. 
\end{itemize}

At the quantitative level, in section \ref{bra} we use the results of this 
section to derive detailed formulae for the electromagnetic and linearized 
gravitational radiation from arbitrary sources in general dimensions, once 
again with a focus on the large $D$ limit.

\subsection{Greens function in frequency space} \label{gffreq}

Consider the retarded Greens function $G(x_\mu, x'_\mu)$ defined by the 
equation 
\begin{equation}\label{gfdef}
-\Box~ G(x-x')= \delta^D(x-x'),
\end{equation}
together with the boundary condition that $G$ vanishes if $x$ lies 
outside the future lightcone of $x'$. In \eqref{gfdef} the d'Alembertian
\footnote{Throughout this paper we employ the mostly 
positive sign convention.} is taken is taken w.r.t the coordinate $x$. 
$G$ may be thought of as the causal response at the point $x$ to a unit 
normalized delta function source at $x'$.

Although the equation \eqref{gfdef} is Lorentz invariant, our Greens function cannot be thought 
of as a function only of $x^2$ (this is a consequence of retarded boundary conditions). In order to solve for the Greens function 
(and to understand its properties) we found it most convenient to sacrifice 
manifest Lorentz invariance. We choose a particular rest frame and so a particular 
time coordinate. In this section we further  locate the  source point $x'$ of our Greens 
function at the origin of spatial coordinates and Fourier transform
w.r.t. time
\begin{equation} \label{ftgf} 
G({\omega,\vec r} )= \int G(t,{\vec r}) e^{i \omega t} dt.
\end{equation}
It follows from \eqref{gfdef} that  $G(\omega,\vec{r})$ obeys the equation 
\begin{equation} \label{gfft}
-\left( \omega^2 + {\vec \nabla}^2 \right) G(\omega,\vec{r})=\delta^{D-1}({\vec r}).
\end{equation}
As $G(\omega,{\vec r})$ is spherically symmetric it 
is convenient to work in polar coordinates, i.e. in coordinates 
in which the Minkowskian metric is given by
$$ ds^2 = -dt^2 + dr^2 + r^2 d \Omega_{D-2}^2,$$
\eqref{gfft} simplifies to 
\begin{equation}\label{gfftn}
\omega^2 G(\omega,r) + 
\frac{1}{r^{D-2}} \partial_r \left( r^{D-2} \partial_r G(\omega,r) \right)
= - \delta^{D-1}({\vec r}).
\end{equation}  
The boundary conditions on $G(r,t)$ require $G(\omega,r)$ to be purely outgoing (i.e. $\propto e^{i \omega r}$)
at infinity. The unique solution to \eqref{gfftn} subject to these 
boundary conditions is 
\begin{equation}\label{Gofs}
G(\omega,r)= \frac{i}{4}\left(\frac{\omega}{2\pi r} \right)^\frac{D-3}{2}
H_\frac{D-3}{2}^{(1)}(\omega r).
\end{equation}
Here $H_n^{(1)}(x)$ is the $n^{th}$ Hankel function of first kind, whose small and large 
argument asymptotics are given by  
\begin{equation}\label{hfa} \begin{split}
&H_n^{(1)}(x) \approx  ~~~-i\left(\frac{ 2}{x}\right)^n \frac{1}{\pi} \Gamma (n)\left(1+\frac{x^2}{4(n-1)} + {\cal O} \left( x^4/ n^2 \right) \right) \quad \mbox{for} \quad x^2\ll  n,\\ 
&H_n^{(1)}(x) \approx  ~~~(1-i)e^{-\frac{in\pi}{2}}e^{ix}\frac{1}{\sqrt{\pi x}}\left(1+i\frac{4n^2-1}{8x}+ {\cal O} \left( n^4/x^2 \right) \right) \quad \mbox{for} \quad x\gg  n^2 .\\
\end{split}
\end{equation}
Using \eqref{hfa} it follows that our Greens function is given by 
\begin{equation}\label{gfaa} \begin{split}
&G(\omega,r) \approx \frac{1}{(D-3)\Omega_{D-2}}\frac{1}{r^{D-3}}\left(1+\frac{\omega^2r^2}{2(D-5)} 
+ {\cal O}(\omega^4 r^4/ D^2) \right)
\quad \quad \mbox{for} ~r^2\omega^2 \ll D, \\ 
&G(\omega,r) \approx -\left(2i\right)^{-\frac{D}{2}}\left(\frac{\omega}{\pi r}\right)^{\frac{D-2}{2}}\frac{e^{i\omega r}}{\omega}\left(1+i\frac{(D-2)(D-4)}{8\omega r} + {\cal O}(D^4/r^2\omega^2) \right)\\
 & \mbox{for} ~r\omega\gg D^2.\\
\end{split}
\end{equation}

\subsubsection{Lightcone structure of the retarded Greens function} \label{cov}

In the previous subsubsection we presented an exact result for the retarded Greens function 
as a function of $\omega$ and $r$.  In Appendix \ref{rps} we evaluate the Fourier 
transform of the expressions of the previous subsection and obtain a formula for 
the retarded Greens function directly in position space. In this brief subsection 
we simply report the final results of Appendix \ref{rps}. 

When $D$ is even we find 
 \begin{equation}\label{ge} 
 G(x, x')=\frac{\theta(X^0)}{2}\left(\frac{1}{\pi}\right)^{\frac{D-2}{2}}\delta^{\left(\frac{D-4}{2}\right)} \left(-X_M X^M \right),
 \end{equation}
where 
$$ X^M=x^M-(x')^M , ~~~\delta^n(X) = \partial_X^n \delta(X).$$
When $D$ is odd, on the other hand we find 
\begin{equation} \label{go}
G(r,t)= \frac{\Omega_{D-3}}{(2\pi)^{D-4}} (\partial_M \partial^M )^{\frac{D-3}{2}}\left(\frac{\theta(t-r)}{\sqrt{-x_Mx^M} } \right),
\end{equation}
where $\Omega_{n}$ is the volume of the unit $n$ sphere
\begin{equation}\label{omegan}
\Omega_{n}= \frac{2 \pi^{\frac{n+1}{2}}}{\Gamma\left( \frac{n+1}{2} \right)}.
\end{equation}
In either case the Greens function is given by linear sums of finite 
numbers of derivatives acting on expressions that vanish outside the 
future lightcone; it follows that these Greens functions never propagate
signals faster than light. 
\footnote{Note, however, that the number of derivatives that appears 
in the expression for the Greens functions increases without bound in the 
large $D$ limit. This allows naive large $D$ approximations of the 
Greens function to mimic apparently acausal behaviour in some situations. 
When used 
correctly, however, the Greens function is causal in every $D$. }

Although the expressions \eqref{ge} and \eqref{go} are exact, they are 
not particularly well suited for taking the large $D$ and obscure various 
features of the Greens functions in this limit. In the rest of this paper 
we will revert to working with the non manifestly Lorentz invariant but highly 
explicit representation Greens functions \eqref{Gofs}. We will now 
proceed to estimate the expression \eqref{Gofs} in the large $D$ limit; 
we find that the large $D$ limit is smooth and can be taken without 
differentiating between odd and even $D$.

\subsection{Large $D$ expansion through WKB} \label{ldgf}

In this section we will use the WKB approximation to determine the large 
$D$ limit of the retarded Greens function. The main conclusions of this 
subsection are 
\begin{itemize}
\item Upto a scaling (see \eqref{gfwf}) the scalar Greens function is given
by the solution to the one dimensional Schrodinger equation listed in 
\eqref{schfor}
\item In the large $D$ limit the potential in this Schrodinger equation 
exceeds the energy when  $\omega r < 2D$ and is less than the energy 
when $\omega r>2D$. The wave function that yields the 
Greens function describes a process of tunneling through a wide
potential barrier. The exponential tunneling suppression ensures that the 
oscillating solution that emerges when $ \omega r>2D$ is very small. 
This explains the smallness of radiation at large $D$.
\item All Greens functions (e.g. retarded, advanced, Feynman) are all 
essentially identical for $\omega r \ll 2 D$. In particular when $\omega r$ 
is of order unity, the differences between different Greens functions are 
of order $1/D^{D}$.
\end{itemize}
In the rest of this subsection we will explain these points in some more 
detail relegating detailed derivations to appendices. 

The transformation 
\begin{equation}\label{gfwf} 
G(\omega,r)=\frac{1}{r^{\frac{D-2}{2}}}\psi(\omega,r),
\end{equation} 
recasts the equation \eqref{gfftn} into 
\begin{equation}\label{schfor}
-\partial_r^2\psi(\omega,r) +\frac{(D-2)(D-4)}{4 r^2}\psi(\omega,r) = 
\omega^2 \psi(\omega,r),
\end{equation}
\footnote{For the purposes of this discussion we stay away from the point $r=0$ 
and so ignore the term proportional to the $\delta$ function.}
i.e. a one dimensional Schrodinger 
equation with potential $V$ and energy $E$ given by 
$$V(r)=\frac{{D^*}^2}{4 r^2}, ~~~E=\omega^2~~~~{\rm where}~D^* = 
\sqrt{(D-4)(D-2)} \approx D-3 + {\cal O}(1/D).$$ 
This potential divides the $r$ axis into the classically allowed and 
disallowed regions  
$$ 2 \omega r > D^* :~{\rm allowed};$$ 
$$ 2 \omega r < D^* : ~{\rm disallowed} .$$
In Appendix \ref{wkb} we demonstrate that WKB approximation of the solutions 
to this equation are exact in the large $D$ limit 
away from the turning points. \footnote{Although we do not go beyond leading 
order in this paper, higher order corrections to the WKB approximation 
generate a systematic expansion of the Greens function in a power series in 
$1/D$.}

Let us first consider the classically disallowed region. We define 
\begin{equation}\label{kappa}
\kappa(r)= \left(\frac{D^{*2}}{4r^2}-\omega^2\right)^\frac{1}{2}.
\end{equation}
The WKB solution to $\psi(\omega, r)$ takes the form 
\begin{equation}\label{psiform}
\psi(\omega,r) =\frac{1}{\sqrt{\kappa(r)}}
\left( A \left( \frac{e\omega}{D} \right)^{D-3} e^{\int \kappa(r)dr } + B e^{-\int \kappa(r) dr} \right) ,
\end{equation}
(where $e$ is Euler's number $2.7182...$) 
for some constants $A$ and $B$. In \eqref{psiform} we have chosen to multiply 
$A$ by the constant factor $\left( \frac{e \omega}{D} \right)^{D-3}$ for 
future convenience. Note that this factor is of order $1/D^D$.

At small $r$ and with an appropriate choice of integration constants  
we have 
$$ \int \kappa(r) dr \approx \frac{D^*}{2} \ln r  - \frac{r^2 \omega^2}{2 D^*} 
+ {\cal O}(r^4 \omega^4/(D^*)^3),$$
so that 
$$e^{ \int \kappa(r) dr} \approx r^{ D^*/2} \left( 1-\frac{r^2 \omega^2}{2 D^*}  
+ \ldots \right).  $$
\footnote{Note, in particular, that the correction to the leading order 
small $r$ behaviour in $e^{ \int \kappa(r) dr} \approx r^{ D^*/2}$ is 
negligible provided $\frac{r^2 \omega^2}{2 D^*} \ll 1$, in agreement with 
the estimate for the validity of the small argument expansion 
of the exact formula for the Greens function presented in \eqref{gfaa}.}  
It follows that at small $r$ 
\begin{equation}\label{gfsg}
G(\omega,r) = A \left( \frac{e\omega}{D} \right)^{D-3} + \frac{B}{r^{D-3}},
\end{equation}
where we have accounted for the proportionality factor between 
$G(\omega, r)$ and $\psi(\omega,r)$ (see \eqref{gfwf}).\footnote{In fact we choose the integration constants in \eqref{psiform} to ensure that \eqref{gfsg} is valid. The constants 
The combination of the equations \eqref{psiform} and \eqref{gfsg} give a complete definition of the constants $A$ and $B$. }

Now the equation 
$$\nabla^2 G(\omega, r) = -\delta(r); $$
leaves $A$ undetermined but fixes the constant $B$ to 
\begin{equation}\label{bval}
B= \frac{1}{(D-3)\Omega_{D-2}},
\end{equation}
($\Omega_n$, the volume of the unit $n$ sphere, is listed in \eqref{omegan}). 
The constant $A$ is determined by matching with the solution
in the classically allowed region as we explain below. 

In the classically allowed region we have $k(r)= \sqrt{\omega^2-\frac{D^{*2}}{4r^2}}$. 
The usual formulae of the WKB approximation yield 
\begin{equation}\label{wkbappf}
\psi(\omega,r) = \frac{1}{\sqrt{k(r)}} \left( 
C e^{i \int k(r)-\frac{iD\pi}4}  + Ee^{-i \int k(r)+\frac{iD\pi}4}   \right)
\approx \frac{1}{\sqrt{\omega}} \left( C  e^{i \left(\omega r-\frac{D\pi}4\right)} + E e^{-i \left(\omega r-\frac{D\pi}4\right)}
\right) .
\end{equation}
The last expression in \eqref{wkbappf} holds in the limit 
$2\omega r \gg D^*$. \footnote{The integration constants in the integrals 
in the first expression in \eqref{wkbappf} are determined by the 
requirement that it reduce to the second expression in the same equation 
at large $r$.}

For the special case of the retarded Greens function the wave function must 
be outgoing at infinity so that $E=0$. The constants $A$ and $C$ are both 
determined by matching across the turning
point; in Appendix \ref{wkb} we use standard WKB matching formulae to find 
\begin{equation} \label{matching} \begin{split} 
C&= \frac{(1+i)}{\sqrt{2}} B \sqrt{\frac{D^*}2}\left( \frac{D^*}{\omega} \right) ^{-\frac{D^*}{2}} e^{\frac{D^*}{2}}  =\frac{(1+i)}{\sqrt{2}}\left(2\right)^{-\frac{D}{2}}\frac{\omega^\frac{D-3}{2}}{\pi^\frac{D-2}{2}},\\
A&= \frac{iB}{2} = \frac{i}{2(D-3)\Omega_{D-2}}.
\end{split}
\end{equation}

The parametric dependences of these results may be understood as follows. 
At the turning point we expect the two terms in \eqref{psiform} to be 
of comparable magnitude. Using the WKB approximation to evolve the solution 
inwards to small $r$ we obtain the following estimate.  The ratio of the 
decaying to the growing solution at the point $r$ should approximate 
$e^{2 \int_{r}^\frac{D^*}{2 \omega}  \kappa(r) dr}$. At large $D$ and when 
$r \ll \frac{D^*}{2 \omega}$ we find 
$$ 2 \int_{r}^\frac{D^*}{2 \omega}  \kappa(r) dr \approx 
D^* \ln \frac{D^*}{e \omega r}.$$
Comparing with \eqref{gfsg} it follows that 
\begin{equation}\label{aapprox}
A \sim B ,
\end{equation}
in approximate agreement with the more precise formulae \eqref{matching}. 
Using similar logic we can use \eqref{wkbappf} to estimate 
the value of $\psi(\omega,r)$ when we approach the turning point from 
the large $r$ limit. Matching this estimate with the value of the wave 
function when the turning point is approached from the small $r$ limit 
we find  
\begin{equation}\label{capprox}
C \sim B \left( \frac{2 \omega}{D^*} \right)^{\frac{D-2}{2}}, 
\end{equation}
an estimate that is once again in agreement with the 
precise result \eqref{matching}. 

The utility of the rough approximations \eqref{aapprox} and \eqref{capprox}
is that they are equally valid for other Greens functions (e.g. the 
retarded Greens function or the Feynman Greens function). It follows that 
for all these Greens functions the term in \eqref{psiform} proportional
to $B$ dominates over the term proportional to $A$ when $r\omega \ll D/2$. 
When $r\omega$ is of order unity, in particular, the term proportional 
to $A$ (which is sensitive to the precise nature of the Greens function) 
is subdominant to the term proportional to $B$ (which is universal) at 
relative order $1/D^D$. It follows that different reasonable Greens functions
\footnote{We call a Greens function `reasonable' if the large $r$ boundary 
condition that defines it ensures that the ratio of the 
decaying and growing solutions at the turning point is of order unity. 
The retarded, advanced and Feynman Greens functions are all reasonable 
by this criterion. It is possible to rig up Greens functions whose 
boundary conditions are finely tuned (in a $D$ dependent way) so as 
to violate the conclusions of this paragraph. Such Greens functions 
are unphysical for our purposes, and will be ignored through the rest of this 
paper.}
 differ from each other only at order $1/D^D$ when $\omega r$ is of order unity.

The fact that $C/B$ is of order $1/D^{D/2}$ captures the smallness of 
radiation in the large $D$ limit.

Let us end this subsubsection with a brief discussion of a subtle point. 
In the limit that $r^2 \omega^2 \ll D$ the Greens function $G(\omega, r)$ is effectively independent of $\omega$. Upon Fourier transforming, this observation 
suggests that the Greens function in this limit is time independent but nonlocal in space 
(in fact the spatial dependence of the propagator is exactly that of the Euclidean propagator for 
$\nabla^2$ in $D-1$ Euclidean dimensions). This suggests that the retarded propagator mediates 
instantaneous action at a distance and so is acausal. Of course the 
exact formulae of subsubsection \ref{cov} make it clear that this conclusion 
is erroneous. While we have not carefully tracked down the fallacy in the
naive argument, we believe it has its roots in the following fact. 
In order to really argue for acausality one should turn on a source that is 
sharply localized in time and detect a response outside the lightcone of 
this source. Such a source is necessarily non analytic and so 
always has significant support at arbitrarily high $\omega$. It follows 
that the approximations of the previous paragraph, which work for $\omega$
of order unity  cannot really be consistently used to argue for acausality. 
It would be interesting to understand this point better but we leave it 
for future work.

\section{Review of Background Material: the stress tensor and 
conserved currents on codimension one membranes}\label{ccom}

In this section we study conserved currents and stress tensors localized 
on codimension one surfaces in space time. 

Consider the flat space $R^{D-1, 1}$. Consider a function $\rho$ defined on 
this spacetime, and consider a membrane whose world volume is given by the 
solutions to the equation $\rho-1=0$. The normal to the membrane world volume
is given by the equation  
\begin{equation}\label{normal}
n_M = \frac{\partial_M \rho} {|\partial \rho|}, ~~~
|\partial \rho|=\sqrt{\partial_M \rho \partial^M \rho}
\end{equation}
and is assumed to be everywhere spacelike. 

\subsection{Scalar sources localized on a membrane}

As a warm up consider the minimally coupled scalar equation
\begin{equation}\label{pheommntxt}
 -\Box \phi = \mathcal{S}.
\end{equation}
 Consider 
a situation in which the source ${\cal S}$ of that equation is given by 
the distributional valued field ${\cal S}_{ST}$ localized on the membrane 
\begin{equation}\label{mls}
{\cal S}_{ST}=  |\partial \rho| \delta(\rho -1)
{\cal S},
\end{equation}
where $\cal S$ is a smooth function on the membrane. Integrating \eqref{mls} over 
a pillbox whose faces are just above and just below the membrane we conclude that 
\begin{equation}\label{phon}
{\vec n}\cdot\partial \phi_{out} - {\vec n}\cdot\partial \phi_{in}= -{\cal S},
\end{equation}
where ${\vec n}$ is the outward pointing unit normal to the membrane 
(i.e. from `in' to `out'), $\phi_{out}$ is the 
scalar field just outside the membrane and $\phi_{in}$ is the scalar field just 
inside the membrane. 

The source ${\cal S}$ can also be given the following interpretation.
Let $\phi_0$ be the value of the field $\phi$ on the membrane world volume. 
Let $S_{out}[\phi_0]$ represent the action of the outer part of the solution 
as a functional of $\phi_0$, the value of the field $\phi$ on the membrane 
\footnote{If the external region 
of spacetime has an additional boundary, the action would also depend 
on the value of the field $\phi$ on this additional boundary. This 
dependence plays no role in what follows and is suppressed in the notation. 
Similar remarks hold for the internal solution.}. Using 
$$ S_{out}= - \frac{1}{2} \int (\partial \phi)^2,$$ 
it follows that 
\begin{equation}\label{hj}
 \delta S_{out}= \int \delta \phi_{out} \partial^2 \phi_{out} -
\int \partial_M \left( \delta \phi_{out} \partial ^M \phi_{out}\right)=
 \int \delta \phi_{out}( n\cdot\partial) \phi_{out} .
 \end{equation}
The first two integrals on the RHS of \eqref{hj} are taken over 
the bulk spacetime to the exterior of the membrane. The last integral is 
taken over the membrane world volume. In the final step in 
\eqref{hj} we have used the scalar equation of motion and Stokes theorem. 

It follows from \eqref{hj} that  
\begin{equation} \label{outeraction} 
({\vec n}\cdot\partial) \phi_{out}= \frac{ \delta S_{out}} {\delta \phi_0}
\end{equation}
(this is simply the Hamilton Jacobi equation: the 
LHS is evaluated on the membrane approached from the outside). 
In a similar manner, making similar definitions we have
\begin{equation} \label{inaction} 
({\vec n}\cdot\partial) \phi_{in}= -\frac{ \delta S_{in}} {\delta \phi_0}.
\end{equation}
The difference in sign between \eqref{inaction} and \eqref{outeraction} stems
from the fact that the normal $n$ is outward pointing from the point of 
view of the inside, but inward pointing from the point of view of the outside. 
It follows that \eqref{phon} can be rewritten as 
\begin{equation}\label{phonn}
{\cal S}=-\frac{ \delta S_{in}} {\delta \phi_0} - 
\frac{ \delta S_{out}} {\delta \phi_0}.
\end{equation} 

It is not difficult to present explicit expressions for the actions 
$S_{out}$ and $S_{in}$ in terms of integrals over the membrane 
of $\phi_0$ and the normal derivatives of $\phi$ on the 
outer and inner solutions respectively on the membrane.  
\begin{equation}\label{formsin}
S_{in}[\phi_0]= - \frac{1}{2} \int (\partial \phi)^2 =
\left( - \frac{1}{2} \int \partial_M \left( \phi \partial^M \phi \right) +\frac{1}{2} \int \phi \partial^2 \phi \right)= 
 - \frac{1}{2} \int \phi_0(n\cdot\partial)\phi_{in}.
 \end{equation}
The integral in the last expression in \eqref{formsin} is taken over the membrane world volume; 
all other integrals are taken over the region of bulk spacetime that lies to the interior of the 
membrane; in obtaining the last equality we have used the bulk equation of motion and Stokes theorem. 
 In a similar manner  
\begin{equation} \label{formsout} 
S_{out}[\phi_0]= \frac{1}{2} \int \phi_0(n\cdot\partial) \phi_{out} .
\end{equation}

\subsection{Membrane Charge current}
Let us now study the Maxwell equation. Consider the action for the bulk gauge field $A_M$ coupled to a current ${\cal J}^M$
\begin{equation}\label{aischrg}
\text{Action}=-\int 
\left(  \frac{F_{MN} F^{MN} }{4}
+{\cal J}^M A_{M} \right),
\end{equation}
where 
\begin{equation}\label{fmnchrg}
F_{MN} = \partial_M A_N -\partial_N A_M.
\end{equation}
The equation of motion that follows from this action
\begin{equation}\label{mmaxeqmn}
\partial^MF_{MN} = {\cal J}^N.
\end{equation}
Let the charge current 
that is tangent to and localized on the membrane .
\begin{equation}\label{cc}
{\mathcal J}^M =  |\partial \rho|  \delta(\rho -1)
J^M,
\end{equation}
where $J^M$ is a smooth vector field tangent to the membrane (i.e. $J^M n_M=0$).
Integrating \eqref{maxeq} over a pillbox that encloses the membrane we conclude 
that 
\begin{equation}\label{maxo}
n_M F^{MN}_{(out)} - n_MF^{MN}_{(in)}= J^N,
\end{equation}
where $n$ is the outward pointing normal to the membrane.

As in the previous subsection, \eqref{maxo} may be rewritten as 
\begin{equation}\label{maxon}
J^N= \frac{\delta S_{out}[(A_0)_N]}{\delta (A_0)_N}+ 
\frac{\delta S_{in}[(A_0)_N]}{\delta (A_0)_N},
\end{equation}
$S_{out}[(A_0)_N]$ is the action of the outer part of the solution as a
functional of the gauge field restricted to the membrane.

As in the previous subsection it is not difficult to present explicit 
expressions for the actions 
$S_{out}$ and $S_{in}$ in terms of integrals over the membrane 
of $(A_0)_M$ and the normal 
derivatives of the gauge field in  the outer and inner solutions respectively.  
\begin{equation}\label{formsing}
\begin{split} 
S_{in}[A_0]&=  -\frac{1}{2} \int (A_0)_N n_M F^{MN}_{(in)},\\ 
S_{out}[A_0]&= \frac{1}{2} \int (A_0)_N n_M F^{MN}_{(out)}. \\
\end{split}
\end{equation}

We will now demonstrate that the divergence of ${\mathcal J}_{ST}^M$,  viewed as a distributional 
current in spacetime, vanishes provided $J^M$ is a conserved current on the 
membrane. 

In order to see this we note that 
\begin{equation}\label{cccons} \begin{split}
\partial_M {\cal J}^M &=  \delta(\rho -1)|\partial \rho|
\left[ \partial_M \left( \ln \left(\sqrt{\partial_M \rho \partial^M\rho}  
\right)  \right)J^M + \partial_M J^M \right] \\
& =  \delta(\rho -1) |\partial \rho| 
\left[ J^N(n\cdot\partial) n_N  + \partial_M J^M \right] \\
& = \delta(\rho -1) |\partial \rho| 
\left[ {\Pi}^N_M \partial_N J^M \right]. 
\end{split}
\end{equation}
Here $\Pi_{MN}= \eta_{MN} - n_M n_N$. 
In the first line of \eqref{cccons} have used  
$(\partial_M \rho) J^M=0$. In order to obtain the second line of the equation 
we have used $\partial_M \partial_N \rho = \partial_N \partial _M \rho$ 
and $n_M J^M=0$. In order to obtain the third line we have 
used $n_M J^M=0$ to conclude that $n^N n^M \partial_M J^N= - J^N (n\cdot\partial) n_N$. 
As ${\Pi}^N_M \partial_N J^M$ is simply the divergence of $J^N$ viewed as 
a vector field on the membrane, it follows from \eqref{cccons} that 
the ${\cal J}^M$ is conserved in spacetime if and only if the 
$J^M$ is conserved on the membrane world volume. 

\subsection{Membrane localized stress tensor}

Let us now turn to a study of the Einstein equation .  the action for the bulk gauge field $g_{MN}$ coupled to a current ${\cal T}^{MN}$
\begin{equation}\label{aisstr}
\text{Action}= \frac{1}{16 \pi} \int \sqrt{-g} 
R - \left( \frac{1}{2}\right)\int 
h^{MN} {\cal T}_{MN} .
\end{equation}
Consider a membrane localized stress tensor given by  
\begin{equation}\label{stm}
{\cal T}^{MN} =  |\partial \rho| \delta(\rho -1)
T^{MN}.
\end{equation}
The equation of motion that follows from this action
\begin{equation}\label{eineqmn} \begin{split}
& R_{MN}- \left(\frac{R}{2}\right) g_{MN}=8 \pi {\cal T}_{MN},\\
\end{split}
\end{equation} 
where $T_{MN}$ \cite{Eric:2004ep}is a symmetric tensor that is tangent to and smooth on the membrane. 
By integrating Einstein's equations over a pill box that surrounds the membrane 
one can show that 
\begin{equation}\label{eedis}
\left( {\cal K}^{(out)}_{MN}-{\cal K}^{(out)} (g_0)_{MN} \right) -  
\left( {\cal K}^{(in)}_{MN}-{\cal K}^{(in)} (g_0)_{MN} \right)
= -8 \pi T_{MN},
\end{equation}
where $(g_0)_{MN}$ is the space-time metric restricted to the  membrane.  $ {\cal K}^{(out)}_{MN}$ and $ {\cal K}^{(in)}_{MN}$ are the extrinsic curvature computed from `outside' and `inside' the membrane respectively.\\
In other words the discontinuity of the Brown- York stress tensor across the membrane 
is proportional to $T_{MN}$. 

As in the previous subsection, \eqref{eedis} may be rewritten as 
\begin{equation}\label{einon}
T^{MN}= -\left[\frac{\delta S_{out}[(g_0)_{MN}]}{\delta ((g_0)_{MN})}+ 
\frac{\delta S_{in}[(g_0)_{MN}]}{\delta ((g_0)_{MN})}\right],
\end{equation}
$S_{out}[(g_0)_{MN}]$ is the action of the outer part of the solution as a
functional  $(g_0)_{MN}$, the space-time metric, restricted to the membrane.

As in the previous subsection it is not difficult to present explicit 
expressions for the actions 
$S_{out}$ and $S_{in}$ in terms of integrals over the membrane 
of $(g_0)_{MN}$ and the normal derivatives of the metric in  the outer and 
inner solutions respectively.  The action is given entirely by the Gibbons
Hawking term and takes the form
\begin{equation} \label{memacto}
S=S_{out}+ S_{in}
\end{equation}
where
\begin{equation}\label{formein} \begin{split}
S_{in}&= -\frac{1}{8 \pi} \int \sqrt{-g_{(ind)}}~ {\cal K}_{in},\\
S_{out}&= \frac{1}{8 \pi} \int \sqrt{-g_{(ind)}} ~{\cal K}_{out},\\
\end{split}
\end{equation}
where the integral is taken over the world volume of the membrane, viewed 
as a boundary of the internal and external solutions respectively. The 
difference in signs in the two equations above is because ${K}$ is 
defined as the trace of the extrinsic curvature of the normal vector $n$ 
which always runs from in to out.

We emphasize that $T^{MN}$ is assumed tangent to 
the membrane, i.e. $T^{MN} n_M=0$. We will now demonstrate that ${\cal T}^{MN}$ 
is conserved in spacetime if and only if 
\begin{itemize}
\item $T^{MN}$ is a conserved stress tensor on the membrane world volume
\item $T^{MN}{\cal K}_{MN}=0$, where ${\cal K}_{MN}$ is the extrinsic curvature
on the membrane.
\end{itemize}

Unlike the equation for charge conservation, the equation for the conservation 
of the spacetime stress tensor has a free index. We get the first condition 
above when the free index in this equation is in the membrane world volume, 
and the second condition when the free index is chosen proportional to the 
membrane normal.
 
Let us first consider the equation for stress tensor conservation 
projected tangent to the membrane world volume: 
\begin{equation}\label{stons} \begin{split}
{\mathfrak p}_N^P\nabla_M {\cal T}^{MN} &= 
\delta(\rho -1)|\partial \rho|
 {\mathfrak p}_N^P \left[  \nabla_M \left( \ln \left(\sqrt{\partial_M \rho \partial^M
\rho}  \right)  \right)T^{MN} + \nabla_M T^{MN} \right] \\
& =  \delta(\rho -1)|\partial \rho| 
{\mathfrak p}_N^P  \left[ (n\cdot\nabla) n_M T^{MN} + \nabla_M T^{MN} \right] \\
& = \delta(\rho -1) |\partial \rho|
 \left[  {\mathfrak p}_N^P {\mathfrak p}^Q_M \nabla_Q T^{MN} \right]. 
\end{split}
\end{equation}
The manipulations in \eqref{stons} are essentially identical to those in 
\eqref{cccons}. Note that ${\mathfrak p}_N^P {\mathfrak p}^Q_M \nabla_Q T^{MN}$ is the membrane 
world volume divergence of the membrane stress tensor $T^{MN}$. 

On the other hand 
\begin{equation}\label{norm}
n_N \nabla_M {\cal T}^{MN}
= -\left(\nabla_M n_N \right) {\cal T}^{MN} = -{\cal K}_{MN} {\cal T}^{MN}
= -\delta(\rho -1)|\partial \rho|
{\cal K}_{MN} T^{MN},
\end{equation}
(in going from the first to the second expression in \eqref{norm} we have used 
${\cal T}^{MN}n_N=0$). 
It follows that the normal component of the stress tensor conservation equation 
is satisfied if and only if ${\cal K}_{MN} T^{MN}=0$. 

\subsection{The stress tensor for a Nambu-Goto membrane}

In order to gain some intuition for membrane stress tensors is useful 
to consider a simple example. Consider a relativistic membrane whose only 
degree of freedom is its shape and whose dynamics is governed by the 
relativistic Nambu-Goto action 
\begin{equation}\label{memact}
S= -\sigma \int \sqrt{-g_{(ind)}},
\end{equation}
where
$g_{(ind)}$ is the determinant of the metric $g^{(ind)}_{\mu\nu}$ induced on the world volume
of the membrane and $\sigma$ is the tension of the membrane. It is easily 
verified that the equation of motion that follows from this action is simply 
\begin{equation}\label{eom}
{\cal K}=0,
\end{equation}
where ${\cal K}$ is the trace of the extrinsic curvature of the membrane 
world volume. The spacetime stress tensor for this system may be obtained by 
varying the action w.r.t the spacetime metric. The stress tensor is easily 
verified to take the form \eqref{stm} with 
\begin{equation}\label{stst}
T^{MN}= -\sigma ~{\mathfrak p}^{MN}.
\end{equation}
Note that $T^{MN}$ is proportional to the world volume metric; it follows that 
$T^{MN}$ - viewed as membrane world volume stress tensor - is trivially 
conserved. On the other hand the requirement that  $T_{MN} {\cal K}^{MN}=\sigma 
{\cal K} =0$ is nontrivial and yields the membrane 
equation of motion. 

In the simple example reviewed above the conservation of the membrane 
stress tensor was trivial in the world volume directions as a consequence 
of diffeomorphism invariance in these directions. On the other hand the 
conservation of the stress tensor in the normal direction was nontrivial 
and yields the equations of motion - a relativistic version of Newton's laws
in the normal direction. Below we will see that the large $D$ gravitational 
membranes of interest to this paper behave in an orthogonal fashion. 
In that case the equation of stress tensor conservation in the normal direction
is obeyed in a relatively trivial manner, while the equation for world 
volume conservation of the stress tensor yields the membrane equations of 
motion.

\section{Membrane Currents from Linearized solutions: Description 
of the Map} \label{mstcc}

In this section and the next we study the minimally coupled scalar, Maxwell and linearized 
Einstein equation in the vicinity of the world volume of a codimension one 
membrane. We assume that our membrane is embedded in a flat $D$ dimensional 
spacetime and work in the large $D$ limit. 

Let us suppose we are given a solution to the exterior of the membrane 
world volume that decays rapidly towards infinity. \footnote{As we will 
see later, the true exterior solution also has small constant modes
with coefficient of order $\frac{1}{D^D}$ (see \eqref{gfsg} for an example). 
At distances of order unity from the membrane - where we work in this 
section - the constant modes (the mode proportional to $A$ in \eqref{gfsg}) 
are nonperturbatively smaller than the decaying piece, and so 
are invisible to the large $D$ analysis of this section. However 
the details of this constant piece shape the nature of the radiation  
far away; see e.g. the discussion under 
\eqref{capprox}. }
We then search for a corresponding {\it regular}  solution in the interior region of the membrane 
subject to the requirement that the scalar field, tangential components of 
field strengths and curvatures are continuous across the membrane while 
allowing for first derivatives of these quantities to be discontinuous 
across the membrane. Our continuity requirement 
effectively imposes a Dirichlet type boundary condition for the (as yet unknown) solution in the interior of the membrane. This boundary condition, together with the requirement of regularity, turns out to be sufficient to uniquely - and practically -
determine the interior solution order by order in the $1/D$ expansion. 
\footnote{The fact that these boundary and regularity conditions uniquely determine our 
solution is true only in the $1/D$ expansion and is certainly untrue at finite $D$. 
As an example consider the minimally coupled scalar equation $\Box \phi=0$  
with the membrane manifold taken to be $S^{D-2} \times {\rm time}$ and the Dirichlet boundary condition
that $\phi$ vanish on the membrane. One solution with these boundary conditions is $\phi=0$, but 
this solution is clearly not unique. In the $l=0$ sector, for instance, we also have solutions 
of the form $\phi= \sum_n a_n e^{-i \omega_n t} \left( \frac{\omega_n}{r} \right)^{\frac{
D-3}{2}} J_{\frac{D-3}{2}}(\omega_n r)$ where $\omega_n$ run over the set of zeroes of $J_{\frac{D-3}{2}}(\omega_n)$. Note however that at large $D$ the first zero of this Bessel function occurs at a value 
of order $D/2$. It follows that the frequencies $\omega_n$ are all of order $D$ or higher at large 
$D$. In the large $D$ limit we disallow solutions with such high frequencies. In this extremely 
simple toy example it follows that the unique allowed interior solution is 
simply $\phi=0$.} 

Though the interior and exterior solutions are continuous across the membrane they are not 
analytic continuations of each other. In particular normal derivatives of fields are generically 
discontinuous across the membrane. The discontinuities in these normal derivatives 
determine an effective source for the wave equations that is localized on the 
membrane (see  \eqref{phon}, \eqref{maxo}
and \eqref{eedis} ). As explained in those equations, this 
 source is the difference between an `exterior' current 
(the exterior normal derivative) and `interior' current (the interior normal derivative).  
\footnote{As explained in the introduction, the interior current is neatly encoded 
in the action of the interior 
solution as a function of the metric, gauge field or scalar field on the membrane.}

To recap,  the procedure described in this section and the next allows us to constructively establish a one to one map between decaying linearized solutions to the exterior of a membrane and an auxiliary 
solution (which has no physical reality). The auxiliary solution agrees with the decaying solution 
- upto corrections of order $1/D^D$ - to the exterior of the membrane. It is constructed to ensure 
that it is regular everywhere in the interior of the membrane.
The auxiliary solution solves the free uncharged equations everywhere to the exterior and interior of the membrane. The auxiliary solution also solves the bulk equations precisely on the 
membrane provided the membrane is assumed to 
carry a charge; in this section and the next we find precise formulae for this charge as a functional of the 
prescribed external solution. The discussion of this section and the next is precise 
(even conceptually) only in the $\frac{1}{D}$ expansion. 

The starting point of the discussion of this section was a decaying external solution which 
was assumed to be known in the neighbourhood of the membrane surface. This original solution is - 
in general - not known far away from the membrane. However the analysis of this section 
- together with one additional piece of information - allows us to determine this asymptotic behaviour 
as we now explain. 

Recall that the auxiliary solution obeys the linearized bulk equation, with a known charge, all over 
spacetime. It follows that the auxiliary solution is given {\it all over spacetieme}  by the convolution of the membrane current with a Greens function. This statement does not, as yet, completely determine 
the auxiliary solution as all of the linearized equations of motion we study admit an infinite number 
of inequivalent Greens functions (e.g. advanced, retarded, Feynman etc). We now add an additional 
condition on the auxiliary solution; we demand that it is (e.g.) purely outgoing at infinity. 
This condition uniquely singles out one particular Green's function (e.g. the retarded 
Green's function) and yields a well defined - and practically useful - formula for the auxiliary 
solution all over spacetime. 
\footnote{
The fact that the auxiliary solution is given by the convolution of a membrane current with the Green's function depends crucially on the fact that the auxiliary solution was defined to be regular 
in the interior of the membrane. Had we defined the auxiliary solution differently- perhaps by 
allowing prescribed singularities in the interior of the membrane - we would have obtained 
an integral formula for this solution given by the convolution of the Greens function with 
all sources - those located at singularities together with those on the membrane. }

Recall, however, that the original external solution agrees with the auxiliary solution in an exterior neighbourhood of the membrane. If physical considerations inform us that the external solution obeys (e.g.) outgoing boundary conditions at infinity, it then follows that the external solution agrees with the auxiliary solution - to non perturbative 
accuracy - everywhere outside the membrane. It follows that the external solution is also given 
everywhere outside the membrane by the integral formula described in the previous paragraph. 

In summary let us suppose we are given a linearized external solution in the neighbourhood of the membrane 
world volume that is known to be purely outgoing at infinity. The following two step procedure 
can be used to continue this solution to large $r$. In the first 
step we determine the `membrane current' corresponding to our external solution. This determination is 
the topic of this section and the next. In the second step we convolute this current against a Greens 
function - this is the topic of section \ref{bra}.  The resultant expression is the continuation of the 
external solution to large $r$. In the external neighbourhood of the membrane this expression 
is guaranteed to agree with the configuration we started out with, upto nonperturbative corrections. 
The large $r$ behaviour of this solutions yeilds the radiation field that our external solution 
continues to at infinity.

\subsection{Minimally coupled scalar}

We start with the case of a minimally coupled scalar equation 
\begin{equation} \label{mcse}
\Box \phi= -{\cal S},
\end{equation}
with the source ${\cal S}$ assumed to be delta function localized on the membrane. 

Given the decaying part of the solution to \eqref{mcse} in the exterior, we wish 
to construct the matching interior solution. Our tactic for achieving this is 
very straightforward. We first construct the most general decaying solution to 
\eqref{mcse} in the vicinity of the exterior of the membrane. We then construct 
the most general regular solution to the same equation in the vicinity of the 
interior of the membrane. By matching solutions in the exterior with those in the 
interior we produce the most general solution to \eqref{mcse} that is 
continuous across the membrane. Our construction 
- which uniquely pairs any external solution with an internal solution - turns out 
to depend on one free function on the membrane. This function can be thought of as the 
value of $\phi$ on the membrane or equally as the source `current' ${\cal S}$. 
The construction 
thus gives us 
\begin{itemize}
\item{1.} An explicit classification and construction of all consistent decaying external 
solutions.
\item{2.} A one to one map between such solutions and corresponding interior solutions.
\item{3.} Consequently a one to one map between decaying external solutions and a 
source function ${\cal S}$ localized on the membrane.
\end{itemize}
Our construction of the exterior and interior solutions takes the form of 
a power series expansion in the distance $s$ away from the membrane. The 
radius of convergence of this expansion is of order $D/{K}$ and 
so this expansion is useful, from a practical point of view, only when 
$s \ll D/{K}$. The coefficients in this power series expansion are 
each individually determined in a power series expansion in $\frac{1}{D}$. 

Given that \eqref{mcse} is a second order equation, the reader may 
wonder how it is possible that exterior and interior solutions to this 
equation are parametrized 
by one rather than two functions on the membrane. The key point here is the restriction that 
the exterior solution rapidly decay away from the membrane and that the interior solution 
be regular (in particular not grow arbitrarily large as $D$ is taken to 
infinity at any point reliably captured by our approximations). 
These two conditions cut down the set of 
exterior and interior solutions each to solutions parametrized by 
a single function on the membrane; 
upon imposing continuity across the membrane we find a set of sewn solutions
parametrized by a single function on the membrane.

As this point is very important, we now explain it again in a more precise 
and much more detailed manner. 

The full set of solutions to the equation $\Box \phi=0$ - either to the 
exterior or in the interior of the membrane - is indeed parametrized by 
two functions on the membrane world volume. Let us denote these two  
functions by $\alpha$ and $\beta$. It follows from linearity that the 
most general solution of the equation $\Box \phi=0$ away from the 
membrane is given by 
\begin{equation}
\label{formsolns} 
 \phi= F_1[\alpha(x)]+ F_2[\beta(x)],
\end{equation}
where $F_{1,2}$ are linear maps from the space of functions on the membrane 
to functions in the flat spacetime in which the membrane is embedded. 
Later in this section we will 
explicitly construct the two functionals $F_1$ and $F_2$ (in a Taylor series 
expansion in distance away from the membrane) \footnote{We determine the 
coefficients of this expansion order by order in $1/D$.} with the following 
two properties. 
\begin{itemize}
\item
First, on the membrane 
$F_1[\alpha]=\alpha$ and 
$F_2[\beta]=\beta$. In other words $\alpha$ and $\beta$ are the values of $\phi$ restricted to the membrane. 
$F_1[\alpha]$ and $F_2[\beta]$ are two different continuations of the scalar field on the membrane into the 
bulk.  
\item 
Second $F_1$ decays rapidly (over a distance scale $1/D$) 
to the 
exterior of the membrane, and grows rapidly over the same distance scale on the 
interior of the membrane, while $F_2$ neither grows nor decays as we move 
distances of order $1/D$ away from the membrane. Instead the variation of 
$F_2$, as we move away from the membrane, occurs over length scales of order 
unity. \footnote{The functionals $F_1$ and $F_2$ are effectively local functions 
of $\alpha$ and $\beta$ in the following sense: it is possible to foliate spacetime around the 
membrane into tubes each of which cuts the membrane and is labeled by the point 
$x_0$ at which it does so. To any given order in $1/D$, $F_1$ and $F_2$ at 
any $x_0$ depend only on the distance from the membrane (which is assumed small in 
units of the local radius of extrinsic curvature of the membrane), the extrinsic 
geometry of the membrane at $x_0$ and a finite number of derivatives of 
$\alpha(x_0)$ or $\beta(x_0)$. 
The reason for this locality is simply that the boundary conditions of decay in the exterior 
and lack of blow up in the interior can each effectively be imposed at distances of order 
$1/D$ away from the membrane. The thinness of the region enclosed by our boundary 
conditions is the underlying reason for the locality of our expansion.}
\end{itemize}

We will now use the two functionals $F_1$ and $F_2$ to construct solutions $\phi(x)$ of \eqref{mcse}
that are of the form described in the previous subsection, or, more specifically 
have the following properties
\begin{itemize}
\item 
$\phi(x)$ reduces to an arbitrarily prescribed function 
$\phi_0(x)$ on the membrane world volume.
\item 
$\phi(x)$ is continuous across the membrane but its normal derivative is 
across this surface
\item $\phi(x)$ decays to the exterior of the membrane, and stays regular (does not 
blow up) in the interior.
\end{itemize}
A moment's thought will convince the reader that the required solution is given by 

\begin{equation}\label{unsols}\begin{split} 
\phi(x)&= F_1[\phi_0]~~~{\rm outside},\\
\phi(x)&=F_2[\phi_0]~~{\rm inside}.\\
\end{split}
\end{equation}
As mentioned above, in the next section we will explicitly determine the functionals 
$F_1$ and $F_2$ in a power series expansion in $1/D$. 

Note that the solutions \eqref{unsols} are parameterized by a {\it single} membrane's function 
worth of data - which can be thought of either as $\phi_0(x)$ or the source function 
${\cal S}$ on the membrane. This fact can also be 
understood in the following terms. Suppose we are given a source ${\cal S}$ localized 
on the world volume of the membrane. Clearly the most general solution to \eqref{mcse} in the 
presence of this source takes the form 
\begin{equation}\label{phsc} \begin{split}
\phi (x)&= \int dy G(x-y) {\cal S}(y),\\
\end{split}
\end{equation}
where $G$ is a Greens function for the operator $\Box$ and the integral 
over $y$ is taken over the membrane world volume. At finite 
$D$ \eqref{phsc} does not define a unique solution to the problem, because 
the Greens function, $G$, is not unique. As we have explained in 
subsection \ref{ldgf}, however, all reasonable Greens functions are 
identical (upto differences of 
order $1/D^D$) at distances of order unity around the source.
It follows that the formula 
\eqref{phsc} does unambiguously define a unique solution to \eqref{mcse} in
the neighbourhood of the membrane an expansion in $1/D$. \eqref{unsols} is 
this unique solution; i.e. \eqref{phsc} can be identified with \eqref{unsols}
in the neighbourhood of the membrane for every reasonable choice of the 
Greens function $D$, even though the expressions \eqref{phsc} begin to depend 
sensitively on the choice of Greens function at large $r$ (i.e. distances of 
order $D$). As we have explained in detail above, the `correct' choice of 
Greens functions is determined by physical considerations for the problem at hand; 
the relevant Greens function for this paper will always prove to be the retarded 
Greens function.

\subsection{Maxwell Equation}

Although it is possible to solve the Maxwell equations in a gauge invariant manner,
we will find it convenient to proceed by fixing a gauge. We first define a foliation of spacetime 
into surfaces of constant $\rho$, chosen so that the surface $\rho=1$ is the membrane. 
We choose the function $\rho$ to obey the equation $\Box \left( \frac{1}{\rho^{D-3}}\right) =0$ 
(see subsection \ref{fost} below). We then choose to work in a gauge in which 
$A^\rho$ vanishes, 
i.e. the gauge $d\rho. A=0$. 

With this choice of foliation, the Maxwell equations can be divided up into the constraint equations (Maxwell equations dotted with $d \rho$) and the dynamical 
equations. More precisely, by a slight misuse of terminology, we will refer 
to the equations 
\begin{equation}\label{dyndef}
\Pi_C^A \partial_BF^{BC}=0,
\end{equation}
as dynamical equations where 
\begin{equation}\label{projdef} \begin{split}
\Pi_{CA}&=\eta_{CA} -n_A n_C, \\
n_A&=\frac{\partial_A \rho}{\sqrt{\partial_D \rho \partial^D \rho}}.\\
\end{split}
\end{equation} 
On the other hand we refer to 
\begin{equation}\label{constdef} \begin{split}
{\cal M}& =0,\\
{\cal M}&\equiv n_C \partial_BF^{BC},
\end{split}
\end{equation}
as the constraint Maxwell equation

We proceed by first solving the dynamical equations defined above and 
then turn later to the constraint equation. 
The dynamical equations are very similar in character to the minimally coupled scalar equation 
discussed in the previous subsubsection. As in the previous subsubsection 
we find in general that the solutions to the dynamical Maxwell equations 
take the form 
\begin{equation}
\label{formsolnv} 
A= F_1[C_\mu (x)]+ F_2[B_\mu(x)],
\end{equation}
where $A$ is the oneform gauge field in spacetime and $C_\mu$ and $B_\mu$ are worldvolume gauge 
fields on the membrane. $F_{1,2}$ are now linear maps from gauge fields on the membrane 
to oneform gauge fields in flat spacetime. These functional share the following properties 
with their scalar counterparts. First, on the membrane 
$F_1[C_\mu]=C_\mu$ and $F_2[B_\mu]=B_\mu$ (it makes sense to equate a spacetime gauge field 
with a world volume gauge field precisely because $d\rho . A$ vanishes).  
As for scalars $F_1$ decays rapidly (over a distance scale $1/D$) 
to the exterior of the membrane, and grows rapidly over the same distance scale on the 
interior of the membrane, while $F_2$ neither grows nor decays as we move 
distances of order $1/D$ away from the membrane. Instead the variation of 
$F_2$, as we move away from the membrane, occurs over length scales of order 
unity.

As in the case of scalars above, the boundary condition that our spacetime gauge 
field decays in the exterior, is regular and bounded in the interior and that the 
field strength restricted to the membrane is continuous on the membrane, and 
that it takes the value $(A_0)_\mu$ on the membrane  leaves us 
with the solutions
\begin{equation}\label{unsolv}\begin{split} 
A(x)&= F_1[(A_0)_\mu]~~~{\rm outside},\\
A(x)&=F_2[(A_0)_\mu]~~{\rm inside}.\\
\end{split}
\end{equation}

We have completed our programme of solving the dynamical equations. 
What remains is to solve the Maxwell constraint equations. It is a well 
known property of Maxwell's equations that if the dynamical equations 
are obeyed everywhere and the constraint equation is obeyed on a single 
slice then the constraint equation is obeyed everywhere. Our definition 
of dynamical and constraint equations are different from the usual ones 
(which are adapted to a foliation of spacetime into coordinate systems 
including $\rho$ as a special coordinate) and it is instructive to work 
our our version of this standard statement. This is easily done. Note that
\begin{equation}\label{manipconst}
\partial_A \left( \Pi^{A}_B \partial_C F^{CB} \right) 
= \partial_C \partial_B F^{CB} - 
\partial_A \left( n^A n_B\partial_C F^{CB} \right) =
-n.\partial \left( n_B\partial_C F^{CB} \right) -{K} 
\left( n_B\partial_C F^{CB} \right) ,
\end{equation} 
(where we have used the antisymmetry of $F^{AB}$ in the last step). 
It follows that 
\begin{equation}\label{constevi} 
(n.\partial ){\cal M} = - {K} {\cal M} - 
\partial_A \left( \Pi^{A}_B \partial_C F^{CB} \right) ,
\end{equation}
(see \eqref{constdef} for a definition of ${\cal M}$).
Now the last term on the RHS of \eqref{constev} is the divergence of the 
dynamical equations and so vanishes once those equations are solved. 
On solutions of the dynamical equations it thus follows that
\begin{equation}\label{constev} 
(n.\partial ) {\cal M} = - {K} {\cal M} .
\end{equation}
Integrating \eqref{constev} along flow lines of the vector field $n$ it 
follows that 
\begin{equation}\label{meq} 
{\cal M}(\rho) = {\cal M}_0 e^{ - \int_1^\rho {K} d s },
\end{equation}
where ${\cal M}_0$ is the value of ${\cal M}$ at $\rho=1$ (i.e. on the 
membrane) and $ds$ is the proper distance from the membrane along the integral curves of the vector field $n$.\\
 Note that ${K}$, the extrinsic curvature of slices of 
constant $\rho$ is positive and of order $D$ (see subsection \ref{fost} below).

Let us assume that ${\cal M}_0$ is nonzero. It follows that 
${\cal M}(\rho)$ decays rapidly to zero (over a length 
scale of order $1/D$) as we move away from the membrane towards the exterior. 
But it also follows that ${\cal M}(\rho)$ blows up rapidly - over a length 
scale of order $1/D$ - as we move away from the membrane towards the 
interior. 

Let us now apply these results to the two special solutions $F_1[(A_0)_\mu]$
and $F_2[(A_0)_\mu]$ defined above. The solution $F_1[(A_0)_\mu]$ is defined 
so that it decays rapidly to the exterior of the membrane and blows up 
rapidly in the interior of the membrane. The fact that ${\cal M}$ also 
has the same behaviour comes as no surprise for this solution. On the 
other hand the solution $F_2[(A_0)_\mu]$ is defined so that it {\it does not}
blow up in the interior of the membrane. It is thus impossible for 
${\cal M}$ to blow up in the interior - in the manner determined by 
\eqref{meq}. It follows that ${\cal M}_0$ must in fact vanish on the solution 
$F_2[(A_0)_\mu]$.

In summary we have demonstrated that the solution $F_2[(A_0)_\mu]$ is very 
special; it is the solution on which the constraint equation is automatically
satisfied - without the need to impose any further constraint on 
$(A_0)_\mu$. On the other hand the configuration $F_2[(A_0)_\mu]$ is 
a solution of the full Maxwell equations not for all $(A_0)_\mu$ but only 
for those that are constrained to obey a further condition (which we will 
interpret below as the condition of conservation of the membrane current). 

Matching the solutions $F_1$ and $F_2$ as in \eqref{unsolv} yields a 
class of solutions of Maxwell's equations parametrized by $(A_0)_\mu$ 
subject to the single constraint just described above. The solution 
\eqref{unsolv} is a solution to Maxwell's equations with a current of the 
form \eqref{cc} with the function $J^M$ given in \eqref{maxo}. This 
current may be rewritten as 
\begin{equation}\label{currentmaxn}
J_M=J^{(out)}_M-J^{in}_M, ~~~J_M^{(out)}=n^NF^{(out)}_{NM}, ~~~
J_M^{(in)}=n^NF^{(in)}_{NM}.
\end{equation}
Note that the conservation of this current follows immediately from the 
constraint equations applied to the external and internal solutions 
respectively. As we have explained above this conservation is automatic
for the internal solution, but imposes a constraint on the data $(A_0)_\mu$
in the case of the external solution. 

The interior current $J_M^{(in)}$ is most compactly presented by evaluating the 
action of the interior solution $S_{in}[A_0]$. The current $J_M^{(in)}$ is 
then given by varying this action w.r.t $A_0$ using
\begin{equation}\label{hjeq}
\delta S_{in}[A_0] = \int  \delta (A_0)_M J^{(in) M},
\end{equation}
(see \eqref{maxon}). As the interior solution 
$F_2[A_\mu(x)]$ is well defined for every value of the boundary 
gauge field $(A_0)_\mu(x)$, $S_{in}[A_0]$, is a gauge invariant functional 
of this boundary gauge field that also turns out to be local in the 
large $D$ limit. 
\footnote{Recall that $(A_0)_\mu$ is also the gauge field on the membrane 
viewed from the outside and so is known.}

 On the other hand the external contribution to the current is simply 
evaluated from the definition \eqref{currentmaxn}, where the quantity on the 
RHS of that equation is evaluated on the external solution which is 
assumed to be known.

Let us summarize. Solutions of Maxwell's equations that obey our boundary conditions are parametrized 
by the membrane gauge field subject to a single constraint (the conservation of the exterior contribution
to the membrane current). The full membrane current is given by adding the exterior contribution to the interior contribution which, in turn, is obtained from the variation of a gauge invariant `counterterm' boundary action. In order 
to compute the current associated with a given external solution the 
{\it only} remaining nontrivial step is the determination of the 
counterterm action associated with the interior solution. 

\subsection{Linearized Einstein Equation}

Let the metric be given by $\eta_{MN} +H_{MN}$. As in the previous subsection we work with a particular gauge choice; we impose the gauge $n^N H_{NM}=0$.

In parallel with the previous subsection it is convenient to decompose Einstein's equations
into dynamical and constraint equations. Let us define 
\begin{equation} \label{dyndefe1} 
{\cal E}_{MN}= R_{MN} - \frac{R}{2} g_{MN} - 8 \pi T_{MN}.
\end{equation}
The Einstein equations take the form  
\begin{equation} \label{dyndefen} 
{\cal E}_{MN}=0.
\end{equation}
The dynamical equations are defined to be 
\begin{equation} \label{dyndefe} 
\Pi_A^M {\cal E}_{MN} \Pi_B^N= 0.
\end{equation}
The constraint Einstein equations are 
\begin{equation} \label{consteindef}\begin{split} 
{\cal C}^E_M &\equiv n^A {\cal E}_{AM}, \\
{\cal C}^E_A&=0.
\end{split}
\end{equation}

As in the previous subsection we first solve the dynamical Einstein equations to find a structure very similar 
to that for the minimally coupled scalar. The most general solution is given 
by
\begin{equation}
\label{formsolng} 
H= F_1[\mathfrak{h_{\mu\nu}} (x)]+ F_2[\mathfrak{g_{\mu\nu}}(x)],
\end{equation}
where the $G= \eta + H$ is the spacetime metric and $\mathfrak{h_{\mu\nu}}(x)$ and $\mathfrak{g_{\mu\nu}}(x)$ are induced metrics on the membrane.  $F_{1,2}$ are now maps from the 
induced metric on the membrane to linearized metric fluctuations in flat 
spacetime. Note that the induced metric is nontrivial even in the 
absence of the fluctuation $H_{MN}$. The maps $F_1$ and $F_2$ linearly 
map {\it changes} in this induced metric to linearized fluctuations of the
bulk.

As in the previous 
section $F_1$ decays rapidly (over a distance scale $1/D$) 
to the exterior of the membrane, and grows rapidly over the same distance scale in the 
interior of the membrane, while $F_2$ neither grows nor decays as we move 
distances of order $1/D$ away from the membrane. 

Following the previous subsection we proceed to solve the dynamical equations 
subject to the boundary conditions that $g_{MN}$ reduces to 
$ \left[g^{(ind)}_{\mu\nu}=g^{(ind,f)}_{\mu\nu} + h^{(0)}_{\mu\nu}\right]$ 
on the membrane where $g^{(ind,f)}_{\mu\nu}$ is the induced metric on the 
membrane viewed as a submanifold of the spacetime with metric $\eta_{MN}$ 
and $h^{(0)}_{\mu\nu}$ is arbitrary but small. Through this section we work to 
linearized order in $h^{(0)}_{\mu\nu}$. 

 Imposing the boundary conditions of fall off to the exterior 
and regularity in the interior and the continuity of the induced 
metric on the membrane as we pass from outside to inside, 
 we find that the unique solutions to our 
equations are 
\begin{equation}\label{unsolg}\begin{split} 
H&= F_1[g^{(ind)}_{\mu\nu}]~~~{\rm outside},\\
H&=F_2[g^{(ind)}_{\mu \nu}]~~~{\rm inside},\\
\end{split}
\end{equation}
where $H$ is a spacetime symmetric two tensor (we have omitted its indices 
for brevity). 

As with the study of the Maxwell equation the main qualitative difference between the solutions of the linearized Einstein equations and the minimally coupled scalar equation lies in the 
constraint equations. However the Einstein constraint equations are of two 
varieties. Let 
$$X_N \equiv{\cal C}^{E}_M \Pi^M_N .$$
We refer to the equation $X_M=0$ as the momentum constraint equations. Moreover let 
$$Y \equiv{\cal C}^{E}_M n^M .$$
We refer to the equation $Y=0$ as the Hamiltonian constraint equation. 

In Appendix \ref{eec} we use the identity 
$$\nabla_M  \left( {\cal E}^{MN} \right)=0,$$ 
to demonstrate that the momentum and Hamiltonian Einstein constraint equations obey 
the equations 
\begin{equation}\label{eecc} \begin{split}
&\Pi_B^C(n\cdot\nabla) X_C = -{K}~ X_B  -X^AK_{AB} -Y (n\cdot\nabla)n_B\,,\\
& n\cdot\nabla Y =- {K}~Y - \nabla\cdot X +X^C(n\cdot\nabla)n_C \,.\\
\end{split}
\end{equation}
As in the previous subsection, these equations determine the $\rho$ dependence of 
the constraint equations in terms of their value at $\rho=1$. Let us first consider the 
momentum constraint equations. The first term on the RHS of the first line of \eqref{eecc} 
is of order $D$ while the last two terms on the RHS of this equation are of order unity 
and can be ignored. It follows that, as in the previous subsection, the constraint equations 
$X_C$ grow exponentially as we move away from the membrane in the interior region, but decay 
exponentially in the exterior. As in the previous subsection this means that the constraint 
equations $X_C$ must simply vanish for the interior solution, $F_2$ in \eqref{unsolg}. Once 
this result has been 
established for $X_C$, the second equation in \eqref{eecc} ensures that the same is true of 
the constraint equation $Y$. As in the previous subsection there is no particular reason for
the constraint equations to vanish for the exterior solutions - $F_1$ in \eqref{unsolg}, 
and we will see by explicit computation below 
that they do not. 

It follows that the interior solution $F_2$ is labeled by a boundary metric 
on the membrane. On the other hand the external solution $F_1$ is 
labeled by the same boundary data modulo one constraint. We will later 
interpret this condition as the requirement  
that the membrane stress tensor be conserved.  It follows also that the solution 
\eqref{unsolg} is also labeled by membrane boundary metric subject to a single
constraint.  

We now turn the `Hamiltonian' constraint 
equation
$$ {\cal C}^Mn_M=0.$$
Recall that in section 
\ref{ccom} we demonstrated that a stress tensor of the form 
\eqref{stm} is conserved in spacetime provided that 
\begin{itemize} 
\item $T^{MN}$, viewed as a tensor on the membrane world volume is conserved.
\item $T_{MN}{\cal K}^{MN}=0$. 
\end{itemize}
We have just argued that the  `momentum' constraint equations guarantee that the first condition is satisfied. We will now use the 
`Hamiltonian' constraint equations to show that the second condition is also satisfied.  

It is well known that the Hamiltonian constraint equation can be rewritten 
in terms of the membrane extrinsic curvature and intrinsic membrane curvatures 
as follows (see e.g. eqn 10.2.30.  page 259, of \cite{Wald:1984rg})\footnote{In \cite{Wald:1984rg}, the eqn 10.2.30 is derived for a  spacelike hypersurface where the normal is timelike. But in our case the normal is spacelike and this is why the sign in the first term of our equation \eqref{quote} is different from what it is there in \cite{Wald:1984rg}. See appendix (\ref{identity}) for a derivation.}

\begin{equation}\label{quote}
\begin{split}
0=n^A n^B E_{AB} = \frac{1}{2}\left(-{\cal R} + {\cal K}^2 -{\cal  K}_{AB} {\cal K}^{AB}\right),
\end{split}
\end{equation}
where $E_{AB}=$ is the Einstein Tensor, ${\cal R}$ is the intrinsic Ricci 
scalar on $(\rho=const)$ slices and ${\cal K}^{AB}$ is the extrinsic curvature of 
the same slices. All indices in \eqref{quote} 
are raised or lowered using the induced metric on $\rho=const$ slices, 
embedded in full space-time. As Einstein's equations are obeyed both just 
outside and just inside the membrane, it follows in particular that 
\begin{equation}\label{quoten}
\begin{split}
&\frac{1}{2}\left(-{\cal R}_{(out)} + {\cal K}_{(out)}^2 - {\cal K}^{(out)}_{AB} {\cal K}_{(out)}^{AB}\right)=0,\\
&\frac{1}{2}\left(-{\cal R}_{(in)} + {\cal K}_{(in)}^2 - {\cal K}^{(in)}_{AB} {\cal K}_{(in)}^{AB}\right)=0,\\
\end{split}
\end{equation}
where all quantities with the subscript `out' are evaluated on the 
special slice $\rho=1$ (we refer to this slice as the membrane)
as approached from the outside, while all quantities with the subscript 
`in' are evaluated on the membrane when approached from the interior. 

Recall that the membrane world volume - viewed as a submanifold of flat space 
- has a nontrivial Ricci curvature tensor $R_{\mu\nu}$ and a nontrivial 
extrinsic curvature tensor ${\cal K}_{MN}$; the trace of ${\cal K}_{MN}$ is ${\cal K}$. 
Now $R^{(out)}_{\mu\nu}$, ${\cal K}^{(out)}_{MN}$ and ${\cal K}_{(out)}$ 
refer to the same quantities - but evaluated with the membrane regarded 
as a submanifold of $\left[g_{MN}=\eta_{MN} + h^{(out)}_{MN}\right]$. Similar remarks apply 
to the inside. It follows that - for instance ${\cal K}^{(out)}_{MN}$ differs 
from $K_{MN}$ at first order in the fluctuation field $h_{MN}$. Let us now 
subtract the two equations in \eqref{quoten} above. Using the fact that  
${\cal R}_{(out)}={\cal R}_{(in)}$ (this follows because ${\cal R}$ is a function
only of the induced metric on the membrane and not its normal derivative) 
we find 

\begin{equation}\label{mani1}
\begin{split}
0=&~n^A n^B E_{AB}|_{out} - n^A n^B E_{AB}|_{in}\\
 =&~   {K} \left({{\cal K}}_{out} - {{\cal K}}_{in}\right) - K_{AB} 
\left({\cal K}_{out}^{AB} -{\cal K}_{in}^{AB} \right)\\
 =&~- K_{AB} \left[\left({\cal K}^{AB}_{out} -{\cal K}^{AB}_{in}\right)  -\left( {{\cal K}}_{out}- {{\cal K}}_{in}\right)~ \Pi^{AB}\right]\\
 =&~ 8 \pi K_{AB} T^{AB}.
\end{split}
\end{equation}
In the second line of this equation we have worked to linear order in 
$h_{AB}$. The third line is an algebraic rearrangement of the second line 
and in the fourth line we have used the definition of the membrane 
stress tensor given in \eqref{eedis}

Notice that, as in the previous subsection it is useful to define  
\begin{equation}\label{tinout}\begin{split}
T^{(out)}_{AB}&= \left({\cal K}^{AB}_{(out)}  -{{\cal K}}_{(out)}~ {\mathfrak p}_{(out)}^{AB} \right),\\
T^{(in)}_{AB}&= \left({\cal K}^{AB}_{(in)}  -{\cal K}_{(in)}~ {\mathfrak p}_{(in)}^{AB}  \right),\\
\text{where}&\\
{\mathfrak p}_{(out/in)}^{AB}&=~\text{Projector on the membrane, embedded in outside (inside) metric}.
\end{split}
\end{equation} 
This implies
\begin{equation} \label{tdiff}
T_{AB}=-\left(\frac{1}{8 \pi} \right)\left[T^{(out)}_{AB} - T^{(in)}_{AB}\right].
\end{equation}
In parallel with the previous subsection, the `momentum' Einstein equations 
ensure that $T_{AB}$ is conserved. 
\footnote{More precisely each of $T^{(out)}_{AB}$ and $T^{(in)}_{AB}$ are separately 
conserved when viewed as tensor fields on the membrane with metric induced
from $\eta_{MN}+ h_{MN}$. Note that $T^{(out)}_{AB}$ and $T^{(in)}_{AB}$ each have a 
term that is zeroth order in fluctuations. However this zero order piece 
is common between $T^{(out)}$ and $T^{(in)}$ and so cancels in their difference. 
As a consequence $T_{AB}$ is of first order in fluctuations. It follows 
that $T_{AB}$ is conserved, to first order, even when viewed as a tensor 
field living on the membrane with undeformed induced metric $g^{(ind,f)}_{\mu\nu}$.}

As in the previous subsection, the fact that the interior solution is 
well defined for every value of the induced metric $g^{(ind)}_{\mu\nu}$ without
restriction allows us determine $T_{AB}^{(in)}$ by first evaluating the 
action $S_{in}$ using \eqref{formein} and obtaining the current using 
\eqref{einon}. Note that $S_{in}$ is a gauge invariant function of 
$g^{(ind)}_{\mu\nu}$ which will also turn out to be local in the large $D$ limit. 

\subsubsection{Counterterm Action for $T^{(in)}_{AB}$ at first order}

As we have seen above, the interior solution $F_2$ that appears in \eqref{unsolg} is labeled by 
a metric on the boundary of the membrane. As we have explained in the previous section, the interior 
contribution to this stress tensor may be obtained as follows. We first compute the boundary action 
\begin{equation}\label{intac}
\begin{split}
S_{(in)} = -\left(\frac{1}{8\pi}\right)\int \sqrt{-g^{(ind)}}~~ {\cal K}^{(in)},
\end{split}
\end{equation}
of this solution. This action should be viewed as a functional of the membrane metric that parameterizes
solutions of the functional $F_2$. Varying the action \eqref{intac} w.r.t this boundary metric then 
yields the contribution of the interior stress solution to the membrane stress tensor 
(see \eqref{einon}).  

It turns out that, upto first order in the expansion in $\frac{1}{D}$,  the action \eqref{intac} 
is easily evaluated as a functional of the metric on the membrane using the Gauss Codacci formalism
For any Ricci-flat space, the intrinsic quantities could be related to extrinsic quantities in the following way \cite{Wald:1984rg} (see Appendix (\ref{identity}) for derivation).
\begin{equation}\label{G2mntxt}
\begin{split}
&0= {\cal R}^{\mu\nu} -{\cal K} {\cal K}^{\mu\nu} +{\cal K}^{\mu\alpha} {\cal K}_\alpha^\nu +
e^\mu_A e^\nu_B R^{ACBC'}~n_C n_{C'}, \\
&0= {\cal R} -{\cal K}^2 + {\cal K}_{\mu\nu}{\cal K}^{\mu\nu} ,
\end{split}
\end{equation}
where ${\cal R}^{\mu\nu}$ and ${\cal R}$ is the intrinsic Ricci tensor and Ricci scalar of the membrane, $R^{ACBC'}$ is the Riemann tensor of the full space-time and and $n_C$ is the unit normal to the membrane. $e^\mu_A$ is the matrix that relates coordinates along the membrane ($\{x^\mu\}$) to the full space-time coordinate ($\{X^A\}$) as
$$x^\mu = e^\mu_A ~X^A.$$
The following scalings with $D$ apply to the various quantities that in equation \eqref{G2mntxt} 
when evaluated on the interior solution $F_2$
\begin{equation}\label{scalD}
\begin{split}
&{\cal R}\sim {\cal O}(D^2),
~~~~{\cal R}_{\mu\nu}\sim {\cal O}(D),\\
 &{\cal K}^{(in)}\sim {\cal O}(D),
~~~~{\cal K}^{(in)}_{\mu\nu}\sim {\cal O}(1),\\
&e^\mu_A e^\nu_B R^{ACBC'}~n_C n_{C'}\sim 
{\cal O}(1),
\end{split}
\end{equation}
(the derivation of these scalings use the fact that in the interior solution $F_2$ the metric varies in the $\rho$ direction on length scale unity - rather than length scale $1/D$ (as is the case for the exterior solution $F_1$).

The nature of these scalings allow us determine  ${\cal K}$ in terms of intrinsic Riemann curvature tensor by solving equation \eqref{G2mntxt} order by order in $\left(\frac{1}{D}\right)$ expansion.
\begin{equation}\label{solinmn}
\begin{split}
{\cal K}^{(in)} &= \sqrt{{\cal R}^{(in)}}+\frac{1}{2}\left[\frac{{\cal R}_{\mu\nu}{\cal R}^{\mu\nu}}{{\cal R}_{(in)}^\frac{3}{2}}\right] + {\cal O}\left(\frac{1}{D}\right),\\
{\cal K}^{(in)}_{\mu\nu}&=\frac{{\cal R}_{\mu\nu}}{\sqrt{{\cal R}}}+ {\cal O}\left(\frac{1}{D}\right).\\
\end{split}
\end{equation}
Note that the last term in the first equation of \eqref{G2mntxt} has not contributed to this order. In 
order to evaluate this complicated term we would need the full details of the solution $F_2$ developed 
in the next section. As this term does not contribute, however, the computation we have presented 
is identical to the computation of the counter term on a curved membrane surface embedded in flat-Minkowski space \footnote{Note that if we are considering the outside solution, the equation \eqref{G2mntxt} is still applicable, but the scaling rules described in equation \eqref{scalD} are not valid. In that case, ${\cal K}^{(in)}_{\mu\nu}$ also scales like order ${\cal O}(D^2)$ and therefore the solution that we have presented in equation \eqref{solinmn} is not valid for the space-time outside the membrane.}.\\
 Substituting the first equation of \eqref{solinmn} in equation \eqref{intac} we get the form of the counter term action in terms of membrane's intrinsic curvature:
 \begin{equation}\label{couteraction}
\begin{split} 
S_{counter}=-8\pi S_{(in)} = \int \sqrt{g^{(ind)}}~ \left[\sqrt{{\cal R}}+\frac{1}{2}\left(\frac{{\cal R}_{\mu\nu}{\cal R}^{\mu\nu}}{{\cal R}^\frac{3}{2}}\right) + {\cal O}\left(\frac{1}{D}\right)\right].
\end{split}
\end{equation}
In Appendix \ref{cact} we have demonstrated that the stress tensor 
$$-8\pi  \sqrt{g^{(ind)}} T^{(in)}_{\mu\nu} = g^{(ind)}_{\mu\alpha}\left[ \frac{\delta S_{counter}}{\delta g^{(ind)}_{\alpha\beta}}\right]g^{(ind)}_{\nu\beta},$$
obtained from this action is given by 
\begin{equation}\label{tabin}
\begin{split}
(-8\pi) T^{(in)}_{\mu\nu} &= -\left(\frac{{\cal R}_{\mu\nu} }{2\sqrt{{\cal R}}}\right) + \left(\frac{g^{(ind)}_{\mu\nu}}{2}\right)\left[\sqrt{{\cal R}}+\frac{1}{2}\left(\frac{{\cal R}_{\alpha\beta}{\cal R}^{\alpha\beta}}{{\cal R}^\frac{3}{2}}\right)\right] + {\cal O}\left(\frac{1}{D}\right).
\end{split}
\end{equation}

\section{Membrane currents from Linearized Solutions: Detailed Construction} \label{mclsdc}

In this detailed technical section we present an explicit construction of the 
functionals $F_1$ and $F_2$  defined in the previous section, separately  
for the scalar, vector and linearized gravity theories (\eqref{formsolns}, 
\eqref{formsolnv}, \eqref{formsolng}). As explained behind we construct 
these functionals in a power series expansion in $\rho-1$. Each Taylor 
series coefficient in this expansion is computed in an expansion in $1/D$.

The results of this section will be used in the next section to read off 
the current and stress tensor carried by the large $D$ gravitational 
membrane. The only aspect of the internal solution that will be needed 
for this purpose is its action; as explained in the previous section 
the action is given by a surface integral of the solution and its 
first normal derivative at the membrane. For the purposes of computing this 
action we are thus specially interested in the first Taylor series expansion
coefficient of our solution. 

As explained above we present our solutions in terms of a Taylor series 
in $\rho-1$. Before proceeding to the explicit constructions we thus
need to pause to give a precise definition of the function $\rho$ and to 
briefly explore its properties. 

\subsection{A membrane adapted foliation of spacetime} 
\label{fost}

Consider a function $\rho$ defined in flat Minkowski space by the 
following conditions. 
\begin{itemize}
\item $\rho$ takes the value unity on the membrane world volume.
\item $\rho$ obeys the equation
\begin{equation}\label{rhof}
\Box \left( \frac{1}{\rho^{D-3}} \right)=0,
\end{equation}
everywhere outside the membrane.
\item $\frac{1}{\rho^{D-3}}$ decays at infinity and is purely outgoing there. 
\end{itemize}
The conditions above uniquely define the function $\rho$ to the exterior of 
the membrane 
at any $D$. \footnote{This may be understood as follows. Any solution of the second order 
differential equation \eqref{rhof} is uniquely specified by two boundary conditions. In the 
present context the two boundary conditions are the requirement that $\rho=1$ on the membrane 
world volume and the requirement that the solution is outgoing at infinity.} Once we have 
the solution for $\rho$ to the exterior of the membrane, we define it in the interior 
of the membrane by an analytic continuation. The interior solution $\rho$ 
defined in this manner continues to obey the equation \eqref{rhof} 
in the interior except at positions of potential singularities of 
$\frac{1}{\rho^{D-3}}$. We will see below that such singularities 
- which are always present - do not occur at distances 
$ \ll \frac{D}{{K}}$ away from the membrane and will play no role in 
our analysis below. 

While the requirements above uniquely determine the function $\rho$ in principle, an explicit 
determination of $\rho$ as a functional of the membrane world volume is a difficult job  
at finite $D$. The situation in this regard is much better at large $D$. In this subsection 
we explicitly determine the function $\rho$ in a Taylor series expansion in distance away from 
the membrane \footnote{This expansion is good at distances $ \ll \frac{D}{ {K}}$ 
away from the membrane}. The coefficients of this expansion are determined in a Taylor series expansion 
in $\frac{1}{D}$. The key simplification at large $D$ is that, in this limit, the function 
$\rho$ turns out to be locally determined by the shape of the membrane world volume (see later 
in this subsection for the precise version of this statement). \footnote{A related fact is that 
we do not need to use the boundary condition that $\rho$ is outgoing at infinity in order to 
determine $\rho$ in the large $D$ limit. If, in other words we were to define a new 
function ${\tilde \rho}$ by the conditions listed in this subsection, with the one replacement 
that ${\tilde \rho}$ is required to be ingoing rather than outgoing at infinity, then 
in the $\frac{1}{D}$ expansion ${\tilde \rho}$ would have the same Taylor series expansion around
the horizon as $\rho$. It turns out that the two functions $\rho$ and ${\tilde \rho}$ differ only 
at order $\frac{1}{D^D}$ at distances of order unity away from the black hole. The two functions
begin to differ substantially from each other only at distances of order $D$ away from the 
membrane. All these remarks are, of course, tightly connected to the properties of 
Greens functions at large $D$ discussed in section \ref{bgm}. }

Consider a point in flat spacetime with coordinates $x^M$. To every 
such point we can associate a point ${\hat x}^\mu$ on the membrane by 
the requirement that the straight line between $x^M$ and ${\hat x}^\mu$ 
is collinear with the normal at ${\hat x}^\mu$. Let $s(x^M)$ denote the 
distance between $x^M$ and ${\hat x}^\mu(x^M)$ measured along this straight line. 
In Appendix \ref{rhoconstruct} we demonstrate that 
\begin{equation}\label{rhodef} \begin{split}
& \rho(x)=1+ s(x) \left( \frac{K}{D-2} +
\frac{2}{K} \left(\frac{1}{2 K} {\hat \nabla}^2 
\left( \frac{ K}{D-2} \right)  + \frac{ K^2}{2 (D-2)^2}  + 
\frac{K_{MN} K^{\mu\nu}}{ K} \right) +{\cal O} \left( \frac{1}{D^2} \right) \right)
+ \\ & s(x^\mu)^2 \left(\frac{1}{2 K} {\hat \nabla}^2 
\left( \frac{ K}{D-2} \right) + \frac{ K^2}{2 (D-2)^2}  + 
\frac{K_{\mu\nu} K^{\mu\nu}}{ K}  + {\cal O}\left(\frac{1}{D} \right) \right) + {\cal O} \left(s^3 \right),
\end{split} 
\end{equation}
where all intrinsic membrane quantities (like $K$, $K_{\mu\nu}$ etc) are evaluated at the membrane point 
${\hat x}(x)$. The quantity ${\hat \nabla}$ 
represents the covariant derivative along the world volume of the membrane. 
\footnote{The structure of the equations we encounter in evaluating the function 
$\rho(x^M)$ in the large $D$ expansion is as follows. At leading order in perturbation theory 
we are able to obtain the ${\cal O}(1)$ part of the coefficient of $s$. At next leading order 
we find the ${\cal O}(1/D)$ piece in the coefficient of $s$ together with the ${\cal O}(1)$ 
part of the coefficient of $s^2$. At third order we would find the  ${\cal O}(1/D)^2$ contribution 
to the coefficient of $s$, the  ${\cal O}(1/D)$ contribution to the coefficient of $s^2$ and the 
${\cal O}(1)$ part of the coefficient of $s^3$, and so on. In other words if we 
specialize to the case that $s(x^\mu)$ is of order $1/D$ then our perturbative expansion 
evaluates $\rho$ in an expansion in $\frac{1}{D}$. In  
\eqref{rhodef} have reported the result of our expansion upto second order. In the special 
case that $s \sim {\cal O}(1/D)$ we have  
\begin{equation}\label{rhodefn} \begin{split}
& \rho(x^\mu) -1=
 s(x^\mu) \frac{K(\hat{r}^\mu)}{D-2} +  \\
& \left( \frac{2 s(x^\mu)}{K} + s(x^\mu)^2 \right) 
\left(\frac{1}{2 K} {\hat \nabla}^2 
\left( \frac{ K}{D-2} \right)  + \frac{ K^2}{2 (D-2)^2}  + 
\frac{K_{MN} K^{MN}}{ K} \right)\\
&+  {\cal O}
\left( \frac{1}{(D-2)^3} \right),
\end{split} 
\end{equation}
where we have arranged terms so that the first and second lines in this  \eqref{rhodefn} are respectively of order $1/D$ and $1/D^2$ }

Later in this subsection will need to take derivatives of the function $\rho$. As 
we have expressed $\rho$ as a function of $s$, it is useful to first compute relevant derivatives
of the function $s$. It is possible to verify that  
\begin{equation} \begin{split} \label{spt}
&\partial_M s= n_M, \\
&\Box s = K + s K_{MN} K^{MN}, \\
&+ {\cal O}(1/D) + s \times {\cal O}(1) + s^2 {\cal O}(D),\\
\end{split} 
\end{equation}
where $n_M$ is the vector $\partial_M \rho$ rescaled to have unit norm. 
\footnote{The second equation in \eqref{spt} may be understood as follows. 
As $\partial_\mu s= n_\mu$, it follows that $\Box s$ equals $K$
of the constant $\rho$ slice at that point. To the 
appropriate order in $1/D$, $K(x^M)$ can be re-expressed in terms 
of curvature invariants at the corresponding ${\hat x}$ point, 
yielding the second equation of \eqref{spt}} 
Using these results it may be verified that 
\begin{equation} \label{rhonorm}
\begin{split}
&N^2\equiv |\partial \rho|^2 \equiv \partial_M \rho \partial^M \rho = 
\left(\frac{K}{D-2} \right)^2\\
&+ \frac{4}{D-2} 
\left( 1 + K s \right) 
\left( \frac{2}{K} \left(\frac{1}{2 K} {\hat \nabla}^2 
\left( \frac{ K}{D-2} \right)  + \frac{ K^2}{2 (D-2)^2}  + 
\frac{K_{MN} K^{MN}}{ K} \right) \right)\\
&+{\cal O}(1/D^2)  + s \times {\cal O}(1/D) + s^2 \times {\cal O}(1). \\
\end{split}
\end{equation}

\subsection{Membrane solutions of the minimally coupled scalar}

In this subsection we will construct the solution \eqref{unsols} (see the previous section)  
both for $\rho>1$ and $\rho<1$.  We obtain 
our solution in a Taylor series expansion in $\rho-1$. The coefficients in this expansion are
obtained in a power series expansion in  $\frac{1}{D}$.
\footnote{As in the previous subsection, at leading order in our expansion we find the coefficient of the constant term in the Taylor series 
expansion at order unity in the expansion in $\frac{1}{D}$. At next order we find the ${\cal O}(1/D)$ 
correction to this constant together with the order unity (i.e leading) contribution to the coefficient 
of $(\rho-1)$. We stop our expansion at this point. 
Had we gone to one higher order in the perturbative expansion we would have obtained the ${\cal O}(1/D^2)$
correction to the constant, the ${\cal O}(1/D)$ correction to the coefficient of $\rho-1$ and the 
order unity correction to the coefficient of $(\rho -1)^2$. In other words our expansion reduces 
to an honest expansion in $\left(\frac{1}{D}\right)$ provided $(\rho-1)$ is of order $\left(\frac{1}{D}\right)$ .}

Recall that the solution \eqref{unsols} is labeled by the value 
$\phi_0({\hat x})$ of the scalar field on the membrane. In the special case that $\phi_0({\hat x})$ is a constant $\alpha$, it follows immediately that the solution of interest is given by $\phi_a=\frac{\alpha}{\rho^{D-3}}$ (for $\rho >1$) and 
$\phi=\alpha$ (for $\rho<1$). Note that in the exterior region $\phi$ varies on the length scale $1/D$ in the direction normal to the membrane. If 
$\phi_0({\hat x})$ is a function that varies on length scale
unity, the relative slowness of this variation suggests the following.  
Let $\alpha(x)$ in \eqref{phsol} 
be any smooth extension of the membrane function $\phi_0({{\hat x}})$ into the bulk. Then
\begin{equation}\label{phsol}\begin{split}
\phi_a(x)&= \frac{\alpha(x)}{\rho^{D-3}} ~~~(\rho \geq 1),\\
\phi_a(x)&=  \alpha(x)~~~(\rho \leq 1),
\end{split}
\end{equation}
\footnote{The subscript $a$ in $\phi_a$ stands for `ansatz'; $a$ is not a spacetime vector index.}
also solves the minimally coupled scalar equations; not exactly (as was the case when 
$\alpha$ was constant), but at leading order in the expansion in $\frac{1}{D}$. We will 
check below that this expectation is indeed correct. 

In order to proceed with our computation we need to make a particular choice 
for the extension of the membrane valued function $\phi_0({\hat x})$ to the 
bulk function $\alpha(x)$. In the rest of this section we choose, arbitrarily, to extend the function 
$\phi_0({{\hat x}})$ into the bulk in such a way that it obeys the 
`subsidiary condition' 
\begin{equation}\label{auxcond}
d\rho\cdot d \alpha=0.
\end{equation}
This requirement together with the condition that $\alpha(x)$ agrees with $\phi_0({\hat x})$ on the membrane. 
\footnote{The subsidiary condition \eqref{auxcond} is simply one convenient way of extending 
$\alpha$ away from the membrane surface in a smooth, $D$ independent way. The auxiliary 
condition \eqref{auxcond} is convenient but essentially arbitrary. We could, for example, 
also have used the condition $\alpha(x^\mu)= \alpha({\hat x}^\mu( x^\mu))$. This condition 
would also have served our purposes in principle but proves less convenient for actually 
solving the problem in practice.}
completely determines the bulk field in terms of the membrane valued field $\alpha(x)$. 

$\phi_a(x)$ in \eqref{phsol} is a function of order unity which varies on length scale $\left(\frac{1}{D}\right)$. 
We would thus expect that the action of $\Box$ on a configuration of this sort should 
yield an expression of order ${\cal O}(D^2)$. Using \eqref{rhof}, however, it is easily verified that 
\begin{equation}\label{phsole}\begin{split}
\Box \phi_a(x)&= \frac{\Box \alpha(x)}{\rho^{D-3}} ~~~(\rho \geq 1),\\
\Box \phi_a(x)&=  \Box\alpha(x)~~~(\rho \leq 1).
\end{split}
\end{equation}
Recall from the introduction that even though the function $\alpha$ varies over length scale unity, 
$\Box \alpha$ is generically of order ${\cal O}(D)$. It follows that the ansatz \eqref{phsol} satisfies
the minimally coupled scalar equation at order $D^2$ - the order at which we might at first expect this 
equation to be violated,

\subsubsection{Systematic procedure to correct the ansatz $\phi_a$}

In order to proceed, we search for a systematic correction of 
\eqref{phsol}. The corrections should have the property that they are 
subleading compared to $\phi_a(x)$ presented above when $(\rho-1)$ is of order 
${\cal O}\left(\frac{1}{D}\right)$, and also that they are capable of canceling the RHS of \eqref{phsole}. 
An ansatz that obviously satisfies the first criterion and turns out to
satisfy the second is 
\begin{equation}\label{phsoled}\begin{split}
\phi(x)&= \frac{\sum_{n=0}^\infty \alpha_n(x) (\rho-1)^n}{\rho^{D-3}} ~~~(\rho \geq 1),\\
\phi(x)&=  \sum_{n=0}^\infty \beta_n(x)(\rho-1)^n~~~(\rho \leq 1),\\
\alpha_0(x)&=\beta_0(x)= \alpha(x),\\
\end{split}
\end{equation}
with
\begin{equation}\label{auxcondb}
n\cdot\partial \alpha_n= n\cdot\partial \beta_n=0.
\end{equation}

Assuming the expansion \eqref{phsoled} and focusing on the region $\rho>1$, a straightforward
algebraic exercise demonstrates that 
\begin{equation}\label{phsoledd}
\begin{split}
\Box \phi(x)&= \sum_{n=1}^\infty A_n \frac{(\rho-1)^n}{\rho^{D-3}} ,\\
A_n&=\left( \Box \alpha_n + \left( (n+1)(D-2) -2(D-3) \right) 
\frac{(d\rho\cdot.d\rho)~  \alpha_{n+1}}{\rho} + (n+2)(n+1) (d\rho\cdot.d\rho)
~\alpha_{n+2} \right) .\\
\end{split}
\end{equation}
\footnote{We have used the fact that 
\begin{equation}\label{rhoharm} 
(D-2) \partial_\mu \rho \partial^\mu \rho= \rho \Box \rho,
\end{equation} 
(this is an expansion of the equation $\Box \frac{1}{\rho^{D-3}}=0$) to 
simplify the RHS of \eqref{phsoledd}.}

When $\rho-1<1$, on the other hand, we find  
\begin{equation}\label{phsolf}
\begin{split}
\Box \phi(x)&= \sum_{n=1}^\infty B_n (\rho-1)^n, \\
B_n&=\left( \Box \alpha_n + \left( (n+1)(D-2) \right) 
\frac{d\rho.d\rho  \alpha_{n+1}}{\rho} + (n+2)(n+1) (d\rho\cdot.d\rho)
\alpha_{n+2} \right). \\
\end{split}
\end{equation}

The coefficients $A_n$ and $B_n$ in the expansion above can themselves be 
expanded in a power series in $(\rho-1)$. Let 
\begin{equation}\label{abexp} \begin{split}
A_n&= \sum_m A_n^m (\rho-1)^m,\\
B_n&= \sum_m B_n^m (\rho-1)^m,\\
\end{split}
\end{equation}
where 
\begin{equation}\label{anbn}
n.\nabla A_n^m= n.\nabla B_n^m=0.
\end{equation} 
The equations \eqref{abexp} and \eqref{anbn} define the expansion functions 
$A_n^m$ and $B_n^m$. The expressions for $\Box \phi$ can be rewritten in terms of 
these expansion coefficients as 
\begin{equation}\label{phsolg}
\begin{split}
\Box \phi(x)&= \sum_{n=1}^\infty {\tilde A}_n \frac{(\rho-1)^n}{\rho^{D-3}}, 
~~~(\rho>1) \\
\Box \phi(x)&= \sum_{n=1}^\infty {\tilde B}_n (\rho-1)^n, ~~~(\rho<1)\\
{\tilde A}_n&= \sum_{m=0}^n A^m_{n-m},\\
{\tilde B}_n&= \sum_{m=0}^n B^m_{n-m},\\
&n\cdot\partial {\tilde A}_n= n\cdot\partial {\tilde B}_n=0.\\
\end{split}
\end{equation}
The condition that $\phi$ is harmonic then simply reduces to the condition  
${\tilde A}_n={\tilde B}_n=0$. We will now demonstrate that these equations 
are very easily solved in a power series expansion in $1/D$. 

\subsubsection{Explicit solution at low orders for $\rho>1$}

In this subsection we construct the functional $F_1$ defined in \eqref{formsolns}.

Let us consider the special case $n=0$. ${\tilde A}_0=0$ implies that 
$A_0=0$ i.e. that 
$$\Box \alpha_0 -(D-4)   
\frac{d\rho\cdot d\rho  }{\rho}\alpha_{1} + 2(d \rho\cdot d \rho)~
\alpha_{2}=0.$$
This equation is practically solvable in the large $D$ limit because 
the term proportional to $\alpha_2$ is subleading at large $D$ compared 
to the other terms in this equation. Ignoring this term in the equation we 
obtain the equation 
\begin{equation}\label{defboo}
\alpha_1= \frac{\rho \Box \alpha_0}{(D-4)(d\rho\cdot d\rho)}.
\end{equation}
More precisely $\alpha_1$ is given by \eqref{defboo} on the membrane and determined elsewhere 
by subsidiary conditions $n\cdot\partial \alpha_1=0$. \footnote{ To see why this is so recall that 
\eqref{defboo} was obtained by equating the coefficient
of $(\rho-1)^0$ in \eqref{phsolg} to zero. Clearly \eqref{defboo} is not 
the unique solution to this condition; if we add $(\rho-1) G$ to the 
solution for $\alpha_1$ presented in \eqref{defboo} the coefficient of $(\rho-1)^0$ in \eqref{phsolg} 
continues to vanish. In other words \eqref{defboo} is too strong; the correct statement is  
\begin{equation}\label{defboor}
\alpha_1= \frac{\rho \Box \alpha_0}{(D-4)(d\rho\cdot d\rho)} +{\cal O}(\rho-1).
\end{equation}
The ambiguity of extending $\alpha_1$ off the membrane is then resolved by the condition 
$n.\nabla \alpha_1=0$. } 

 At any event we are most interested in $\alpha_1$  evaluated on  
membrane surface. The solution we have presented for $\alpha_1$ on the membrane is given in terms of the spacetime d'Alembertian of $\alpha$. 
This result may be reworded in terms of the membrane d'Alembertian acting on 
the membrane valued function $\phi_0$ using  
\begin{equation}\label{relbetlap}
\Box \alpha = \overset{\widetilde{\phantom{A}}}{\smash{\Box}}(\phi_0) - \frac{{\hat \nabla} K\cdot {\hat \nabla} \phi_0 }{K} + {\cal O}\left(\frac{1}{D}\right),
\end{equation} 
(here $\Box$ in \eqref{defboo} is the full spacetime d'Alembertian operator, 
 $\overset{\widetilde{\phantom{A}}}{\smash{\Box}}$ is the d'Alembertian on the membrane world volume and \eqref{relbetlap} is 
 derived using the subsidiary condition $n\cdot\partial \alpha=0$). The dot product in the 
last term on the RHS of \eqref{relbetlap} is taken in the membrane world volume metric $\Pi_{MN}= \eta_{MN}-n_M n_N$. Note that the second term on the RHS of 
\eqref{relbetlap} is of order unity in the $1/D$ expansion, and so 
is subleading compared to the first term in that equation. On the membrane (i.e. on the 
surface $\rho=1$ and at leading order 
$$\Box \alpha = {\overset{\widetilde{\phantom{A}}}{\smash{\Box}}} \phi_0.$$

Using \eqref{rhonorm} it then 
follows that on the membrane surface $\rho=1$  
\begin{equation} \label{scefl}
\begin{split}
n\cdot\partial \phi = &\left[\frac{ K}{D-2}\right]
\left[- (D-3) \alpha +\left( \frac{D}{K^2} \right)\Box \alpha_0(x^\mu) \right]\\
=& - K\alpha_0 \left(1-\frac{1}{D} \right) + \frac{\Box \alpha_0(x^\mu)}{K} 
+ {\cal O}\left(\frac{1}{D}\right)\\
=&- K\alpha \left(1-\frac{1}{D} \right) + \frac{\overset{\widetilde{\phantom{A}}}{\smash{\Box}}(\phi_0)}{K}
+ {\cal O}\left(\frac{1}{D}\right).\\
\end{split}
\end{equation}
Recall from the introduction that $\Box \alpha_0$ and $K$ are both of order $D$. The RHS 
of \eqref{scefl} has terms of order $D$ and order unity. 

The procedure outlined here can be generalized to all orders. 
The equation ${\tilde A}_1=0$ will now allow us to determine 
$\alpha_2$ to leading order. Plugging this result into the equation 
${\tilde A}_0=0$ then allows us to determine the first subleading correction
to $\alpha_1$ in the $1/D$ expansion. In a similar manner the equation 
${\tilde A}_2=0$
allows us to determine $\alpha_3$ to leading order; which in turn permits
the determination of $\alpha_2$ to first subleading and $\alpha_1$ to 
second subleading order in $1/D$, and so on.

\subsubsection{Explicit solution at low orders when $\rho<1$}

In this subsection we construct the functional $F_2$ defined in \eqref{formsolns} at lowest 
nontrivial order. In order to do this we focus on the special case $n=0$. ${\tilde B}_0=0$ implies that 
$B_0=0$ i.e. that 
$$\Box \beta_0 + (D-2)   
\frac{\left(d\rho\cdot d\rho\right)  \beta_{1}}{\rho} + (n+2)(n+1) \left(d \rho\cdot d \rho \right)
\beta_{2}=0.$$
Once again the term proportional to $\beta_2$ is subleading at large 
$D$ compared 
to the other terms in this equation. It follows that 
\begin{equation}\label{defboi}
\beta_1= -\frac{\rho \Box \alpha_0}{(D-2)(d\rho\cdot d\rho)}.
\end{equation}
Once again \eqref{defboi} is reliable only on the membrane; $\beta_1$ is 
extended off the membrane using the condition $n\cdot\partial \beta_1=0$.

We are particularly interested in this coefficient evaluated on the 
membrane surface. Using \eqref{relbetlap} it follows that on the membrane
\begin{equation} \label{scefli}
\begin{split}
n\cdot\partial \phi \vert_{\rho=1}
=& -\frac{\Box \alpha_0(x^\mu)}{K} +{\cal O}\left(\frac{1}{D}\right)\\
=&-\left(\frac{1}{D}\right)\left[ \overset{\widetilde{\phantom{A}}}{\smash{\Box}}(\phi_0) - \frac{{\hat \nabla} K\cdot {\hat \nabla} \phi_0 }{K}\right]
+ {\cal O}\left(\frac{1}{D}\right)\\
=& - {\hat \nabla}_\mu \left( \frac{{\hat \nabla}^\mu \phi_0}{{K}} \right) + {\cal O}\left(\frac{1}{D}\right).\\
\end{split}
\end{equation}

According to \eqref{mls}, the contribution of the internal solution to the current on 
the membrane is given by the spacetime source
\begin{equation}\label{stsource}
\begin{split}
{\mathcal S}&= \left(\sqrt{ d \rho\cdot d\rho} \right)\delta( \rho -1) (n\cdot\partial \phi_{in}) \\
&=-\left(\sqrt{ d \rho\cdot d\rho} \right)\delta( \rho -1)\hat \nabla_\mu \left( \frac{\hat\nabla^\mu \phi}{{K}} \right) .
\end{split}
\end{equation}
This current can be derived from the variation of the action for the internal 
solution w.r.t. $\phi_0$ using the equation \eqref{inaction} once we identify 
\begin{equation}\label{sintph}
S_{in} = \frac{1}{2} \int \frac{ (\hat\nabla \phi_0)^2}{K},
\end{equation}
\eqref{sintph} can also be obtained from \eqref{formsin} using 
\eqref{scefli}.

\subsubsection{Current}

Using the results of the previous two subsubsections it is easily 
verified that 
$$n\cdot\partial \phi|_{out}- n\cdot\partial \phi|_{in}=
- K\alpha_0(x^\mu) \left(1-\frac{1}{D} \right) +\left( \frac{2}{K} \right)\Box \alpha_0(x^\mu).$$
In other words our field $\phi$ obeys the equation \eqref{mls} (which we repeat here for convenience)
\begin{equation} \label{source}
\Box \phi= \left[ \left(\sqrt{ d\rho\cdot  d \rho}\right) \delta( \rho-1) \right]{\cal J},
\end{equation} 
\begin{equation}\label{jform}
{\cal J}= 
 - K\phi_0 \left(1-\frac{1}{D} \right) + 
\frac{2}{K} \left(\overset{\widetilde{\phantom{A}}}{\smash{\Box}}(\phi_0) - \frac{{\hat \nabla} K\cdot {\hat \nabla} \phi_0 }{K} \right) .
\end{equation}


\subsection{Membrane solutions of the Maxwell Equations}

We will now imitate the analysis of the previous subsection to demonstrate that 
the most general solution of the Maxwell equations is parametrized by a conserved current living on the membrane, and explicitly construct the solution 
generated by any particular current. 

\subsubsection{$\rho>1$}\label{linearout}

In this subsubsection we find the solution $F_1[A_0]$ (see \eqref{formsolnv}). We will 
find it convenient to slightly change notation as compared to the previous section; in particular 
the data for our solution - referred to as $(A_0)_\mu$ in the previous section will be taken to be 
$G^{(0)}_M$ below. As we explain in detail below, $G^{(0)}_M$ is a bulk spacetime gauge field whose 
restriction onto the membrane equals $(A_0)_\mu$ of the previous section.

Following  previous subsections we assume that the gauge 
field ${\cal A}_A$ can 
be expanded outside the membrane as
\begin{equation}\label{out1}
\begin{split}
{\cal A}_A &= \rho^{-(D-3)} G_A,\\
G_A& =\sum_{k=0}^\infty (\rho -1)^k G_A^{(k)},
\end{split}
\end{equation}
where each of $G_A^{(k)}$ admits further expansion in $\left(\frac{1}{D}\right)$.

As in the previous subsection, the leading term $G^{(0)}_B$ in this expansion will turn out to 
be the data of our solution (which we will later be able to trade for a conserved current). 
Below we will outline 
the procedure that  determines all the remaining coefficient functions in 
terms of $G_A^{(0)}$.

In order to set up the problem we work in the gauge 
${\cal A}_A n^A=0$. Of course this is simply a convenient device; the gauge 
invariant content in our expansion lies in the field strengths. 
 This 
particular gauge is convenient as our problem has a special oneform - $\partial \rho$ - 
at each point in spacetime. By using this oneform to fix gauge we obtain a parametrization that 
keeps all the symmetries of the physical problem manifest. 

Our gauge condition implies
\begin{equation}\label{outgauge}
n^AG_A=0,
\end{equation}
\begin{equation}\label{outgaugec}
n^A G_A ^{(k)} =0 ,~~\text{for every}~k
\end{equation}
where $n_A$ is the unit normal to the $\rho = constant$ surfaces, 
defined by 
$$\partial_A \rho = N ~n_A,~~~N = \sqrt{(\partial_A\rho )(\partial^A\rho)},$$
(recall that $N$ was evaluated in \eqref{rhonorm} and equals $\frac{{K}}{D-2}$ 
to leading order).  
As in the previous subsection, we impose a subsidiary condition on the coefficient functions $G_A$
to give our expansion meaning. The condition we impose is 
\begin{equation} \label{subsidonGa}
\Pi_C^A (n.{\partial}){G_A ^{^{(k)}}} = 0,~~\text{for every}~k
\end{equation}
where 
$$\Pi_{AB} = \eta_{AB} - n_A n_B.$$ 

From \eqref{outgauge} it follows that 
\begin{equation}\label{evn}
n^A (n.{\partial})G_A = -G_A \big[(n.{\partial})n^A \big].
\end{equation}
Similarly from \eqref{outgaugec} it follows that 
\begin{equation}\label{evnc}
n^A (n.{\partial})G_A^{(k)} = -G_A^{(k)} \big[(n.{\partial})n^A \big],
\end{equation}
(the last two equations are consistent because of \eqref{outgaugec})\footnote{Here all lowering, raising and contraction of indices have been done using the flat metric $\eta_{AB}$.}.

Our discussion above has been presented in a particular gauge. However the functions
$G_A$ actually have a simple gauge invariant significance as we now explain. Note that 
\begin{eqnarray}\label{fieldstrength}
F_{AB} &=& \partial _A (\rho^{-(D-3)} G_B) - \partial _B (\rho^{-(D-3)} G_A) \nonumber \\
&=& ({\partial}_A \rho^{-(D-3)})G_B - ({\partial}_B \rho^{-(D-3)})G_A + \rho^{-(D-3)} ({\partial}_A G_B - {\partial}_B G_A). \nonumber
\end{eqnarray}
Now using
\begin{eqnarray}
-n_A \partial_B G^A &=& -\partial_B ( n_A G^A) + (\partial_B n_A)G^A \nonumber \\
&=& \eta_B^C (\partial_C n_A) G^A \nonumber \\
&=& \left( \Pi_B^C + n^C n_B  \right) (\partial_C n_A) G^A \nonumber \\
&=& K_{BA} G^A + n_B G^A {(n.\partial)n_A} \nonumber \\
&=& K^A _B G_A -{n_B n^A  (n.\partial) G_A}, 
\end{eqnarray}
where the projector $\Pi_{AB}= \eta_{AB} - n_A n_B$.\\ 
It follows that 
\begin{equation}\begin{split} \label{nfs}
& n_A {F^A}_B =\frac{-N(D-3) G_B}{\rho^{D-2}} + \frac{1}{\rho^{D-3}}  \left[(n.\partial)G_B - n_A \partial_B G^A \right]  \\
&=  \frac{-(D-3) N G_B}{\rho^{D-2}}  +  \frac{1}{\rho^{D-3}}  K^A _B G_A .\\ 
\end{split}
\end{equation}
Here in the last line we have used the subsidiary condition \eqref{subsidonGa}.\\
Moreover 
\begin{equation}\begin{split} \label{fst}
& \Pi_A^{ A'} F_{A'B'} \Pi_{B}^{B'} =  \left(\frac{1}{ \rho^{D-3}} \right) \Pi_{ A}^{A'} \left( \partial_{A'} G_{B'}-\partial_{B'} G_{A'}    \right)  \Pi_{B}^{B'}.
\end{split}
\end{equation}
Equations  \eqref{fieldstrength} (and in particular \eqref{nfs} and \eqref{fst}) are presentations of the gauge invariant significance of the functions $G_A$. 

We now proceed to use the Maxwell equations to determine $G_A^{k}$  (for $k \geq 1$) in terms of $G_A^{(0)}$. Our analysis proceeds in analogy with that of the 
previous subsection (scalar field) with one crucial difference. While there 
are $(D-1)$ unknown functions $G_A$ we have $D$ Maxwell equations. In order
to solve for $G_A$ we will use only the $(D-1)$ dynamical Maxwell equations 
\eqref{dyndef}. \footnote{As we have explained in detail in the previous section, the 
remaining constraint equation \eqref{constdef} constrains the 
data $G_A^{(0)}$ (which we referred to as $(A_0)_\mu$ in the previous section) 
that parametrizes general solutions of the Maxwell equation.}
In Appendix \ref{Maxeq} we have presented all the algebraic details of our 
computation of $G_A$. Here we simply 
present our results. 

Let us define 
\begin{equation}\label{auxfs}
F_{AB}^{(m)} = \partial_A G_B^{(m)} - \partial_B G_A^{(m)}.
\end{equation}
At first subleading order in $\frac{1}{D}$ we find 

\begin{equation}\label{10mt}
\begin{split}
G_B^{(1)} &= \frac{\Pi_B^C~\partial^A F_{AC}^{(0)}}{2(D-3) N^2 -N{K}}+ {\cal O}\left(\frac{1}{D}\right)\\\\
&=\left(\frac{\Pi_B^C~\partial^A F_{AC}^{(0)}}{{N K}}\right) + {\cal O}\left(\frac{1}{D}\right).\\
\end{split}
\end{equation}
Here, in the second line, we have used the fact that ${K} = D N + {\cal O}(1)$.\\
Note that  $\Pi_B^C~\partial^A F_{AC}^{(0)} $ could be re-expressed completely in terms of quantities and covariant derivatives that are defined only along the membrane.
\begin{equation}\label{membmid}
\begin{split}
\Pi_B^C~\partial^A F_{AC}^{(0)}& =\Pi_B^C\partial^A\left[n_A n^{A'} F^{(0)}_{A'C} - n_C n^{A'} F^{(0)}_{A'A} + \Pi^{A'}_A F^{(0)}_{A'C'} \Pi^{C'}_C\right]\\
& ={K}~n^{A}F^{(0)}_{AC}\Pi_B^C +\Pi_B^C \partial^A\left[\Pi^{A'}_A F^{(0)}_{A'C'} \Pi^{C'}_C\right]+{\cal O}\left(1\right)\\
& ={K}~n^{A}(\partial_C G^{(0)}_{A})\Pi_B^C +\Pi_B^C \partial^A\left[\Pi^{A'}_A F^{(0)}_{A'C'} \Pi^{C'}_C\right]+{\cal O}\left(1\right)\\
& ={K}~K^A_BG^{(0)}_{A} +\Pi_B^C \partial^A\left[\Pi^{A'}_A F^{(0)}_{A'C'} \Pi^{C'}_C\right]+{\cal O}\left(1\right).\\
\end{split}
\end{equation}
In \eqref{membmid} all free indices  are projected on the membrane and also all contracted indices and derivatives run along the membrane directions only. Similarly, because of our gauge condition, $G_A^{(k)}$  for every value of $k$ could also be considered as a vector 
field ($G_{\mu}^{(k)}$) defined only along the membrane. Therefore It follows that $G_\mu^{(1)}$ - the first Taylor coefficient 
in the expansion of the gauge field off the membrane but viewed  as a vector field along the membrane - can be rewritten 
entirely in terms of intrinsic quantities on the membrane as
\begin{equation}\label{10mti}
\begin{split}
G_{\mu}^{(1)}=\left(\frac{1}{N}\right)\left( K^\nu_\mu G^{(0)}_{\nu} + \frac{\hat\nabla^\nu\hat F_{\nu\mu}}{K}\right)+{\cal O}\left(\frac{1}{D}\right),
\end{split}
\end{equation} 
where $\hat F_{\mu\nu}$ is the field strength along the surface and $\hat\nabla_\mu$ is the covariant derivative on the membrane surface, with respect to the intrinsic metric of the membrane. Also all raising lowering and contraction of indices have been done using the intrinsic metric of the membrane as embedded in flat space.

Restricting attention to the surface $\rho=1$ we have in particular 
\begin{equation}\label{outcurrentexact}
n_A {F^A}_B|_{\rho=1}=J^{(out)}_B = -(D-3) N G^{(0)}_B + N G_B^{(1)} + K_B^A G_A^{(0)}.
\end{equation}
Using the same argument as given above and substituting equations \eqref{10mti}  in equation \eqref{outcurrentexact} we get the outside current as vector field along the membrane (upto first subleading order)
\begin{equation}\label{outcurrentsub}
J^{(out)}_\mu = -(D-3) N G^{(0)}_\mu + \frac{\hat\nabla^\nu\hat F_{\nu\mu}}{{K}} +2 K_\mu^\nu G_\nu^{(0)}+{\cal O}\left(\frac{1}{D}\right),
\end{equation}
where  $\hat F_{\mu\nu}$ is the field strength along the surface

 As explained in the previous section, the constraint Maxwell equation 
asserts that 
$$\hat\nabla_\mu J_{(out)}^\mu=0,$$
(where $\hat\nabla_\mu$ is the covariant derivative on the membrane surface) yielding
an effective constraint on the data $G^{(0)}_\mu$ of the solution.

\subsubsection{$\rho<1$}\label{linearin}

In this subsection we proceed to construct the functional $F_2$ defined in \eqref{formsolnv}. As in the
previous subsection, the data for this solution will be taken to be the spacetime 
gauge field ${\tilde G_A^{(0)}}$ whose restriction onto the membrane defines $A_0$ of the previous section.

In order to proceed with our computation we proceed assuming that the solution in the 
region  $\rho<1$ can be expanded as
\begin{equation}\label{in1}
\begin{split}
{\tilde G}_A&= \sum_{k=0}^\infty (\rho -1)^k \tilde G_A^{(k)}.
\end{split}
\end{equation}

In order that the gauge field is continuous across the membrane we will require that the restriction 
of $\tilde G^{(0)}_B$ to the surface $\rho=1$ agree with the restriction of  $G^{(0)}_B$ on the same 
surface. As in the previous subsection we will use 
Maxwell's equations to determine the higher order terms in the expansion of the gauge field in terms of $G^0_A$. As in the previous subsection we adopt the 
gauge  $ n^A {\tilde G}_A=0$ which implies that .
\begin{equation}\label{ingauge}
n^A \tilde G_A ^{(k)} =0,~~\text{for every}~k  .
\end{equation}
As in the previous subsubsection we also demand that 
$$ \Pi^C_B  n^A\partial_A \tilde G_C^{(k)} =0.$$
Again as in the previous subsubsection it follows that 
$$ (n.{\partial})\tilde G^{(k)}_A = -n_A~ \tilde G^{(k)}_B \big[(n\cdot{\partial})n^B \big].$$
The quantities $\tilde G_A^{(k)}$ have the following gauge invariant significance: 

\begin{equation} \begin{split}
\tilde F_{AB}&=\partial_A {\tilde G}_B-\partial_B {\tilde G}_A,\\
\tilde F_{AB} &= \sum_{k=0}^\infty k(\rho - 1 )^{k-1} N \left[ n_A \tilde G_{B}^{(k)} - n_B \tilde G_{A}^{(k)} \right] + \sum_{k=0}^\infty (\rho - 1 )^{k} \left[ \partial_A \tilde G_B^{(k)} - \partial_B\tilde G_A^{(k)} \right]. \\
\end{split}
\end{equation}

Solving the equation $(\partial_A \tilde F^{AB} =0)$ at first subleading order we find
(see Appendix \ref{Maxeq})

\begin{equation}\label{insolutionexact}
\begin{split}
\tilde G_B^{(1)} &= -\frac{\Pi_B^C~\partial^A F_{AC}^{(0)}}{N{K}}+ {\cal O}\left(\frac{1}{D}\right)\\
&= -\left(\frac{1}{N}\right)\left(K^A_BG^{(0)}_{A} +\frac{\Pi_B^C \partial^A\left[\Pi^{A'}_A F^{(0)}_{A'C'} \Pi^{C'}_C\right]}{K}\right)+ {\cal O}\left(\frac{1}{D}\right),
\end{split}
\end{equation}
where $$\tilde F_{AB}^{(m)} = \partial_A \tilde G_B^{(m)} - \partial_B \tilde G_A^{(m)},~~K_{AB} = \text{Extrinsic curvature}, ~~ {K}= \eta^{AB}K_{AB}.$$
In the last line we have used equation \eqref{membmid}.

As in previous subsection we could also express $G^{(1)}$ as a vector field defined intrinsically on the membrane
\begin{equation}\label{insolution2}
\begin{split}
\tilde G_\mu^{(1)} &=
 -\frac{\hat\nabla^\nu \hat F_{\nu\mu}}{N{K}}-\frac{K^\nu_\mu G^{(0)}_{\nu}}{N} + {\cal O}\left(\frac{1}{D}\right),\\
\end{split}
\end{equation}
where $\hat F_{\mu\nu}$ is the field strength along the surface and $\hat\nabla_\mu$ is the covariant derivative on the membrane surface, with respect to the intrinsic metric of the membrane. Also all raising lowering and contraction of indices have been done using the intrinsic metric of the membrane as embedded in flat space.

\eqref{insolution2} is our result for the first 
Taylor coefficient of the internal solution expressed entirely in terms 
of the gauge field $G^{(0)}_A$ restricted to the membrane (which we denote here as $G^{(0)}_\mu$). 

According to \eqref{maxo} and \eqref{currentmaxn} we have  
\begin{equation}\label{incurrentexact}
J^{(in)}_B =n^A \tilde F_{AB}\vert_{\rho=1}=   N\tilde G^{(1)}_B + K_B^A {{G}}_A^{(0)}.
\end{equation}
Substituting equation \eqref{insolutionexact} in equation \eqref{incurrentexact} to first subleading order we find
\begin{equation}\label{incurrentsub}
\begin{split}
J^{(in)}_B&=-\frac{\Pi_B^C \partial^A\left[\Pi^{A'}_A F^{(0)}_{A'C'} \Pi^{C'}_C\right]}{K} + {\cal O}\left(\frac{1}{D}\right)\\
&=-\Pi_B^C~ \Pi^{A''A}~\partial_A\left[\frac{\Pi^{A'}_{A''} F^{(0)}_{A'C'} \Pi^{C'}_C}{K}\right] + {\cal O}\left(\frac{1}{D}\right).
\end{split}
\end{equation}
It follows from \eqref{incurrentsub} and \eqref{cc} 
that the contribution of the internal 
solution to the 
current on the membrane is given by the spacetime source 
\begin{equation}\label{stsourcema}
\begin{split}
{\mathcal J}^{in}_B&=-\left(\sqrt{ \nabla \rho\cdot \nabla \rho}\right) \delta( \rho -1)J_B^{in}\\
&=\left(\sqrt{ \nabla \rho\cdot \nabla \rho}\right) \delta( \rho -1)\bigg[\Pi_B^C~ \Pi^{A''A}~\partial_A\left(\frac{\Pi^{A'}_{A''} F^{(0)}_{A'C'} \Pi^{C'}_C}{K}\right) + {\cal O}\left(\frac{1}{D}\right)\bigg].\\
\end{split}
\end{equation}
As before we could also view the current as a vector defined only along the membrane.
\begin{equation}\label{stsourcema2}
\begin{split}
J^{(in)}_\mu&=-\frac{\hat\nabla^\nu F_{\nu\mu}}{K} + {\cal O}\left(\frac{1}{D}\right).\\
\end{split}
\end{equation}
This current is consistent with \eqref{maxon} if we define
\begin{equation}\label{sintma}
S_{int} = -\frac{1}{4} \int \frac{F_{\mu\nu} F^{\mu\nu}}{K} ,
\end{equation}
where the integration is now taken only over the membrane world volume. 
It may be verified using \eqref{formsing} and \eqref{insolutionexact} 
that \eqref{sintma} is indeed the action of the interior solution. 
As explained in the previous section, the fact that the interior current 
is identically conserved follows immediately from the gauge invariance of 
the action \eqref{sintma}.

\subsubsection{Membrane Current}

Let us summarize. We have constructed the most general decaying solution to the 
linearized Maxwell equations in the exterior neighbourhood of a membrane 
surface. This solution is parametrized by one vector field $G^{(0)}_B$ 
on the membrane world volume, or equivalently a conserved current on the 
membrane world volume. The conserved current is given in terms of $G^{(0)}_B$ 
by the formula 
\begin{equation}\label{finalcurrent}
\begin{split}
J^B & = J^B_{(out)}  - J^B _{(in)}\\
&=\left[-(D-3) N G^{(0)}_B + N G_B^{(1)} + K_B^A G_A^{(0)}\right]-\left[ N\tilde G^{(1)}_B + K_B^A {{G}}_A^{(0)}\right]\\
&=-(D-3) N G^{(0)}_B + N\left[ G_B^{(1)} -\tilde G_B^{(1)} \right]\\
&=-(D-3) N G^{(0)}_B +\left(\frac{2~\Pi^C_B}{{K}}\right) \partial^A\left[\partial_A G_C^{(0)} -\partial_C G_A^{(0)}\right] + {\cal O}\left(\frac{1}{D}\right) + {\cal O}\left(\frac{1}{D}\right).
\end{split}
\end{equation}\\

Expressed as  current as a vector intrinsic to the membrane, we find
\begin{equation}\label{finalcurrent2}
\begin{split}
J^\mu & = J^\mu_{(out)}  - J^\mu _{(in)}\\
&=\left[-(D-3) N G^{(0)}_\mu + N G_\mu^{(1)} + K_B^A G_\mu^{(0)}\right]-\left[ N\tilde G^{(1)}_\mu + K_\mu^\nu {{G}}_\nu^{(0)}\right]\\
&=-(D-3) N G^{(0)}_\mu + N\left[ G_\mu^{(1)} -\tilde G_\mu^{(1)} \right]\\
&=-(D-3) N G^{(0)}_\mu + \frac{2\hat\nabla^\nu F_{\nu\mu}}{K}  + {\cal O}\left(\frac{1}{D}\right).
\end{split}
\end{equation}\\

\subsection{Membrane solutions of the linearized Einstein Equations}

In this subsection we will find the most general solution of the 
Einstein equation linearized around flat space-time
\begin{equation}\label{linein}
\begin{split}
g_{AB}&= \eta_{AB} + h_{AB},\\
R_{AB} &= \frac{1}{2} \left( {\partial _C}{\partial _A}h^C_B + {\partial _C}{\partial _B}h^C_A -\square h_{AB} - {\partial _A}{\partial _B}h^C_C \right) + {\cal O}(h^2) =0.
\end{split}
\end{equation}

As explained in the previous section we proceed by first solving the {\it dynamical} Einstein equations
\eqref{dyndefe} to determine the functionals $F_1$ and $F_2$ defined in \eqref{formsolng}. We construct 
these two functionals - to lowest nontrivial order - in the next two subsubsections. As in the previous 
subsection, in this subsubsection we find it convenient to use the bulk metrics 
$\eta_{MN} + h^{(0)}_{MN}$ and $\eta_{MN} + {\tilde h}^{(0)}_{MN}$ (see below) as the data in terms of which 
we write our solutions. The restrictions of these metrics to the membrane defines the intrinsic 
metric $g^{(ind)}_{\mu\nu}$ used as the data for the functionals $F_1$ and $F_2$ used in \eqref{formsolng}.

As explained in the previous section, once we have solved the dynamical equations, the constraint 
equation is automatic for the inner solution. For the outer solution it is simply the requirement 
that the Brown York stress tensor is conserved on the membrane approached from the outside. Below we will 
find explicit expressions for the Brown York Stress tensor on the membrane approached from both the 
outside and the inside in our solutions.

\subsubsection{$\rho>1$:}\label{einlinearout}
Let us first study the external region $\rho>1$. In analogy with previous subsections 
the solution in this region takes the form 
\begin{equation}\label{hexp}
{{h}}_{AB}= \left[ \rho^{-(D-3)}\sum_{m=0}^\infty (\rho -1)^m h_{AB}^{(m)}  \right].
\end{equation} 

As in the previous subsection we adopt a gauge condition adapted to the foliation 
of spacetime in slices of constant $\rho$ 
\begin{equation}\label{metoutgauge}
n^A h_{AB}^{(m)} = 0.
\end{equation}
As in the previous subsection we impose the subsidiary conditions 
\begin{equation}\label{outsubsidiary}
\Pi^{C'}_B \Pi^C _A~ (n.\partial) h_{CC'} ^{(m)} =0.
\end{equation}
on the expansion coefficients of \eqref{hexp}. These conditions together with the gauge conditions 
\eqref{outgauge} make \eqref{hexp} a well defined expansion of the metric function. 

As in the previous subsection the functions $h_{AB}^{(0)}$ may be thought of as the basic data of the 
solutions. The dynamical Einstein equations determine the higher order coefficients in \eqref{hexp}
in terms of $h_{AB}^{(0)}$. We present the details for how this works in Appendix \ref{Eineq}. 
To first order in the expansion in $(\rho-1)$ and at leading order in $(1/D)$ we find 
 \begin{equation}\label{h10out}
 \begin{split}
h_{AB}^{(1)}
 =& -\Pi^{C'}_B \Pi^C _A \left[ \frac{\partial _{C} \partial ^M h_{M{C'}}^{(0)} + \partial _{C'} \partial ^M h_{MC}^{(0)} - \square h_{CC'}^{(0)}  + (D-3) N h^{(0)}  K_{CC'}+\partial_C\partial_{C'} h^{(0)}}{2(D-3)N^2-N {K}} \right] \\
 &+{\cal O}\left(\frac{1}{D}\right)\\
 =& -\Pi^{C'}_B \Pi^C _A \left[ \frac{\partial _{C} \partial ^M h_{M{C'}}^{(0)} + \partial _{C'} \partial ^M h_{MC}^{(0)} - \square h_{CC'}^{(0)}  + (D-3) N h^{(0)}  K_{CC'}+\partial_C\partial_{C'} h^{(0)}}{NK} \right] \\
 &+{\cal O}\left(\frac{1}{D}\right),\\
\text{where}&~~ h^{(0)} = \eta^{AB}h^{(0)}_{AB}.
\end{split}
\end{equation}

As explained in the previous section, (see around \eqref{tinout}), the Einstein constraint equation is 
simply the condition that the Brown York stress tensor 
\begin{equation}\label{byst}
T^{(out)}_{AB}={\cal K}^{(out)}_{AB} - {\cal K}^{(out)}~ {\mathfrak p}_{AB},
\end{equation}
is conserved on the membrane, w.r.t the induced metric on the membrane. Here ${\cal K}_{AB}$ is the 
extrinsic curvature of the $\rho=1$ slice, 
${\cal K}$ is its trace, ${\mathfrak p}_{AB}$ is the projector on the $\rho=1$ slice. 

 At leading nontrivial order the stress 
tensor evaluated at $\rho=1$ turns out to be 
\begin{equation}\label{stressout}
\begin{split}
T^{AB}_{(out)}
=~& \left( \tilde K^{AB} -{\tilde K}\tilde \Pi^{AB} \right)  
+ \frac{N}{2} \left( {h}^{AB} _{(1)} -{h}^{(1)} \Pi^{AB}  \right)\\
& -\frac{N}{2} (D-3) \left(  h^{AB} _ {(0)}  - h^{(0)} \Pi^{AB}\right).
\end{split}
\end{equation}
Here  $\tilde K^{AB}$ and $\tilde \Pi^{AB} $ denote  the extrinsic curvature and the projector  respectively on the membrane embedded in the metric $\left[\eta_{AB} + h^{(0)}_{AB}\right]$ and $\tilde{K}$ is the trace of $\tilde K^{AB}$.
\begin{equation}\label{inthex1}
\begin{split}
&\tilde \Pi^{AB}= \eta^{AB} - n^A n^B - h_{(0)}^{AB},\\
&\tilde K^{AB} = K^{AB} -\frac{1}{2} \left[K_{AC}h_{(0)}^{CB} +K^B_{C}h_{(0)}^{CA}\right],\\
&\tilde{K} =\tilde\Pi_{AB} \tilde K^{AB} =\left[\Pi_{AB}+ h^{(0)}_{AB}\right]\tilde K^{AB}= K + {\cal O}(h^2),
\end{split}
\end{equation}
where $K_{AB}$ and $\Pi_{AB}$ are respectively the 
extrinsic curvature and projectors on the $\rho=1$ slice as embedded in flat Minkowski space-time, 
$K\equiv \eta^{AB}K_{AB}$. 

As explained in the previous section, this `internal' stress tensor can be rewritten more 
elegantly in terms of purely intrinsic geometrical quantities on the membrane (see \eqref{tabin}). 
The less elegant expression \eqref{stressin2} will, however,  prove practically more useful to us 
in the next subsection, as the cancellations with the outer stress tensor \eqref{stressout} are more
manifest in this form.

Plugging \eqref{h10out} into \eqref{stressout} yields an expression for the Brown York stress tensor
purely in terms of $h^{AB} _{(0)}$. The requirement that this stress tensor is conserved on the 
membrane yields an effective constraint on $h^{AB} _{(0)}$.
\footnote{
Note that the stress tensor \eqref{stressout} is non vanishing even when $h_{AB}=0$, i.e. when the 
spacetime metric is flat. The conservation of this zero order stress tensor w.r.t. the zero order metric (i.e. the induced metric on the surface 
$\rho=1$ viewed as a submanifold of the flat bulk spacetime with metric 
$\eta_{AB}$.)
on the membrane is a trivial identity. The conservation of \eqref{stressout}, when expanded to first order in $h_{AB}$ is 
nontrivial. If we expand the stress tensor \eqref{stressout} as $T_{AB}= T^0_{AB} + T^1_{AB}$ 
and the world volume metric on the membrane as 
$P_{AB}=P^{0}_{AB} + P^1_{AB}$ (where superscripts denote the order of expansion 
in $h_{AB}$) then the conservation equation, expanded to first order 
takes the schematic form 
\begin{equation}\label{schcons}
 (\nabla^1)^M T^0_{MN}+ (\nabla^0)^M T^1_{MN}=0
\end{equation}
(here we have expanded the covariant derivative as 
$\nabla= \nabla^0+ \nabla^1$; as above superscripts keep track of the order of 
$h_{AB}$ and have used the fact that $(\nabla^0)^M T^0_{MN}$ vanishes 
identically.)
Note that the equation \eqref{schcons} asserts that $T^1_{MN}$ is not quite 
a conserved stress tensor on the membrane. The lack of perfect conservation 
of $T^1_{MN}$ is a direct consequence of the nonvanishing of $T^0_{MN}$. }


\subsubsection{$\rho<1$:}\label{einlinearin}
As above, in the interior of the membrane we expand the metric as 
\begin{equation}\label{ein}
\begin{split}
\text{Bulk metric}=& g_{AB} = \eta_{AB} +\tilde h_{AB},\\
&\tilde h_{AB} = \left[ \sum_{m=0}^\infty (\rho-1)^m\tilde  h_{AB}^{(m)}  \right].
\end{split}
\end{equation}
As above we use the gauge condition 
\begin{equation}\label{metingauge}
n^A \tilde h_{AB}^{(m)} = 0.
\end{equation}
As above we require the coefficients of the expansion \eqref{ein} to obey the additional subsidiary 
constraints 
\begin{equation}\label{insubsidiary}
\Pi^{C'}_B \Pi^C _A ~(n.\partial) \tilde h_{CC'} ^{(m)} =0.
\end{equation}
As above $h_{AB}^0$ may be regarded as data of the solutions. The dynamical Einstein equations 
determine all other terms in the expansion in terms of data. At leading order we find (see Appendix 
\ref{Eineq} for details) 
 \begin{equation}\label{h10in}
\tilde h_{AB}^{(1)} = \Pi^{C'}_B \Pi^C _A \left[ \frac{\partial _{\bar{C}} \partial ^M h_{M{C'}}^{(0)} + \partial _{C'} \partial ^M h_{MC}^{(0)} - \square h_{CC'}^{(0)}-\partial_C\partial_{C'} h^{(0)} }{N {K}} \right] +{\cal O}\left(\frac{1}{D}\right).
\end{equation}

As explained in the previous section, the momentum constraint equations in the interior of the 
membrane assert the conservation of the stress tensor 
\begin{equation}\label{stressin}
\begin{split}
8 \pi T^{AB}_{(in)}
=~& {\cal K}_{(in)}^{AB} - {\cal K}_{(in)} ~{\mathfrak p}_{(in)}^{AB} ,
\end{split}
\end{equation}
where ${\cal K}_{(in)}^{AB}$ and ${\mathfrak p}_{(in)}^{AB} $ are  the extrinsic curvature and the projector on the membrane embedded in the metric $\left[\eta_{AB} + \tilde h_{AB}\right]$ . $ {\cal K}_{(in)} $ is the trace of ${\cal K}_{(in)}^{AB}$. Using the expansion equation \eqref{ein} we find 
\begin{equation}\label{stressin2}
\begin{split}
T^{AB}_{(in)}
=~& \left( \tilde K^{AB} -\tilde {K}\tilde \Pi^{AB} \right) 
+ \frac{N}{2} \left( \tilde{h}^{AB} _{(1)} -\tilde{h}^{(1)} \Pi^{AB}  \right).
\end{split}
\end{equation}
As described before, here  $\tilde K^{AB}$ and $\tilde \Pi^{AB} $ denote  the extrinsic curvature and the projector  respectively on the membrane embedded in the metric $\left[\eta_{AB} + h^{(0)}_{AB}\right]$ and $\tilde{K}$ is the trace of $\tilde K^{AB}$.
\begin{equation}\label{inthex}
\begin{split}
&\tilde \Pi^{AB}= \eta^{AB} - n^A n^B - h_{(0)}^{AB},\\
&\tilde K^{AB} = K^{AB} -\frac{1}{2} \left[K_{AC}h_{(0)}^{CB} +K^B_{C}h_{(0)}^{CA}\right],\\
&\tilde{K} =\tilde\Pi_{AB} \tilde K^{AB} =\left[\Pi_{AB}+ h^{(0)}_{AB}\right]\tilde K^{AB}= K + {\cal O}(h^2).
\end{split}
\end{equation}

\subsubsection{The conserved membrane stress tensor}

The full membrane stress tensor is given by 
\begin{equation}\label{stressfinal}
\begin{split}
8 \pi T^{AB}  =~& -\left( T^{AB}_{(out)}-T^{AB}_{(in)}\right)\\
=~& \frac{N}{2} (D-3) \left(  h^{AB} _ {(0)}  - h^{(0)} \Pi^{AB}\right)- \frac{N}{2} \left[h^{AB}_{(1)}-\tilde{h}^{AB} _{(1)} -(h^{(1)}-\tilde{h}^{(1)} )\Pi^{AB}  \right].
\end{split}
\end{equation}
Now from equation \eqref{h10out} and \eqref{h10in} it follows that 
 $$ \tilde h^{(1)}_{AB} =-h^{(1)}_{AB}-\left(\frac{D}{K}\right) h^{(0)} K_{AB} + {\cal O}\left(\frac{1}{D}\right).$$
Substituting we find
\begin{equation}\label{stressfinal12}
\begin{split}
8 \pi T^{AB} 
=~& \frac{N}{2} (D-3) \left(  h^{AB} _ {(0)}  - h^{(0)} \Pi^{AB}\right)- \frac{K}{D} \left[{h}^{AB} _{(1)} 
-{h}^{(1)} \Pi^{AB}  \right]\\
&+\left(\frac{h^{(0)}}{2} \right)K_{AB}+ {\cal O}\left(\frac{1}{D}\right).
\end{split}
\end{equation}
In the next section we shall see that for our particular solution $h^{(0)}\sim{\cal O}\left(\frac{1}{D}\right)$. In that case the expression for the final stress tensor simplifies further and we find.

\begin{equation}\label{stressfinal2}
\begin{split}
8 \pi T^{AB} 
=~& \frac{N}{2} (D-3) \left(  h^{AB} _ {(0)}  - h^{(0)} \Pi^{AB}\right)-\left( \frac{K}{D}\right) \left[{h}^{AB} _{(1)} 
-{h}^{(1)} \Pi^{AB}  \right]\\
&+ {\cal O}\left(\frac{1}{D}\right).
\end{split}
\end{equation}

\section{The Charge Current and Stress Tensor for the large D black hole membrane}
\label{ccst}

\subsection{Review of the nonlinear large $D$ charged black hole membrane solutions} \label{rcp}

As reviewed in some detail in the introduction, the authors of 
 \cite{Bhattacharyya:2015dva,Bhattacharyya:2015fdk, Dandekar:2016fvw}
found a class of asymptotically 
flat solutions to the Einstein Maxwell equations. 
The solutions obtained in \cite{Bhattacharyya:2015dva,Bhattacharyya:2015fdk, 
Dandekar:2016fvw} are in one to one correspondence with the configuration (shape, velocity and charge density) of a membrane in flat space, and describe the dynamics of black holes in a large 
number of dimensions at time and distance scales of order unity. 

The spacetime metric ${\cal G}_{MN}$ and gauge field ${\mathfrak a}_M$ of 
\cite{Bhattacharyya:2015dva,Bhattacharyya:2015fdk, Dandekar:2016fvw}
take the schematic form 
\begin{equation}\label{schematic} \begin{split}
&{\cal G}_{MN}= \eta_{MN} + {\mathfrak g}_{MN},~~~ {\mathfrak g}_{MN}= \sum_{n=1}^\infty \frac{G^n_{MN}(\rho-1)}{\rho^{n(D-3)}},\\
&{\mathfrak a}_{N}=\sum_{n=1}^\infty \frac{A^n_{N}(\rho-1)}{\rho^{n(D-3)}}.\\
\end{split}
\end{equation}
The functions $G^n_{MN}(\rho-1)$ and $A^n_N(\rho-1)$ each admit a power series 
expansion in $\rho-1$. Schematically 
\begin{equation}\label{ha}
  G^{n}_{MN}(\rho-1)= \sum_{k=0}^\infty G^{nk}_{MN}~(\rho-1)^k, ~~~
 A^n_{N}(\rho-1)= \sum_{k=0}^\infty A^{nk}_{N}~(\rho-1)^k .
\end{equation} 
The coefficients $G^{nk}_{MN}$ and  $A^{nk}_{MN}$ are all finite in the limit 
$D\to \infty$ and each themselves admit a power series expansion in $\frac{1}{D}$, whose 
coefficients are various
 derivatives of the shape, velocity and charge density fields of the membrane. 

The authors of \cite{Bhattacharyya:2015dva,Bhattacharyya:2015fdk, Dandekar:2016fvw} have developed 
a systematic perturbative procedure to determine the coefficients $G_{MN}^{nk}$ and $A_M^{nk}$. 
The $m^{th}$ iteration of the perturbative procedure of 
\cite{Bhattacharyya:2015dva,Bhattacharyya:2015fdk, Dandekar:2016fvw}
simultaneously determines the coefficients $G^{nk}_{MN}$ $A^{nk}_{N}$ upto order 
$\frac{1}{D^{m-k}}$ (simultaneously for all $n$). 

It follows that the $m^{th}$ iteration allows 
systematic determination of the metric and gauge field to order 
$\frac{1}{D^m}$ for those values of $\rho$ for which $\rho-1$ is of order $\frac{1}{D}$. 
This was, in fact, the method adopted in 
\cite{Bhattacharyya:2015dva,Bhattacharyya:2015fdk, Dandekar:2016fvw}. The authors of those papers 
work with a scaled coordinate $R= D(\rho-1)$ and then, in the $m^{th}$ order of perturbation theory, 
systematically determine the gauge field and metric to order $1/D^m$. The fact that the authors 
of \cite{Bhattacharyya:2015dva,Bhattacharyya:2015fdk, Dandekar:2016fvw} found solutions of the full 
nonlinear Einstein Maxwell equations is reflected in the fact that the perturbative procedure works uniformly at every value of $n$ in the expansion \eqref{schematic}.

Note that \eqref{ha} reduces to the 
expansions \eqref{hexp}and \eqref{out1} when \eqref{schematic} is truncated 
to the term with $n=1$. This observation makes perfect sense; the terms in 
\eqref{schematic} with $n \geq 2$ are all highly subdominant compared to the 
leading term when $\rho -1 \gg \frac{1}{D}$. As explained in the introduction 
this is precisely the matching region in which we expect the general nonlinear 
solution of \cite{Bhattacharyya:2015dva,Bhattacharyya:2015fdk, Dandekar:2016fvw}
to reduce to a particular linearized solution of the Einstein Maxwell
equations. In fact the attentive reader will have noticed that the 
structure of the perturbative expansion described in the previous 
paragraph is {\it precisely} the structure employed to obtain the general 
solution to the linearized Einstein Maxwell solutions in section \ref{mclsdc}. 
In other words the solution of \cite{Bhattacharyya:2015dva,Bhattacharyya:2015fdk, Dandekar:2016fvw} is guaranteed to reduce to a special case of the 
construction of section \ref{mclsdc} when we truncate \eqref{schematic} 
to $n=1$. 

In this section we will see how this works in detail in a particular example. 
Our starting point is the {\it first order} solution of the perturbative 
procedure of \cite{Bhattacharyya:2015dva,Bhattacharyya:2015fdk, Dandekar:2016fvw}
presented in \cite{Bhattacharyya:2015fdk}. \footnote{The results presented in \ref{mclsdc}
have since been generalized to one higher order in  \cite{Dandekar:2016fvw}
for the special case of uncharged membranes. As this generalization has not
yet been performed for the case of charged membranes, in this paper we 
restrict our attention to metrics and gauge fields at first order in 
the derivative expansion, leaving the extension to second order results
to future work.} In the rest of this section we massage the explicit solution 
of \cite{Bhattacharyya:2015fdk} to put it in the form \eqref{schematic} and 
\eqref{ha}. We then drop all terms with $n\geq 2$ in this expansion, identify the 
effective solution of section \ref{mclsdc} that we are left with and thereby read off 
the membrane charge current and stress tensor of the solution.

In the rest of this subsection we simply recall the final result for the membrane metric and gauge field determined in \cite{Bhattacharyya:2015fdk} in some detail. This solution
 is parametrized by the shape of a metric 
in flat space, a velocity field $u_M$ on the membrane and a charge density 
field $Q$ on the membrane As in earlier sections in this paper, the 
symbols $n_M$ denotes the normal of the flat space membrane while 
$K_{NM}$ is its extrinsic curvature and ${K}$ is the trace of $K_{MN}$ 
in flat space. We follow  \cite{Bhattacharyya:2015fdk} to define 
$$O_M= n_M-u_M.$$ 

In terms of all these quantities the metric and gauge field, 
presented in  \cite{Bhattacharyya:2015fdk} is given by 
\begin{equation}\begin{split}\label{metgform}
{\cal G}_{MN} &= \eta_{MN} + {\mathfrak g}_{MN}\\
{\mathfrak g}_{MN} &= F(\rho) O_M O_N +{\mathfrak g}^{(T)}_{MN} + 2O_{(M}{\mathfrak g}_{N)}^{(V)} + {\mathfrak g}^{(S)} O_M O_N + {\mathfrak g}^{(Tr)} P_{MN},\\
\sqrt{16 \pi} ~{\mathfrak a}_M&= \sqrt{2} Q ~\rho^{-(D-3)}~ O_M +\left( {\mathfrak a}^{(S)}O_M + {\mathfrak a}^{(V)}_M\right),\\
\text{where}&\\
{ P}_{MN}&= \eta_{MN}-O_M n_N -O_N n_M + O_M O_N,\\
{ P}^{MN}& {\mathfrak g}_{N}^{(V)}={ P}^{MN} {\mathfrak a}_{N}^{(V)}=0, ~~~{ P}^{MN} {\mathfrak g}^{(T)}_{MQ} =0, ~~~{ P}^{MN}{\mathfrak g}^{(T)}_{MN}=0,
\end{split}
\end{equation}
The factor of $\sqrt{16 \pi}$ in the third line of \eqref{metgform} is a consequence of the differences 
in the conventions used for the gauge field in \cite{Bhattacharyya:2015fdk} and the current paper 
(see around \eqref{translate}). 
The various free functions appearing in equations \eqref{metgform} are given by
\begin{equation}\label{gaugegeo}
\begin{split}
{\mathfrak a}^{(V)}_M=&~-\left(\frac{\sqrt{2}}{D}\right)Q\rho^{-D}\bigg[D(\rho-1){V}^{(1)}_M - Q^2 [1+\log(1-\rho^{-D}Q^2)]{V}^{(2)}_M\bigg]\\
& + {\cal O}\left(\frac{1}{D}\right)^2,\\
{\mathfrak a}^{(S)} =&~\left(\frac{1}{D}\right)\bigg[ \sqrt{2}~Q~D (\rho-1)~\rho^{-D} {S}^{(1)} + 2\sqrt{2}\left(\frac{Q^3}{1-Q^2}\right)\rho^{-D}~\Upsilon_A(\rho)~{S}^{(2)}\bigg] + {\cal O}\left(\frac{1}{D}\right)^2.
\end{split}
\end{equation}
\begin{equation}\label{metricgeo1}
\begin{split}
{\mathfrak g}^{(T)}_{MN}=&~\left(\frac{2}{D}\right)\log(1-Q^2\rho^{-D})~\tau_{MN} + {\cal O}\left(\frac{1}{D}\right)^2,\\
{\mathfrak g}^{(V)}_M =&~
\left(\frac{1}{D}\right)\bigg[Q^2\left[(F(\rho)-\rho^{-(D-3)}) +(F(\rho)-1)\log(1-Q^2\rho^{-D})\right]{V}^{(2)}_M\\
&~~~~~~~~~~~-D(\rho-1) F(\rho)~{V}^{(1)}_M\bigg]+ {\cal O}\left(\frac{1}{D}\right)^2.\\
\end{split}
\end{equation}
\begin{equation}\label{metricgeo2}
\begin{split}
{\mathfrak g}^{(S)}=&~-\sqrt{2} Q~ \rho^{-D}{\mathfrak a}^{(S)} +\left(\frac{1}{D}\right)\bigg[\rho^{-(D-3)} - F(\rho)\bigg]\\
& + \left(\frac{2}{D}\right)\rho^{-D}\left[Q^2~ D(\rho-1) ~ {S}^{(1)}
+\Upsilon_H(\rho){S}^{(2)}\right] +{\cal O}\left(\frac{1}{D}\right)^2,\\
\\
{\mathfrak g}^{(Tr)}=& ~ {\cal O}\left(\frac{1}{D}\right)^3,
\end{split}
\end{equation}

The different functions and the derivative structures that appear in equations \eqref{gaugegeo},  \eqref{metricgeo1} and \eqref{metricgeo2} are defined as\footnote{Here our basis for the independent  boundary data ( the derivatives of velocity and the shape of the membrane) is little different from what has been used in \cite{Bhattacharyya:2015fdk} . The basis we have used turns out to be more convenient for our analysis later in this paper.  } 

\begin{table}[ht]
\vspace{0.5cm}
\centering 
\begin{tabular}{|c| c| }
\hline
 &$S^{(1)} =   \left( \frac{D}{K^2} \right)  \bar\nabla^2 Q$ \\
Scalars&\\
&$S^{(2)}= \left( \frac{D}{{K}}\right)\left[ u^A u^BK_{AB} - \frac{(u\cdot\partial){K}}{{K}}\right]$\\
\hline
&$V^{(1)}_M =\left( \frac{D}{{K}}\right)\left[\frac{\bar\nabla^2 u_N} {{K}} +u^C K_{CN}\right]~{ P}^N_M$ \\
Vectors&\\
&$V^{(2)}_M=  \left( \frac{D}{{K}}\right)\left[\frac{\partial_N {K}} {K} - (u\cdot\partial)u_N\right]~{ P}^N_M $ \\
\hline
&\\
Tensor& $\tau_{MN} =  { P}_M^{Q_1}~\left(\frac{D}{{K}}\right)\left[\frac{\partial_{Q_1} O_{Q_2}+\partial_{Q_2} O_{Q_1}}{2}-\eta_{Q_1 Q_2}\left(\frac{\partial\cdot O}{D-2}\right) \right]~{ P}_N^{Q_2}$\\
\hline
\end{tabular}\vspace{.5cm}
\caption{A listing of the `first order' quantities that appear in the 
formula for the metric and gauge field, taken from  
\cite{Bhattacharyya:2015fdk}. }
\label{table:geometric} 
\end{table}
\noindent

\begin{equation} \label{fdef}
\begin{split}
F(\rho) &= \left[(1+Q^2)\rho^{-(D-3)} -Q^2 \rho^{-2(D-3)}\right],\\
\Upsilon_A(\rho) &= \int_0^{D(\rho-1)} dx~\log(1-Q^2e^{-x}),\\
\Upsilon_H(\rho)&=\left[(\rho^D-Q^2)\log(1-Q^2\rho^{-D})-(1-Q^2)\log(1-Q^2) + Q^2\left(\frac{1+Q^2}{1-Q^2}\right)\Upsilon_A(\rho)\right].\\
\bar\nabla^2 Q &=\Pi^A_B \partial_A\left[\Pi^{BC}\partial_C Q\right],~~\bar\nabla^2u_A= \Pi_{AA'}\Pi^{B}_{C}\partial_{B}\left[\Pi^{CC'}\Pi^{A'A''}(\partial_{C'}u_{A''})\right].
\end{split}
\end{equation}

\subsection{The Membrane Charge Current}

From equation \eqref{gaugegeo} it is not difficult to read off the corresponding value of 
$A_M^1$ (see \eqref{schematic}). Recall that $A^1_M$ is guaranteed to be a solution for the linearized 
Maxwell equations around flat space. We find 
\begin{equation}\label{out1m}
\begin{split}
\sqrt{16 \pi} { A}^1_B &\equiv M_B=\sum_{k=0}^\infty (\rho -1)^k M_B^{(k)},\\
\end{split}
\end{equation}
with
\begin{equation} \label{expm}
\begin{split}
M_B^{(0)}&=\sqrt{2} Q~ O_B + \left(\frac{\sqrt{2}}{D}\right) Q^3\left(\frac{D}{K} \right)\left(\frac{\partial_A{K}}{{K}} - (u\cdot\partial)u_A\right)P^A_B + {\cal O}\left(\frac{1}{D}\right)^2,\\
M_B^{(1)} &= \sqrt{2} \left(\frac{D}{K}\right)\left(\frac{\bar\nabla^2 Q}{K}\right) O_B -\sqrt{2} \left(\frac{D}{K}\right)\left[\frac{\bar\nabla^2 u_A}{K} + u^C K_{CA}\right]P^A_B+ {\cal O}\left(\frac{1}{D}\right),\\
\text{where}&~O_B = n_B -u_B,~~P_{AB} = \eta_{AB} - n_A n_B + u_A u_B = \Pi_{AB}+ u_A u_B,\\
&~\bar\nabla^2 Q =\Pi^A_B \partial_A\left[\Pi^{BC}\partial_C Q\right],~~\bar\nabla^2u_A= \Pi_{AA'}\Pi^{B}_{C}\partial_{B}\left[\Pi^{CC'}\Pi^{A'A''}(\partial_{C'}u_{A''})\right],
\end{split}
\end{equation}
(for notational convenience we have renamed $A_A^{1k}$ of \eqref{ha} 
as $M_A^{(k)}$; we have dropped the superscript unity as we will only 
concern ourselves with the linearized part of the solution from now on).

As we have emphasized above, the configuration \eqref{out1m} is guaranteed to 
be a linearized solution of the form presented in subsection \ref{linearout}. 
As we have explained around that subsection, every such solution may is associated
with a membrane current. This current is given by $J_M=J^{(out)}_M -J^{(in)}_M$ where 
$J^{(out)}_M$ is simply $n^N F_{NM}$ where $F_{NM}$ is the field strength evaluated 
on the solution \eqref{out1m} above and $J^{(in)}_M$ is given by \eqref{incurrentsub} 
where the field strength in that expression is once again evaluated on the configuration
\eqref{out1m} using the solution \eqref{out1m}. The algebra required to evaluate these
two components of the current is straightforward; in Appendix \ref{memour} we demonstrate
that
\begin{equation}\label{jext}
\begin{split}
\sqrt{16 \pi} J^{out}_B  
=&~\sqrt{2}\left[Q\left({K} + \frac{\bar\nabla^2{K}}{{K}^2}-\frac{2K}{D}\right)+(u\cdot\partial) Q - \left(\frac{\bar\nabla^2 Q +Q(u\cdot \partial){K}}{K}\right) + Q(u^C u^{C'} K_{CC'})\right] u_B\\
&-\sqrt{2}Q\left[\left(\frac{\partial_A Q}{Q}\right) + (u\cdot\partial)u_A \right]P^A_B
+ {\cal O}\left(\frac{1}{D}\right),\\
\end{split}
\end{equation}
while
\begin{equation}\label{jint}
\begin{split}
\sqrt{16 \pi} J^{(in)}_B 
&= \sqrt{2}\bigg[ \left(\frac{\bar\nabla^2 Q}{K}  + Q~ u^C u^{C'} K_{CC'}\right)u_B + Q ~P^A_B \left(\frac{\bar\nabla^2 u_A}{K}\right)-Q~K^C_A u_C\bigg]+ {\cal O}\left(\frac{1}{D}\right).\\
\end{split}
\end{equation}
Here we have used the following short-hand notation for derivatives projected along the membrane.
\begin{equation}\label{notderi}
\begin{split}
&\bar \nabla^2 K \equiv \Pi^{AB}\partial_A\left[\Pi^{B'}_B\partial_{B'} K\right],~~\bar \nabla^2 Q \equiv \Pi^{AB}\partial_A\left[\Pi^{B'}_B\partial_{B'}Q\right],\\
&\bar\nabla_A\bar u_B \equiv \Pi_A^{A'}\Pi_B^{B'}\partial_{A'}u_{B'},~~~\bar\nabla^2u_A \equiv \Pi^{CB}\partial_C\left[\Pi^{A'}_A\Pi^{B'}_B\partial_{B'} u_{A'}\right] ,
\end{split}
\end{equation}
\eqref{jext} and \eqref{jint} are our final results for the internal and external contributions to the 
membrane current. Putting them together we find 
$$J_B = J_B^{(out)} - J_B^{(in)}.$$
Subtracting equation \eqref{jext} from equation \eqref{jint} we find
\begin{equation}\label{memcur}
\begin{split}
\sqrt{16 \pi} J_B 
=&~\sqrt{2}\left[Q\left({K} + \frac{\bar\nabla^2{K}}{{K}^2}-\frac{2K}{D}\right)+(u\cdot\partial) Q - \left(\frac{2\bar\nabla^2 Q +Q(u\cdot \partial){K}}{K}\right) \right] u_B\\
&-\sqrt{2}Q\left[\left(\frac{\partial_A Q}{Q}\right) + (u\cdot\bar\nabla)u_A+ \left(\frac{\bar\nabla^2 u_A}{K}\right) -K^C_A u_C\right]P^A_B
+ {\cal O}\left(\frac{1}{D}\right).
\end{split}
\end{equation}
Note that by construction $J_B$ is a vector tangent to the membrane and also all the derivatives that appears in the expression of $J_B$ are all along the membrane. All these derivatives could re expressed as covariant derivatives with respect to the intrinsic metric of the membrane. In terms of the coordinates intrinsic to the membrane we write the current as
\begin{equation}\label{memcurnt}
\begin{split}
\sqrt{16 \pi} J^\mu
=&~\sqrt{2}\left[Q\left({K} + \frac{\hat\nabla^2{K}}{{K}^2}-\frac{2K}{D}\right)+(u\cdot\hat\nabla) Q - \left(\frac{2\hat\nabla^2 Q +Q(u\cdot\hat \nabla){K}}{K}\right) \right] u^\mu\\
&-\sqrt{2}Q\left[\left(\frac{\hat\nabla_\nu Q}{Q}\right) + (u\cdot\hat\nabla)u_\nu+ \left(\frac{\hat\nabla^2 u_\nu}{K}\right) -K^\alpha_\nu u_\alpha\right]p^{\nu\mu}
+ {\cal O}\left(\frac{1}{D}\right),\\
\text{where}&~~
p_{\mu\nu}= g^{(ind,f)}_{\mu\nu} + u_\mu u_\nu,~~\hat\nabla_\mu = \text{Covariant derivative w.r.t $g^{(ind,f)}_{\mu\nu}$},\\
g^{(ind,f)}_{\mu\nu}=&~\text{Induced metric on  membrane, embedded in flat space-time}\\
(\cdot )&~\text{denotes contraction w. r. t ~~$g^{(ind,f)}_{\mu\nu}$}.
\end{split}
\end{equation}

\subsection{A consistency check} \label{acc}

In the previous subsection we obtained the results for the membrane charge current assuming 
that the configuration \eqref{out1m} is indeed a particular case of a solution of the 
general solution presented in subsection \ref{linearout}. While this must be the case on 
logical grounds, it is, of course, reassuring to have a direct algebraic check of this claim.
We have performed such a direct check; in this subsection 
we present a brief explanation of the check we have 
 here relegating most details to Appendix \ref{memour}.

 In subsection \ref{linearout} we argued that the most general linearized solution to the Maxwell equation 
is parametrized by the single function $G^{(0)}_A$, the gauge field on the 
membrane. The Taylor series coefficients of this gauge field off the membrane
are completely determined in terms of $G^{(0)}_A$. In particular, to first 
order, $G^{(1)}_A$ is given in terms of $G^{(0)}_A$ by \eqref{10mt}. We will now 
verify that \eqref{expm} is consistent with \eqref{10mt}. 

Roughly speaking, $G^{(0)}_A$ is simply $M_B^{(0)}$ while $G^{(1)}_A$ is $M_B^{(1)}$ 
(see \eqref{expm}). However this is not completely accurate for two reasons
\begin{itemize}
\item{} The analysis of the previous section was performed with the choice 
of gauge $n^B. G_B=0$. Unfortunately the solution \eqref{out1m} is 
presented in a different gauge. In order to compute $G^{(0)}_M$ and $G^{(1)}_M$, 
consequently, we must either compute gauge invariants or perform a 
gauge transformation that puts the solution \eqref{out1m} into the 
gauge $n^B. G_B=0$. We found it more convenient to actually perform 
the gauge transformation.
\item{} The statement that $G^{(0)}_A$ is given by \eqref{expm} evaluated 
at $\rho=1$ is unambiguous. However the statement that $G^{(1)}_M$ is the 
part of \eqref{expm} proportional to $(\rho-1)$ is meaningful only once 
we have agreed on a set of subsidiary conditions on the coefficients 
of the expansion in $(\rho-1)$. In the analysis of the previous section 
we assumed that all coefficient functions obeyed the subsidiary conditions 
\eqref{subsidonGa}. The coefficient functions in \eqref{expm} turn out 
not to obey these subsidiary conditions (the coefficients in 
\eqref{expm} obey the subsidiary conditions employed in \cite{Bhattacharyya:2015fdk}, which are 
slightly different from \eqref{subsidonGa}). Consequently they have 
to be re-expanded in terms of quantities that do obey \eqref{subsidonGa} 
before we can read off $G^{(1)}_A$.  
\end{itemize}
In Appendix \ref{memour} we have carefully dealt with both these issues, 
and verified that the solution \eqref{expm} does indeed take the 
general form presented in subsection \ref{linearout} with 

\begin{equation}\label{goo}
\begin{split}
\sqrt{16 \pi}G_B^{(0)}&=- \sqrt{2} Q ~u_B  + \frac{\sqrt{2}Q^3}{D}\left(\frac{D}{K}\right)\left(\frac{\partial_A{K}}{K} - (u\cdot\partial)u_A\right)P^A_B\\
& +\sqrt{2}\Pi^A_B\left[\frac{\partial_A Q}{K} - \frac{Q\partial_A K}{K^2}\right]+ {\cal O}\left(\frac{1}{D}\right)^2,\\
\sqrt{16 \pi}G_B^{(1)}&=\left[\tilde M_B^{(1)} + C_B^{(0)}\right]=-\sqrt{2} \left(\frac{D}{K}\right) \left(\frac{\bar\nabla^2 Q}{K} \right)u_B - \sqrt{2}Q \left(\frac{D}{K}\right) \left(\frac{P^B_A\bar\nabla^2 u_B}{K} \right)+ {\cal O}\left(\frac{1}{D}\right).
\end{split}
\end{equation}

\subsection{Membrane equation of motion from conservation of the charge current}

In section \ref{mclsdc} we have argued that any membrane  constructed out of  the general 
linearized solution of the Maxwell equations presented in that section must be automatically 
conserved. Earlier in this section we have used the formalism of section \ref{mclsdc} to 
 explicitly determine a charge current for the membrane spacetimes of 
\cite{Bhattacharyya:2015dva,Bhattacharyya:2015fdk, Dandekar:2016fvw}. Our final result, presented 
in \eqref{memcur} is given in terms of the curvatures, charge and velocity derivatives of the large
$D$ black hole membrane. If the analysis presented in this paper is self consistent it must turn out 
that the charge current \eqref{memcur} - which can simply be algebraically determined in terms of 
membrane curvatures, velocity and charge derivatives - must automatically vanish using only
constraints between these derivatives that were already determined in \cite{Bhattacharyya:2015dva,Bhattacharyya:2015fdk, Dandekar:2016fvw}. In this subsection we explain how 
this works in detail.

At leading order in the large $D$ limit, the current \eqref{memcur} takes the form
\begin{equation}\label{locur}
\sqrt{16 \pi}J^\mu=\sqrt{2} Q {K} u^\mu
\end{equation}
and is of order $D$ \footnote{This scaling is because ${K}$ is of order $D$ as explained 
in \cite{Bhattacharyya:2015fdk} - see the introduction.}
. The divergence of a current of order ${\cal O}(D)$ is generically of order ${\cal O}(D^2)$. In the current 
context the naively order ${\cal O}(D^2)$ term in the divergence of the leading order current is given by 
\begin{equation}\label{concurd2}
\begin{split}
& \sqrt{16 \pi} \hat\nabla_\mu J^\mu = \sqrt{2} Q{K} \left(\hat\nabla_\mu u^\mu\right) + {\cal O}(D).\\
\end{split}
\end{equation}
This expression is naively of order ${\cal O}(D^2)$ because ${K}$ is of order ${\cal O}(D)$ and $\left(\nabla_B u^B\right)$ would also be of order ${\cal O}(D)$ if $u$ were an unrestricted arbitrary velocity field. The fact that the 
divergence of the charge current must vanish tells us that $u$ {\it cannot} be an unrestricted 
velocity field; it must, in fact, be chosen to ensure that  
\begin{equation}\label{divf}
\left(\hat\nabla_\mu u^\mu\right) ={\cal O}(1).
\end{equation}
The requirement \eqref{divf} is the first of \eqref{memeom} and was, in fact, the starting point of the membrane construction of 
 \cite{Bhattacharyya:2015dva,Bhattacharyya:2015fdk, Dandekar:2016fvw}.
 
In this paper we have systematically determined the large $D$ membrane 
charge current \eqref{memcur} upto ${\cal O}(1)$. \footnote{The determination of the 
charge current to order ${\cal O}(1/D)$ requires knowledge of the gauge field in the solutions 
of \cite{Bhattacharyya:2015dva,Bhattacharyya:2015fdk, Dandekar:2016fvw} at order 
${\cal O}(1/D)$ which has not yet been worked out. } As the operation of taking the 
divergence generically increases the order of $D$ of a current by one power, our knowledge 
of the charge current \eqref{memcur} is sufficient to determine the divergence of this current only 
to order ${\cal O}(D)$. We have already explained that the condition \eqref{divf} ensures that the divergence
of the charge current vanishes at order ${\cal O}(D^2)$. We will now explore the requirement that this 
divergence also vanishes at order ${\cal O}(D)$.
 
Apart from the expression listed in \eqref{locur}, every term in \eqref{memcur} is of ${\cal O}(1)$ rather than 
order ${\cal O}(D)$. While a generic term in a current of order unity has a divergence of order $D$, it follows from 
\eqref{divf} that any term of order unity proportional to $u^M$ has a divergences of order unity. 
It follows that such terms do not contribute to the divergence of the charge current at order ${\cal O}(D)$. 
Dropping all such terms we find the simplified current 
\begin{equation}\label{memcurrel}
\begin{split}
\sqrt{16 \pi} J_\mu^{(simp)}
=&~\sqrt{2}Q\bigg\{{K} u_\alpha
-\left[\left(\frac{\hat\nabla_\mu Q}{Q}\right) + (u\cdot\hat\nabla)u_\mu+ \left(\frac{\hat\nabla^2 u_\mu}{K}\right) -K^\nu_\mu u_\nu\right]p_\alpha^\mu\bigg\}
+ {\cal O}\left(\frac{1}{D}\right),
\end{split}
\end{equation}
whose divergence is given by 
\begin{equation}\label{divcur}
\begin{split}
&\sqrt{16 \pi} \hat\nabla_\mu J^\mu_{(simp)}\\ 
=&
-\hat\nabla_\mu\bigg\{\left[\hat\nabla_\nu Q  + Q(u\cdot\hat\nabla)u_\nu+ Q\left(\frac{\hat\nabla^2 u_\nu}{K}\right) -QK^\alpha_\nu u_\alpha\right]p^{\nu\mu}\bigg\}\\
&+{K}Q(\hat\nabla\cdot u)+{K} (u\cdot\hat\nabla) Q +  Q(u\cdot\hat\nabla){K}  + {\cal O}(1)\\
\\
=&~ {K}\bigg\{Q (\hat\nabla\cdot u)+(u\cdot\hat\nabla) Q +  Q\left[\frac{(u\cdot\hat\nabla){K}}{ K}\right]-\left[\frac{\hat\nabla^2 Q}{{K}} \right]- Q~(u^\mu u^\nu K_{\mu\nu})\bigg\} + {\cal O}(1).
\end{split}
\end{equation}
In computing equation \eqref{divcur} we have used the identities \eqref{I1} and \eqref{I2}.
\footnote{We emphasize that it is permissible to replace the full charge current $J_\mu$ by 
$J_\mu^{simp}$ only for the purposes of computing its divergence and not for the purposes of computing 
radiation.}

In the analysis of \cite{Bhattacharyya:2015dva,Bhattacharyya:2015fdk, Dandekar:2016fvw} it turns out that 
$\nabla\cdot u= {\cal O}(1/D)$. Moreover the `charge' equation of motion of \cite{Bhattacharyya:2015fdk}
asserts that 
\begin{equation}\label{ceom}
(u\cdot\hat\nabla) Q +  Q\left[\frac{(u\cdot\hat\nabla){K}}{ K}\right]-\left[\frac{\hat\nabla^2 Q}{{K}} \right]- Q~(u^\mu u^\nu K_{\mu\nu})={\cal O}(1/D).
\end{equation}
It follows that the last line of \eqref{divcur} - and so the divergence of the charge current 
\eqref{memcur} - does indeed vanish at order $D$. 

In summary, the charge current computed in \eqref{memcur} is indeed divergence free; the fact that this 
is the case is, in fact, a restatement of the `charge' equation of motion of \cite{Bhattacharyya:2015fdk}.

\subsection{The Membrane Stress Tensor and its conservation}

In the rest of this section we imitate the analysis already presented for the membrane 
charge current in order to obtain and analyse the large $D$ black hole  membrane stress tensor. 
As the logic of our construction proceeds in close analogy with the case of the 
charge current we 
keep our explanations brief. 
 
Expanding the metric presented in \eqref{metgform},\eqref{metricgeo1} and \eqref{metricgeo2})
in the form \eqref{schematic}, it is not difficult to show that the function $G^{1}_{MN}$ in 
\eqref{schematic} (which, for notational convenience, we refer to below as $M_{AB}$) is given by
\begin{equation}\label{expmet}
G^{1}_{MN}\equiv M_{AB}
 =\sum_n (\rho-1)^n M_{AB}^{(n)},
\end{equation}
where
\begin{equation}\label{metricread}
\begin{split}
M_{AB} ^{(0)} &= (1+Q^2)O_A O_B 
+ 2Q^4 \left( O_A V_B ^{(2)} + O_B V_A ^{(2)} \right) -Q^2 O_A O_B -2Q^2 {\tau}_{AB} \\
& + {\cal O}\left(\frac{1}{D}\right)^2,\\
M_{AB}^{(1)} &= 2Q^2 S^{(1)} O_A O_B - (1+Q^2) \left[  V_A ^{(1)} O_B + O_A  V_B ^{(1)} \right]    \\
& + {\cal O}\left(\frac{1}{D}\right),
\end{split}
\end{equation}
with\footnote{In equation \eqref{expmet} and \eqref{metricread} we simply renamed $G_{AB}^{1k} $ as $M_{AB}^{(k)}$ to avoid confusion.}
\begin{equation}\label{metnot}
\begin{split}
&V_A ^{(1)} = \left( \frac{D}{{K}}\right)\left[\frac{\bar\nabla^2 u_B} {{K}} +u^C K_{CB}\right]~{ P}^B_A, \\
&V_A ^{(2)} =  \left( \frac{D}{K} \right)   \left[ \frac{\partial _C{K} }{K} - (u\cdot\partial)u_C \right]P_A^C , \\
&S^{(1)} =  \left( \frac{D}{K^2} \right)  \bar\nabla^2 Q ,\\
&\tau_{AB} =  { P}_A^{A'}~\left(\frac{D}{{K}}\right)\left[\frac{\partial_{A'} O_{B'}+\partial_{B'} O_{A'}}{2}-\eta_{A' B'}\left(\frac{\partial\cdot O}{D-2}\right) \right]~{ P}_B^{B'},\\
&\text{where}\\
&\bar\nabla^2 Q =\Pi^A_B \partial_A\left[\Pi^{BC}\partial_C Q\right],~~\bar\nabla^2u_A= \Pi_{AA'}\Pi^{B}_{C}\partial_{B}\left[\Pi^{CC'}\Pi^{A'A''}(\partial_{C'}u_{A''})\right].
\end{split}
\end{equation}

The metric \eqref{expmet} is a particular example of the general linearized solution to the Einstein
equations presented in subsection \eqref{einlinearout}.  As in the previous subsection we have also verified in detail that the 
solution \eqref{expmet} and \eqref{metricread} after appropriate transformation agrees with the general structure listed in 
subsection \ref{einlinearout} provided we identify 
\begin{equation}\label{readpmntxt}
\begin{split}
h_{AB}^{(0)} &= (1+ Q^2)~ u_A u_B\\
&+\left(\frac{1}{D}\right)\bigg[-2Q^4 \left( u_A V_B^{(2)} + u_B V_A ^{(2)}  \right)  - Q^2 u_A u_B- 2Q^2~ {\tau}_{AB}  \\
& ~~~~~~~~~~~~~+  \Pi^C_A  \left[\partial _C \zeta _{C^ \prime} + \partial _{C^\prime} \zeta _C \right] \Pi^{C'}_B \bigg]+{\cal O}\left(\frac{1}{D}\right)^2,\\
h_{AB} ^{(1)} &= \left( \frac{D}{{K}^2} \right) \bigg[ 2Q \bar\nabla ^2 Q ~u_A u_B + (1+ Q^2)~ \Pi^C_B\left(u_{A}\bar\nabla ^2 u_{C}+u_{C}\bar\nabla ^2 u_{A} \right) \bigg]
+ {\cal O}\left(\frac{1}{D}\right) , \\
\end{split}
\end{equation}
where
\begin{equation}\label{notmetmntxt}
\begin{split}
\zeta _A &= (1+Q^2)\left( \frac{D}{K} \right)\left( \frac{n_A}{2} -u_A \right).   \\
\end{split}
\end{equation}
We have, in particular, verified that the results quoted in \eqref{readpmntxt} are consistent 
with \eqref{h10out}.

From the first equation of \eqref{readpmntxt} it follows that the   trace of $h^{(0)}_{AB}$($= \Pi^{AB} h^{(0)}_{AB} $) is of order $ {\cal O}\left(\frac{1}{D}\right)$, which justifies our expression of stress tensor as given in equation \eqref{stressfinal2} of previous section.

According to the general analysis of that 
subsection, any such solution is associated with a stress tensor, which is given by the difference 
between the Brown York stress tensor evaluated on the metric \eqref{expmet} and the expression 
\eqref{stressin2} evaluated on the same solution. Also in equation \eqref{stressfinal2}, we have an expression for the final stress tensor explicitly in terms of $h^{(0)}_{AB}$ and $h^{(1)}_{AB}$. Substituting equation \eqref{readpmntxt} in equation \eqref{stressfinal2} we find the explicit expression for the stress tensor for metric \eqref{metricread}.

At this stage to simplify our calculation  of stress tensor we shall use a trick. We shall define $T_{AB}^{(NT)}$ as 
$$T_{AB}^{(NT)} = \frac{N}{2}(D-3) h^{(0)}_{AB} -\left(\frac{K}{D}\right) h^{(1)}_{AB}.$$
Then from equation \eqref{stressfinal2} we could clearly see that $T_{AB} - T_{AB}^{(NT)}\propto\Pi_{AB}$. We write the proportionality factor as $\Delta$. 
With this notation the stress tensor could be written as 
\begin{equation}\label{mntxtstress}
\begin{split}
&8 \pi T_{AB}= 8 \pi \left[T_{AB}^{(NT)}+\Delta ~\Pi_{AB}\right].\\
\end{split}
\end{equation}
Now we shall determine $\Delta$ using the condition that $K_{AB}T^{AB}=0$ (see equation \eqref{norm}).
$$K_{AB}T^{AB}=0\Rightarrow \Delta = -\frac{K^{AB}T_{AB}^{(NT)}}{K}.$$

Now collecting all these pieces together we finally get the explicit expression for  the stress tensor
\begin{equation}\label{stnt}
\begin{split}
8 \pi T_{AB}^{(NT)}=& \left( \frac{K}{2} \right) (1+Q^2) u_A u_B + \left( \frac{1-Q^2}{2}  \right)K_{AB} - \left(   \frac{\bar\nabla_A u_B +\bar \nabla_B u_A}{2} \right) \\
&-\left(\frac{{K}Q^2}{2D} + \frac{2Q\bar\nabla^2Q }{K}+ Q^2 u^C u^{C'} K_{CC'}\right)u_A u_B-\left( u_A{\cal V}_B + u_B{\cal V}_A\right) + {\cal O}\left(\frac{1}{D}\right),\\
\\
{\cal V}_A =&~Q~\bar\nabla_A Q +Q^2(u^CK_{CA} ) +\left(\frac{2Q^4-Q^2-1}{2}\right)\left(\frac{\bar\nabla_A{K}}{K}\right)\\
&-\left(\frac{Q^2 + 2Q^4}{2}\right)(u\cdot\bar\nabla)u_A + \left(\frac{1+Q^2}{K}\right)\bar\nabla^2 u_A
\end{split}
\end{equation}
and 
\begin{equation}\label{mntxtdeltaexp}
\begin{split}
&\Delta=-\bigg[\left(\frac{1+Q^2}{2}\right)\left(u^A u^BK_{AB}\right)+\left(\frac{1-Q^2}{2}\right)\left(\frac{{K}}{D}\right)+{\cal O}\left(\frac{1}{D}\right)\bigg].
\end{split}
\end{equation}
As in the previous subsection, $\bar\nabla_A$ defines the projected derivative along the membrane as embedded in flat Minkowski space. See equation \eqref{notderi} for a more precise definition.Also in our algebra we used the fact that to leading order in $\frac{1}{D}$, 
$$K^{AB}K_{AB}=\frac{{K}^2}{D} +{\cal O}(1).$$\\
In equations \eqref{mntxtstress}, \eqref{stnt} and \eqref{mntxtdeltaexp} all derivatives and all free and contracted indices are along the membrane. Therefore we can as well re-express the stress tensor as a tensor defined completely on the membrane, where all projected derivatives are replaced by covariant derivatives, defined with respect to the membrane's intrinsic metric.
\begin{equation}\label{stmem}
\begin{split}
8\pi T_{\mu\nu}= &~ \left( \frac{K}{2} \right) (1+Q^2) u_\mu u_\nu + \left( \frac{1-Q^2}{2}  \right)K_{\mu\nu} - \left(   \frac{\hat\nabla_\mu u_\nu +\hat \nabla_\nu u_\mu}{2} \right) \\
&-\left(\frac{{K}Q^2}{2D} + \frac{2Q\hat\nabla^2Q}{K} + Q^2 u^\alpha u^\beta K_{\alpha\beta}\right)u_\mu u_\nu-\left( u_\mu{\cal V}_\nu + u_\nu{\cal V}_\mu\right) \\
&~-\bigg[\left(\frac{1+Q^2}{2}\right)\left(u^\alpha u^\beta K_{\alpha\beta}\right)+\left(\frac{1-Q^2}{2}\right)\left(\frac{{K}}{D}\right)\bigg]g^{(ind,f)}_{\mu\nu}\\
&~ + {\cal O}\left(\frac{1}{D}\right),\\
\end{split}
\end{equation}
where
\begin{equation}\label{calv}
\begin{split}
{\cal V}_\mu =&~Q~\hat\nabla_\mu Q +Q^2(u^\alpha K_{\alpha\mu} ) +\left(\frac{2Q^4-Q^2-1}{2}\right)\left(\frac{\hat\nabla_\mu{K}}{K}\right)\\
&-\left(\frac{Q^2 + 2Q^4}{2}\right)(u\cdot\hat\nabla)u_\mu+ \left(\frac{1+Q^2}{K}\right)\hat\nabla^2 u_\mu.
\end{split}
\end{equation}

\subsubsection{Conservation of the stress tensor}

In this subsection we shall compute the divergence of the stress tensor \eqref{mntxtstress} 
and demonstrate that it vanishes at order ${\cal O}(D^2)$ and at order ${\cal O}(D)$ once we impose the 
membrane equations of motion listed in equation 1.1. of \cite{Bhattacharyya:2015fdk}.

As in the case of the charge current, the stress tensor has a leading order piece
\begin{equation}\label{lop}
8 \pi T_{\mu\nu}=\left( \frac{K}{2} \right) (1+Q^2) u_\mu u_\nu,
\end{equation}
which is of order ${\cal O}(D)$. All other terms in \eqref{mntxtstress} are of order ${\cal O}(1)$. As in our 
analysis of the charge current the divergence of \eqref{lop} is naively of order ${\cal O}(D^2)$; 
the requirement that the divergence vanish at this order reimposes the condition \eqref{divf}. 
As in the case of the charge current we must now impose the condition that the divergence 
of the stress tensor vanishes also at order ${\cal O}(D)$. The order ${\cal O}(D)$ part of this divergence receives 
contributions only from those ${\cal O}(1)$ terms in \eqref{mntxtstress} whose divergences is 
of order ${\cal O}(D)$. This criterion excludes all order ${\cal O}(1)$ terms proportional to $g^{(ind,f)}_{\mu\nu}$ in 
\eqref{mntxtstress}.
\footnote{In order to see this recall that $\hat\nabla_\mu g_{(ind,f)}^{\mu\nu}=0$ identically.
Therefore 
\begin{equation}\label{tracediv}
\begin{split}
\hat\nabla^\mu\left[\Delta ~g^{(ind,f)}_{\mu\nu}\right] =\hat\nabla_\nu\Delta = {\cal O}(1).
\end{split}
\end{equation} }
In order to compute the divergence of the stress tensor at order $D$, it follows that we can replace 
the stress tensor in \eqref{mntxtstress} by the simpler effective stress tensor $T_{\mu\nu}^{(eff)}$. 
\begin{equation}\label{effst}
\begin{split}
T_{\mu\nu}^{(eff)} =&~\left( \frac{K}{2} \right) (1+Q^2) u_\mu u_\nu + \left( \frac{1-Q^2}{2}  \right)K_{\mu\nu} \\
&- \left(   \frac{\hat\nabla_\mu u_\nu +\hat \nabla_\nu u_\mu}{2} \right) 
-\left( u_\mu{\cal V}_\nu + u_\nu{\cal V}_\mu\right), 
\end{split}
\end{equation}
where ${\cal V}_\mu $ is defined in equation \eqref{calv}.\\
The divergence of $T_{\mu\nu}^{(eff)}$ has a free index and so can be decomposed into the part orthogonal to 
$u^\mu$ and the part in the direction of $u^\mu$. We will find it convenient to give these two different 
pieces names. Let 
$$E^\mu\equiv p^\mu_\nu~\hat\nabla_\alpha T_{(eff)}^{\alpha\nu}$$
and let 
$$E\equiv u_\nu\hat\nabla_\alpha T_{(eff)}^{\alpha\nu}.$$
We will first demonstrate that the requirement that $E^\mu$ vanish at order ${\cal O}(D)$ is simply a 
restatement of the motion presented in equation (1.1) in \cite{Bhattacharyya:2015fdk}. On the
other hand the requirement that $E$ vanish at order $D$ tells us $(\hat\nabla\cdot u)$ is of order ${\cal O}\left(\frac{1}{D}\right)$ or 
smaller (this is a strengthening of the condition \eqref{divf}). As both these conditions 
were independently met in \cite{Bhattacharyya:2015fdk}, it follows that the stress tensor dual 
to the large $D$ membrane of \cite{Bhattacharyya:2015fdk} is indeed conserved. 

We now turn to a demonstration of the first of these assertions. 
\begin{equation}\label{ebman}
\begin{split}
E^\mu 
= &-\left(\frac{K}{2}\right)(1+Q^2) (u\cdot\hat\nabla)u^\mu - \left(\frac{1-Q^2}{2}\right) p^{\mu\nu}\hat\nabla_\alpha K^\alpha_\nu\\
 &+p^{\mu\alpha}\left(\frac{\hat\nabla^2 u_\alpha + \hat\nabla_\nu\hat \nabla_\alpha u^\nu}{2}\right)
+ {\cal O}(1)\\
= &-\left(\frac{K}{2}\right)\bigg[(1+Q^2) (u\cdot\hat\nabla)u^\mu + (1-Q^2) p^{\mu\nu}\left(\frac{\hat\nabla_\nu {K}}{K}\right)\\
 &~~~~~~~~~~~~-p^{\mu\nu}\left(\frac{\hat\nabla^2 u_\nu}{K} +\frac{ \hat\nabla_\alpha \hat\nabla_\nu u^\alpha}{K}\right)\bigg]
 +{\cal O}(1)\\
= &-\left(\frac{K}{2}\right)\bigg[(1+Q^2) (u\cdot\hat\nabla)u^\mu + (1-Q^2) p^{\mu\nu}\left(\frac{\hat\nabla_\nu {K}}{K}\right)\\
&~~~~~~~~~~~~ -p^{\mu\nu}\left(\frac{\hat\nabla^2 u_\nu}{K} + K_{\nu\alpha}u^\alpha\right)\bigg]
+ {\cal O}(1).\\
\end{split}
\end{equation}
In the last step we have used identities \eqref{I3} and \eqref{I4} and also \eqref{divf}. 

We now turn to the quantity $E$. After a little bit of algebra (see appendix (\ref{fincal}) we 
are able to show that
\begin{equation}\label{emanmntxt}
\begin{split}
E\equiv&~u_\nu\hat\nabla_\alpha T_{(eff)}^{\alpha\nu}\\
&= \left(\frac{K}{2}\right)(1+Q^2)(\hat\nabla\cdot u) -(1+2Q^2) (u\cdot\hat\nabla){K} + \frac{K}{2} (u^\mu u^\nu K_{\mu\nu}) \left( 1 + 3 Q^2 + 2Q^4 \right)  \\
&+ \frac{1}{2} \left( \frac{D^2 {K}}{K} \right) \left(1+ Q^2 - 2Q^4  \right)+{\cal O}(1) \\
&=  \frac{(1+2Q^2)}{2}  \left[ -2 (u\cdot\hat\nabla){K} + {K}{(1+Q^2)} (u^\mu u^\nu K_{\mu\nu})+ (1-Q^2)  \left(  \frac{\hat\nabla^2 {K}}{K} \right)  \right]  \\
&~~~~+\left(\frac{K}{2}\right)(1+Q^2)(\hat\nabla\cdot u)+{\cal O}(1)\\
&=\left(\frac{K}{2}\right)(1-Q^2)(\hat\nabla\cdot u) -\left(\frac{1+2Q^2}{{K}} \right)(\hat\nabla_\mu E^\mu)+{\cal O}(1).
\end{split}
\end{equation}
We have already argued above that all the $E_\mu$ are of order unity or smaller. It follows that 
$\nabla_\mu E^\mu$ is of order $D$ or smaller and so \eqref{emanmntxt} implies that 
\begin{equation}\label{nablau}
\begin{split}
&\left(\frac{K}{2}\right)(1+Q^2)(\hat\nabla\cdot u)= {\cal O}(1)\\
\Rightarrow~&(\hat\nabla\cdot u)={\cal O}\left(\frac{1}{D}\right),
\end{split}
\end{equation}
as we claimed above.

\subsection{Stress Tensor and current conservation imply the membrane equations of motion}
We have already demonstrated above that the membrane equations of motion, equation 1.1 of 
\cite{Bhattacharyya:2015fdk}, are sufficient to ensure that the charge current and stress tensors 
dual to the solutions constructed in \cite{Bhattacharyya:2015fdk} are automatically conserved. 
In this brief subsection we point out that the relationship between the membrane equations of motion 
and current conservation can be reversed. Just as the equations of motion imply current and 
stress tensor conservation, the conservation equations in turn imply the membrane equations of motion. 

The argument is immediate. The first equation in 1.1 of \cite{Bhattacharyya:2015fdk} is simply 
\eqref{ebman}, which we have already derived as a consequence of conservation. Plugging 
\eqref{nablau} in equation \eqref{divcur} then yields the second equation in 1.1 of  \cite{Bhattacharyya:2015fdk}. In other words the conservation equations directly imply 
\begin{equation}\label{EOMF}
\begin{split}
&(1+Q^2) (u\cdot\hat\nabla)u^\mu + (1-Q^2) (p^{\mu\nu}\nabla_\nu {K})
 -P^{\mu\nu}\left(\frac{\hat\nabla^2 u_\nu}{K} + K_{\nu\alpha}u^\alpha\right)= {\cal O}(1),\\
 &(u\cdot\hat\nabla) Q +  Q\left[\frac{(u\cdot\hat\nabla){K}}{ K}\right]-\left[\frac{\hat\nabla^2 Q}{{K}} \right]- Q~(u^\mu u^\nu K_{\mu\nu})= {\cal O}(1),\\
\end{split}
\end{equation}
the two membrane equations of motion listed in equations in (1.1) of \cite{Bhattacharyya:2015fdk} .

\subsection{Qualitative discussion of the  uncharged membrane stress tensor and resulting equation of motion}

In this subsection we focus our attention to the relatively simple case of an uncharged membrane. 
In this special case we re-discuss the structure of the membrane stress tensor and resulting equation of 
motion emphasizing qualitative features. The purpose of this subsection is to help the reader 
develop some intuition for the structure of the large $D$ membrane.

Let us first note that the expression for the membrane stress tensor, \eqref{stcc}, simplifies considerably 
when we specialize to the study of uncharged membranes. We find
\begin{equation}\label{stuc} \begin{split}
T_{\mu\nu}=&\left(\frac{1}{8 \pi}\right) \bigg[ \left( \frac{K}{2} \right)u_\mu u_\nu + \left( \frac{1}{2}  \right)K_{\mu\nu} - \left(   \frac{\hat\nabla_\mu u_\nu + \hat\nabla_\nu u_\mu}{2} \right) \\
&~~~~~~~~~~~~+  \left( \frac{u_\mu\hat\nabla_\nu K+u_\nu\hat\nabla_\mu K}{2K}  \right) 
-~\bigg(\frac{u^\alpha u^\beta K_{\alpha\beta}}{2}+\frac{K}{2D}\bigg)g_{\mu\nu}^{(ind,f)}\bigg]\\
&~~~~~~~~~~~~+{\cal O}\left(\frac{1}{D}\right).
\end{split}
\end{equation}
At leading order in the large $D$ limit \eqref{stuc} simplifies to 
\begin{equation}\label{lost}
T_{\mu\nu}= \frac{{K}}{16 \pi} u_\mu u_\nu .
\end{equation} 
This term is of order ${\cal O}(D)$ because ${K}$ is of order ${\cal O}(D)$; all other terms in \eqref{stuc} are of 
order ${\cal O}(1)$. Note, in particular, that the leading order stress tensor lacks a `surface tension' 
term proportional to $g_{\mu\nu}^{(ind,f)}$. \eqref{lost} appears to assert that the large $D$ black hole 
membrane is made up of a collection of pressure free dust particles with density proportional to 
${K}$. This slogan is misleading, as we now explain. 

The divergence of \eqref{lost} is given by 
\begin{equation}\label{dlost}
\hat\nabla^\nu T_{\mu\nu}= \frac{(\hat\nabla\cdot u ){K}}{16 \pi} u_\mu + \frac{(u\cdot \hat\nabla) ( {K} u^\mu) }{16 \pi} .
\end{equation}
The first term in \eqref{dlost} is or order $D^2$ while the second term is of order ${\cal O}(D)$ 
(recall that $(\hat\nabla\cdot u)$ is of order $D$). Setting the divergence of the stress tensor to zero 
at order ${\cal O}(D^2)$ immediately yields the condition that $(\hat\nabla\cdot u)=0$. We emphasize that even though 
\eqref{lost} is the stress tensor of a collection of pressure free dust particles of (variable) density 
$\frac{{K}}{16 \pi}$, one of the equations of motion that follows from the conservation of 
\eqref{lost} asserts that the velocity flow $u^\mu$ is incompressible The reason for this apparent 
dissonance is that terms involving a derivative of the dust density are subleading in $\left(\frac{1}{D}\right)$ 
compared to the term involving the divergence of the velocity.

It might naively seem from \eqref{dlost} that the remaining equations of motion that follow from
the requirement that the stress tensor is conserved is the equation
\begin{equation} \label{wgde}
p^\mu_\nu \cdot \left( u\cdot \hat\nabla\right) ( {K} u^\nu)=
{K}  \left(  u\cdot\hat\nabla\right) u^\mu =0,
\end{equation}
where $p_{\mu\nu}$ represents the world volume projector orthogonal to the velocity $u^\mu$. 
The equation \eqref{wgde}, if correct, would have been the statement that the `proper acceleration'
of $u^\mu$ vanishes on the membrane world volume in the directions orthogonal to $u^\mu$. 
This statement would have been consistent with the interpretation of $u^\mu$ as the velocity 
field of a pressure free gas of dust. 

The equation \eqref{wgde} is in fact incorrect. This is because 
the expression in \eqref{wgde}, which is of order ${\cal O}(D)$, is of the same order as (parts of) the 
divergence of the ${\cal O}(1)$ terms in the stress tensor \eqref{stuc} that were omitted in  
the leading order expression \eqref{lost}. The corrected version of \eqref{wgde} takes these 
additional terms into account, yielding the membrane equation \eqref{memeom} which can be rewritten for 
the special case of an uncharged membrane as 
\begin{equation}\label{memeomuc}
 {K}~ p^\mu_\nu (u\cdot\hat\nabla) u^\nu = 
p^\mu_\nu  \left( \hat\nabla^2u^\nu + u^\alpha K^\nu_{\alpha}-\hat\nabla^\nu K  \right)=0.
\end{equation}
The equation \eqref{memeomuc} can be thought of as an expression of  Newton's force applied to the 
particles that make up the membrane. The LHS represents `mass density' ({K}) times acceleration 
$(u\cdot\hat\nabla) u$ while the RHS of \eqref{memeomuc} describes the forces that these particles 
are subject to. The first two term on the RHS of \eqref{memeomuc} are an expression of the of the 
force of shear viscosity and have their origin in the last term - the shear viscosity term - 
in the first line of \eqref{stuc}. 
\footnote{The first term on the RHS of \eqref{memeomuc} is the classic expression of a viscous 
force, familiar from the Navier Stokes equations. The second term in 
\eqref{memeomuc} is less familiar because it vanishes when the membrane world volume is flat. 
This term arises because  $\hat\nabla_\mu \hat\nabla_\nu u^\mu$ differs, in general, from  
$\hat\nabla_\nu \hat\nabla_\mu u^\mu$ by terms proportional to the 
curvature of the membrane (see \eqref{I4}). } The final term in the RHS of \eqref{memeomuc} has 
its origin in the second term - the bending or curvature energy term - the on the RHS of the first line of \eqref{stuc} (see \eqref{I3}). Roughly speaking this term reflects the fact membrane has a restoring 
force that tries to smooth out gradients of the membrane extrinsic curvature.

\section{Membrane Entropy current}\label{mec}

In the previous section we have found explicit formulae for the stress tensor and a charge 
current on the world volume of the membrane. In this section we will 
use a pullback of the area form on the event horizon of our the spacetimes
dual to large $D$ black hole membranes to define and determine an entropy current on the 
membrane. The Hawking area increase theorem guarantees that 
the entropy current that we define in this section has a divergence that is point wise 
non negative \cite{Bhattacharyya:2008xc}. 

As in the case of the charge current and stress tensor, in this section we first 
explain the general strategy that we use to construct a membrane entropy current 
at every order in the $\frac{1}{D}$ expansion. We then proceed to implement our construction 
at low orders in this expansion, using explicit results for the spacetimes dual to large 
$D$ membranes.

In previous sections we obtained results for the charge current and stress tensor on the membrane 
using the explicit results of \cite{Bhattacharyya:2015fdk} for the spacetime solutions dual to 
membrane motions accurate to first order in $\frac{1}{D}$. The knowledge of the stress tensor and 
charge current to this order proved sufficient to test one of the most important structural features 
of these currents; namely that the requirement that these currents be conserved is a restatement 
of the membrane equations of motion. In a similar manner it is possible to obtain an entropy current 
to first order in the derivative expansion at first order in $\frac{1}{D}$ using the results of 
\cite{Bhattacharyya:2015fdk}. However the divergence of the current obtained in this manner turns out to 
vanish identically. In other words at this order we are blind to one of the most important general 
properties of the entropy current, namely that it is not conserved, but it's divergence is instead 
point wise positive definite. 

In order to capture this basic qualitative feature of the membrane entropy current, in this section 
we work with {\it second} order (in $\frac{1}{D}$) metrics of \cite{Dandekar:2016fvw} dual to second order 
membrane motions. \footnote{In fact the use of the results of \cite{Dandekar:2016fvw}
(rather than those of  \cite{Bhattacharyya:2015fdk}) proves convenient for another 
unrelated reason. In their construction the authors of \cite{Dandekar:2016fvw} have employed a natural all orders definition of the membrane shape 
and velocity that turn out to significantly simplify their metric in the neighbourhood of its event 
horizon in a manner that proves convenient for the analysis we present below.}
 The disadvantage of our reliance on the results of \cite{Dandekar:2016fvw} is that these 
results apply only to uncharged black holes. Second order spacetimes and gauge fields dual to 
charged large $D$ black holes have not yet been obtained. For this 
reason all the explicit results presented in this section apply only to the case of uncharged 
membranes. The extension of this analysis to charged membranes should be a straightforward exercise
once the charged version of \cite{Dandekar:2016fvw} are available. 

After obtaining our formula for the entropy current we turn our attention to the simplest solution of the 
membrane equations of motion - namely the solution for a static spherical membrane - and compute 
energy, charge and entropy of this solution. We demonstrate that the charges of our solution agree with those 
of exact large $D$ black holes to leading order in the large $D$ limit, demonstrating in particular, 
the consistency of our results for membrane currents with the first law of thermodynamics. 

\subsection{Determination of the entropy current}

Consider the spacetime dual to a membrane configuration. Let the {\it bulk} 
spacetime metric at the event horizon be denoted by $G_{AB}$. \footnote{ We emphasize that 
$G_{AB}$ is the full spacetime metric, not the metric restricted to the 
event horizon. }
The precise definition of the membrane shape function and membrane velocity were chosen  
in \cite{Dandekar:2016fvw} to ensure that the spacetime metric $G_{MN}$ takes the 
following simple form at any point on the event horizon 
\begin{equation}\label{stmoh}
G_{MN}= \eta_{MN}+(n-u)_M(n-u)_N + H^{(T)}_{MN}+ H^{Tr} \frac{ { P}_{MN}}{D-2} .
\end{equation}
Here $n_M$ is the normal oneform on the event horizon normalized so that $$\eta^{MN}n_M n_N=1,$$ 
$u_M$ is the `velocity' field chosen to be orthogonal to $n_M$ (i.e. $\eta^{MN} u_M n_N
=0$) and also to be unit normalized (i.e. $\eta^{MN}u_M u_N=-1$).  Moreover 
\begin{equation}\label{pndef}
{ P}_{MN}=\eta_{MN} + u_M u_N -n_M n_N
\end{equation}
and the `tensor' field $H^{(T)}_{MN}$ is orthogonal to both $u^M$ and $n^M$ and is also traceless
i.e. 
$$H^{(T)}_{MN} n^M = H^{(T)}_{MN} u^M=H^{(T)}_{MN} { P}^{MN}=0,$$
(where all indices are raised using the inverse metric $\eta^{MN}$).

\eqref{stmoh} is a formula for the full $D$ dimensional spacetime metric at any point on the event
horizon. The metric \eqref{stmoh} carries information about the inner product between any two 
vectors in the $D$ dimensional tangent space to the full manifold at any point on the metric. 
In this section we will be primarily interested only in the metric restricted to the event horizon 
itself - i.e. the inner product between any two vectors, both of which lie in the $D-1$ dimensional 
tangent space of the event horizon. The tangent space of the event horizon is a codimension one 
subspace of the tangent space of the full space, consisting of those vectors whose dot product 
with $n_M$ vanishes. It is easily verified that a basis for such 
vectors is given by the tangent vector $u^M= \eta^{MN} u_N$ together with any basis for the $D-2$ dimensional space of vectors orthogonal to both $u_M$ and $n_M$. 

If we are concerned only with the tangent space of the event horizon then the metric \eqref{stmoh} 
is easily verified to be equivalent to  
\begin{equation}\label{stmohi}
G^{eh}_{MN}= H^{(T)}_{MN}+ \left(1+ \frac{H^{Tr}}{D-2} \right) { P}_{MN},
\end{equation}
in the sense that 
$$ j^M k^N G_{MN}=j^M k^N G^{eh}_{MN},$$
where $j^A$ and $k^B$ are arbitrary vectors in the tangent space of the event horizon. 
Note that $G^{eh}_{MN} n^M=G^{eh}_{MN}u^M=0$. It follows that the metric 
\eqref{stmohi} has rank $D-2$, even though the event horizon is a $D-1$ dimensional 
manifold, reflecting the fact that the event horizon is a null manifold. 

We will now define $D-2$ dimensional `area form' on the event horizon. 
Consider any point on the event horizon and consider a `patch' of a 
$D-2$ dimensional sub manifold enclosed by the generalized parallelogram formed out of 
the $D-2$ infinitesimal vectors
$\delta t_1^A \ldots \delta t_{D-2}^A$. Let the $D-2$ volume of this patch - computed using the metric induced 
on this patch by \eqref{stmohi} (or equivalently \eqref{stmohi}) -be given by $\delta V_{D-2}$. 
The $D-2$ area form $A_{B_1 \ldots B_{D-2}}$ on the event horizon is defined by equation  
\begin{equation}\label{dmt}
\delta V_{D-2}= A_{B_1 \ldots B_{D-2}} \delta t_1^{A_1} \ldots \delta t_{D-2}^{A_{D-2}},
\end{equation}
(this equation is required to hold for every choice of the infinitesimal vectors 
 $\delta t_i^A$). 

If one of the boundary vectors, $t_1^A$ is chosen to be $u^A$ then it is clear by inspection that the metric induced by \eqref{stmohi} on the $D-2$ dimensional patch is of rank $D-3$, and 
so $\delta V_{D-2}$ vanishes. It follows from \eqref{dmt} that  $A_{B_1 \ldots B_{D-2}}$ must vanish when 
contracted with $u^A$. Now the area form is only well defined in its action on tangent vectors of the 
event horizon. However we could choose instead to generalize this area form to any $D-2$ form that 
can be contracted with tangent vectors on the full manifold, so long as this form has the correct 
action when acting on tangent vectors of the event horizon. Of course this `uplift' of the 
volume form on the event horizon is not unique; we choose a unique `uplift' by arbitrarily imposing 
the additional requirement that the $D-2$ form vanishes when contracted with $n^M$. With this choice 
the uplifted area form on the event horizon necessarily takes the form
\begin{equation}\label{areaform}
A_{A_1 ... A_{D-2}}= \zeta  \epsilon_{A_1 ... A_{D-2} B_1 B_2}  u^{B_1} n^{B_2},
\end{equation}
where $\epsilon_{A_1 ... A_{D-2} B_1 B_2}$ is the standard volume form in flat $D$ dimensional 
space with metric $\eta_{MN}$ and $\zeta$ is yet to be determined. 

We will now determine  $\zeta$ in \eqref{areaform}. Consider a $D-2$ dimensional parallelepiped constructed out 
of $D-2$ basis vectors $\delta t_1^A \ldots \delta t_{D-2}^A$ where these basis vectors are all chosen to be 
orthogonal to both $u_A$ and $n_A$. As above we will denote the volume of this parallelepiped 
- constructed in the spacetime \eqref{stmohi}- by $\delta V_{D-2}$. 

Let us now consider a different problem. Consider a fictional space time with metric $G'_{AB}$ given by
\begin{equation} \label{Gf}
G'_{AB}= G^{eh}_{MN} + n_A n_B -u_A u_B.
\end{equation} 
Using \eqref{stmohi} and \eqref{pndef}  we find
\begin{equation}\label{stmoheq}
G'_{MN}= \eta_{MN} +H^{(T)}_{MN}+ H^{Tr} \frac{{ P}_{MN}}{D-2} .
\end{equation}
Working with the metric $G'_{MN}$ we now consider the $D$ dimensional parallelepiped bounded the vectors 
$\delta t_1^A \ldots  \delta t_{D-2}^A$ together with the additional two vectors $\delta a~ n^M$ and 
$\delta b~ u^M$. Let $\delta V_D$ denote the volume of this $D$ dimensional parallelepiped. A little thought 
will convince the reader that (upto a sign we will not keep track of)
\begin{equation}\label{ddv}
\delta V_{D}= \delta a ~\delta b ~\delta V_{D-2}.
\end{equation}
However $\delta V_D$ is easily independently determined. Using the fact that the 
volume form of the non degenerate $D$ dimensional metric 
$G'_{AB}$ is simply given by $\sqrt{-G'} \epsilon_{A_1  \ldots A_D}$ we conclude that 
\begin{equation}\label{cddv}
\delta V_{D}= \sqrt{-G'} ~\delta a \delta b ~\epsilon_{A_1 ... A_{D-2} B_1 B_2}  u^{B_1} n^{B_2}
\end{equation}
Comparing \eqref{cddv}, \eqref{ddv} and \eqref{areaform} we conclude that (upto a 
sign) 
\begin{equation} \label{detzet}
\zeta= \sqrt{-G'},
\end{equation}  
so that 
\begin{equation}\label{areaformf}
A_{A_1 ... A_{D-2}}= \sqrt{-G} \epsilon_{A_1 ... A_{D-2} B_1 B_2}  u^{B_1} n^{B_2},
\end{equation}
\eqref{areaformf} is our final result for the area form on the world volume of the membrane. 

At least for the case of uncharged black holes, it was demonstrated in 
\cite{Dandekar:2016fvw} that $H^{(T)}_{MN}$ and $H^{Tr}$ both vanish at leading and first subleading 
order in $\frac{1}{D}$ and are nonzero only at order ${\cal O}(1/D^2)$. As $H^{(T)}_{MN}$ is traceless, 
it follows that the contribution of this term to the determinant $G'$ starts at order $\frac{1}{D^4}$. 
On the other hand the trace of $\frac{{\cal P}_{MN}}{D-2}$ is unity. It follows that upto order 
$\frac{1}{D^2}$
\begin{equation}\label{lol} \begin{split}
\sqrt{-G'}&= 1 +\frac{H^{Tr}(\rho=1)}{2}=  1 -\frac{C}{2}+ {\cal O}{(1/D^3)},\\
C&=\frac{2}{{K}^2}(u\cdot K-u.\nabla u)^2,\\
\end{split}
\end{equation}
where we have used the explicit result for $H^{Tr}(\psi=1)$ at order 
$\frac{1}{D^2}$ (see Equation 4.16 of \cite{Dandekar:2016fvw}) to obtain the explicit value for 
$C$. Note that $C$ is of order $\frac{1}{D^2}$. \footnote{We emphasize again that the explicit result 
for $C$ reported in the second line of \eqref{lol} is accurate only for uncharged black holes; 
the computations required to determine $C$ have not yet been performed for charged black holes. 
We leave the determination of $C$ with nonzero charge as a task for the future. }

The entropy current on the membrane is obtained by dualizing the area $D-2$ form and dividing by $4$ 
\cite{Bhattacharyya:2008xc}. 
We obtain 
\begin{equation} \label{entcur}
J_S^\mu=\sqrt{-G'} \frac{u^\mu}{4} \approx \left( 1 - \frac{C}{2} + {\cal O}\left( {1/D^3} \right) \right)  \frac{u^\mu}{4}.
\end{equation}
Note in particular that at leading order in $1/D$
\begin{equation} \label{entcurl}
J_S^\mu= \frac{u^\mu}{4}.
\end{equation}
The first correction to this leading order result occurs at order $1/D^2$.

The divergence of this entropy current \eqref{entcur} is easily computed; 
\begin{equation} \label{entcurd}
\hat\nabla_\mu J_S^\mu= \frac{\hat\nabla_\mu u^\mu}{4} - \frac{u^\mu\hat\nabla_\mu C}{8} +{\cal O}\left( 1/D^3 \right),
 \end{equation} 
$\nabla_A u^A$ was evaluated in \cite{Dandekar:2016fvw} with the result
\begin{equation}\label{divofvel}
\hat\nabla_\mu u^\mu= \frac{p^{\mu\nu} p^{\alpha\beta} \hat\nabla_{(\mu} u_{\alpha)}\hat \nabla_{(\nu} u_{\beta)}}{2K} +{\cal O}(1/D^2).
\end{equation}
Note in particular that $\nabla_A u^A$ is of order $\frac{1}{D}$. As $C$ is of order $\frac{1}{D^2}$, 
if follows from \eqref{entcurd} that 
\begin{equation} \label{entcure}
\hat\nabla_\mu J_S^\mu= \frac{p^{\mu\nu} p^{\alpha\beta} \hat\nabla_{(\mu} u_{\alpha)}\hat \nabla_{(\nu} u_{\beta)}}{8K}
+{\cal O}(1/D^2).
 \end{equation} 
 Note that the RHS of \eqref{entcure} is positive definite. As we have explained earlier in this section 
 this positivity could have been anticipated on general grounds using 
 the Hawking area increase theorem \cite{Bhattacharyya:2008xc}. 
  
 \subsection{Thermodynamics of spherical membranes}
 
 The simplest solution of the membrane equations of motion \eqref{memeom} is a static 
spherical bubble of radius $r_0$ with $u=-dt$ and $Q=Q_0={\rm const}$. In this brief subsection we 
compute the charges of this solution and match these with the thermodynamic charges of 
black holes.

At leading order in the large $D$ limit it follows from \eqref{stcc} that $T_{00}$ for this solution 
is given by 
$$T_{00}= \frac{(D-2)(1+Q_0^2)}{16 \pi r_0}.$$
It follows that the mass $m$ of this solution is given by 
\begin{equation}\label{bhmass} 
m= \Omega_{D-2} r_0^{D-2} T_{00} = \frac{ \Omega_{D-2} (D-2) r_o^{D-3}(1+ Q_0^2)}{16 \pi} .
\end{equation}
Note that $m$ in \eqref{bhmass} agrees with the mass of the black hole \eqref{erns}
at large $D$ (recall that $c_D=1$ in \eqref{mandq} the large $D$ limit). 

The static membrane solution described above has a gravitational tail at infinity. 
It follows from \eqref{npe} and \eqref{Gstatl} that the curvature of this tail is given by 
\begin{equation}\label{curavturetail}
R_{0i0j}=-\frac{8\pi}{(D-2)\Omega_{D-2}}\nabla_i \nabla_j \left( \frac{m}{r^{D-3}} \right), 
\end{equation}
in agreement with \eqref{curvlin}, supporting our identification of $\int T_{00}$ with the mass 
of the membrane. 

 It may be verified that \eqref{curavturetail} agrees with the curvature of the 
black hole solution \eqref{erns} at large $r$ and large $D$. 

In a similar manner the charge density of our solution is given by 
$$J^0 = \frac{Q_0(D-2)}{2 \sqrt{2\pi} r_0} .$$ It follows that the charge of our membrane 
configuration is given by 
\begin{equation}\label{bhcharge} 
q= \Omega_{D-2} r_0^{D-2} J^0 = \frac{\Omega_{D-2} (D-2) Q_0 r_o^{D-3}}{2 \sqrt{2 \pi} } .
\end{equation}
Once again the $q$ in \eqref{bhcharge} agrees with the charge of the black hole \eqref{erns}
at large $D$ (see \eqref{mandq}). 

At large $r$ our charged static membrane solution sources an electric field given by  
\begin{equation}\label{chargetail}
E^i=F^{i0}=-\frac{1}{(D-2)\Omega_{D-2}}\nabla_i \left( \frac{q}{r^{D-3}} \right),
\end{equation}
in agreement with \eqref{flin}.

It follows from this analysis of metric and field strength tails at infinity that 
the spherical membranes studied in this section are dual to
static black holes \eqref{erns} of mass $m$ and charge $Q_0$. 

Finally it follows from \eqref{entcurl} that the entropy $S$ 
of our static solution is given by the area of 
the membrane divided by 4, i.e. 
\begin{equation}\label{entbhm}
S=\frac{\Omega_{D-2} r_0^{D-2}}{4} ,
\end{equation}
in agreement with the entropy \eqref{entbh} of a black hole with the same mass and charge. 

In the study of black hole physics we define the black hole temperature and chemical potential 
via the formulae  \eqref{tempbh} and \eqref{mubh}. These definitions ensure that black holes obey 
the first law of thermodynamics \eqref{thermoalg}. As the spherical membranes of this subsection 
are dual to the corresponding black holes, it is natural to assign them the same temperature 
and chemical potentials 
\begin{equation} \label{tempident} \begin{split}
T&= \frac{(1-Q_0^2) {K}}{4 \pi},\\
\mu&= \frac{Q}{\sqrt{8 \pi}}.
\end{split}
\end{equation}
With these definitions the equation \eqref{thermoalg} can be viewed as the assertion that the 
spherical membranes of this subsection obey the first law of thermodynamics. 

In the spirit of the equations of hydrodynamics, the identification \eqref{tempident} can be made 
locally for any membrane configuration, allowing us to discuss the 
evolution of the local black hole temperature and chemical potential in the course of a 
dynamical evolution.

\section{Radiation in general dimensions} \label{bra}

Earlier in this paper we have determined the explicit form of the stress tensor and charge current
carried by a large D black hole membrane. As the membrane undergoes a dynamical motion these currents 
source electromagnetic and gravitational radiation. The resultant radiation 
field is determined by plugging these currents into radiation formulae: the formulae that determine 
radiation fields in terms of currents. In this section we review radiation formulae in 
arbitrary dimensions.

For completeness - and clarity of presentation - we begin this section with a discussion of the 
formulae for the radiation response of a massless minimally coupled field to a scalar source, even 
though this theory is not needed in order to analyse the black hole membrane. We then 
turn to the analysis of the cases of real interest;  the radiation response of a Maxwell field to an 
arbitrary conserved current and the analysis of the radiation response of the linearized gravitational 
field to an arbitrary conserved stress tensor. In the next section we will apply the formulae developed 
in this section to a particular situation, namely to the study of small fluctuations about a 
static membrane. 

In a certain abstract sense the radiation formulae are extremely simple; they take the schematic
form
\begin{equation}\label{radiform}
 R(x)= \int d^{D}x' G(x-x') J(x'),
\end{equation}
where $J$ is the source, $R$ the radiation response and $G$ a retarded Greens function. 
From this point of view the theory of radiation ends with the computation of the appropriate 
Greens function, a topic we have already discussed in section \ref{bgm}. 

Let us now, however, specialize to situations in which the `centre of mass' of the sources 
is at rest and localized in a shell of radius $R$ about a particular spatial point $x'$. 
If we are interested in the radiation response at points $x$ whose distance
from $x'$ is much larger than $R$, the resultant radiation formulae will clearly be most transparent 
when expressed 
in spherical polar coordinates with $x'$ as origin. In this coordinate system the 
sources and radiation fields are both naturally expanded in a basis of scalar, vector and tensor
spherical harmonics. \eqref{radiform} then turns into an integral transform that expresses 
radiation fields a particular symmetry property (say, e.g. 
radiation fields in the $l^{th}$ vector spherical harmonic) as an integral over sources 
in the same representation. The resultant final expressions are much more explicit - and 
so much more transparent - than \eqref{radiform}.

The starting point for the derivation of the derivation of the formulae presented in this section is 
the expansion of the retarded scalar Greens function of section \ref{bgm} in spherical coordinates. 
In \eqref{Gofs} the Greens function was already presente in polar coordinates in the special case 
of the source at the origin. In Appendix \ref{ocgf} we demonstrate 
that when the source is displaced away from 
the origin, the generalization of \eqref{Gofs} is given by (we assume  $|{\vec r}|> |{\vec r}'|$)
\begin{equation}\label{gfexp}
G(\omega,|{\vec r}-{\vec r}'|)= \frac{i\pi}{2} \sum_{l=0}^\infty
\frac{1}{(r'r)^{\frac{D-3}{2}}}
H_\frac{D-3+2 l}{2}^{(1)} (\omega r) 
J_\frac{D-3+2 l}{2} (\omega r') {\mathcal P}_l(\theta, \theta'),
\end{equation}
where 
${\mathcal P}_l(\theta, \theta')$ is the projector onto the space of functions whose angular dependence is 
a linear combinations of $l^{th}$ scalar spherical harmonics; see around  \eqref{scspproj} 
in the Appendix for more details) and $\theta$ and $\theta'$ are the angular locations of 
${\vec r}$ and ${\vec r}'$ respectively.
\footnote{
While the formulae developed in this section are 
standard extensions of textbook treatments of radiation to arbitrary dimensions, we were unable 
to locate a reference with all formulae presented in a clear and systematic manner and so 
chose to undertake the exercise ourselves. All the formulae developed in this section 
are derived for arbitrary values of $D$ ; however we also emphasize special simplifications 
that occur at large $D$.  }

As mentioned above, all the results of this section are presented in terms of 
 scalar, vector and tensor spherical harmonics. We define and study spherical harmonics in 
 arbitrary dimensions in Appendix \ref{sphericalharmonics}.

\subsection{Scalar Radiation}

Consider a minimally coupled scalar $\phi$ which is zero at early times. The scalar is
subsequently kicked out of its `vacuum' state by coupling to a source according to the equation 
\begin{equation}\label{pheom}
 -\Box \phi = \mathcal{S}.
\end{equation}
where $\mathcal{S}$ is an arbitrary function of space and time. 
\footnote{This equation of motion 
follows from the action 
\begin{equation} \label{mincact}
S= \int d^Dx \sqrt{-G} \left( -  \frac{1}{2}(\nabla \phi)^2 + {\cal S} \phi \right)  .
\end{equation}
}
We further assume that the sourcce $\mathcal{S}(x)$ is spatially localized about a particular 
point in space at all times in a particular Lorentz frame. We choose to work in this 
Lorentz frame, and choose the this point as the origin of our spatial coordinates.  At large 
enough distance from the origin the equation of motion for $\phi$ simplifies
to $-\Box \phi=0$.

\subsubsection{Spherical Expansion of outgoing radiation}

 The most general solution to the minimally coupled scalar equation that is outgoing radiation 
at infinity takes the form (see \eqref{plrs})
\begin{equation} \label{phrad}
\phi(\omega,\vec{x})= \sum_l  \alpha_l(\omega,\theta) \frac{H_{\frac{D+ 2 l-3}{2}} (\omega r)}{r^{\frac{D-3}2}}.
\end{equation}
The functions $\alpha_l(\omega, \theta)$ are angular functions in the $l^{th}$ 
spherical harmonic sector for scalars, i.e. they obey the equation 
$${\mathcal P}_l\alpha_{l'}= \delta_{l, l'} \alpha_l. $$
\footnote{This is an abbreviated form of the equation
\begin{equation}\label{projintdef}
\int d\Omega_{D-2}'P_l(\theta,\theta') \alpha_{\omega,l'} (\theta')= \delta_{l, l'} \alpha_{\omega,l}(\theta).
\end{equation}
}

\subsubsection{Radiation in terms of sources}

The response of the field $\phi$ to the source function ${\mathcal S}$ is given by the formula 
\begin{equation}\label{radscform}
\phi(\omega, {\vec x})= \int d^{D-1} {\vec x}' G(\omega, |{\vec x} - {\vec x}'|) {\mathcal S}(\omega, 
{\vec x}') .
\end{equation}
Here $G(\omega, |{\vec x} - {\vec x}'|)$ is the retarded Greens function determined in 
\eqref{Gofs} and 
\begin{equation} \label{jft}
\mathcal{S}(\omega,{\vec x}')= \int e^{i \omega t} \mathcal{S}({x'}^\mu) dt.
\end{equation}
In other words $\mathcal{S}(\omega,{\vec x}')$ is the source function Fourier transformed in time. 

It is useful to decompose the source into its distinct angular momentum components
\begin{equation} \label{jos}
\mathcal{S}(\omega,\vec{x})= \sum_l \mathcal{S}_l(\omega,r',\theta'),
\end{equation}
where 
\begin{equation}\label{jsphd}
\mathcal{P}_l\mathcal{S}_{l'}= \delta_{l,l'} \mathcal{S}_l.
\end{equation}
In other words $\mathcal{S}_l(\omega, r', \theta')$ is the part 
of $\mathcal{S}(\omega,{\vec x}')$ that transforms in the $l^{th}$ spherical 
harmonic representation.
Inserting the expansion \eqref{gfexp} for the Greens function in \eqref{radscform} and specializing 
that formula to large $r$, it is easily verified that \eqref{radscform} reduces to \eqref{phrad} 
with 
\begin{equation} \label{aval}
\alpha_l(\omega, \theta)= \frac{i \pi}{2} \int dr' J_\frac{D-3+2 l}{2} (\omega r')~r'^{\frac{D-1}{2}}\mathcal{S}_l(\omega,r', \theta) ,
\end{equation} 
\eqref{aval} is our final formula for scalar radiation. In the rest of this subsection we will study 
limits and properties of the formula \eqref{aval}.

\subsubsection{The static limit}

Recall that 
\begin{equation}\label{saj}
J_n(x) \approx \frac{ \left( \frac{x}{2} \right)^n}{\Gamma(n+1)}, ~~~
(x^2 \ll n).
\end{equation}
This observation may be used to simplify \eqref{aval} in two different 
physical situations. The first is the static limit $ \omega \to 0$ 
taken at finite $D$. The second - of particular interest to this paper 
- is the limit in which $\omega R$ is held fixed as $D$ is taken to 
infinity (here $R$ is an estimate of the spatial size of the support for 
the scalar source ${\mathcal S}$ which is assumed to be of finite extent). 
In either limit we obtain the simplified formula 

\begin{equation} \label{avalsimp}
\alpha_l(\omega,\theta)= {\tilde \alpha}_l\omega^{l+\frac{D-3}2}
\int dr'(r')^{l+D-2}
\mathcal{S}_l(\omega,r', \theta) ,
\end{equation}  
with 
\begin{equation}\label{tal}
\begin{split}
{\tilde \alpha_l} &= \frac{i \pi }{ 2^{l+\frac{D-1}2}}\frac{1}{\Gamma\left(l+\frac{D-1}{2}\right)}.\\
\end{split}
\end{equation}

In this subsubsection we study the static limit, postponing our stdy of the large $D$ limit to the next 
subsubsection.

In the static limitthere is a further simplification. 
As $\omega$ is taken to zero the Hankel function 
in \eqref{phrad} may be approximated by its small argument expansion 
\eqref{hfa} and we find 
\begin{equation} \label{scastat}
\phi(0,r,\theta) = \frac{1}{r^{D-3}}\sum_{l=0}^{\infty}\left(\frac{1}{2}\right)\frac{2}{2l+D-3}\int dr'r'^{D-2}\left(\frac{r'}{r}\right)^l S_l(0,r',\theta),
\end{equation}
\eqref{scastat} is simply the multipole expansion of the solution to the Euclidean equation 
$\nabla^2 \phi= -\mathcal{S}$ in $D-1$ Euclidean dimensions, and may directly be obtained by 
inserting \eqref{greenstat} into \eqref{radscform}. Note that the 
static field falls off much faster at large $r$ than the radiation field does; this is the 
generalization of the familiar fact that the Coulomb field in $D=4$ falls off like $\frac{1}{r^2}$ 
while a radiation field decays more slowly, like $\frac{1}{r}$.

\subsubsection{Large D limit}

Let us now turn to the  large $D$ limit at finite $\omega$. The 
Sterling approximation allows us to 
simplify the expression ${\tilde \alpha_l}$; we find 
\begin{equation}\label{alpha}
{\tilde \alpha_l} \approx \frac{i\sqrt{\pi}}{2 D^{\frac{D}2}}\left(\frac{e^{l+\frac{D-3}2}}{D^{l-1}}\right).
\end{equation}

We would now like to estimate how fast our system loses `charge' via radiation at large $D$. In order 
to make this 
question precise, let us slightly generalize the discussion of this subsection to the case 
of a complex scalar  field $\phi$ and source ${\mathcal S}$. All the formulae derived above continue 
to apply. The advantage is that our scalar field now carries a current given by 
$J_M=i ( \partial_M \phi^* \phi-\phi^* \partial_M \phi)$. 
Let us assume that the source function is 
nonzero only in a shell of radius of order $R$ and vanishes outside the shell. In the external region 
the current $J_M$ is conserved. We will now estimate first the integrated density of this charge 
contained in the field $\phi$ to the exterior of the shell of ${\mathcal S}$ and second 
the rate of loss of charge to infinity by radiation. The ratio of these two quantities 
will give us an estimate of the rate of loss of charge due to radiation per unit time.

Using \eqref{phrad} we see that the charge carried by our configuration 
in the $l^{th}$ mode is of order 
$$ \frac{R D^D}{(\omega R)^{2l + D-4}}\int_{S^{D-2}} |\alpha_l(\theta)^2|,$$
where the source is assumed to be of size $R$ we have retained only leading 
order terms in the limit of large $D$. 
On the other hand the rate of energy lost due to radiation 
is of order
$$ \int_{S^{D-2}} |\alpha_l(\theta)^2| .$$
It follows that the fractional loss of charge by radiation per unit time is 
of order $ \frac{ (\omega R)^{2 l + D-4}}{R D^D}$, and so is extremely small 
at large $D$.

\subsection{Electromagnetic Radiation}

In this section we will find the solution to the Maxwell equation 
\begin{equation}\label{maxeq} 
\nabla^M F_{MN} = \mathcal{J}_{N},
\end{equation} 
sourced by an arbitrary localized charge current ${\mathcal J}_M$.
It follows from \eqref{maxeq} that 
\begin{equation} \label{lapmax}
\Box F_{MN} = -(\nabla_N \mathcal{J}_M - \nabla_M \mathcal{J}_N).
\end{equation}
In particular the electric field defined by 
\begin{equation}\label{ef}
{ E}_i= F_{0i},
\end{equation}
obeys the equation
\begin{equation}\label{nablasqv} \begin{split}
\Box {\vec E}& =-{\vec {\mathcal J}}_{eff},\\
{\vec {\mathcal J}}_{eff}&={\vec \nabla} {\mathcal J }_0 - \partial_0 {\vec {\mathcal J }}.
\end{split}
\end{equation}
 
In order to determine the radiation response to a current it is sufficient
to determine the electric field at large distances; all 
other components of the field strength may be obtained rather simply from the 
electric field using the Bianchi identity. To see how this works recall 
that the Bianchi identity with free indices $0, i, j$ takes the form
\begin{equation} \begin{split} 
&\partial_0 F_{ij} = \nabla_i E_j - \nabla_j E_i, ~~~{\rm i.e}\\
& F_{ij}(\omega,\vec{x}) = 
\frac{i( \nabla_i E_j(\omega,\vec{x}) - \nabla_j E_i (\omega,\vec{x}))}{\omega}.
\end{split}
\end{equation}
It follows that the $F_{ij}$ is completely determined in terms of 
${\vec E}$ at every nonzero $\omega$. 

\subsubsection{Free outgoing solutions of Maxwell's equations} 

At large distances where the source ${\mathcal J}^M$ vanishes, 
\eqref{nablasqv} reduces to 
\begin{equation}\label{lapelv}
\Box {\vec E}=0.
\end{equation}
Now ${\vec E}$ is a vector field in spacetime. As we have explained
in Appendix \ref{dvf}, any such field can be written in terms of 
two scalar fields and one divergenceless, purely tangential (to the 
sphere) vector field. This tangential divergenceless vector field can 
be expanded in vector spherical harmonics while the two scalars
are expanded in scalar spherical harmonics. A useful basis for 
this decomposition is listed in \eqref{arbitvector} in the Appendix. 
In this basis the action of $\nabla^2$ is diagonal and is listed in 
\eqref{actnablav}. 

It follows immediately from \eqref{actnablav} that the most general solution to
\eqref{lapelv} is given by a vector of the form \eqref{arbitvector} 
where the radial dependence of the coefficients $\alpha_l$, ${\vec \gamma}_l$ and 
$\beta_l$ respectively is the same as that of the $(l-1)^{th}$, $l^{th}$ and 
$(l+1)^{th}$ angular momentum component of the modes in \eqref{phrad}. 
In other words the most general harmonic solution for ${\vec E}$ is given by 
\begin{equation} \label{emad} \begin{split}
{\vec E} (\omega,\vec{x}) &= 
\sum_{l=0}^{\infty} \left( \frac{H_{\frac{D+ 2 l-5}{2}} (\omega r)}{r^{\frac{D-3}2}}
 {\vec {\cal A}}^{-} [
S^-_l(\omega,\theta) ] +  \frac{H_{\frac{D+ 2 l-1}{2}} (\omega r)}{r^{\frac{D-3}2}}
 {\vec {\cal A}}^{+} [
S^+_l(\omega,\theta) ] \right)\\
& +  \sum_{l=1}^{\infty} \left( \frac{H_{\frac{D+ 2 l-3}{2}} (\omega r)}{r^{\frac{D-3}2}}
 {\vec V}_l \right) ,
\end{split}
\end{equation}
where $S^{\pm}_l$ are arbitrary $r$ independent 
scalar functions in the $l^{th}$ scalar 
spherical harmonic sector and ${\vec V}_l$ is an arbitrary 
vector function in the 
$l^{th}$ vector spherical harmonic sector, normalized so that each of the 
Cartesian components of  ${\vec V}_l$ is a function only of the angles and 
is independent of $r$. Here ${\vec {\cal A}}^{-} [
S^-_l(\omega,\theta)$ and ${\vec {\cal A}}^{+} [
S^+_l(\omega,\theta) ]$ are the maps from scalar to vector functions in $R^{D-1}$ 
defined in \eqref{lcs}.

The equation  
\begin{equation}\label{de}
{\vec \nabla}\cdot{\vec E}=0,
\end{equation}
(which also holds in the absence of sources) 
further constrains radiation fields. Using 
\eqref{divbasis} and appropriate recursion relations for Hankel functions 
we demonstrate in subsection \ref{consistem} below that 
\begin{equation} \label{rbs}
lS_l^- = (l+D-3) S_l^+,
\end{equation}
\eqref{emad} with the constraint \eqref{rbs} is the most general solution 
to the source free Maxwell equations. 

At very large distances,  $\omega r \gg D^2$,  \eqref{emad} simplifies to 
\begin{equation} \label{emadlr} \begin{split}
{\vec E} (\omega,\vec{x}) &= 
\sqrt{\frac{2}{\pi \omega}}\frac{e^{i r \omega}}{r^{{D-2}\over2}}\sum_{l=0}^{\infty} e^{\frac{-i(D+2l)\pi}{4}}\left({\vec {\cal A}}^{+} [
S^+_l(\omega,\theta) ]
 -{\vec {\cal A}}^{-} [
S^-_l(\omega,\theta) ] 
  \right)\\
& +  i\sqrt{\frac{2}{\pi \omega}}\frac{e^{i r \omega}}{r^{{D-2}\over2}}\sum_{l=1}^{\infty} e^{\frac{-i(D+2l)\pi}{4}}  
 {\vec V}_l \\
&= \sqrt{\frac{2}{\pi \omega}}\frac{e^{i r \omega}}{r^{{D-2}\over2}}\sum_{l=0}^{\infty} e^{\frac{-i(D+2l)\pi}{4}}\left(\hat{r}\left(lS_l^--(l+D-3)S_l^+\right)-r\vec{\tilde{\nabla}}\left(S_l^-+S_l^+\right)
  \right)\\
& +  i\sqrt{\frac{2}{\pi \omega}}\frac{e^{i r \omega}}{r^{{D-2}\over2}}\sum_{l=1}^{\infty} e^{\frac{-i(D+2l)\pi}{4}}  
 {\vec V}_l.
\end{split}
\end{equation}
Where we have used \eqref{lcs} in the last step. Note, in particular, that the radiation electric 
field is orthogonal to ${\hat r}$ - and so is finally well approximated by a local plane wave
- at these distances.

\subsubsection{Radiation in terms of sources}

In order to determine the radiation field sourced by an arbitrary current
we expand the effective source ${\vec {\mathcal J}}_{eff}$ in the 
form \eqref{arbitvector}. In particular let 
\begin{equation} \label{arbitscalarJ}
{\vec {\mathcal J} }_{eff} =
\left( {\vec {\cal A}}^-[\mathfrak{a}] +
{\vec {\cal A}}^+[\mathfrak{b}]+ {\vec {\mathfrak c}  } \right),
\end{equation}
where $\mathfrak{a}$, $\mathfrak{b}$ and ${\vec {\mathfrak c}}$
respectively 
play the role of $\alpha$, $\beta$ and ${\vec \gamma}$ in \eqref{arbitvector}.
In Appendix \ref{segf} we have determined the action of the retarded Greens function 
on an arbitrary vector field expanded in the basis employed in \eqref{arbitscalarJ}. 
Using \eqref{sov}, \eqref{sovv} and \eqref{vrgf} of the Appendix we find  that 
the electric field at large $r$ takes the form \eqref{emad} with 
\begin{equation} \label{avvv} \begin{split}
S^-_l&= \frac{i \pi}{2} \int dr' J_{\frac{D-3+2 (l-1)}{2}} (\omega r')~
r'^{\frac{D-1}{2}}\mathfrak{a}_l(\omega,r', \theta) ,\\
S^+_l&= \frac{i \pi}{2} \int dr' J_{\frac{D-3+2 (l+1)}{2}} (\omega r')~
r'^{\frac{D-1}{2}}\mathfrak{b}_l(\omega,r', \theta), \\
{\vec V}_l&= \frac{i \pi}{2} \int dr' J_{\frac{D-3+2 l}{2}} (\omega r')~
r'^{\frac{D-1}{2}}{\vec {\mathfrak c}}_l(\omega,r', \theta) .\\
\end{split}
\end{equation}
The conservation of the electromagnetic current ${\vec {\mathcal J}}$
can be used to show that the effective current obeys the following 
equation
\begin{equation}\label{fcc}
{\vec \nabla}\cdot{\vec {\mathcal J}}^{eff}
= \Box {\mathcal J}_0.
\end{equation}
This relation can be used to verify that the coefficients \eqref{avvv} 
obey \eqref{rbs}.
\footnote{At the formal level it is obvious that this had to work.  
\begin{equation} \label{boxe}
\nabla\cdot E= -{\mathcal J}_0.
\end{equation}
This is simply the Gauss law, and ensures that the Electric field is 
divergence free in the absence of a source. The fact that the actual 
formulae \eqref{avvv} obey \eqref{rbs} may be regarded as a check on our 
algebra.}

\subsubsection{Special limits}

As in the previous subsection \eqref{saj} may be used to simplify 
\eqref{avvv} in both the static and the the large 
$D$ limits. In either limit we obtain the simplified formula 

\begin{equation} \label{emradcompsimpv}
\begin{split}
S^-_l(\omega,\theta) &= {\tilde S}^-_l\omega^{l+\frac{D-5}2}
\int dr'(r')^{l+D-3}
\mathfrak{a}(\omega,r',\theta),\\
S^+_l(\omega,\theta) &= {\tilde S}^+_l\omega^{l+\frac{D-1}2}
\int dr'(r')^{l+D-1}
\mathfrak{b}(\omega,r',\theta),\\
V_l(\omega,\theta)_i &= {\tilde V}_l\omega^{l+\frac{D-3}2}
\int dr'(r')^{l+D-2}
\mathfrak{c}(\omega,r',\theta),
\end{split}
\end{equation}  
with 
\begin{equation}\label{coef}
\begin{split}
{\tilde S}^-_l &= \frac{i\pi}{ 2^{l+\frac{D-3}2}}\frac{1}{\Gamma\left(l+\frac{D-3}{2}\right)},\\
{\tilde S}^+_l &= \frac{i\pi}{ 2^{l+\frac{D+1}2}}\frac{1}{\Gamma\left(l+\frac{D+1}{2}\right)},\\
{\tilde V}_l &= \frac{i\pi}{ 2^{l+\frac{D-1}2}}\frac{1}{\Gamma\left(l+\frac{D-1}{2}\right)}.
\end{split}
\end{equation}

As in the previous subsection in the large 
$D$ limit at fixed $\omega$ we use the Sterling approximation to further 
simplify ${\tilde \alpha_l}$; we find 
\begin{equation}\label{coefld}
\begin{split}
{\tilde S}^-_l \approx \frac{i\sqrt{\pi}}{2 D^{\frac{D}2}}\left(\frac{e^{l+\frac{D-5}2}}{D^{l-2}}\right),\\
{\tilde S}^+_l \approx \frac{i\sqrt{\pi}}{2 D^{\frac{D}2}}\left(\frac{e^{l+\frac{D-1}2}}{D^{l}}\right),\\
{\tilde V}_l \approx \frac{i\sqrt{\pi}}{2 D^{\frac{D}2}}\left(\frac{e^{l+\frac{D-3}2}}{D^{l-1}}\right).
\end{split}
\end{equation}

As in the previous subsection \eqref{coefld} does not apply in the 
static limit at fixed $D$. In this limit, however, the small argument 
expansion of the Hankel function leads to simplifications. In Appendix 
\ref{consist} we demonstrate that in the limit $\omega \to 0$ the radiation 
formulae \eqref{avvv} yield results consistent with the familiar 
formulae of electrostatics
\begin{equation}\label{staticvec} \begin{split}
{\vec E}&=- \nabla \Phi_E,\\
\nabla^2 \Phi_E&= {\mathcal J}_0(r'),\\
F_{ij}&=\partial_i A_j -\partial_j A_i,\\
\nabla^2 {\vec A} &= {\vec {\mathcal J}}.\\
\end{split}
\end{equation}

\subsection{Gravitational Radiation}

In this section we will find the unique purely outgoing 
solution to the linearized 
Einstein equation; i.e. the linearized version of 
\begin{equation}\label{eineq} 
R_{MN} = 8\pi \mathcal{T}_{MN}, 
\end{equation} 
as a functional of an arbitrarily specified conserved 
${\mathcal T}_{MN}$.

It follows from \eqref{eineq} that, to linear order in an expansion 
around flat space
\begin{equation}\label{lapein}
\begin{split}
\Box R_{MNPQ} =& 8 \pi \left(\partial_M\partial_PT_{NQ}-\partial_M\partial_QT_{NP}-\partial_N\partial_PT_{MQ}+\partial_N\partial_QT_{MP}\right)\\
&-\frac{8\pi}{D-2}\left(\eta_{MP}\partial_N\partial_QT-\eta_{MQ}\partial_N\partial_PT-\eta_{NP}\partial_M\partial_QT+\eta_{NQ}\partial_M\partial_PT\right).
\end{split}
\end{equation}
In particular
\begin{equation}\label{nablasqg} \begin{split}
\Box R_{0i0j}& =-({\mathcal T}_{eff})_{ij},\\
(\mathcal{T}_{eff})_{ij} &= 8 \pi \left( 
\omega^2\mathcal{T}_{ij}-
i\omega(\partial_i\mathcal{T}_{0j}+\partial_j\mathcal{T}_{0i})- 
\partial_i\partial_j\left(\mathcal{T}_{00}+\frac{\mathcal{T}}{D-2}\right)
-\eta_{ij}\omega^2\frac{\mathcal{T}}{D-2} \right) .
\end{split}
\end{equation}

As in the previous subsection it is sufficient to consider $R_{0i0j}$ as 
all other curvature components are easily obtained from this one by 
use of the Bianchi identity. \footnote{ The Bianchi identity yields
 \begin{equation}\label{Biang}
 \begin{split}
 R_{0ijk}&=\frac{-i}{\omega}(\partial_kR_{0i0j}-\partial_jR_{0i0k}),\\
 R_{ijpq}&=\frac{-i}{\omega}(\partial_qR_{0pij}-\partial_pR_{0qij}).
 \end{split}
 \end{equation} }
One way to understand this statement is to work in the $h_{0M} = 0$. In this gauge and in Fourier space
\begin{equation}\label{metRrel}
h_{ij} = \frac{-2}{\omega^2}R_{0i0j}.
\end{equation}
As all gauge invariants can be built out of $h_{ij}$, it follows that all gauge invariant information
is also contained in $R_{0i0j}$ except in the special limit $\omega \to 0$.

\subsubsection{Parametrization of vacuum solutions}
When all source currents vanish \eqref{nablasqg} reduces to 
\begin{equation}\label{lapg}
\Box R_{0i0j}=0.
\end{equation}

As in the previous subsection the most general tensor field 
$R_{0i0j}$ can
be decomposed into four scalars, two divergence free tangential vector
fields and one divergence free, traceless tangential tensor field - the later
can be decomposed in tensor spherical harmonics. The form of this expansion 
is given in \eqref{arbitscalarg}. Away from all sources the equation 
\eqref{lapg} determines the radial dependence of all the 
coefficient functions in \eqref{arbitscalarg}. It follows 
from \eqref{actnabla} that the radial dependence of $\kappa_l$, $\gamma_l$ 
and $\chi^{ij}_l$ (in the decomposition \eqref{arbitscalarg} applied to $R_{0i0j}$)
is precisely that of the coefficient of the mode 
$\alpha_l$ in the equation \eqref{phrad}. On the 
other hand the radial dependence of $\alpha_l$, ${\vec \phi}_l$, 
${\vec \psi}_l$ and $\beta_l$ is that of the modes with angular momentum 
$l-2$, $l-1$, $l+1$ and $l+2$ respectively in \eqref{phrad}. It thus follow
that away from all sources
\begin{equation} \label{gmad} \begin{split}
R_{0i0j}(\omega,\vec{x}) &= 
\sum_{l=0}^{\infty} \bigg( \frac{H_{\frac{D+ 2 l-7}{2}} (\omega r)}{r^{\frac{D-3}2}}
 (\mathcal{C}^{-})_{ij} [
S^-_l(\omega,\theta) ] +  \frac{H_{\frac{D+ 2 l+1}{2}} (\omega r)}{r^{\frac{D-3}2}}
 (\mathcal{C}^{+})_{ij} [
S^+_l(\omega,\theta) ] \\&+ \frac{H_{\frac{D+ 2 l-3}{2}} (\omega r)}{r^{\frac{D-3}2}}
 \left((\mathcal{C}^{0})_{ij} [
S^0_l(\omega,\theta) ]
+ \delta_{ij}S^{Tr}_l(\omega,\theta)\right)\bigg)\\
&+ \sum_{l=1}^{\infty}\left( \frac{H_{\frac{D+ 2 l-5}{2}} (\omega r)}{r^{\frac{D-3}2}}
 (\mathcal{B}^{-})_{ij} [
V^-_l(\omega,\theta) ] +  \frac{H_{\frac{D+ 2 l-1}{2}} (\omega r)}{r^{\frac{D-3}2}}
 (\mathcal{B}^{+})_{ij} [
V^+_l(\omega,\theta) ]\right)\\
& +  \sum_{l=2}^{\infty} \left( \frac{H_{\frac{D+ 2 l-3}{2}} (\omega r)}{r^{\frac{D-3}2}}
 (X_l)_{ij} \right) ,
\end{split}
\end{equation}
where $S^{\pm}_l$, $S^0_l$ and $S^{Tr}_l$ are arbitrary $r$ independent 
scalar functions in the $l^{th}$ scalar 
spherical harmonic sector, ${\vec V}^{\pm}_l$ is an arbitrary 
vector function in the 
$l^{th}$ vector spherical harmonic sector, normalized so that each of the 
Cartesian components of  ${\vec V}^{\pm}_l$ are functions only of the angles and 
are independent of $r$,  and $X_l$ is an arbitrary 
symmetric, divergencelsess, traceless tensor function in the 
$l^{th}$ vector spherical harmonic sector, normalized so that each of the 
Cartesian components of  $X_l$ are functions only of the angles and 
are independent of $r$, and all functionals (e.g. $({\mathcal C})_{ij}$) 
were defined in \eqref{lct}. 

\eqref{gmad} is the most general solution to the linearized dynamical 
Einstein equations; however the general solution \eqref{gmad} does not 
automatically solve the Einstein constraint equations.
Using 
\begin{equation}\label{fccgp}
\nabla^i({\mathcal T}_{eff})_{ij} = 8\pi\Box\left(i\omega\mathcal{T}_{0j}+\partial_j\left(\mathcal{T}_{00}+\frac{\mathcal{T}}{D-2}\right)\right),
\end{equation}
we find the linearized gravity analogue of the electromagnetic 
Gauss law of the previous subsection 
\begin{equation}\label{fccg}
{\nabla}^i R_{0i0j} = 
-8\pi \left(i\omega\mathcal{T}_{0j}+\partial_j\left(\mathcal{T}_{00}+\frac{\mathcal{T}}{D-2}\right)\right).
\end{equation}
In particular, in the absence of sources we have
\begin{equation}\label{deg}
\nabla^i R_{0i0j}=0.
\end{equation}
Using \eqref{deg}, \eqref{divbasis} and appropriate recursion relations for 
Hankel functions we find 
\begin{equation} \label{grbs}
\begin{split}
&(l-1)S^-_l = \frac{(l+D-3)S^0_l}{2(2l+D-3)}\left((2l+D-3)-\frac{4l}{D-1}\right),\\
&(l+D-2)S^+_l = \frac{lS^0_l}{2(2l+D-3)}\left((2l+D-3)-\frac{4(l+D-3)}{D-1}\right),\\
&(l-1)\vec{V}_l^- = (l+D-2) \vec{V}_l^+.
\end{split}
\end{equation}

Moreover it is easily verified that
\begin{equation}\label{traceofscal}
({\mathcal T}_{eff})_i^i = 8\pi\Box \left(\mathcal{T}_{00}+\frac{\mathcal{T}}{D-2}\right),
\end{equation}
so that
\begin{equation}\label{tor}
R^i_{~0i0}= -8\pi\left(\mathcal{T}_{00}+\frac{\mathcal{T}}{D-2}\right).
\end{equation}
The equation \eqref{tor} implies that $R_{0i0j}$ is traceless in the 
absence of sources; this sets 
\begin{equation}\label{trfu}
S^{Tr}_l = 0.
\end{equation} 
\eqref{gmad} with the constraint \eqref{grbs} and \eqref{trfu} 
is the most general source free solution of the linearized Einstein equations.  
Notice that the general radiation field is parametrized by a single 
scalar function, a single divergence free 
vector function and a single traceless divergence free tensor 
function on the unit sphere. \\
In the large distance limit $\omega r \gg D^2$ \eqref{gmad} simplifies to 
\begin{equation} \label{gmadlr} \begin{split}
R_{0i0j} (\omega,\vec{x}) &= 
i\sqrt{\frac{2}{\pi \omega}}\frac{e^{i r \omega}}{r^{{D-2}\over2}}\sum_{l=0}^{\infty} e^{\frac{-i(D+2l)\pi}{4}}\left({\cal C}_{ij}^{0} [
S^0_l(\omega,\theta) ]-{\cal C}_{ij}^{+} [
S^+_l(\omega,\theta) ]
 -{\cal C}_{ij}^{-} [
S^-_l(\omega,\theta) ] 
  \right)\\
&+ \sqrt{\frac{2}{\pi \omega}}\frac{e^{i r \omega}}{r^{{D-2}\over2}}\sum_{l=1}^{\infty} e^{\frac{-i(D+2l)\pi}{4}}\left({\cal B}_{ij}^{+} [
{\vec V}^+_l(\omega,\theta) ]
 -{\cal B}_{ij}^{-} [
{\vec V}^-_l(\omega,\theta) ] 
  \right)
\\
& +  i\sqrt{\frac{2}{\pi \omega}}\frac{e^{i r \omega}}{r^{{D-2}\over2}}\sum_{l=2}^{\infty} e^{\frac{-i(D+2l)\pi}{4}}  
 (X_l)_{ij} \\ 
&= i\sqrt{\frac{2}{\pi \omega}}\frac{e^{i r \omega}}{r^{{D-2}\over2}} \sum_{l=0}^{\infty} e^{\frac{-i(D+2l)\pi}{4}}\\&\bigg(\hat{r}_i\hat{r}_j\left(l(l+D-3)\left(\frac{D-3}{D-1}\right)S^0_l- (l+D-3)(l+D-2)S^+_l-l(l-1)S^-_l\right)\\
&+ r\hat{r}_i\tilde{\nabla}_j\left(\frac{D-3}2S_l^0 + (l+D-2)S_l^+ - (l-1)S_l^-\right)+\{i\leftrightarrow j\}\\
&- r^2\tilde{\nabla}_{ij}\left(S_l^0+S_l^++S_l^-\right)\\
&-\Pi_{ij}\left(\frac{2l(l+D-3)}{D-1}S_l^0+(l+D-3)S_l^+-lS_l^-\right)\bigg)\\
&+\sqrt{\frac{2}{\pi \omega}}\frac{e^{i r \omega}}{r^{{D-2}\over2}}\sum_{l=0}^{\infty} e^{\frac{-i(D+2l)\pi}{4}}\bigg(\hat{r}_i\left((l-1)(V_l)_j^--(l+D-2)(V_l)_j^+\right)\\ &-r\tilde{\nabla}_i\left((V_l)_j^-+(V_l)_j^+\right)
  +\{i\leftrightarrow j\}\bigg)\\
& +  i\sqrt{\frac{2}{\pi \omega}}\frac{e^{i r \omega}}{r^{{D-2}\over2}}\sum_{l=1}^{\infty} e^{\frac{-i(D+2l)\pi}{4}}  
 (X_l)_{ij}.
\end{split}
\end{equation}
Where we have expanded the expressions of $\mathcal{C}$s and $\mathcal{B}$s as given in \eqref{lct} using \eqref{lcv} in the last step. Note that the radiation field is polarized orthogonal to the line of 
sight from the observation to the source point in this limit.

\subsubsection{Radiation in terms of sources}

These scalar, vector and tensor functions on the unit sphere that 
characterize radiation may be determined in 
terms of the $({\mathcal T}_{eff})_{ij}$ as follows. The tensor field 
$({\mathcal T}_{eff})_{ij}$
may be decomposed along the lines of \eqref{arbitscalarg} as 
\begin{equation} \label{arbitten}
\begin{split}
(\mathcal{T}_{eff})_{ij} =&
\left( {\cal C}_{ij}^-[\mathfrak{a}] + {\cal C}_{ij}^+[\mathfrak{b}] + {\cal C}_{ij}^0[\mathfrak{c}] + \delta_{ij}\mathfrak{d}\right)\\ 
&+ \left({\cal B}_{ij}^-[{\vec {\mathfrak u}} ] + 
{\cal B}_{ij}^+[{\vec {\mathfrak v}}]\right)
+ \mathfrak{z}_{ij},
\end{split}
\end{equation}
where ${\cal C}_{ij}^-[\mathfrak{a}]$,  ${\cal C}_{ij}^+[\mathfrak{b}]$ and  ${\cal C}_{ij}^0[\mathfrak{c}]$
are the maps from scalars to tensors in $R^{D-1}$ defined in \eqref{lct} in the Appendix. 

In what follows we  use obvious notation to denote the $l^{th}$ spherical harmonic 
components of the scalar, tensor and vector functions on the unit sphere 
that appear in \eqref{arbitten}.  
For instance  $\mathfrak{a}_l$ denotes the projection of the scalar function 
${\mathfrak a}$ to the $l^{th}$ scalar spherical harmonic sector, while 
$({\mathfrak z}_l)_{ij}$ denotes the projection of the tensor field 
${\mathfrak z}_{ij}$ to the $l^{th}$ tensor harmonic sector. As in the 
previous subsection, the action of the retarded Greens function on an arbitrary tensor 
field \eqref{arbitten} takes the for, \eqref{sott} and \eqref{trgf}
and we obtain 
\begin{equation} \label{avvvg} \begin{split}
S^-_l&= \frac{i \pi}{2} \int dr' J_{\frac{D-3+2 (l-2)}{2}} (\omega r')~
r'^{\frac{D-1}{2}}\mathfrak{a}_l(\omega,r', \theta) ,\\
S^+_l&= \frac{i \pi}{2} \int dr' J_{\frac{D-3+2 (l+2)}{2}} (\omega r')~
r'^{\frac{D-1}{2}}\mathfrak{b}_l(\omega,r', \theta), \\
S^0_l&= \frac{i \pi}{2} \int dr' J_{\frac{D-3+2 l}{2}} (\omega r')~
r'^{\frac{D-1}{2}}\mathfrak{c}_l(\omega,r', \theta), \\
{\vec V}^-_l&= \frac{i \pi}{2} \int dr' J_{\frac{D-3+2 (l-1)}{2}} (\omega r')~
r'^{\frac{D-1}{2}}{\vec {\mathfrak u}}_l(\omega,r', \theta), \\
{\vec V}^+_l&= \frac{i \pi}{2} \int dr' J_{\frac{D-3+2 (l+1)}{2}} (\omega r')~
r'^{\frac{D-1}{2}}{\vec {\mathfrak v}}_l(\omega,r', \theta), \\
(X_l)_{ij}&= \frac{i \pi}{2} \int dr' J_{\frac{D-3+2 l}{2}} (\omega r')~
r'^{\frac{D-1}{2}}({\mathfrak z}_l)_{ij}(\omega,r', \theta),\\
S^{Tr}_l&= \frac{i \pi}{2} \int dr' J_{\frac{D-3+2 l}{2}} (\omega r')~
r'^{\frac{D-1}{2}}\mathfrak{d}_l(\omega,r', \theta)=0.
\end{split}
\end{equation}

Although it may not be apparent from a casual glance, the 
solution \eqref{avvvg} obeys the constraints \eqref{trfu} and \eqref{grbs}. 
\eqref{trfu} is obeyed after a partial integration 
simply because $\mathfrak{d} $ equals the operator $\Box$ acting on another 
function (see \eqref{fccgp}). Moreover the expressions for  
$\mathfrak{a}_l$, $\mathfrak{b}_l$ and $\mathfrak{c}_l$ 
in \eqref{avvvg} may also be shown to obey \eqref{grbs} by 
using \eqref{Tdivbasis}, integrating by parts and using an appropriate 
recursion relation (see subsection \ref{consistem} for details). 

\subsubsection{Special limits}

As in the previous subsection \eqref{saj} may be used to simplify 
\eqref{avvvg} in both the static and the the large 
$D$ limits. In either limit we obtain the simplified formula 

\begin{equation} \label{emradcompsimp}
\begin{split}
S^-_l&= {\tilde S}^-_l\omega^{l+\frac{D-7}2}
\int dr'(r')^{l+D-4}\mathfrak{a}_l(\omega,r', \theta), \\
S^+_l&= {\tilde S}^+_l\omega^{l+\frac{D+1}2}
\int dr'(r')^{l+D}\mathfrak{b}_l(\omega,r', \theta), \\
S^0_l&= {\tilde S}^0_l\omega^{l+\frac{D-3}2}
\int dr'(r')^{l+D-2}\mathfrak{c}_l(\omega,r', \theta) ,\\
{\vec V}^-_l&= {\tilde V}^-_l\omega^{l+\frac{D-5}2}
\int dr'(r')^{l+D-3}
\vec{\mathfrak{u}}(\omega,r',\theta),\\
{\vec V}^+_l&= {\tilde V}^+_l\omega^{l+\frac{D-1}2}
\int dr'(r')^{l+D-1}
\vec{\mathfrak{v}}(\omega,r',\theta) ,\\
(X_l)_{ij}&= {\tilde X}_l\omega^{l+\frac{D-3}2}
\int dr'(r')^{l+D-2}
(\mathfrak{z})_{ij}(\omega,r',\theta),
\end{split}
\end{equation}  
with
\begin{equation}\label{coefg}
\begin{split}
{\tilde S}^-_l &= \frac{i\pi}{ 2^{l+\frac{D-5}2}}\frac{1}{\Gamma\left(l+\frac{D-5}{2}\right)},\\
{\tilde S}^+_l &= \frac{i\pi}{ 2^{l+\frac{D+3}2}}\frac{1}{\Gamma\left(l+\frac{D+3}{2}\right)},\\
{\tilde S}^0_l &= \frac{i\pi}{ 2^{l+\frac{D-1}2}}\frac{1}{\Gamma\left(l+\frac{D-1}{2}\right)},\\
{\tilde V}^-_l &= \frac{i\pi}{ 2^{l+\frac{D-3}2}}\frac{1}{\Gamma\left(l+\frac{D-3}{2}\right)},\\
{\tilde V}^+_l &= \frac{i\pi}{ 2^{l+\frac{D+1}2}}\frac{1}{\Gamma\left(l+\frac{D+1}{2}\right)},\\
{\tilde X}_l &= \frac{i\pi}{ 2^{l+\frac{D-1}2}}\frac{1}{\Gamma\left(l+\frac{D-1}{2}\right)}.
\end{split}
\end{equation}

As in the previous subsection in the large 
$D$ limit at fixed $\omega$ we use the Sterling approximation to further 
simplify ${\tilde \alpha_l}$; we find 
\begin{equation}\label{coefldg}
\begin{split}
{\tilde S}^-_l &\approx \frac{i\sqrt{\pi}}{2 D^{\frac{D}2}}\left(\frac{e^{l+\frac{D-7}2}}{D^{l-3}}\right),\\
{\tilde S}^+_l &\approx \frac{i\sqrt{\pi}}{2 D^{\frac{D}2}}\left(\frac{e^{l+\frac{D+1}2}}{D^{l+1}}\right),\\
{\tilde S}^0_l&\approx \frac{i\sqrt{\pi}}{2 D^{\frac{D}2}}\left(\frac{e^{l+\frac{D-3}2}}{D^{l-1}}\right),\\
{\tilde V}^-_l &\approx \frac{i\sqrt{\pi}}{2 D^{\frac{D}2}}\left(\frac{e^{l+\frac{D-5}2}}{D^{l-2}}\right),\\
{\tilde V}^+_l &\approx \frac{i\sqrt{\pi}}{2 D^{\frac{D}2}}\left(\frac{e^{l+\frac{D-1}2}}{D^l}\right),\\
{\tilde X}_l &\approx \frac{i\sqrt{\pi}}{2 D^{\frac{D}2}}\left(\frac{e^{l+\frac{D-3}2}}{D^{l-1}}\right).
\end{split}
\end{equation}

As in the previous subsection \eqref{coefldg} does not apply in the 
static limit at fixed $D$. In this limit, however, the small argument 
expansion of the Hankel function leads to simplifications. In Appendix 
\ref{statgrav} we demonstrate that the radiation formulae \eqref{gmad}, 
in this limit yield results consistent with the equations
\begin{equation}\label{statgrav1} \begin{split}
R_{0i0j}&=\nabla_i\nabla_j\Phi^G ,\\
\nabla^2 \phi^G&= -8\pi\left(\mathcal{T}_{00}+\frac{\mathcal{T}}{D-2}\right),\\
R_{0ijk} &= -\nabla_i\left(\nabla_jA^G_k-\nabla_kA^G_j\right),\\
\nabla^2 A^G_i &= -8\pi \mathcal{T}_{0i},\\
R_{ijkl} &= \nabla_i\nabla_k\mathfrak{T}_{jm}+\nabla_j\nabla_m\mathfrak{T}_{ik}-\nabla_j\nabla_k\mathfrak{T}_{im}-\nabla_i\nabla_m\mathfrak{T}_{jk},\\
\nabla^2 \mathfrak{T}_{ij}^G &= 8\pi\mathcal{T}_{ij}.
\end{split}
\end{equation}

\section{Radiation from linearized fluctuations about spherical membranes}\label{rfl}

\subsection{Electromagnetic Radiation}

As we have explained in the previous section, the simplest solution of the charged membrane equations 
of motions is a static spherical membrane whose world volume is $S^{D-2} \times$ time. 
This solution is dual to a static charged black hole. The spectrum of linearized 
membrane fluctuations about this simple solution was determined in \cite{Bhattacharyya:2015fdk}. 
These linearized solutions are dual to the light quasinormal modes around
the dual stationary black holes. In this section we will compute 
the radiation sourced by these linearized membrane modes. The radiation 
fields we compute have the bulk interpretation as the `outgoing' pieces 
of the corresponding quasinormal modes. 

We begin this section by briefly recalling the linearized solutions 
of \cite{Bhattacharyya:2015fdk}. As in \cite{Bhattacharyya:2015fdk} 
we choose our background solution to be a charged black hole of unit 
radius (as explained in \cite{Bhattacharyya:2015fdk}, the scale invariance
of the Einstein Maxwell equations ensures that this choice involves no
loss of generality). We work to linearized order about this static solution. 
In other words the membrane configurations we study are 
\begin{equation}\label{pert}
\begin{split}
r &= 1 + \delta r(t,\theta),\\
Q &= Q_0 + \delta Q(t,\theta),\\
u &= -dt + \delta u_{\mu}(t,\theta)dx^{\mu}.
\end{split}
\end{equation}
As we have demonstrated earlier in this paper, the charge current 
associated with any membrane configuration is given, in terms of 
arbitrary coordinates on the membrane world volume, by 
\begin{equation}\label{sfcurr}
\begin{split}
J^\mu
=&~\left(\frac{Q}{2\sqrt{2\pi}}\right)\left[{K} u^\mu -\left(\frac{p^{\nu\mu}\hat\nabla_\nu Q}{Q}\right) - (u\cdot\hat\nabla)u^\mu- \left(\frac{\hat\nabla^2 u^\mu}{K}\right) +K^{\alpha\mu} u_\alpha\right]\\
&~+ {\cal Q}~ u^\mu + {\cal O}\left(\frac{1}{D}\right),\\
\end{split}
\end{equation}
where 
\begin{equation}
{\cal Q} =~  \left(\frac{Q}{2\sqrt{2\pi}}\right)\left[ \frac{\hat\nabla^2{K}}{{K}^2}-\frac{2K}{D} -\frac{(u\cdot\hat\nabla) K}{K} -\left(\frac{2\hat\nabla^2 Q +K(u\cdot\hat \nabla) Q}{Q~K}\right) +\left(u^\alpha u^\beta K_{\alpha\beta}\right)\right].
\end{equation}
We will now evaluate the current $J_\mu$ listed in \eqref{sfcurr} for the 
special case of small fluctuations around the spherical membrane 
(\eqref{pert}) to first order in fluctuations. For this purpose we use 
the angular coordinates on the unit $S^{D-2}$ and time as coordinates on 
the membrane world volume. Note that, to linear order in fluctuations, 
the metric on the membrane world volume is given by 
\begin{equation}\label{metonmem}
ds^2=-dt^2 + (1+2 \delta r) dr^2.
\end{equation}
All covariant derivatives in \eqref{sfcurr} must be evaluated on this metric. 
However, following \cite{Bhattacharyya:2015fdk}, we will find it most 
convenient to view our fluctuation fields $\delta r$ and $u_\mu$ as living 
on the undeformed unit sphere. In all formulae below the symbol $\nabla_a$ 
will refer to the covariant derivative on this round sphere ($a$ are the 
angular directions on the sphere). \footnote{In order to present all our 
formulae in terms of covariant derivatives w.r.t the unit sphere, we 
sometimes need to rewrite covariant derivatives w.r.t. the metric 
\eqref{metonmem} in terms of covariant derivatives on the unit sphere. }
Adopting these conventions the formulae 
\begin{equation} \label{componmem}
\begin{split}
n_r &= 1,\\
n_{\mu} &= -\partial_{\mu}\delta r,\\
K_{tt} &= -\partial_t^2\delta r,\\
K_{ta} &= -\partial_t\nabla_a\delta r,\\
K_{ab} &= -\nabla_a\nabla_b\delta r + (1 + \delta r)g_{ab},\\
\delta u_t &= 0, ~~~~~~~~~~(u\cdot u =-1)\\
(u\cdot K)_t &= K_{tt} = -\partial_t^2\delta r,\\
(u\cdot K)_a &= -\partial_t\nabla_a\delta r + \delta u_a,\\
K &= {K_A}^A = D\left(1-\left(1+\frac{\nabla^2}{D}\right)\delta r\right),
\end{split}
\end{equation}
(which we have borrowed from \cite{Bhattacharyya:2015fdk}) allow us to 
explicitly evaluate all components of the membrane world volume current
in terms of the linearized fluctuations in \eqref{pert}; we find 
\begin{equation}\label{flccurRN}
\begin{split}
J_t =&\frac{1}{2\sqrt{2\pi}}\bigg( -D Q_0+ \bigg(DQ_0\left(1+\frac{\nabla^2}{D}\right)\delta r -D\delta Q-\partial_t \delta Q-Q_0\partial^2_t \delta r+\frac{\nabla^2}{D}\delta Q\bigg)\bigg),\\
J_i=& \frac{1}{2\sqrt{2\pi}}\bigg( Q_0\delta u_i-Q_0\partial_t\delta u_i-\partial_i \delta Q-Q_0\partial_i\partial_t\delta r-Q_0\frac{\nabla^2}{D}\delta u_i\bigg).
\end{split}
\end{equation}
Note that we have presented our current with lower indices, i.e. as a 
oneform field. This oneform field lives of the membrane world volume 
whose metric is given by \eqref{metonmem}. 

Recall that the membrane current is conserved,  i.e.
\begin{equation}
\nabla\cdot J=0,
\end{equation}
Explicitly evaluating this conservation equation for the current 
\eqref{flccurRN} we obtain the equation
\begin{equation} \label{qlin}
\left(\frac{\nabla^2}{D}-\partial_t\right) \delta Q = Q_0\left(\partial_t^2-\partial_t\left(\frac{\nabla^2}{D}+1\right)\right)\delta r 
+{\cal O}(1/D).
\end{equation}
Note that \eqref{qlin} is precisely the linearized `charge' membrane 
equation presented in \cite{Bhattacharyya:2015fdk}. We view this 
agreement as a consistency check on the algebra that led to 
\eqref{flccurRN}

The expression \eqref{flccurRN} is the current evaluated on the membrane 
world volume in our particular choice of world volume coordinates. 
Radiation is sourced by the current viewed as a distributional vector
field in spacetime. We obtained the spacetime current as follows. We first
converted the oneform field $J$ into a vector field on the membrane world 
volume using its metric \eqref{metonmem} \footnote{However the term 
proportional to $\delta r$ does not contribute to leading order in fluctuations
 as $J_a$ vanishes for the stationary membrane. In effect, thus, 
consequently we raise all indices using the metric of the unit sphere}. 
We then converted the vector field on the membrane to a 
vector field in spacetime using the formulae
\begin{equation}\label{jmsj} \begin{split}
J_{ST}^a&= \delta(r+ \delta r -1) J^a,\\
J_{ST}^t&= \delta(r+ \delta r -1) J^t,\\
J_{ST}r&=\delta(r+ \delta r -1) \left( 
J^t\partial_t \delta r + J^a\partial_a \delta r  \right)
=\delta(r+ \delta r -1)  
J^t\partial_t \delta r .
\end{split}
\end{equation}
The equality in the last line of this equation holds to linear order 
in fluctuations as $J^a$ vanishes on the static membrane. Note that 
we have also omitted the measure factor $\sqrt{1 + (\nabla \delta r)^2}$
in the spacetime current (see e.g. \eqref{cc}) as this term is unity 
to linear order. We find the following expression for the spacetime 
current 
\begin{equation}\label{jmstact} \begin{split}
J_{ST}^a&= \frac{1}{2\sqrt{2\pi}}\delta(r+ \delta r -1) \left(  D Q_0\delta u^a-Q_0\partial_t\delta u^a-\partial^a \delta Q-Q_0\partial^a\partial_t\delta r-Q_0\frac{\nabla^2}{D}\delta u^a  \right), \\
J_{ST}^t&= \frac{1}{2\sqrt{2\pi}}\delta(r+ \delta r -1) \bigg(  D Q_0- \bigg(DQ_0\left(1+\frac{\nabla^2}{D}\right)\delta r -D\delta Q\\&-\partial_t \delta Q-Q_0\partial^2_t \delta r+\frac{\nabla^2}{D}\delta Q\bigg)   \bigg),  \\
J_{ST}^r&= \frac{1}{2\sqrt{2\pi}} \delta(r+ \delta r -1) \left( DQ_0\partial_t\delta r  \right), \\
\end{split}
\end{equation}

As was explained in \cite{Bhattacharyya:2015fdk} the linearized solutions \
take the form 
\begin{equation}\label{genfllnr}
\begin{split}
\delta r&=\sum_{l,m}a_{lm}Y_{lm}e^{-i\omega^r_l t},\\
\delta Q&=\sum_{l,m}a_{lm} 
\frac{i \omega^r_l Q_0 \left( l-1-i\omega^r_l \right)}{l-i\omega^r_l}
Y_{lm}e^{-i\omega^r_l t} + 
\sum_{l,m}q_{lm}Y_{lm}e^{-i\omega^Q_l t},\\
\delta u_i&=\sum_{l,m} \frac{-i \omega^r_l}{l} a_{lm} \nabla_i Y_{lm}e^{-i\omega^r_l t} + \sum_{l,m}b_{lm}V_i^{lm}e^{-i\omega^v_l t},\\
\end{split}
\end{equation} 
(the summation over $l$ involving $a_{lm}$ in the last line excludes 
$l=0$). The coefficients $b_{lm}$ parametrize the `velocity fluctuations'
of \cite{Bhattacharyya:2015fdk}; note that these fluctuations affect only 
the velocity field. The coefficients $q_{lm}$ parametrize the `charge 
fluctuations' of  \cite{Bhattacharyya:2015fdk}; note that they affect only 
the charge field. The coefficients $a_{lm}$ parametrize the `shape'
fluctuations of  \cite{Bhattacharyya:2015fdk}. These are the most 
complicated quasinormal modes, as they affect the shape $\delta r$, the 
charge $\delta Q$ and the velocity $\delta u$. In the rest of this 
subsection we will determine the radiation field sourced by each of these
fluctuations in turn. 

\subsubsection{Radiation from Charge Fluctuations}

Let us first restrict our attention to the $l^{th}$ spherical harmonic 
mode of `charge' fluctuations (i.e. mode in \eqref{genfllnr} that is 
proportional to $q_{lm}$). In this special case the spacetime current 
\eqref{jmstact} reduces to 
\begin{equation}\label{stcurcf}
 \begin{split}
J_{ST}^a&= -\frac{1}{2\sqrt{2\pi}}\delta(r -1) q_{lm}  e^{-i\omega^Q_lt }\left(   \partial^a  Y_{lm}\right), \\
J_{ST}^t&= \frac{D}{2\sqrt{2\pi}}\delta(r -1)  \left( q_{lm} Y_{lm} e^{-i\omega^Q_lt }  \right),  \\
J_{ST}^r&=  0 .\\
\end{split}
\end{equation}
It follows that the quantities  $\mathfrak{b}$ and 
$\mathfrak{c}$ relevant for \eqref{arbitscalarJ} are given by 
\begin{equation}\label{stcurcf2}
 \begin{split}
\mathfrak{b}&= \frac{l}{2\sqrt{2\pi}}\delta(r -1)  q_{lm} Y_{lm}e^{-i\omega^Q_lt} - \frac{1}{2\sqrt{2\pi}}q_{lm}Y_{lm}e^{-i \omega_l^Qt}\partial_r\delta(r-1), \\
\mathfrak{c}&=  0 ,\\
\end{split}
\end{equation}
(at $\omega= \omega^Q_l=-il$). In writing our result \eqref{stcurcf} we have 
taken the large $D$ limit and retained only leading order terms.
It follows from \eqref{emradcompsimpv} that 
\begin{equation}\label{radformq1}
 \begin{split}
S_l^+&= \frac{D}{2\sqrt{2\pi}}\tilde{S}_l^2\omega^{l+\frac{D-1}{2}}q_{lm}Y_{lm}e^{-i\omega_l^Q t},  \\
{\vec V}_l&=  0 ,\\
\end{split}
\end{equation}
where the coefficients  $\tilde{S}_l^+$ are defined in \eqref{coef}. The $S_l^+$ can be computed from the constraint equation \eqref{rbs} which in the large $D$ limit reduces to 
$$S_l^-=\frac{D}{l}S_l^+.$$
 The electromagnetic 
radiation field associated with the $l^{th}$ 
`charge' fluctuation quasinormal mode is given by plugging these results into \eqref{emad}.

\subsubsection{Radiation from Velocity Fluctuations}

Let us now  restrict our attention to the $l^{th}$ spherical harmonic 
mode of `velocity' fluctuations (i.e. mode in \eqref{genfllnr} that is 
proportional to $b_{lm}$). In this special case the spacetime current 
\eqref{jmstact} reduces to 
\begin{equation}\label{stcurcfu}
 \begin{split}
J_{ST}^a&= \frac{D Q_0}{2\sqrt{2\pi}}\delta(r -1) b_{lm} V_{lm}^a e^{-i\omega^v_lt }, \\
J_{ST}^t&=  0,\\
J_{ST}^r&=  0. \\
\end{split}
\end{equation}
It follows that the quantities  $\mathfrak{b}$ and 
$\mathfrak{c}$ relevant for \eqref{arbitscalarJ} are given by 
\begin{equation}\label{stcurcfs}
 \begin{split}
\mathfrak{b}&= 0,\\
\mathfrak{c}&=  \frac{iDQ_0\omega^v_l}{2\sqrt{2\pi}} \delta(r-1)b_{lm} V_{lm}^a e^{-i\omega^v_lt }, \\
\end{split}
\end{equation}
(at $\omega= \omega^v_l=\frac{-i (l-1)}{1+Q_0^2}$).
It follows from \eqref{avvv} that 
\begin{equation}\label{radformu}
 \begin{split}
{\vec V}_l&= \frac{i DQ_0}{2\sqrt{2\pi}} (\omega_l^v)^{\frac{D-1}{2}+l}\tilde{V}_lb_{lm}\vec{V}_{lm}e^{-i\omega_l^vt} ,\\
\end{split}
\end{equation}
where the coefficients $\tilde{V}_l$ is defined in \eqref{coef}.
The electromagnetic radiation field associated with the $l^{th}$ 
`velocity' fluctuation quasinormal mode is obtained by plugging \eqref{radformu} into \eqref{emad}.

\subsubsection{Radiation from Shape Fluctuations}

Let us first restrict our attention to the $l^{th}$ spherical harmonic 
mode of `shape' fluctuations (i.e. mode in \eqref{genfllnr} that is 
proportional to $a_{lm}$). The radiation due to the shape fluctuation is little complicated compared to the 'charge ' fluctuation  or the 'velocity ' fluctuation , since the small perturbation in the shape turns on  both the charge and the velocity fluctuation\eqref{genfllnr} . In this special case the spacetime current 
\eqref{jmstact} reduces to 
\begin{equation}\label{stcurcfshp}
 \begin{split}
J_{ST}^a&=-i\frac{\omega_l^r}{l} \frac{DQ_0}{2\sqrt{2\pi}}\delta(r -1) a_{lm} \nabla^aY_{lm} e^{-i\omega^r_lt } ,\\
J_{ST}^t&= \delta(r-1)\frac{DQ_0}{2\sqrt{2\pi}}   \left(   a_{lm} 
\frac{i \omega^r_l  \left( l-1-i\omega^r_l \right)}{l-i\omega^r_l} Y_{lm}e^{-i\omega^r_lt } \right) +\frac{D Q_0}{2\sqrt{2\pi}} a_{lm}Y_{lm}\partial_r\delta(r-1)e^{-i \omega^r_lt}\\&-\frac{DQ_0(-l+1)}{2\sqrt{2\pi}}\delta(r-1)a_{lm}Y_{lm}e^{-i \omega^r_lt}, \\
J_{ST}^r&=  -\frac{i \omega^r_lDQ_0}{2\sqrt{2\pi}}\delta(r-1) a_{lm} Y_{lm}e^{-i\omega^r_lt}. \\
\end{split}
\end{equation}
It follows that the quantities  $\mathfrak{b}$ and 
$\mathfrak{c}$ relevant for \eqref{arbitscalarJ} are given by 
\begin{equation}\label{stcurcfsh}
 \begin{split}
\mathfrak{b}&= \frac{Q_0}{2\sqrt{2\pi}}a_{lm}Y_{lm}\partial^2_r\delta(r-1)e^{-i \omega^r_lt}, \\
\mathfrak{c}&=  0, \\
\end{split}
\end{equation}
(at $\omega= \omega^r_l$).
It follows from \eqref{avvv} that 
\begin{equation}\label{radform}
 \begin{split}
S_l^+&=  \tilde{S}_l^+\omega^{l+\frac{D-1}{2}}\frac{D^2Q_0}{2\sqrt{2\pi}}a_{lm}Y_{lm}e^{-i\omega^r_lt}, \\
{\vec V}_l&=  0 ,\\
 \end{split}
\end{equation}
where the coefficient $\tilde{S}_l^+$ are defined in \eqref{coef}.  The $S_l^+$ can be computed from the constraint equation \eqref{rbs} which in the large $D$ limit reduces to 
$$S_l^-=\frac{D}{l}S_l^+.$$
The electromagnetic radiation field associated with the $l^{th}$ 
`shape' fluctuation quasinormal mode is obtained by plugging \eqref{emad} ).

\subsection{Gravitational Radiation}
In this subsection  we compute the gravitational radiation emitted by the quasinormal modes described 
earlier in this section. As we demonstrated earlier in this paper, the stress tensor on the world 
volume of the large $D$ black hole membrane  is given by
\begin{eqnarray}\label{stcc1}
\begin{split}
T_{\mu\nu}= &~\left(\frac{1}{8\pi}\right)\bigg[ \left( \frac{K}{2} \right) (1+Q^2) u_\mu u_\nu + \left( \frac{1-Q^2}{2}  \right)K_{\mu\nu} - \left(   \frac{\hat\nabla_\mu u_\nu +\hat \nabla_\nu u_\mu}{2} \right) \\
&-\left(\frac{{K}Q^2}{2D} + \frac{2Q\hat\nabla^2Q}{K} + Q^2 u^\alpha u^\beta K_{\alpha\beta}\right)u_\mu u_\nu-\left( u_\mu{\cal V}_\nu + u_\nu{\cal V}_\mu\right) \\
&~-\bigg[\left(\frac{1+Q^2}{2}\right)\left(u^\alpha u^\beta K_{\alpha\beta}\right)+\left(\frac{1-Q^2}{2}\right)\left(\frac{{K}}{D}\right)\bigg]g^{(ind,f)}_{\mu\nu}\bigg]\\
&~ + {\cal O}\left(\frac{1}{D}\right),\\
\end{split}
\end{eqnarray}
where 
\begin{equation}
\begin{split}
{\cal V}_\mu =&~Q~\hat\nabla_\mu Q +Q^2(u^\alpha K_{\alpha\mu} ) +\left(\frac{2Q^4-Q^2-1}{2}\right)\left(\frac{\hat\nabla_\mu{K}}{K}\right)\\
&-\left(\frac{Q^2 + 2Q^4}{2}\right)(u\cdot\hat\nabla)u_\mu+ \left(\frac{1+Q^2}{K}\right)\hat\nabla^2 u_\mu.\\
\end{split}
\end{equation}
The stress tensor is conserved upto the membrane equation of motion \eqref{memeom} and the divergencelessness of the velocity field.
\begin{equation}
\nabla_\mu T^{\mu}_\nu=0.
\end{equation}
The most general form of the fluctuation of the stress tensor  about the RN background takes the form 
\begin{equation}\label{strstnfl}
\begin{split}
-8\pi T_{tt}&=-\frac{D}{2}(1+Q_0^2)+\frac{Q_0^2}{2}-\bigg(-(1+Q_0^2)\left(1+\frac{\nabla^2}{D}\right)\left(\frac{D}{2}+\partial_t\right)\delta r+(D+1)Q_0\delta Q\nn\\
&+Q_0^2\left(1+\frac{\nabla^2}{D}\right)(Q_0^2\partial_t-1/2)\delta r+2Q_0\left(\frac{\nabla^2}{D}-\partial_t\right)\delta Q+2Q_0^2\partial_t^2\delta r\bigg)\\&+\frac{1+Q_0^2}{2}\partial_t^2\delta r-\frac{1-Q_0^2}{2}\left(1-\left(1+\frac{\nabla^2}{D}\right)\delta r\right)+\frac{1-Q_0^2}{2}\partial_t^2\delta r,\\
-8\pi T_{ta}&=\bigg(\frac{D(1+Q_0^2)-Q_0^2}{2}\delta u_a+\frac{1-Q_0^2}{2}\partial_t\nabla_a\delta r+\frac{1+Q_0^4}{2}\partial_t\delta u_a-\frac{1-Q_0^2-Q_0^4}{2}(1+\frac{\nabla^2}{D})\nabla_a\delta r\nn\\
&-Q_0\partial_a\delta Q-\left(\frac{1+Q_0^2}{2}\right)\frac{\nabla^2}{D}\delta u_a+2Q_0^2\partial_t\nabla_a\delta r-2Q_0^2\delta u_a\bigg),\\
-8\pi T_{ab}&=-\frac{1+Q_0^2}{2} \partial_t^2\delta r g_{ab}+\bigg(\frac{1-Q_0^2}{2}\left(\nabla_a\nabla_b\delta r-g_{ab}\delta r\right)+Q_0g_{ab}\delta Q+\frac{\nabla_a\delta u_b+\nabla_b\delta u_a}{2}+g_{ab}\partial_t\delta r\bigg)\\
&-\frac{1-Q_0^2}{2} \left(1+\frac{\nabla^2}{D}\right)\delta rg_{ab}.
\end{split}
\end{equation}
The velocity fluctuation and the shape fluctuation along the angular direction  follows the following second order differential equation 
\begin{equation} \label{vecsc}
\begin{split}
\bigg(\left(1+\frac{\nabla^2}{D}\right)-(1+Q_0^2) \partial_t\bigg)\delta u_a =& - \left((1-Q_0^2)\nabla_a\left(1+\frac{\nabla^2}{D}\right)-\partial_t\nabla_a\right)\delta r,
\end{split}
\end{equation}
and along the t direction the velocity fluctuation and the shape fluctuation follows the constraint equation{\footnote{The expression below can be obtained either by use the expansion of the divergence in terms of the Christoffel symbol or use the form $$(\nabla.T)_i=\frac{1}{\sqrt{g}}\partial_k(\sqrt{g}T^k_i)-\frac{1}{2}\partial_ig_{kl}T^{kl}.$$}}
\begin{equation}
\nabla_a\delta u^a=-(D-2)\partial_t\delta r\approx -D\partial_t\delta r.
\end{equation}
We have, so far, been working with tensor fields living on the world volume of the membrane. In order to
determine the source for radiation we are really interested in the stress tensor viewed as a distributional 
tensor field living in spacetime. Apart from the delta functions that localize the spacetime quantities 
to the membrane (and which we will explicitly present in later formulae) the relationship between these 
two structures is given in general by the following translation formulae between the membrane 
field $A^{\mu\nu}$ and the spacetime field $A_{ST}^{MN}$
\begin{equation}
\begin{split}
A^{tt}_{ST}&=A^{tt},\qquad A^{ab}_{ST}=A^{ab},\nonumber\\
A^{rr}_{ST}&={\cal{O}}(\epsilon^2),\qquad A^{ t a}_{ST}=A^{t a},\nonumber\\
A^{tr}_{ST}&=\left(A^{tt}\partial_t\delta r+A^{ta}\partial_a\delta r\right),A^{ar}_{ST}=\left(A^{at}\partial_t\delta r+A^{ab}\partial_b\delta r\right).
\end{split}
\end{equation}
It follows that to linear order in fluctuations
\begin{equation}\label{jmst} \begin{split}
T_{ST}^{tt}&= \delta(r+ \delta r -1) T^{tt},\\
T_{ST}^{ab}&=  \delta(r+ \delta r -1)T^{ab},\\
T_{ST}^{ta}&= \delta(r+ \delta r -1)T^{ta},\\
T_{ST}^{tr}&=\delta(r+ \delta r -1) \left( T^{tt}\partial_t \delta r+T^{ta}\partial_a\delta r\right)=\delta(r+ \delta r -1)  T^{tt}\partial_t \delta r,\\
T_{ST}^{rr}&= 0,\\
T_{ST}^{ar}&=0.        
\end{split}
\end{equation}
Explicitly we find 
\begin{equation}\label{strjmstact} \begin{split}
-8\pi T_{ST}^{tt}&= \delta(r+ \delta r -1)\bigg(-\frac{D}{2}(1+Q_0^2)-\bigg(-(1+Q_0^2)\left(1+\frac{\nabla^2}{D}\right)\left(\frac{D}{2}+\partial_t\right)\delta r\nn\\&+(D+1)Q_0\delta Q+\frac{1-Q_0^2}{2}\partial_t^2\delta r\nn\\
&+Q_0^2\left(1+\frac{\nabla^2}{D}\right)(Q_0^2\partial_t-1/2)\delta r+2Q_0\left(\frac{\nabla^2}{D}-\partial_t\right)\delta Q+2Q_0^2\partial_t^2\delta r\bigg)\\&+\frac{1+Q_0^2}{2}\partial_t^2\delta r-\frac{1-Q_0^2}{2}\left(1-\left(1+\frac{\nabla^2}{D}\right)\delta r\right)\bigg) , \\
-8\pi T_{ST}^{ta}&= -\delta(r+ \delta r -1)  \bigg(\frac{D(1+Q_0^2)-Q_0^2}{2}\delta u_a+\frac{1-Q_0^2}{2}\partial_t\nabla_a\delta r+\nn\\
&\frac{1+Q_0^4}{2}\partial_t\delta u_a-\frac{1-Q_0^2-Q_0^4}{2}(1+\frac{\nabla^2}{D})\nabla_a\delta r-Q_0\partial_a\delta Q-\left(\frac{1+Q_0^2}{2}\right)\frac{\nabla^2}{D}\delta u_a+2Q_0^2\partial_t\nabla_a\delta r\\&-2Q_0^2\delta u_a\bigg) ,  \\
-8\pi T_{ST}^{ab}&=  \delta(r+ \delta r -1) \bigg( -\frac{1+Q_0^2}{2} \partial_t^2\delta r g_{ab}+\bigg(\frac{1-Q_0^2}{2}\left(\nabla_a\nabla_b\delta r-g_{ab}\delta r\right)+Q_0g_{ab}\delta Q\\&+\frac{\nabla_a\delta u_b+\nabla_b\delta u_a}{2}+g_{ab}\partial_t\delta r\bigg)
-\frac{1-Q_0^2}{2} \left(1+\frac{\nabla^2}{D}\right)\delta rg_{ab} \bigg) \\
-8\pi T_{ST}^{tr}&=  \delta(r+ \delta r -1) \bigg( -\frac{D}{2}(1+Q_0^2)\partial_t\delta r \bigg), \\
-8\pi T_{ST}^{rr}&=0,\\
-8\pi T_{ST}^{ar}&=0.
\end{split}
\end{equation}

 \subsubsection{Gravitational Radiation from Charge fluctuation}
 Let us first restrict our attention to the $l^{th}$ spherical harmonic 
 mode of `charge' fluctuations (i.e. mode in \eqref{genfllnr} that is 
 proportional to $q_{lm}$). In this special case the stress tensor
 \eqref{strjmstact} reduces to 
 \begin{equation}\label{strchrg}
  \begin{split}
  -8\pi T^{tt}_{ST}&=-\delta(r-1)DQ_0\delta Q,\\
  -8\pi T^{ta}_{ST}&=\delta(r-1)Q_0 \partial^a\delta Q,\\
 -8\pi T^{ab}_{ST}&=\delta(r-1)Q_0g^{ab}\delta Q,\\
 -8\pi T^{tr}_{ST}&=0,\\
 -8\pi  T^{rr}_{ST}&=0,\\
 -8\pi   T^{ar}_{ST}&=0,\\
  \end{split}
  \end{equation}
  where $\delta Q$ is given by the part of \eqref{genfllnr} proportional to $q_{lm}$. 
  
 It follows that we can read of the relevant quantity $\mathfrak{b}$  from the effective stress tensor \eqref{nablasqg} and \eqref{arbitten}  is given by 
 \begin{equation}\label{strchradfldq}
  \begin{split}
 \mathfrak{b}&= \frac{Q_0}{D}(\partial_{r}^2\delta(r-1))q_{lm}Y_{lm}e^{-i\omega^Q_lt}\\
 \end{split}
 \end{equation}
 (at $\omega= \omega^Q_l$).
 It follows from \eqref{avvv} that 
 \begin{equation}\label{radformq2}
  \begin{split}
 S_l^+ = \tilde{S}_l^+DQ_0(\omega^Q_l)^{l+\frac{D+1}2}q_{lm}Y_{lm}e^{-i\omega^Q_lt},
 \end{split}
 \end{equation}
 where $\tilde{S}_l^+$ is given by \eqref{coefldg}.
 The other  components can be read of using the constraint equation \eqref{grbs},which in the large $D$ limit can be simplified as 
\begin{equation}\label{rfo}
S_l^0 = \frac{2D}l S_l^+,~~~S_l^- = \frac{D^2}{l(l-1)} S_l^+.
\end{equation}
 The contribution to the radiation due to the charge fluctuation to the vector sector and the tensor sector vanishes in the linear order.
 
The explicit formula for gravitational radiation from the charge fluctuations is given by 
plugging \eqref{radformq2} and \eqref{rfo} into \eqref{gmad}.
 
 \subsubsection{Gravitational Radiation from Velocity fluctuation}
We now turn our attention to the $l^{th}$ spherical harmonic 
 mode of `velocity' fluctuations (i.e. mode in \eqref{genfllnr} that is 
 proportional to $b_{lm}$). In this special case the spacetime current 
 \eqref{jmstact} evaluates to 
 \begin{equation}\label{strchrg1}
  \begin{split}
 -8\pi T^{tt}_{ST}&=0,\\
 -8\pi T^{ta}_{ST}&=\delta(r-1)Q_0 \frac{D(1+Q_0)^2}{2}\delta u^a,\\
-8\pi  T^{ab}_{ST}&=\delta(r-1)\left(\frac{\nabla^a\delta u^b+\nabla^b\delta u^a}{2}\right),\\
-8\pi  T^{tr}_{ST}&=0,\\
 -8\pi  T^{rr}_{ST}&=0,\\
 -8\pi   T^{ar}_{ST}&=0,\\
  \end{split}
  \end{equation}
  where $\delta u^a$ is obtained from the part of \eqref{genfllnr} proportional to $b_{lm}$. 
  We can read of the relevant quantity $\vec{\mathfrak{v}}$   from the effective stress tensor \eqref{nablasqg} and 
  \eqref{arbitten}. We find 
  \begin{equation}\label{strchradfldu}
   \begin{split}
  \vec{\mathfrak{v}} = \frac{-i\omega^v_l( 1+Q_0^2)}{2}(\partial_{r}\delta(r-1))b_{lm}V_i^{lm}e^{-i\omega^v_lt},\\
  \end{split}
  \end{equation}
  (at $\omega= \omega^v_l$).
  It follows from \eqref{avvv} that 
  \begin{equation}\label{radformq3}
   \begin{split}
  \vec{V}_l^+ = iD\tilde{V}_l^+(\omega^u_l)^{l+\frac{D-1}2}\frac{1+Q_0^2}{2}b_{lm}V_i^{lm}e^{-i\omega^v_lt},
  \end{split}
  \end{equation}
  where $\tilde{V}_l^+$ is given by \eqref{coefldg}.
  The other  components can be read of using the constraint equation \eqref{grbs},which in the large $D$ limit can be simplified as 
\begin{equation}\label{rarel}
\vec{V}_l^- = \frac{D}{(l-1)} \vec{V}_l^+.
\end{equation}  
  The contribution to the radiation due to the velocity fluctuation to the scalar sector and the tensor sector vanishes in the linear order. The gravitational radiation associated with the $l^{th}$ 
  `velocity' fluctuation quasinormal mode is given by plugging \eqref{radformq3} and \eqref{rarel}
  into \eqref{gmad}.

 \subsubsection{Gravitational Radiation from Shape fluctuation}
Finally, we turn to the  $l^{th}$ spherical harmonic 
 mode of `shape' fluctuations (i.e. mode in \eqref{genfllnr} that is 
 proportional to $a_{lm}$). In this special case the spacetime current 
 \eqref{jmstact} reduces to 
 \begin{equation}\label{strchrg3}
  \begin{split}
 -8\pi T^{tt}_{ST}&=\delta(r-1)\left(\frac{1+Q_0^2}{2}\left(1+\frac{\nabla^2}{D}\right)\frac{D}{2}\delta r-DQ_0 \delta Q\right)-\frac{D(1+Q_0^2)}{2}\delta r\partial_r\delta(r-1),\\
-8\pi  T^{ta}_{ST}&=\delta(r-1)Q_0 \frac{D(1+Q_0)^2}{2}\delta u^a,\\
-8\pi  T^{ab}_{ST}&=\delta(r -1) \bigg( -\frac{1+Q_0^2}{2}\partial_t^2\delta r g_{ab}+\bigg(\frac{1-Q_0^2}{2}\left(\nabla_a\nabla_b\delta r-g_{ab}\delta r\right)+Q_0g_{ab}\delta Q\\&+\frac{\nabla_a\delta u_b+\nabla_b\delta u_a}{2}+g_{ab}\partial_t\delta r\bigg)
  -\frac{1-Q_0^2}{2} \left(1+\frac{\nabla^2}{D}\right)\delta rg_{ab} \bigg)\\
-8\pi  T^{tr}_{ST}&=\delta(r -1) \bigg( -\frac{D}{2}(1+Q_0^2)\partial_t\delta r \bigg),\\
 -8\pi  T^{rr}_{ST}&=0,\\
 -8\pi   T^{ar}_{ST}&=0,\\
  \end{split}
  \end{equation}
  where all fluctuation fields are obtained from the part of \eqref{genfllnr} proportional to $a_{lm}$.
  It follows that we can read of the relevant quantity  $\mathfrak{b}$  from the effective stress tensor \eqref{nablasqg} and \eqref{arbitten} and is given by 
  \begin{equation}\label{strchradfldr}
   \begin{split}
  \mathfrak{b} = \frac{1+Q_0^2}{2D}\left(\partial_{r}^2\delta(r-1)a_{lm}Y_{lm}e^{-i\omega^r_lt}-\partial_{r}\delta(r-1)a_{lm}Y_{lm}e^{-i\omega^r_lt}\right),
  \end{split}
  \end{equation}
  (at $\omega= \omega^r_l$).
  It follows from \eqref{avvv} that 
  \begin{equation}\label{radformq4}
   \begin{split}
 S_l^+ = \tilde{S}_l^+D^2\frac{1+Q_0^2}{2}(\omega^r_l)^{l+\frac{D+1}2}a_{lm}Y_{lm}e^{-i\omega^r_lt},
  \end{split}
  \end{equation}
  where $\tilde{S}_l^+$ is given by \eqref{coefldg}.
  The other  components can be read of using the constraint equation \eqref{grbs},which in the large $D$ limit can be simplified as 
  $$S_l^0 = \frac{2D}l S_l^+,~~~S_l^- = \frac{D^2}{l(l-1)} S_l^+.$$
  The contribution to the radiation due to the charge fluctuation to the vector sector and the tensor sector vanishes in the linear order.The radiation field associated with the $l^{th}$ 
  `shape' fluctuation quasinormal mode is obtained by plugging these results into  \eqref{gmad}.
 
\section{Discussion}\label{disc}

In this paper we have obtained explicit formulae for the stress tensor, charge current and entropy 
current that live on the world volume of the large $D$ black hole membrane of \cite{Bhattacharyya:2015dva,Bhattacharyya:2015fdk, Dandekar:2016fvw}. We have demonstrated that the 
membrane stress tensor and charge current are conserved. When written in terms of membrane variables, the 
requirement of conservation is simply a restatement of the membrane equations of motion of \cite{Bhattacharyya:2015dva,Bhattacharyya:2015fdk, Dandekar:2016fvw}. In contrast to the charge 
current and the stress tensor, the entropy current on the membrane world volume is not conserved; 
$\nabla_M J^M_S$ is nonvanishing at order $\frac{1}{D}$. We have used the Hawking area increase theorem to demonstrate that the divergence of this entropy current is point wise positive definite. 
At lowest nontrivial order (order $\frac{1}{D})$ we have demonstrated that this divergence 
is proportional to the square of the shear tensor.

In this paper we have also derived explicit formulae for linearized radiation response of the 
metric and the electromagnetic field to an arbitrary stress tensor and a charge current.
Plugging the our explicit results for the membrane stress tensor and charge current into 
these general formulae yields a formula for the radiation emitted from a large $D$ 
black hole membrane as it undergoes any particular 
solution of the large $D$ membrane equations.  A central qualitative result of this 
paper is that the fractional energy lost to radiation as the large $D$ black hole membrane moves, 
oscillates and vibrates around is of order $\frac{1}{D^D}$. The smallness of radiation is a simple kinematical consequence of the nature of Greens functions in large $D$, and results in the decoupling of 
membrane motion from asymptotic low energy gravitons at large $D$. It also ensures that 
the `radiation reaction' on large $D$ black hole membranes can be ignored when working to 
any fixed order in $\frac{1}{D}$.

The results of this paper could be generalized in many ways. First, the membrane stress tensor has been derived in this paper at first subleading order in $\frac{1}{D}$. It should be straightforward to 
use the explicit results of \cite{Dandekar:2016fvw} to generalize this stress tensor to second order 
in the large $D$ expansion, and verify that the conservation of this improved stress tensor leads to the 
second order membrane equations of motion derived in \cite{Dandekar:2016fvw}. Second it would be 
interesting to generalize the construction of \cite{Dandekar:2016fvw} to the study of charged membranes
and thereby obtain the formula for the leading entropy production for charged membranes. 

It was demonstrated in \cite{Dandekar:2016jrp} that the `black brane' equations of 
Emparan, Suzuki and Tanabe (EST) and collaborators are a special scaling limit of the membrane equations 
of \cite{Bhattacharyya:2015dva,Bhattacharyya:2015fdk, Dandekar:2016fvw}. It should be straightforward to take the same scaling limit of the stress tensor derived in 
this paper and compare the result with the `black brane stress tensor' constructed 
by EST and collaborators. 

It would be interesting (and may be possible) to use the formulae derived in this paper - especially the 
formula for the divergence of the entropy current - to classify all stationary 
solutions of the membrane equations of motion. 

The membrane equation of motion \eqref{memeom} and the formula for the membrane stress tensor 
\eqref{stcc} apply, at first order in $\frac{1}{D}$, note only to membranes in flat space but 
also to membranes propagating in any slowly varying solution of the vacuum Einstein equations 
$R_{MN}=0$, e.g. a gravitational wave.
 Using this fact the  membrane equations of motion 
together with the formulae for the membrane stress tensor and charge current of this paper can be used 
to study how external gravitational waves `polarize' large $D$ black holes. The induced polarization 
will set the black hole oscillating, and the black oscillation will in turn radiate gravitational and 
electromagnetic waves in accordance with the formulae derived in this paper. It should be straightforward
to work out the details of this process in order to compute the $\omega$ dependent analogues of the 
`Love Numbers' for black holes described, for instance,  in \cite{Kol:2011vg}.
\footnote{It is interesting to understand how energy conservation works when a gravitational 
wave is incident on a black hole at large $D$. Consider, for instance, a spherical wave of amplitude $A$ incident on a black 
hole. Only a fraction $\epsilon A$ of this amplitude reaches the membrane (where $\epsilon$ is a small 
number of order $\frac{1}{D^{\frac{D}{2}}}$). This part of the wave excites the membrane into a motion of 
amplitude proportional to $\epsilon A$ and so of energy proportional to $(\epsilon A)^2$. The 
membrane oscillation set up by this process result in radiation, and so a back scattered wave of 
amplitude of order $\epsilon^2 A$. The interference of this back scattered wave with the initial incident 
wave reduces the energy of the initial wave by an amount proportional to $\epsilon^2 A \times A= 
\epsilon^2 A^2$, accounting for the energy deposited into membrane vibrations.}

It would be interesting to generalize the construction of the membrane entropy current to higher 
to the large $D$ black hole membrane for higher derivative theories of gravity. The study of this 
subject should make contact with ongoing attempts to establish the second law of thermodynamics 
in higher derivative theories of gravity.

The RHS of the formula \eqref{entcure} for the divergence of the entropy is of order $\frac{1}{D}$. 
At least naively, this fact suggests that the fractional rate of entropy production in black hole 
motion is of order $\frac{1}{D}$. This conclusion appears to lead to a paradox. 

Consider the head 
on collision of two non rotating black holes, each of which is moving at a substantial 
fraction of the speed of light. If the energy lost as radiation in this collision process is 
very small - as suggested by the discussion of this paper - then almost all of the initial 
energy of this configuration must find its way into the black hole that is formed out of this 
collision. It follows that the mass of this daughter black hole is substantially 
larger than the sum of masses of the initial colliding black holes implying that 
the entropy of the final black hole is also substantially larger than the sum of the 
entropies of the original colliding black holes. 
In other words the collision of two black holes at large $D$ appears to lead to 
fractional entropy production of order unity, in apparent contradiction with the 
claim of the previous paragraph. 

We do not have a clear resolution to the puzzle described above. Note, however, that 
there is a time period of order $\frac{1}{D}$ when the colliding black holes first 
come very near to each other, when the membrane description of 
\cite{Bhattacharyya:2015dva,Bhattacharyya:2015fdk, Dandekar:2016fvw} breaks down. 
It is possible that the solution over this time period is a rather violent one, 
leading to the emission of a substantial amount of radiation over the short time 
scale of order $\frac{1}{D}$, invalidating the claim the energy lost in radiation at 
large $D$ is rather small. This discussion suggests that the solution describing the 
collision of two black holes may be rather interesting when the black holes first touch. 
It is possible that the details of this solution are amenable to an analytical analysis 
of some sort. We hope to return to this fascinating question in the future. 

\vskip .8cm
\section*{Acknowledgments}
We would like to thank R. Emparan,  B. Kol and D. Stanford for useful discussions. 
We would especially like to thank  Y. Dandekar, S. Mazumbdar and A. Saha for several 
extremely useful discussions over the course of this project. 
S.M. would like to acknowledge the hospitality of the Institute for 
Advanced Study, Princeton, while this work was in progress.
The work of S.B. was supported by an India Israel (ISF/UGC) 
joint research grant. The work of M.M, S.M and S.T was supported 
by a separate India Israel (ISF/UGC) grant, as well as the Infosys Endowment 
for the study of the Quantum Structure of Space Time. We would all also like 
to acknowledge our debt to the people of India for their steady and 
generous support to research in the basic sciences.

\appendix

\section{Conventions and notation}\label{exactlywhatwemean}

\begin{table}[!htb] 
    \caption{Different indices}
    \centering 
       \begin{tabular}{|r|l|}
            \hline
  Minkowski Spacetime indices &   Capital Latin    ($A$, $B$, $M$, $N$) \\
  \hline
  Indices in the  membrane & Small Greek ($\alpha$, $\beta$, $\mu$, $\nu$)\\
  \hline
  Cartesian Space indices & Small Latin   ($i$, $j$, $k$, $m$)\\
  \hline
  Angle indices    on $S^{D-2}$  &   Small Latin ($a$, $b$, $c$, $d$)\\
  \hline
  
        \end{tabular}
        \end{table}

        \begin{table}[!htb]
          \caption{Gauge fields}
        \centering
         \begin{tabular}{|r|l|}
           \hline
    Full (nonlinear) Gauge field  & ${\mathfrak a}_M$ \\
    (as read off from \cite{Bhattacharyya:2015fdk})&\\
         \hline
    Linearized    part from ${\mathfrak a}_B$ &  ${\cal A}_B = \rho^{-(D-3)}M_{B}$ \\
   ( not satisfying  gauge conditions of this paper) &  \\
   \hline
   Coefficient of $k$th term in expansion &$M^{(k)}_{B}$\\
   of $M_B$ around $\rho=1$&\\
   \hline
Linearized  and  outside  the membrane & ${{G}}_{A}$ \\
(satisfying  gauge conditions of this paper) &\\
   \hline
   Coefficient of $k$th term in expansion &$G^{(k)}_{A}$\\
   of $G_A$ around $\rho=1$&\\
   \hline
Linearized and inside  the membrane & ${\tilde{G}}_{A}$ \\
   \hline
   Coefficient of $k$th term in expansion &$\tilde G^{(k)}_{A}$\\
   of $\tilde G_A$ around $\rho=1$&\\
   \hline
         \end{tabular}
        \end{table}

        \begin{table}[!htb]
         \caption{Different metrics}
        \centering
          \begin{tabular}{|r|l|}
          \hline
  Full (nonlinear) metric: & $\mathcal{G}_{AB} = \eta_{AB} + {\mathfrak g}_{AB}$\\
   (as read off from \cite{Bhattacharyya:2015fdk})&\\
    \hline
    Linearized    part from ${\cal G}_{AB}$ &  $\eta_{AB}+ \rho^{-(D-3)}M_{AB}$ \\
   ( not satisfying  gauge conditions of this paper) &  \\
   \hline
   Coefficient of $k$th term in expansion &$M^{(k)}_{AB}$\\
   of $M_{AB}$ around $\rho=1$&\\
   \hline
  Linearized Metric Outside  the membrane: & $g_{AB} = \eta_{AB} + h_{AB}=\eta_{AB} + \rho^{-(D-3)}{\mathfrak h}_{AB}$\\
  \hline
  Linearized Metric Inside  the membrane: & $\tilde{g}_{AB} = \eta_{AB} + {\tilde{ h}}_{AB}$\\
   \hline
   Coefficient of $k$th term in expansion &$h^{(k)}_{AB}$\\
   of $h_{AB}$ around $\rho=1$&\\
    \hline
   Coefficient of $k$th term in expansion &$\tilde h^{(k)}_{AB}$\\
   of $\tilde h_{AB}$ around $\rho=1$&\\
  \hline
  Induced Metric from Full space-time: & $ g^{(ind)}_{\mu\nu}$\\
  \hline
  Induced Metric from   flat space-time: & ${g}^{(ind,f)}_{\mu\nu}$\\
  \hline
          \end{tabular}
    \end{table}

        \begin{table}[!htb]
          \caption{Differential operators}
          \centering
           \begin{tabular}{|r|l|}
    \hline
w.r.t induced metric on the membrane   &  $ \hat{\nabla}$  \\
 \hline
 w.r.t full space-time  metric  &  $ \nabla$  \\
 \hline
   w.r.t  Minkowski metric    &  $\partial$  \\
  \hline 
  d'Alembertian   & $\Box $ \\
  \hline
d'Alembertian  w.r.t $g_{\mu\nu}^{(ind,f)}$  & $\tilde\Box $ \\
  \hline
    \end{tabular}
        \end{table}

        \begin{table}[!htb]
          \caption{Different projectors}
        \centering
         \begin{tabular}{|r|l|}
            \hline
  On the membrane as& \\ embedded in flat space-time & $\Pi_{AB}$ = $\eta_{AB}-n_An_B$.\\
  \hline
 On the membrane as embedded in & \\  space time with metric $ g_{AB} = {\eta}_{AB} + h_{AB}$   &   $ \mathfrak{p }^{(out)}_{AB}  $\\
 \hline
 On the membrane as embedded in & \\  space time with metric $ \tilde g_{AB} = {\eta}_{AB} + \tilde h_{AB}$  &   $ \mathfrak{p }^{(in)}_{AB}  $\\
  \hline
  Projector orthogonal to both the normal as&  $P_{AB}$ = $\eta_{AB}-n_An_B+u_A u_B$.\\
  embedded in flat space and the velocity  &\\
  \hline
  Projector orthogonal to velocity along the as&  $ p_{\mu \nu} = g^{(ind,f)}_{\mu \nu} + u_{\mu} u_{\nu}$.\\
 membrane as  embedded in flat space &\\
   \hline
  Projector orthogonal to membrane as&  $ \tilde \Pi_{AB} = \eta_{AB} + h^{(0)}_{AB} -n_A n_B$.\\
embedded in  space with metric $\eta_{AB} + h^{(0)}_{AB}$&\\
 \hline
  Projector on the membrane as & \\dependence of the &$\mathcal{P}_l$\\ $l^{th}$spherical harmonic&\\
  \hline
  \end{tabular}
        \end{table}

        \begin{table}[!htb]
         \caption{Extrinsic curvature}
        \centering
          \begin{tabular}{|r|l|}
       \hline
  when embedded in $g_{AB}=\eta_{ AB} +h_{AB}$ & ${\cal K}^{(out)}_{AB}$\\
  \hline
  when embedded in $\tilde g_{AB}=\eta_{AB} +\tilde h_{AB}$ & ${\cal K}^{(in)}_{AB}$\\
  \hline
  when embedded in $\eta_{AB} + h^{(0)}_{AB}$ & $\bar K_{AB}$\\
   \hline
  when embedded in $\eta_{AB}$ & $K_{AB}$\\
  \hline
 $g^{AB}{\cal K}^{(out)}_{AB}$& ${\cal K}^{(out)}$\\
 \hline
 $\tilde g^{AB}{\cal K}^{(in)}_{AB}$& ${\cal K}^{(in)}$\\
  \hline
 $\left[\eta^{AB}- h_{(0)}^{AB}\right]\bar K_{AB}$& $\bar K$\\
  \hline
  $\eta^{AB} K_{AB}$&$K$\\
  \hline
       \end{tabular}
    \end{table}

        \begin{table}[!htb]
         \caption{Intrinsic curvature and Field strength}
        \centering
          \begin{tabular}{|r|l|}
       \hline
  Riemann Tensor  for full space time & $R_{ABCD}$\\
  (for general analysis)&\\
  \hline
Ricci Tensor  for $g_{AB} =\eta_{AB} +h_{AB}$ & $R^{(out)}_{AB}$\\
\hline
Ricci Tensor  for $\tilde g_{AB} =\eta_{AB} +\tilde h_{AB}$ & $R^{(in)}_{AB}$\\
 \hline
Ricci Scalar  for $g_{AB} =\eta_{AB} +h_{AB}$ & $R^{(out)}$\\
\hline
Ricci Scalar  for $\tilde g_{AB} =\eta_{AB} +\tilde h_{AB}$ & $R^{(in)}$\\
\hline
Ricci Tensor  for $g^{(ind)}_{\mu\nu}$ & ${\cal R}_{\mu\nu}$\\
 \hline
Ricci Scalar for $g^{(ind)}_{\mu\nu}$ & ${\cal R}$\\
 \hline
 Field strength for $G_A$ & $F_{AB}$\\
 \hline
  $\partial_AG^{(k)}_B -\partial_BG^{(k)}_A$& $F^{(k)}_{AB}$\\
  \hline
  Field strength for $\tilde G_A$ & $\tilde F_{AB}$\\
   \hline
  $\partial_A\tilde G^{(k)}_B -\partial_B\tilde G^{(k)}_A$& $\tilde F^{(k)}_{AB}$\\
   \hline
  Field strength along the membrane & $\hat F_{\mu\nu}$\\
  \hline
       \end{tabular}
    \end{table}

\begin{table}[!htb]
         \caption{Different Sources}
        \centering
          \begin{tabular}{|r|l|}
      \hline
${\cal T}_{AB}$&Space-time Stress tensor\\
  \hline
  ${\cal J}_A$&Space-time Current\\
  \hline
$T_{AB}$&Defined through ${\cal T}_{AB}= \sqrt{d\rho\cdot d\rho}~ \delta(\rho-1) T_{AB}$\\
  \hline
  $J_A$&Defined through ${\cal J}_A= \sqrt{d\rho\cdot d\rho}~ \delta(\rho-1) J_A$\\
  \hline
 $T_{\mu\nu}$& Stress tensor along the membrane\\
  \hline
  $J_\mu$& Current along the membrane\\
   \hline
$T^{(out/in)}_{AB}$&${\cal K}_{AB}^{(out/in)} -{\cal K}^{(out/in)}{\mathfrak p}_{AB}$\\
  \hline
  $J^{(out)}_A$&$n^BF_{BA} $\\
  \hline
  $J^{(in)}_A$&$n^B\tilde F_{BA} $\\
  \hline
       \end{tabular}
           \end{table}

        \begin{table}[!htb]
         \caption{other notations}
        \centering
          \begin{tabular}{|r|l|}
      \hline
Fourier transform defined as& $\psi(t) = \int e^{-i\omega t} \widetilde{\psi}(\omega)\frac{d\omega}{2\pi}$\\
  \hline
  Outgoing wave  represented by &$e^{-i\omega(t-r)}$.\\
  \hline
  Greens function  defined as & $\Box~G(x,y) = -\delta^D(x-y)$\\
   \hline
$N$ & $\sqrt{d\rho\cdot d\rho}$\\
\hline
$n_A$ & $\frac{\partial_A\rho}{N}$\\
  \hline
       \end{tabular}
           \end{table}

       \begin{itemize}
 \item  For our case because of the continuity of the metric        ${\mathfrak p}_{AB}^{(in)}={\mathfrak p}_{AB}^{(out)}$. So sometimes we have denoted it by just ${\mathfrak p}_{AB}$.
 \item  In all sections we have used $(in)$  and $(out)$ both as superscript and subscript, in a way so that it does not clutter the notation mixing with other raised or lowered indices. The same is true for the superscipt (or sometimes subscript) $(k)$, used to denote the $k$th coefficient in an expansion around $\rho=1$.

   \item In most places $\hat\nabla$ denotes covariant derivative with respect to $g^{(ind,f)}_{\mu\nu}$. But in some sections (e.g., in appendix (\ref{cact})) it denotes covariant derivative with respect to $g^{(ind)}_{\mu\nu}$. What we mean will be clear from the context.\\
      
\item We have used $\nabla$ for  covariant derivative with respect to both $g_{AB}$ and $\tilde g_{AB}$. What we mean, will be clear from the context.\\
      
     \item In section (\ref{rfl})  and section \ref{bra} 
and appendices from (\ref{sphericalharmonics}) to (\ref{segf}), $\nabla_i$ denotes covariant derivative in flat space-time, but not necessarily in Cartesian coordinates and $\hat\nabla_a$ denotes covariant along unit sphere.
     
      \item Throughout this paper we employ the mostly positive sign convention.
       \end{itemize}

\section{Linearized Solutions for point masses and charges}

In this brief Appendix - whose purpose is largely to fix conventions for the normalization 
of mass and charge, we solve the linearized Einstein and Maxwell equations in the 
presence of a point mass and charge at the origin.

\subsection{Conventions for the action and equations of motion employed in this paper}

In this paper we work with the metric and gauge field governed by the 
action 
\begin{equation}\label{actionconv}
S= \frac{1}{16 \pi} \int \sqrt{-g} 
\left( R - 4 \pi F_{MN} F^{MN} \right).
\end{equation}

As explained in the text we will sometimes study this action coupled 
to a classical source - at linearized order. The resultant linearized 
equations can be obtained from the action -
\begin{equation}\label{ais}
\text{Action}= \frac{1}{16 \pi} \int \sqrt{-g} 
\left( R - 4 \pi F_{MN} F^{MN} \right) - \int 
\left( \frac{1}{2}h^{MN} {\cal T}_{MN} + {\cal J}^M A_{M} \right) ,
\end{equation}
where 
\begin{equation}\label{fmn}
F_{MN} = \nabla_M A_N -\nabla_N A_M .
\end{equation}

The linearized equations of motion (about flat space and zero gauge field) 
that follow from this action is 
\begin{equation}\label{eomfa} \begin{split}
& R_{MN}- \frac{R g_{MN}}{2}=8 \pi {\cal T}_{MN} , \\
& \nabla^MF_{MN} = {\cal J}_N\\
\end{split}
\end{equation} 
(the LHS of the first equation in \eqref{eomfa} should be linearized). 

\subsection{Comparison with the conventions used in earlier work}

In contrast with the conventions employed in this paper, 
the paper \cite{Bhattacharyya:2015fdk}
used the action 
\begin{equation}\label{actioncp}
S= \frac{1}{16 \pi} \int \sqrt{-g} 
\left( R - \frac{1}{4}F_{MN} F^{MN} \right) .
\end{equation}

It follows that the gauge fields of this paper are related to the gauge 
fields of \cite{Bhattacharyya:2015fdk} by the map 
\begin{equation}\label{translate}
A^{here}= \frac{ A^{there}}{\sqrt{16 \pi}} .
\end{equation} 

\subsection{Solutions for point sources}

We will now find solutions to the linearized version of \eqref{eomfa} in the presence of 
point sources
$${\cal T}_{00}= m \delta^{D-1} ({\vec x}) , ~~~{\cal J}^0= q \delta^{D-1}({\vec x})$$
(with all other components zero).  By definition, the spacetime that arises in response to 
these sources will be taken to have mass $m$ and charge $q$. 

In order to solve the linearized versions of \eqref{eomfa} we first employ the gauge
$$\nabla^M \left( h_{MN} - \frac{\eta_{MN} h}{2} \right)=0$$
and 
$$\nabla^M A_M=0.$$

We find that the solution to the linearized version of \eqref{eomfa} is given by 
\begin{equation}\label{soln} \begin{split}
& h_{00} = \frac{16 \pi m}{(D-2) \Omega_{D-2} r^{D-3}}   , \\
& h_{ii} = \frac{16 \pi m}{((D-3)(D-2) \Omega_{D-2} r^{D-3}} , \\
&A^0=-A_0= -\frac{q}{(D-3) \Omega_{D-2} r^{D-3}} .\\
\end{split}
\end{equation}

The corresponding linearized metric and gauge field takes the form
\begin{equation} \label{linmet} \begin{split}
ds^2 &= -dt^2 \left( 1-\frac{16 \pi m}{(D-2) \Omega_{D-2} r^{D-3}} \right)
+ dy^i dy^i  \left( 1+\frac{16 \pi m}{(D-3) (D-2) \Omega_{D-2} r^{D-3}} \right) ,\\
A_0&=\frac{q}{(D-3) \Omega_{D-2} r^{D-3}} dt ,
\end{split}
\end{equation} 
where $r^2= y^i y_i$.
It may be verified that the curvature component $R_{0i0j}$ evaluated on this solution is 
given by 
\begin{equation}\label{curvlin}
R_{0i0j}=-\frac{8\pi}{(D-2)\Omega_{D-2}}\nabla_i \nabla_j \left( \frac{m}{r^{D-3}} \right). 
\end{equation}

In a similar manner the field strength $F_{0i}$ evaluated on this solution is 
given by 
\begin{equation}\label{flin}
F_{0i}=-\frac{1}{(D-3) \Omega_{D-2}} \nabla_i \frac{q}{r^{D-3}}.
\end{equation}

The coordinate change 
$$y^i= x^i \left( 1 - \frac{8 \pi m}{(D-3) (D-2) \Omega_{D-2} r^{D-3}} \right) ,$$
turns \eqref{linmet} into 
\begin{equation} \label{linmetn} \begin{split} 
ds^2 &= -dt^2 \left( 1-\frac{16 \pi m}{(D-2) \Omega_{D-2} {\tilde r}^{D-3}} \right)
+ \frac{d {\tilde r}^2}{\left( 1-\frac{16 \pi m}{(D-2) \Omega_{D-2} {\tilde r}^{D-3}} \right)} , \\
A_0&=\frac{q}{(D-3) \Omega_{D-2} r^{D-3}} , \\
\end{split}
\end{equation}
where ${\tilde r}^2= x^i x_i $
\footnote{\eqref{linmetn} is equivalent to \eqref{linmet} under the coordinate 
change listed above only at large $r$. More precisely these two terms are 
equivalent when we keep corrections to the flat space metric of order 
$\frac{1}{r^{D-3}}$ but ignore terms of order $\frac{1}{r^{2(D-3)}}$ }
. The solution in \eqref{linmetn} is presented in the Schwarzschild gauge 
most convenient for comparing with the Reissner Nordstrom black holes of 
the next section. 

\section{Reisnner Nordstorm Black Holes and their thermodynamics}

In this Appendix we present the solutions for Reisnner Naordstorm black holes 
in arbitrary dimensions and also review their thermodynamics. The material 
reviewed here is, of course, well known. Our main purpose is to 
establish conventions. 

The system \eqref{actionconv} admits the following two parameter set of 
exact Reissner Nordstrom solutions 
\begin{equation}\label{erns}
\begin{split}
ds^2 &= -dt^2f(r)+\frac{dr^2}{f(r)}+r^2d\Omega_{D-2}^2 ; \\
f(r) &= \left(1-\frac{(1+c_DQ^2)r_0^{D-3}}{r^{D-3}}+\frac{c_DQ^2r_0^{D-3}}{r^{2(D-3)}}\right) , \\
A &= \frac{Q}{\sqrt{8 \pi}} \left(\frac{r_0}{r}\right)^{D-3}dt ,
\end{split}
\end{equation}
where \footnote{Note that the solution \eqref{erns} agrees with 
the solution reported in \cite{Bhattacharyya:2015fdk} after using 
\eqref{translate}.}
$$c_D= \frac{D-3}{D-2}.$$

The mass and charge of these solutions are easily read off by comparison 
with \eqref{linmetn}; we find 
\begin{equation}\label{mandq}
\begin{split}
m &= \frac{(D-2)(1+c_DQ^2)r_0^{D-3}\Omega_{D-2}}{16\pi} ,\\
q &= \frac{1}{\sqrt{8 \pi}}(D-3)Qr_0^{D-3}\Omega_{D-2} .
\end{split}
\end{equation}

The inverse temperature of these solutions is obtained by continuing to 
Euclidean space and identifying the periodicity of the time circle that 
keeps the solution regular at the outer event horizon; this procedure gives
\begin{equation}\label{tempbh}
T=\frac{(1-c_DQ^2)(D-3)}{4\pi r_0} .
\end{equation}

The chemical potential $\mu$ of this solution is given by $A_0$ evaluated at 
infinity minus $A_0$ evaluated at the outer event horizon and equals 
\begin{equation}\label{mubh}
\mu=- \frac{1}{\sqrt{8 \pi}} Q.
\end{equation}

Finally, the entropy of the black hole is the area of its outer event horizon 
divided by $4$ and is given by 
\begin{equation}\label{entbh}
S =  \frac{\Omega_{D-2}r_0^{D-2}}{4} .
\end{equation}

It is easily verified that 
\begin{equation} \label{thermoalg}\begin{split} 
TdS=& \frac{\Omega_{D-2}(D-2)(D-3)}{16 \pi}(1- c_DQ^2)r_0^{D-4} dr_0 , \\
dm=&  \frac{\Omega_{D-2}(D-2)}{16 \pi} 
\left( (1+ c_D Q^2)(D-3) r_0^{D-4} dr_0 + 2 c_D Q r-0^{D-3} dQ \right) , \\
\mu dq=& - \frac{Q \Omega_{D-2}(D-3)}{8\pi} \left( (D-3) Q r-0^{D-4} dr_0 
+ r_0^{D-3} d Q \right) 
\end{split}
\end{equation} 

It is easily verified that these expressions are consistent with the first 
law of thermodynamics 
\begin{equation}\label{}
T dS= dm + \mu d q .
\end{equation}

\section{Spherical Harmonics} \label{sphericalharmonics}

In this appendix we review various properties of scalar vector and 
tensor spherical harmonics that will prove useful to us in the rest 
of this paper. 

\subsection{Scalar Spherical Harmonics}

Scalar spherical harmonics form a basis for functions on the unit 
$S^{D-2}$. Every scalar spherical harmonic may be obtained as the 
restriction of a polynomial function in $R^{D-1}$ to the 
unit sphere. 
Distinct polynomials that have the same restriction to the unit sphere 
define the same spherical harmonic. In other words spherical harmonics 
may be thought of as equivalence classes of polynomials in $R^{D-1}$. 
In each equivalence class it is possible to find a unique representative 
polynomial which given by a linear combination of monomials of the form 
\begin{equation}\label{cfl}
S_l=C_{i_1 \ldots i_l} x^{i_1} \ldots x^{i_l} ,
\end{equation}
where the coefficients $C_{i_1 \ldots i_l}$ are symmetric and traceless.

Monomials of the form \eqref{cfl} of degree $l$ define a basis for
$l^{th}$ scalar spherical harmonics. Such monomials transform in the 
representation $(l, 0, \ldots 0)$ of $SO(D-1)$, where we label 
$SO(D-1)$ representations by highest weights under rotations in orthogonal 
two planes of $R^{D-1}$. It follows from the tracelessness of 
$C_{i_1 \ldots i_l}$ that 
$\nabla^2 S_l=0$ where $\nabla^2$ is the Laplacian in $R^{D-1}$. 
Transforming this equation to spherical polar coordinates we deduce that 
\begin{equation}\label{nls}
-\nabla^2 Y_{l}= l(D+l-3) Y_{l} , 
\end{equation}
where $Y_l$ is the restriction of $S_l$ onto the unit sphere and 
$\nabla^2$ on the LHS of \eqref{nls} is the Laplacian on the unit sphere.

\subsubsection{Projectors onto spaces of scalar spherical harmonics}

We use notation in which the angles on the unit $S^{D-2}$ are collectively 
denoted by $\theta$. For some purposes it is useful to define 
${\mathcal P}_l$. ${\mathcal P}_l$ acts on the space of functions on the 
unit sphere as a projector onto the $l^{th}$ spherical harmonic sector. 
In other words 
\begin{equation}\label{scspproj}
\int d \Omega'_{D-2} {\mathcal P}_l(\theta, \theta') 
Y_{l'}(\theta')= \delta_{l l'} Y_l(\theta).
\end{equation}

 It is not difficult to find an explicit expression for the projector 
${\mathcal P}_l(\theta, \theta')$. In order to do this first note that 
$${\mathcal P}_l(\theta, \theta') R \left[Y_l'(\theta') \right]
= R \left[ {\mathcal P}_l(\theta, \theta') Y_l'(\theta') \right] , $$
where $R$ is any $SO(D-1)$ rotation operator (this equation follows 
because the action of a rotation operator on any $k^{th}$ spherical harmonic 
is another $k^{th}$ spherical harmonic). It follows, in other words, that 
${\mathcal P}_l(\theta, \theta')$ is invariant under simultaneous rotations 
of $\theta$ and $\theta'$. Let ${\hat r}$ denote the unit vector in the 
direction of $\theta$ and ${\hat r}'$ denote the unit vector in the direction 
of $\theta'$. It follows that 
$${\mathcal P}_l(\theta, \theta')=f_l({\hat r}. {\hat r}') , $$ 
where $f_l(x)$ is an as yet undetermined function of a
single real variable $x$.

In order to determine $f_l(x)$  we note that
\begin{equation} \label{scproj} 
\nabla^2 {\cal P}_l(\theta, \theta')=\nabla'^2 {\cal P}_l(\theta, \theta')= 
-l(D+l-3) {\cal P}_l(\theta, \theta') .
\end{equation}

Let us now specialize to the case that the vector ${\hat r}'$ points along 
the $x^{D-1}$ axis. In this case ${\hat r}. {\hat r}'$ is simply the cosine
of the angle (let us call it $\theta$) that ${\hat r}$ makes with the 
$x^{D-1}$ axis. It follows from \eqref{scproj} that $f_l(\cos \theta)$ is 
an $l^{th}$ spherical harmonic. Notice that $f_{l}(\cos \theta)$ depends 
only on the angle with the $x^{D-1}$ axis and so is rotational invariant 
under $SO(D-2)$ rotations that leave $x^{D-1}$ unchanged. The unique
spherical harmonic with these properties is proportional to the 
unique regular solution to the differential equation 
$$ \frac{1}{\left( \sin (\theta) \right)^{D-3} } \partial_\theta
\left(\left( \sin (\theta) \right)^{D-3} f_l(\cos \theta) \right) 
= -l(D+l-3)f_l( \cos \theta) . $$

Solving the equation we find 
\begin{equation}\label{plsoln}
f_l(\cos \theta) = N_l (\sin \theta)^{-\frac{D-4}{2}}P^{\frac{D}{2}-2}_{\frac{D}{2}+l-2} 
(\cos \theta)  , 
\end{equation}
where $P^{\frac{D}{2}-2}_{\frac{D}{2}+l-2}(x)$ is an associated Legendre function
and $N_l$ is an as yet undetermined constant. 

In order to determine $N_l$ we use the equation \eqref{scspproj} 
for the special case that ${\hat r}$ points along the $x^{D-1}$ axis, and 
the function it acts on ($Y_l'$ in \eqref{scspproj}) is chosen to be 
$ (\sin \theta')^{-\frac{D-4}{2}} P^{\frac{D}{2}-2}_{\frac{D}{2}+l-2}(\cos \theta')$ 
where $\theta'$ is the angle of ${\hat r}'$ 
with the $x^{D-1}$ axis. 
It follows from \eqref{scspproj} that 
\begin{equation}\label{normdet} \begin{split}
\lim_{\theta \to 0} \left(  (\sin \theta')^{-\frac{D-4}{2}} 
P^{\frac{D}{2}-2}_{\frac{D}{2}+l-2}(\cos \theta) \right)  &= 
N_l \Omega_{D-3}\int (sin \theta)^{D-3} 
\left( (\sin \theta)^{-\frac{D-4}{2}} P^{\frac{D}{2}-2}_{\frac{D}{2}+l-2}
(\cos \theta) \right)^2 \\
&= N_l \Omega_{D-3}\int \sin \theta 
\left( P^{\frac{D}{2}-2}_{\frac{D}{2}+L-2}
(\cos \theta) \right)^2    , 
\end{split}
\end{equation}
where $\Omega_{D-3}$ is the volume of the unit $D-3$ sphere. 
The integral on the RHS of \eqref{normdet} is standard in the theory 
of Legendre functions and is given by
$$ \int \sin \theta' \left( P^{\frac{D}{2}-2}_{\frac{D}{2}+l-2} (\cos \theta')
\right)^2= \frac{ 2 (l+D-4)!}{(2l+D-3) l!} .$$

Moreover the limit on the LHS is given by 
$$\lim_{\theta \to 0} \left(  (\sin \theta')^{-\frac{D-4}{2}} 
P^{\frac{D}{2}-2}_{\frac{D}{2}+l-2}(\cos \theta) \right) = 
\left( \frac{-1}{2} \right)^{\frac{D-4}{2}} \frac{(l+D-4)!}{l! 
\Gamma(\frac{D-2}{2}) } . $$ 
These relations determine $N_l$; plugging in the value we obtain  
\begin{equation}\label{pl}
f_l(\cos \theta)= \left( \frac{-1}{2} \right)^{\frac{D}{2}}
\frac{2l+ D-3}{\pi^{\frac{D-2}{2}}} (\sin \theta')^{-\frac{D-4}{2}} 
P^{\frac{D}{2}-2}_{\frac{D}{2}+l-2}(\cos \theta) .
\end{equation}

In particular we have 
\begin{equation} \label{pea}
{\cal P}_l(0)= \lim_{\theta \to 0} f_l(\cos \theta)=
\frac{1}{2^{D-2}\pi^{\frac{D-2}{2}}}\frac{(l+D-4)!}{l!}\frac{(2l+D-3)}{\Gamma\left(\frac{D-2}{2}\right)} .
\end{equation}

\subsection{Vector spherical harmonics}

Vector spherical harmonics form a basis for the set of divergence free 
vector fields on the unit sphere $S^{D-2}$. In this brief section we will 
describe how vector spherical harmonics can be obtained as the restriction 
of polynomial valued vector fields in $R^{D-1}$. We will also use this 
description to compute some of the properties of these harmonics.

Consider a vector field in $R^{D-1}$ of the form 
\begin{equation}\label{alexp}
W^l_i= V_{i, i_1 \ldots i_l}x^{i_1}\ldots x^{i_l}.
\end{equation}
We will be interested in the restriction of this vector field onto the 
unit sphere. As in the previous subsection different expressions of the 
form \eqref{alexp} that restrict to the same vector field on the unit 
sphere will be considered equivalent; in other words vector spherical 
harmonics are identified with equivalence classes of expressions of the form 
\eqref{alexp}. The indices $i_1 \ldots i_l$ are clearly symmetric. 
As the normal component of the vector field $W^l_i$ has no restriction to 
the sphere it is convenient to set this component to zero. 
The requirement $x^i W^l_i=0$ is equivalent to the condition that the 
$V_{i, i_1 \ldots i_l}$ vanishes under symmetrization between $i$ and
(say) $i_1$. As in the previous section one can find a representative
in any equivalence class with the property that the coefficient functions  
$V_{i, i_1 \ldots i_l}$ vanish upon tracing, say, $i_1$ and $i_2$. 
The set of coefficient 
functions with these properties transform in the $(l,1, \ldots 0)$ representation of $SO(D-1)$ (see the previous subsection for an explanation of our labelling of representations).

It follows from all the conditions we have imposed that 
\begin{equation}\label{divvf1}
\nabla. W^l=0  , 
\end{equation}
where the divergence is taken in the embedding $R^{D-1}$. Translating this 
equation to polar coordinates we also find 
\begin{equation}\label{divvf}
\nabla. W^l=0 , 
\end{equation}
where $W^l$ is now thought of as a vector field on the unit sphere and 
$\nabla$ is now regarded as the covariant derivative on the unit sphere.

 The set of vector fields $W^l_i$ - when restricted to the sphere - 
define a basis for $l^{th}$ vector spherical 
harmonics on $S^{D-2}$. We use the symbol $V^\alpha_{l}$ to denote $l^{th}$ 
vector spherical harmonics on $S^{D-2}$. We will sometimes also use the symbol 
$V^\alpha_l$ to denote a vector function in the full embedding $R^{D-1}$ 
defined by 
\begin{equation}\label{vlmd}
V^l_i= V_{i, i_1 \ldots i_l}\frac{x^{i_1}\ldots x^i_l}{r^l} , 
\end{equation} 
where the coefficients  $V_{i, i_1 \ldots i_l}$ are constants independent 
of $r$. With this normalization each Cartesian component of the vector 
field $V^l$ is independent of $r$. 

Note that for any fixed $i$ (where $i$ is a Cartesian coordinate) 
$W^l_i$ is a polynomial of the form 
\eqref{cfl}. It follows that $\nabla^2 W^l_i=0$ (where $\nabla^2$ 
is the Laplacian acting on $R^{D-1}$). In a similar manner for any fixed 
$i$ the function $V^l_i$ defined in \eqref{vlmd} is an $r$ independent
scalar spherical harmonic of degree $l$ and so it follows that 
\begin{equation}\label{nlv}
-\nabla^2 V^{l}_i= \frac{l(D+l-3)}{r^2} V^l_i , 
\end{equation}
where, once again, the Laplacian is taken in the embedding $R^{D-1}$. 

Consider a sphere of radius $r$ centered about the origin of $R^{D-1}$. 
The restriction of $V^l_i$ onto this sphere defines a vector field 
on the sphere. We will now compute the eigenvalue of the Laplacian 
${\hat \nabla^2}$ on this sphere acting on this vector field. 
 Using standard formulae
\begin{equation}
\label{vecsphein}
\begin{split}
&\hat{\nabla}^2 (V_i(\theta) \\&= \partial_m(\Pi_{jn}\partial_m V_{j})\Pi_{in}\\
&=-\frac{1}{r}\Pi_{im}\hat{r}_j\partial_m V_j+\Pi_{ij}\partial_m\partial_n V_j\\
&=\frac{1}{r^2}V_i-\frac{l(l+D-3)}{r^2}V_i\\
&=-\frac{l(l+D-3)-1}{r^2}V_i ,
\end{split} 
\end{equation}
(where $\Pi_{ij}$ denotes the projector onto the unit sphere).
Here we have used some identities
 \begin{equation}\label{projiden}
 \begin{split}
\partial_i \hat{r}_j&=\frac{1}{r}\Pi_{ij} , \\
\partial_{k}\partial_{k}\hat{r}_i&=-\frac{D-2}{r^2}\hat{r}_i , \\
\partial_i \Pi_{ij}&=-\frac{D-2}{r}\hat{r}_j , \\
\partial_k\Pi_{ij}&=-\frac{1}{r}\left(\Pi_{ki}\hat{r}_j+\Pi_{kj}\hat{r}_i\right),\\
\partial_k\partial_k\Pi_{ij}&=-\frac{2}{r^2}\left(\Pi_{ij}-(D-2)\hat{r}_i\hat{r}_j\right) .
 \end{split}
 \end{equation}

Upon setting $r=1$  \eqref{vecsphein} gives the eigenvalue Laplacian (viewed 
as a vector field acting on vectors on the unit sphere) acting on the 
$l^{th}$ vector spherical harmonic.

As in the previous subsection we define the linear operator 
${\mathcal P}^V_l$ which acts on vector fields on the unit sphere and 
projects onto the sector of vector field spanned by $l^{th}$ vector spherical 
harmonics.
\begin{equation}\label{Pvd} \begin{split}
&{\mathcal P}^V_l [V_{l'}]= \delta_{l l'} V_{l'} , \\
&{\mathcal P}^V_l[\partial \chi]=0 .\\
\end{split}
\end{equation}

It is possible to work out an explicit form for the projector 
${\mathcal P}^V_l$; however we will not have need for the explicit expression
in this paper and so will not pause to do so.

\subsection{Tensor Spherical harmonics}

Mimicking the analysis of the previous section, a basis for traceless, 
divergenceless symmetric tensor fields on the unit sphere is given by the 
restriction of the polynomial expressions
\begin{equation}\label{tsfd}
B^l_{ij}=T_{i j, i_1 \ldots i_l}x^{i_1} \ldots x^{i_l} ,
\end{equation}
onto the unit sphere. The coefficient function 
$T_{i j, i_1 \ldots i_l}$ is chosen to have the following properties. 
\begin{itemize}
\item It is symmetric in the indices $i_1 \ldots i_l$ and separately 
in $i,  j$. 
\item It vanishes under tracing any two of the indices. 
\item It vanishes under the symmetrization of (say) $i$ with (say) $i_1$.
\end{itemize}
The coefficient functions $T_{i j, i_1 \ldots i_l}$ 
transform in the $(l,2, 0 \ldots 0)$ representation of $SO(D-1)$.
The restriction of $B^l_{ij}$ to the unit sphere yields a set of symmetric 
traceless, divergenceless tensor fields on the unit sphere that form the 
basis for the set of $l^{th}$ tensor spherical harmonics. 

Note that any fixed Cartesian component of $B^l_{ij}$ is a function 
of the form \eqref{cfl}, and so $\nabla^2 B^l_{ij}=0$,  where $\nabla^2$ 
is the Laplacian on $R^{D-1}$. 
 
As in the previous subsection we sometimes have used $B^l_{ij}$ for a tensor 
spherical harmonic field that is defined in all of $R^{D-1}$. Rather 
than the function $B^l_{ij}$ defined above, we find it convenient to use 
the normalized tensor fields 
\begin{equation}\label{tsf}
T^l_{ij}=T_{i j, i_1 \ldots i_l} \frac{x^{i_1} \ldots x^{i_l}}{r^l}.
\end{equation}

As in the previous section we may restrict $T^{l}_{ij}$ to the surface
of a sphere of radius $r$. The Laplacian of $T^l$ viewed as a tensor 
field on this restricted surface is easily computed; we have 
\begin{equation}
\begin{split}
\hat{\nabla}^2 (T_{ij}(\theta) & = \partial_m(\Pi_{pn}\Pi_{qk}\partial_m T_{pq})\Pi_{in}\Pi_{jk}\\
&=\frac{2}{r^2}T_{ij}-\frac{l(l+D-3)}{r^2}T_{ij}\\
&=-\frac{l(l+D-3)-2}{r^2}T_{ij}.
\end{split}
\end{equation}

As in previous subsections we define the linear operator 
${\mathcal P}^T_l$ , which acts on traceless symmetric tensor on the unit 
sphere and projects onto the sector of tensor fields spanned by 
$l^{th}$ tensor spherical harmonics.
\begin{equation}\label{Pvdt} \begin{split}
&{\mathcal P}^T_l [(T_{ij})_{l'}]= \delta_{l l'} (T_{ij})_{l'} , \\
&{\mathcal P}^T_l[{\rm anything ~else}]=0 ; \\
\end{split}
\end{equation}
where `anything else' refers to tensors formed out of derivatives acting on 
scalar or vector spherical harmonics. 
It should be possible to work out an explicit form for the projector 
${\mathcal P}^T_l$; however we will not have need for the explicit expression
in this paper and so will not pause to do so.

\subsection{Decomposition of the general vector field on $R^{D-1}$ in a spherical basis}
\label{dvf}

As we have mentioned above, the most general vector field on $R^{D-1}$ 
can be constructed out of two scalar fields and one divergenceless purely 
angular vector field. The decomposition takes the form 
\begin{equation}\label{tsov}
{\vec A} = \hat{r} a + \nabla b + {\vec \gamma} , 
\end{equation}
where $a$ and $b$ are arbitrary scalar fields and ${\vec \gamma}$ is an 
arbitrary divergence free, purely angular vector field. We emphasize that 
the scalars $a$ and $b$ and the vector field ${\vec \gamma}$ are arbitrary 
functions of the radial coordinate $r$. 

In \eqref{tsov}
we have arbitrarily chosen a basis for the two scalar fields in the problem; 
of course any two linearly independent linear combinations of $a$ and $b$ 
would form as good a basis. We will now find a geometrically natural 
basis for the problem. Let $\alpha$ and $\beta$ be the two scalar functions and let 
$\alpha_l$ and $\beta_l$ respectively represent the projection of these functions 
into the space of $l^{th}$ spherical harmonics i.e. 
\begin{equation} \label{scaldecomp}
\alpha= \sum_{l=0}^\infty \alpha_l, ~~~\beta= \sum_{l=0}^\infty \beta_l, ~~~
{\cal P}_{l'} \alpha_l= \delta_{l l'} \alpha_l, ~~~
{\cal P}_{l'} \beta_l= \delta_{l l'} \beta_l ,
\end{equation}
where ${\cal P}_{l}$, the projector onto the $l^{th}$ scalar spherical harmonic
was defined in \eqref{scspproj}. 
Let ${\vec \gamma}$ represent the non radial and divergence free vector field 
and let ${\vec \gamma}_l$ represent the projection of this field onto the 
space of $l^{th}$ vector spherical harmonics i.e. let
\begin{equation} \label{scaldecomp1}
{\vec \gamma}= \sum_{l=1}^\infty {\vec \gamma}_l,~~~
{\cal P}^V_{l'} {\vec \gamma}_l= \delta_{l l'} {\vec \gamma}_l,
\end{equation}
where ${\cal P}^V_{l'}$ was defined in \eqref{Pvd}. 
As emphasized above $\alpha_l$ , $\beta_l$ and ${\vec \gamma}_l$ are all 
arbitrary functions of $r$. The most general vector field ${\vec J_{\rm eff} }$ 
can be parametrized in terms of $\alpha$, $\beta$ and ${\vec \gamma}$ by 
\begin{equation} \label{arbitvector}
{\vec {J_{\rm eff}}} =
\left( {\vec {\cal A}}^-[\alpha] +
{\vec {\cal A}}^+[\beta]+ {\vec \gamma} \right) ,
\end{equation}
where{\footnote{ We have defined the projected derivative $\nabla^p$ as follows\begin{equation}\begin{split}&\text{Scalar}:\nabla^p_i\alpha=\Pi_i^j\partial_j\alpha,\\
&\text{Vector}:\nabla^p_i\beta_j=\Pi_i^k\Pi_j^l\partial_k(\Pi_l^m\beta_m).\end{split}\end{equation} and so on for the tensor .}}
\begin{equation}\label{lcs}
\begin{split}
{\vec {\cal A}}^-[\alpha] &= \sum_{l=0}^\infty 
\left( l\hat{r} \alpha_l+r {\vec \nabla}^p \alpha_l \right) , \\
{\vec {\cal  A}}^+[\beta]  &= \sum_{l=0}^\infty 
\left( (l+D-3)\hat{r}_i \beta_l-r {\vec \nabla}^p \beta_l \right) .
\end{split}
\end{equation}

The linear combinations in \eqref{lcs} are special because they 
are `diagonal' under the action of $\mathcal{P}_{l'}$, the projector 
onto scalar spherical harmonics acting separately on each Cartesian 
components. Specifically we have   
\begin{equation} \label{projrel}
\begin{split}
\mathcal{P}_{l'}  \left( {\vec {\cal A}}^-[\alpha] \right)    &={\vec {\cal A}}^-
[\mathcal{P}_{l'+1} \alpha] ,  \\
\mathcal{P}_{l'}  \left( {\vec{ \cal A}}^+[\beta)] \right) &= {\vec {\cal A}}^+
[\mathcal{P}_{l'-1} \beta ] .
\end{split}
\end{equation}
\footnote{This equation can be restated as 
\begin{equation} \label{projreln}
\begin{split}
\mathcal{P}_{l'}({\vec{\cal A}}^-[\alpha_l])  &= \delta_{l',l-1}({\vec {\cal A}}^-[\alpha_l]) , \\
\mathcal{P}_{l'}({\vec {A}}^+[\beta_l]) &= \delta_{l',l+1}({\vec {\cal A}}^+
[\beta_l]).
\end{split}
\end{equation}
}

The action of the scalar projector on individual Cartesian components of vector 
spherical harmonics is automatically diagonal and is very simple 
\begin{equation}\label{povs}
\mathcal{P}_{l} \left( {\vec \gamma} \right)= \mathcal{P}^V_{l} 
\left( {\vec \gamma} \right) 
= {\vec \gamma}_l , 
\end{equation}
where $\mathcal{P}^V_{l}$ represents the projector onto the space of 
$l^{th}$ vector spherical harmonics. 

It is now easy to deduce the action of the $R^{D-1}$ Laplacian $\nabla^2$ 
on the vector field ${\vec J_{\rm eff} }$.  Using the fact the Laplacian 
in Cartesian 
coordinates acts on each component of a vector field as if it were 
a scalar, it follows immediately from \eqref{povs} and \eqref{projrel}
\begin{equation}\label{actnablav} \begin{split}
&\nabla^2 {\vec {\cal A}}^-[\alpha] = {\vec {\cal A}}^-[{\tilde \alpha} ] ,\\
&\nabla^2 {\vec {\cal A}}^+[\beta] = {\vec {\cal A}}^+[{\tilde \beta} ] , \\
&\nabla^2 {\vec \gamma} = {\tilde {\vec \gamma}} , \\
\end{split}\end{equation}
where
\begin{equation}
\begin{split}
&{\tilde \alpha} = \sum_l \left( 
\frac{1}{r^{D-2}} \partial_r \left( r^{D-2} \partial_r \alpha_l \right) - 
\frac{(l-1)(l-1+D-3)}{r^2} \alpha_l  \right) , \\
&{\tilde \beta} = \sum_l \left( 
\frac{1}{r^{D-2}} \partial_r \left( r^{D-2} \partial_r \beta_l\right) - 
\frac{(l+1)(l+1+D-3)}{r^2} \beta_l  \right) , \\
&{\tilde {\vec \gamma}}= \sum_l \left( 
\frac{1}{r^{D-2}} \partial_r \left( r^{D-2} \partial_r {\vec \gamma_l} \right) - 
\frac{(l)(l+D-3)}{r^2} {\vec \gamma}_l  \right) .\\
\end{split}
\end{equation}
In words,  $\nabla^2$ acts on $\alpha_l$ as it would on the $(l-1)^{th}$ 
component of a scalar field. $\nabla^2$ acts on $\beta_l$ as it would on 
the $(l+1)^{th}$ component of a scalar field. $\nabla^2$ acts on 
${\vec \gamma}_l$ as it would on the $(l)^{th}$ component of a scalar field; 
the last statement reflects the fact that the $l^{th}$ scalar and vector 
spherical harmonics have equal eigenvalues under the the action of the 
Laplacian on the unit sphere. 

It is also not difficult to verify that 
\begin{equation}\label{divbasis} \begin{split} 
&{\vec \nabla}\cdot{\vec {\cal A}}^-[\alpha]
= \sum_l   lr^{l-1} \partial_r \left( \frac{\alpha_l}{r^{l-1}} 
\right) , \\
&{\vec \nabla}\cdot{\vec {\cal A}}^+[\beta]
= \sum_l  \frac{l+D-3}{r^{l+D-2}} \partial_r \left( r^{l+D-2} \beta_l 
 \right)  , \\
&{\vec \nabla}\cdot{\vec \gamma}=0.\\
\end{split}
\end{equation}

\subsection{An expansion of the arbitrary tensor field in a spherically adapted basis}

The most general symmetric tensor field in $R^{D-1}$ can be split into 
its trace - which is a decoupled scalar - and a traceless symmetric tensor. 
We ignore the trace part in what follows. The most general traceless 
symmetric tensor field is parametrized by three scalar fields, two 
angular divergence free vector fields and one angular divergence free 
tensor field. As in the previous subsection, it is dynamically convenient
to choose a basis for the vectors and the scalars that diagonalizes 
the action of $\nabla^2$. The logic and algebra is very similar to 
the previous subsection and we only present final results.

Following the previous 
subsection we use obvious notation to denote the projection of any 
of these quantities to their $l^{th}$ spherical harmonic sector. For instance
$\alpha_l$ represents the projection of $\alpha$ to the $l^{th}$ scalar 
spherical harmonic sector, while ${\vec \phi}_l$ represents the projection 
of ${\vec \phi}$ to the $l^{th}$ vector spherical harmonic sector, etc. 
\footnote{The index $l$ runs from $0$ to $\infty$ in the case of scalars, from 
$1$ to $\infty$ in the case of vectors and from $2$ to $\infty$ in 
the case of tensors} A general tensor field $T_{ij}$ is given in terms 
of this data by the decomposition
\begin{equation} \label{arbitscalarg}
\begin{split}
T_{ij} =&
\left( {\cal C}_{ij}^-[\alpha] + {\cal C}_{ij}^+[\beta] + {\cal C}_{ij}^0[\gamma] + \delta_{ij}\kappa\right)\\ 
&+ \left({\cal B}_{ij}^-[\phi] + {\cal B}_{ij}^+[\psi]\right)
+ \chi_{ij} ,
\end{split}
\end{equation}
where
\begin{equation}\label{lct}
\begin{split}
(\mathcal{C}^{-} )_{ij}[\alpha] =& \mathcal{A}^-_i \left( \mathcal{A}^-_j[\alpha] \right)    = \mathcal{A}^-_j \left( \mathcal{A}^-_i[\alpha] \right)  , \\
(\mathcal{C}^{+} )_{ij}[\beta] =& \mathcal{A}^+_i \left( \mathcal{A}^+_j[\beta] \right)  = \mathcal{A}^+_j \left( \mathcal{A}^+_i[\beta] \right)  , \\
(\mathcal{C}^{0} )_{ij}[\gamma] =
& \frac{1}4\left(\mathcal{A}^-_i \left( \mathcal{A}^+_j[\gamma] \right) 
+\mathcal{A}^-_j \left( \mathcal{A}^+_i[\gamma] \right)  +
\mathcal{A}^+_i \left( \mathcal{A}^-_j[\gamma] \right)   + 
\mathcal{A}^+_j \left( \mathcal{A}^-_i[\gamma] \right)  \right)\\&
-\delta_{ij}\sum_l\left(\frac{2l(l+D-3)}{D-1}+\frac{D-3}2\right)\gamma_l ,\\
\delta_{ij}\kappa =& \sum_l\frac{1}{2(2l+D-3)}\left(\mathcal{A}^-_i 
\left( \mathcal{A}^+_j[\kappa_l] \right)   +
\mathcal{A}^-_j \left( \mathcal{A}^+_i [\kappa_l] \right)  -
\mathcal{A}^+_i \left( \mathcal{A}^-_j[\kappa_l] \right)  
-\mathcal{A}^+_j \left( \mathcal{A}^-_i[\kappa_l] \right) \right) , \\
(\mathcal{B}^{-} )_{ij}[{\vec \phi}] =& \mathcal{A}^-_i[\phi_j]+\mathcal{A}^-_j[\phi_i] , \\
(\mathcal{B}^{+} )_{ij}[{\vec \psi}] =& \mathcal{A}^+_i[\psi_j]+\mathcal{A}^+_j[\psi_i] .
\end{split}
\end{equation}

Quantities like $\mathcal{A}^-_i \left( \mathcal{A}^-_j[\alpha] \right)$ that appear in the equation above 
have the following meaning; the operator $\mathcal{A}^-_i$ acts on each Cartesian component of  
$\mathcal{A}^-_j[\alpha]$ as if it were a scalar (i.e. according to the formula \eqref{lcs}). 
\footnote{Here we explicitly write the expressions for the action of two $\vec{{\mathcal{A}}}'s$ on a scalar
\begin{equation}\label{aaexp}
\begin{split}
\mathcal{A}^-_i\mathcal{A}^-_j[\alpha]&=\left((l-1)\hat{r}_i+r\nabla^p_i\right)\left(l\hat{r}_j+r\nabla^p_j\right)[\alpha]  \\
&=l(l-2)\hat{r}_i\hat{r}_j\alpha+(l-1)(r\hat{r}_i\nabla^p_j+r\hat{r}_j\nabla^p_i)\alpha+l\delta_{ij}\alpha+r^2\nabla_{ij}\alpha , \\
\mathcal{A}^+_i\mathcal{A}^+_j[\alpha]&=\left((l+D-2)\hat{r}_i-r\nabla^p_i\right)\left((l+D-3)\hat{r}_j-r\nabla^p_j\right)[\alpha]\\
&=(l+D-3)(l+D-1)\hat{r}_i\hat{r}_j\alpha-(l+D-2)(r\hat{r}_i\nabla^p_j+r\hat{r}_j\nabla^p_i)\alpha-(l+D-3)\delta_{ij}\alpha+r^2\nabla_{ij}\alpha ,\\
\mathcal{A}^-_i\mathcal{A}^+_j[\alpha]&=\left((l+1)\hat{r}_i+r\nabla^p_i\right)\left((l+D-3)\hat{r}_j-r\nabla^p_j\right)[\alpha]\\
&=l(l+D-3)\hat{r}_i\hat{r}_j\alpha-(l+1)r\hat{r}_i\nabla^p_j\alpha+(l+D-2)r\hat{r}_j\nabla^p_i\alpha+(l+D-3)\delta_{ij}\alpha-r^2\nabla_{ij}\alpha , \\
\mathcal{A}^+_i\mathcal{A}^-_j[\alpha]&=\left((l+D-4)\hat{r}_i-r\nabla^p_i\right)\left(l\hat{r}_j+r\nabla^p_j\right)[\alpha]\\
&=l(l+D-3)\hat{r}_i\hat{r}_j\alpha+(l+D-4)r\hat{r}_i\nabla^p_j-(l-1)r\hat{r}_j\nabla^p_i\alpha-l\delta_{ij}\alpha-r^2\nabla_{ij}\alpha , 
\end{split}
\end{equation}
where we have defined $\nabla_{ij}$ as the complete projected derivative of two $\nabla_i$ and is given by $$\nabla_{ij}\alpha=\Pi_i^k\Pi_j^m\partial_m\left(\Pi_k^n\partial_n\alpha\right).$$
It can be easily shown that $\nabla_{ij}$ is symmetric under the exchange of $i\longleftrightarrow j$.}
More generally the action of the $\mathcal{A}^-_i$ or $\mathcal{A}^+_i$ on any vector field is that 
these operators act on each of the Cartesian components of the corresponding vector field as they 
would on a scalar (i.e. according to \eqref{lcs}). Using the fact that each Cartesian component 
of an $l^{th}$ vector harmonic is an $l^{th}$ scalar harmonic , it follows that the action of these 
operators on tangential divergenceless vector fields (those that can be expanded in vector harmonics) is 
given by 
\begin{equation}\label{lcv}
\begin{split}
{\vec {\cal A}}^-[{\vec \phi}] &= \sum_{l=0}^\infty 
\left( l\hat{r} {\vec \phi}_l+r {\vec \nabla}^p {\vec \phi}_l \right) , \\
{\vec {\cal  A}}^+[{\vec \psi} ]  &= \sum_{l=0}^\infty 
\left( (l+D-3)\hat{r}_i {\vec \psi}_l-r {\vec \nabla}^p {\vec \psi}_l \right) .
\end{split}
\end{equation}
Of course the linear combinations in \eqref{lct} 
are `diagonal' under the action of $\mathcal{P}_{l'}$, the projector 
onto scalar spherical harmonics acting separately on each Cartesian 
component. 
\begin{equation} \label{Tprojrel}
\begin{split}
\mathcal{P}_{l}  \left((\mathcal{C}^{-} )_{ij}[\alpha]\right) &= (\mathcal{C}^{-} )_{ij}[\mathcal{P}_{l+2}\alpha]  , \\
\mathcal{P}_{l}  \left((\mathcal{C}^{+} )_{ij}[\beta]\right) &= (\mathcal{C}^{+} )_{ij}[\mathcal{P}_{l-2}\beta] , \\
\mathcal{P}_{l}  \left((\mathcal{C}^{0} )_{ij}[\gamma]\right) &= (\mathcal{C}^{0} )_{ij}[\mathcal{P}_{l}\gamma] , \\
\mathcal{P}_{l}  \left((\mathcal{B}^{-} )_{ij}[{\vec \phi}]\right) &= (\mathcal{B}^{-} )_{ij}[\mathcal{P}^V_{l+1}{\vec \phi}] , \\
\mathcal{P}_{l}  \left((\mathcal{B}^{+} )_{ij}[{\vec \psi}]\right) &= (\mathcal{B}^{+} )_{ij}[\mathcal{P}^V_{l-1}{\vec \psi}]
\end{split}
\end{equation}
(recall ${\mathcal P}^V_l$ projects onto the subspace of $l^{th}$ vector 
spherical harmonics).

The action of the scalar projector on individual Cartesian components of tensor 
spherical harmonics is automatically diagonal and is very simple 
\begin{equation}\label{pots}
\mathcal{P}_{l} \left(\chi\right)_{ij}= \mathcal{P}^T_{l} \left(\chi\right)_{ij}=\left(\chi_l\right)_{ij} , 
\end{equation}
where $\mathcal{P}^T_{l}$ represents the projector onto the space of 
$l^{th}$ tensor spherical harmonics. Equation \eqref{pots} simply asserts that each  
Cartesian component of modes in the $l^{th}$ tensor spherical harmonic is a scalar spherical 
harmonic of degree $l$. 

The action of the operator $\nabla^2$ is also diagonal - and rather simple 
- in this basis
\begin{equation}\label{actnabla} \begin{split}
&\nabla^2  \left((\mathcal{C}^{-} )_{ij}[\alpha]\right) = (\mathcal{C}^{-} )_{ij}[\tilde{\alpha}]  , \\
&\nabla^2  \left((\mathcal{C}^{+} )_{ij}[\beta]\right) = (\mathcal{C}^{+} )_{ij}[\tilde{\beta}] , \\
&\nabla^2  \left((\mathcal{C}^{0} )_{ij}[\gamma]\right) = (\mathcal{C}^{0} )_{ij}[\tilde{\gamma}] , \\
&\nabla^2  \left((\mathcal{B}^{-} )_{ij}[{\vec \phi}]\right) = (\mathcal{B}^{-} )_{ij}[\tilde { \vec \phi}] , \\
&\nabla^2  \left((\mathcal{B}^{+} )_{ij}[{\vec \psi}]\right) = (\mathcal{B}^{+} )_{ij}[\tilde{\vec \psi}] , \\
&\nabla^2  \chi_{ij} = \tilde{\chi}_{ij} ,\\
&\nabla^2  \kappa = \tilde{\kappa},
\end{split}\end{equation}
where 
\begin{equation}
\begin{split}
&{\tilde \alpha} = \sum_l \left( 
\frac{1}{r^{D-2}} \partial_r \left( r^{D-2} \partial_r \alpha_l \right) - 
\frac{(l-2)(l-2+D-3)}{r^2} \alpha_l  \right) , \\
&{\tilde \beta} = \sum_l \left( 
\frac{1}{r^{D-2}} \partial_r \left( r^{D-2} \partial_r \beta_l\right) - 
\frac{(l+2)(l+2+D-3)}{r^2} \beta_l  \right) , \\
&{\tilde \gamma}= \sum_l \left( 
\frac{1}{r^{D-2}} \partial_r \left( r^{D-2} \partial_r \gamma_l \right) - 
\frac{l(l+D-3)}{r^2} \gamma_l  \right) , \\
&{\tilde {\vec \phi}}= \sum_l \left( 
\frac{1}{r^{D-2}} \partial_r \left( r^{D-2} \partial_r {\vec \phi_l} \right) - 
\frac{(l-1)(l-1+D-3)}{r^2} {\vec \phi}_l  \right) , \\
&{\tilde {\vec \psi}}= \sum_l \left( 
\frac{1}{r^{D-2}} \partial_r \left( r^{D-2} \partial_r {\vec \psi_l} \right) - 
\frac{(l+1)(l+1+D-3)}{r^2} {\vec \psi}_l  \right) , \\
&{\tilde \chi}_{ij} = \sum_l \left( 
\frac{1}{r^{D-2}} \partial_r \left( r^{D-2} \partial_r (\chi_l)_{ij} \right) - 
\frac{l(l+D-3)}{r^2} (\chi_l)_{ij}  \right) , \\
&{\tilde \kappa} = \sum_l \left( 
\frac{1}{r^{D-2}} \partial_r \left( r^{D-2} \partial_r \kappa_l\right) - 
\frac{l(l+D-3)}{r^2} \kappa_l  \right).
\end{split}
\end{equation}

It is also not difficult to verify that 
\begin{equation}\label{Tdivbasis} \begin{split}
&\nabla_i\mathcal{C}_{ij}^-[\alpha]=\mathcal{A}^-_j\left[(l-1)(r)^{l-2}\partial_{r}\left(\frac{\alpha_l}{(r)^{l-2}}\right)\right] , \\
&\nabla_i\mathcal{C}_{ij}^+[\beta]=\mathcal{A}^+_j\left[(l+D-2)\frac{\partial_{r}\left((r)^{l+D-1}\beta_l\right)}{{(r)^{l+D-1}}}\right] , \\
&\nabla_i\mathcal{C}_{ij}^0[\gamma]=\mathcal{A}^+_j\left[\frac{l}{2(2l+D-3)}\left((2l+D-3)-\frac{4(l+D-3)}{D-1}\right)(r)^{l}\partial_{r}\left(\frac{\gamma_l}{(r)^{l}}\right)\right]\\&+\mathcal{A}^-_j\left[\frac{(l+D-3)}{2(2l+D-3)}\left((2l+D-3)-\frac{4l}{D-1}\right)\frac{\partial_{r}\left((r)^{l+D-3}\gamma_l\right)}{{(r)^{l+D-3}}}\right] , \\
&\nabla_i\mathcal{B}_{ij}^-[{\vec \phi}]
= \sum_l   (l-1)r^{l-1} \partial_r \left( \frac{(\phi_l)_j}{r^{l-1}} 
\right) ,  \\
&\nabla_i\mathcal{B}_{ij}^-[{\vec \psi}]
= \sum_l  \frac{l+D-2}{r^{l+D-2}} \partial_r \left( r^{l+D-2} (\psi_l)_j 
 \right)  , \\
&\nabla_i\chi_{ij}=0 .
\end{split}
\end{equation}

\section{Scalar Greens Functions} \label{gf}

\subsection{Retarded Greens Functions in position space} \label{rps}

In this subsection we obtain explicit expressions for the Greens function in position space, starting 
from the exact Fourier space result \eqref{Gofs}. 
\begin{itemize}
\item {\it Even $D$}
\end{itemize}

When $D$ is even the argument of the Hankel function that appears in \eqref{Gofs} is half integral. Now Hankel functions of half integral argument have an amazing property; their large argument expansion truncates at a finite order. In equations
\begin{equation}
 H_{m+1/2}^{(1)}(\omega r)=\sqrt{\frac{2}{\pi\omega r}}(-i)^{m+1}e^{i\omega r}\sum_{k=0}^{m}\frac{(m+k)!}{k!(m-k)!}\frac{i^k}{(2\omega r)^k} ,
\end{equation}
 $m$  (which is $m=\frac{D-4}{2}$ in our context)  is an integer. As this expression takes the form $e^{i \omega r}$ times a polynomial in $\omega$.  
It follows that $G(r,t)$ defined by 
\begin{equation}
 G(r,t)=\int \frac{d \omega}{ 2 \pi}  G_\omega(r) e^{-i\omega t} ,
\end{equation}
is a linear sum of a finite number of derivatives of 
$\delta (r-t)$. We find 
\begin{align} \label{eexp}
 G(r,t) &=\frac{1}{2}\left(\frac{1}{2\pi r}\right)^{m+1}\sum_{k=0}^{m} \frac{(m+k)!}{k!(m-k)!} \int d\omega ~~ \frac{(-i\omega)^{m-k}e^{i\omega(r-t)}}{(2r)^k} \\
 &= -\frac{1}{2}\left(\frac{-1}{2\pi r}\right)^{m+1}\sum_{k=0}^{m}\frac{(m+k)!}{k!(m-k)!}\frac{\partial_r^{m-k}\delta(t-r)}{(-2r)^k} .
\end{align}
It may be verified that \eqref{eexp} resums to  
 \begin{equation}\label{Green_even}
 G(r,t)=\frac{\theta(X^0)}{2}\left(\frac{1}{\pi}\right)^{\frac{D-2}{2}}\delta^{\left(\frac{D-4}{2}\right)}(r^2-t^2)=\frac{\theta(X^0)}{2}\left(\frac{1}{\pi}\right)^{\frac{D-2}{2}}
\left( \frac{X^M\partial_M}{2 X^N X_N} \right)^{\frac{D-2}{2}}
\delta( X^M X_M) .
\end{equation}
\footnote{Recall that $\delta^m(\alpha)$ is the $m^{th}$ derivative of 
the delta function w.r.t. $\alpha$. In the case at hand $\alpha$ is 
$r^2-t^2= X^M X_M$ and partial derivatives w.r.t. $\alpha$ can be converted 
into partial derivatives w.r.t. $X^M$ using the chain rule $\partial_\alpha
= \frac{X^M\partial_M}{2 X^N X_N} $.}
We have checked the equivalence of \eqref{Green_even} and \eqref{eexp} on 
Mathematica for $D \leq 14$.

\begin{itemize}
\item {\it Odd $D$}
\end{itemize}

 In order to obtain an explicit expression for the Greens function in odd 
$D$ we found it convenient to start with the explicit expression for the
 Greens function in $\omega$ and ${\vec k}$. Transforming to polar 
coordinates in ${\vec k}$ space we have 
 \begin{equation} \label{gfcont}
 G_D(r,t)=-\Omega_{D-3}\int \frac{d\omega}{(2\pi)^D}  d\theta (\sin\theta)^{D-3}\frac{k^{D-2}dk }{(\omega+ i \epsilon k)^2-k^2} e^{-i(\omega  t-k r \cos\theta) }
 \end{equation}
(here $\epsilon$ is an infinitesimal dimensionless number and the positive 
factor of $k$ in front of $\epsilon$ has been inserted for future convenience).
For $t<0$ we can close the contour in the upper half plane. As the 
integrand of \eqref{gfcont} is analytic here the integral vanishes, as 
expected for a retarded correlator. On the other hand for $t>0$ we close 
the contour in the lower half plane and pick up contributions from the two 
poles in the integrand. Doing the $\omega$ integral we find
 \begin{eqnarray} \label{gdee}
 G_D(r,t)&=&i\Omega_{D-3}\int\frac{1}{(2\pi)^D} d\theta (\sin\theta)^{D-3} k^{D-2} dk \frac{e^{-ik(t-i\epsilon)}-e^{ik(t+i\epsilon)}}{2k}e^{ik r \cos\theta}\nonumber\\
 &=&\frac{i\Omega_{D-3}}{(2\pi)^{D-3}} \int\frac{1}{(2\pi)^2} (k^2-k^2\cos\theta)^{\frac{D-3}{2}}\frac{e^{-ik(t-i\epsilon)}-e^{ik(t+i\epsilon)}}{2}e^{ik r \cos\theta} d\theta dk\nonumber\\
 &=&\frac{i\Omega_{D-3}}{(2\pi)^{D-3}} (-\partial_t^2+\partial_r^2)^{\frac{D-3}{2}}\int\frac{1}{(2\pi)^2}\frac{e^{-ik(t-i\epsilon)}-e^{ik(t+i\epsilon)}}{2}e^{ik r \cos\theta} d\theta dk\nonumber\\
 &=&\frac{i\Omega_{D-3}}{(2\pi)^{D-3}} (-\partial_t^2+\partial_r^2)^{\frac{D-3}2} G_3(r,t) ,
 \end{eqnarray}
where
\begin{equation} \label{gthree}
G_3(r,t)=\int\frac{e^{-ik(t-i\epsilon)}-e^{ik(t+i\epsilon)}}{2}e^{ik r \cos\theta} d\theta \frac{dk}{(2\pi)^2} .
\end{equation}

We now proceed to explicitly evaluate integral in \eqref{gthree}. 
 Evaluating the integral over $k$ in that expression we find
\begin{eqnarray}
G_3(r,t)&=&\frac{-i}{2}\int \frac{d\theta}{2\pi} \left(\frac{1}{t-r \cos(\theta)-i\epsilon}+\frac{1}{t+r\cos(\theta)+i\epsilon}\right)\nonumber\\
\therefore G_3(r,t)&=& -i\frac{I_1+I_2}{4 \pi i } , \\
\text{where $I_1$ and $I_2$ are defined as : } \nonumber \\
I_1&=&\oint \frac{2idz}{r\left(z-\frac{t+\sqrt{t^2-r^2}}{r}\right)\left(z-\frac{t-\sqrt{t^2-r^2}}{r}\right)} , \\
I_2&=&\oint \frac{-2idz}{r\left(z-\frac{-t+\sqrt{t^2-r^2}}{r}\right)\left(z-\frac{-t-\sqrt{t^2-r^2}}{r}\right)} ,
\end{eqnarray}
where we have defined $e^{i\theta}=z$, and the contour integrals above 
are taken anticlockwise over the unit circle. When $r>t$ the poles 
in $z$ in $I_1$ and $I_2$ both lie on the unit circle. This integral 
can be defined by the principal value and simply vanishes. When $t^2>r^2$, 
on the other hand, the second pole in $I_1$ lies within the unit circle while
the first pole lies outside. The situation is reversed for $I_2$; the 
first pole lies within the unit circle while the second one lies outside. 
Evaluating the integrals by contours we find
\begin{equation} \begin{split}
I_1&=\frac{-i \theta(t^2-r^2)}{\sqrt{t^2-r^2}} , \\
I_2&=\frac{-i \theta(t^2-r^2)}{\sqrt{t^2-r^2}} .\\
\end{split}
\end{equation}

Using the fact that $G_3$ vanishes for negative $t$ it follows that 
\begin{equation}
G_3(r,t)=\frac{-2\pi i\theta(t-r)}{\sqrt{t^2-r^2}} .
\end{equation}
From \eqref{gdee} it follows that 
\begin{equation}
G_D(r,t)=\frac{\Omega_{D-3}}{(2\pi)^{D-4}} (-\partial_t^2+\partial_r^2)^{\frac{D-3}{2}}\left(\frac{\theta(t-r) }{\sqrt{t^2-r^2}}\right) .
\end{equation}

\subsection{Large $D$ expansion of the Greens Function using WKB} \label{wkb}

As we have explained in the main text, the large $D$ limit of the Greens function is given by the solution of an effective Schrodinger equation which takes the form 
\begin{equation}
-\psi''(\omega, r) +\frac{D^{*2}}{4 r^2}\psi(\omega, r) = \omega^2 \psi(\omega, r),~~{\rm where}~D^* = \sqrt{(D-2)(D-4)} .
\end{equation}
\begin{figure}
\begin{center}
  \includegraphics[width=.6\textwidth]{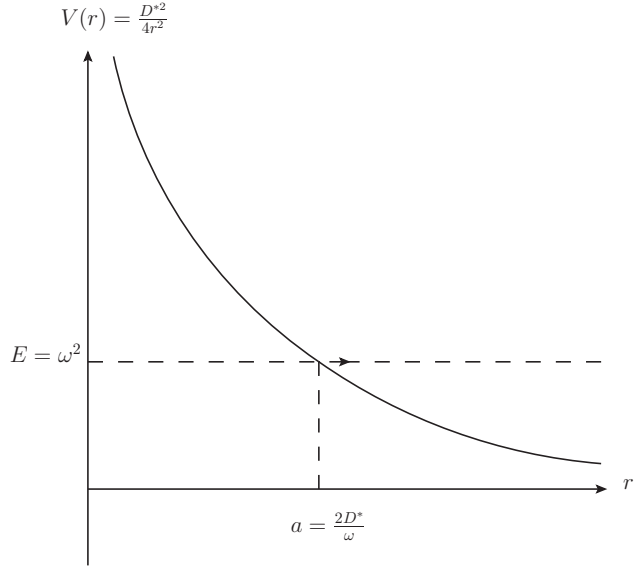}
  \caption{Potential for allowed and disallowed regions .}
  \label{fig:pot}
  \end{center}
\end{figure}

The most general WKB solution to this equation in the classically disallowed region is given by 
\begin{equation}\label{disall}
\begin{split}
\psi(\omega, r)  =~& \frac{\sqrt{D^*}B}{\sqrt{2}\left(\frac{D^{*2}}{4r^2}-\omega^2\right)^\frac{1}{4}}\left(\frac{D^*}{2\omega r}-\sqrt{\frac{D^{*2}}{4\omega^2r^2}-1}\right)^{-\frac{D^*}{2}}   \left( \frac{D^*}{\omega} \right) ^{-\frac{D^*}{2}} e^{\frac{D^*}{2} -\sqrt{\frac{D^{*2}}{4}-\omega^2r^2}}\\
+& \frac{\sqrt{D^*}A}{\sqrt{2}\left(\frac{D^{*2}}{4r^2}-\omega^2\right)^\frac{1}{4}}\left(\frac{D^*}{2\omega r}-\sqrt{\frac{D^{*2}}{4\omega^2r^2}-1}\right)^{\frac{D^*}{2}} \left( \frac{D^*}{\omega} \right) ^{-\frac{D^*}{2}}
e^{ \frac{D^*}{2} +\sqrt{\frac{D^{*2}}{4}-\omega^2r^2} } . 
\end{split}
\end{equation}
In the limit $2r\omega \ll D^*$ this solution reduces to 
\begin{equation}\label{opp}
\psi(\omega, r)  =\frac{B}{ r^{\frac{D-4}{2}}}
+ A \left( \frac{e \omega}{D} \right)^{D-3}  r^{\frac{D-2}{2}} ,
\end{equation}
in agreement with \eqref{gfsg}. In order to obtain \eqref{opp} 
, we have used $D^*= \left( D-3 \right)+ {\cal O}(1/D)$ and have ignored this higher 
order correction. We have also used 
\begin{equation}
\Omega_n= \frac{2\pi^{\frac{n+1}{2}}}{\Gamma(\frac{n+1}{2})} .
\end{equation}

In the classically allowed region, on the other hand, 
\begin{equation}\label{allow}
\psi(\omega, r) = \frac{E e^{i{\frac{D^* \pi}{4}}} e^{-i\frac{D^*}{2}\sin^{-1}\left(\frac{D^*}{2\omega r}\right)}e^{-i\sqrt{\omega^2r^2-\frac{D^{*2}}{4}}}
+ C e^{-i{\frac{D^* \pi}{4}}} e^{i\frac{D^*}{2}\sin^{-1}\left(\frac{D^*}{2\omega r}\right)}e^{i\sqrt{\omega^2r^2-\frac{D^{*2}}{4}}}}{\left(\omega^2-\frac{D^{*2}}{4r^2}\right)^\frac{1}{4}} .
\end{equation}

In the limit $2\omega r \gg D^*$ \eqref{allow} reduces to 
\begin{equation}\label{allowlim}
\psi(\omega, r) = \frac{1}{\sqrt{\omega}} 
\left( E e^{i{\frac{D^* \pi}{4}}}e^{-i \omega r} + C e^{-i{\frac{D^* \pi}{4}}}e^{i \omega r} \right) ,
\end{equation}
in agreement with \eqref{wkbappf}.

The usual WKB crossing formulae relate the four constants $A$ $B$ $C$ $E$. Using Equation 7.35 of   
\cite{Merzbacher:1998rr} 
we have 
\begin{equation} \label{matchingall} \begin{split} 
C&= e^{-\frac{i \pi}{4}} \left(A +\frac{iB}2\right) \sqrt{\frac{D^*}2}\left( \frac{D^*}{\omega} \right) ^{-\frac{D^*}{2}} e^{\frac{D^*}{2}}   , \\
E&= e^{\frac{i \pi}{4}} \left(A- \frac{iB}2\right) \sqrt{\frac{D^*}2}\left( \frac{D^*}{\omega} \right) ^{-\frac{D^*}{2}} e^{\frac{D^*}{2}} .
\end{split}
\end{equation}

As we have explained in the main text, the constant $B$ is universal and is given by $B =\frac{1}{(D-3) \Omega_{D-2}}$. If we now specialize to the case of 
the retarded Greens function we have $E=0$. From \eqref{matchingall} it 
follows that 
\begin{equation} \label{matchinga} \begin{split} 
C&= \frac{(1+i)}{\sqrt{2}} B \sqrt{\frac{D^*}2}\left( \frac{D^*}{\omega} \right) ^{-\frac{D^*}{2}} e^{\frac{D^*}{2}}  =\frac{(1+i)}{\sqrt{2}}\left(2\right)^{-\frac{D}{2}}\frac{\omega^\frac{D-3}{2}}{\pi^\frac{D-2}{2}} , \\
A&= \frac{iB}{2} = \frac{i}{2(D-3)\Omega_{D-2}} .
\end{split}
\end{equation}
\footnote{We have used the large D approximations 
$D^* \approx D-3$ and 
$$ \Omega_{D-2} \approx 2^{-\frac{D-3}{2}}\pi^{\frac{D-2}{2}}e^{\frac{D-3}{2}}D^{-\frac{D-2}{2}}. $$}
It follows that in the classically allowed region 
\begin{equation}\label{wkbapp}
\psi(\omega,r) =-\left(2i\right)^{-\frac{D^*}{2}}\frac{\omega^\frac{D-3}{2}}{\pi^\frac{D-2}{2}}
\frac{ e^{i \left( \sqrt{\omega^2r^2-\frac{D^{*2}}{4}} 
+  \frac{D^*}{2} \sin^{-1} \left( \frac{D^*}{2 \omega r} \right) \right) } }
{\left(\omega^2-\frac{D^{*2}}{4r^2}\right)^\frac{1}{4}} .
\end{equation}
 It is easily verified that \eqref{wkbapp} 
matches both the leading and first subleading terms in \eqref{gfaa} 
when expanded at large $r$.

In the classically disallowed region , where $ D^* > 2{\omega}r $ we find the explicit formula  
\begin{equation}\label{webbpp}
\begin{split}
\psi(\omega,r) =& ~\frac{1}{(D-3)\Omega_{D-2}}\bigg(\frac{e^{\frac{D^*}{2}-\sqrt{\frac{D^{*2}}{4}-\omega^2r^2}}}{\left(\frac{D^{*2}}{4r^2}-\omega^2\right)^\frac{1}{4}}\left(\frac{D^*}{\omega}\left(\frac{D^*}{2\omega r}-\sqrt{\frac{D^{*2}}{4\omega^2r^2}-1}\right)\right)^{-\frac{D^*}{2}}\\
&~~~~~~~~~~~~~~~~~~~~+ \frac{i}2\frac{e^{\frac{D^*}{2}+\sqrt{\frac{D^{*2}}{4}-\omega^2r^2}}}{\left(\frac{D^{*2}}{4r^2}-\omega^2\right)^\frac{1}{4}} \left(\frac{1}{2r}-\sqrt{\frac{1}{4r^2}-\frac{\omega^2}{D^{*2}}}\right)^{\frac{D^*}{2}}  \bigg)\\
=& \frac{1}{(D-3)\Omega_{D-2}}\left(\frac{1}{r^{\frac{D-4}2}}+\frac{i}2\left(\frac{e\omega}D\right)^{D-3}{r^{\frac{D-2}2}}\right) ,
\end{split}
\end{equation} 
where the second expression applies at small $\omega r$. So the Greens Function can be written as
\begin{equation}
G(\omega,r) = \frac{1}{(D-3)\Omega_{D-2}r^{\frac{D-3}2}}\left(\frac{1}{r^{\frac{D-3}2}}+\frac{i}2\left(\frac{e\omega}D\right)^{D-3}{r^{\frac{D-3}2}}\right) .
\end{equation}

It may be verified that this result matches the small $r$ asymptotics of the exact formula \eqref{Gofs} in the following sense. From \eqref{Gofs} the exact
Greens function is given by 
\begin{equation}\label{gn} 
G(\omega, r)= \frac{i}{4} \left( \frac{\omega}{2 \pi r} \right)^{\frac{D-3}{2}}
\left( J_{\frac{D-3}{2}}(\omega r) + i  N_{\frac{D-3}{2}}(\omega r) \right) ,
\end{equation} 
where $N_n$ is the Neumann function. At small $\omega r$ we 
use the small $\omega r$ expansion of the Bessel and Neumann functions to 
obtain 
\begin{equation}\label{gnn}
\begin{split}
G(\omega, r) &= \frac{i}{4}\frac{1}{(D-3)\Omega_{D-2}}\left( \frac{\omega}{2 \pi r} \right)^{\frac{D-2}{2}}
\left(-4i\left(\frac{2\pi}{\omega r}\right)^{\frac{D-3}2}+\frac{2}{e}\left(\frac{2\pi e \omega r}D\right)^{\frac{D-3}2}\right)\\
&= \frac{1}{(D-3)\Omega_{D-2}r^{\frac{D-3}2}}\left(\frac{1}{r^{\frac{D-3}2}}+\frac{i}{2e}\left(\frac{e\omega}D\right)^{D-3}{r^{\frac{D-3}2}}\right) ,
\end{split}
\end{equation} 
in agreement with \eqref{webbpp}. 

We will now explain in what sense the WKB approximation may be thought of as 
the first term in a systematic large $D$ approximation of the Greens function. 
The first correction to any WKB approximation is of order  of 
the fractional change in the wavenumber over a distance scale of order 
one wavelength. In formulae, the first correction to this approximation is
of order $\frac{1}{k(r)}\partial_r \ln k(r)$ where $k(r)$ is the local 
WKB wave number. 
In the classically allowed region $k(r)= \sqrt{\omega^2- \frac{D^2}{4 r^2}}$. So the 
fractional correction, $E(r)$,  to the WKB approximation can be estimated to 
be of order  
$$E(r)=\frac{ \frac{{D^*}^2}{r^3}}{ \left( \sqrt{\omega^2- \frac{{D^*}^2}{4 r^2}} 
\right)^3}.$$  
Provided that $\sqrt{\omega^2- \frac{{D^*}^2}{4 r^2}}$ is of order unity (i.e. 
provided we don't get too near the turning point) it follows that 
$E(r) = {\cal O} \left(1/D \right)$ (recall that  $\omega r> D/2$).
This conclusion works all the way down to 
$\omega r - \frac{D}{2} \sim \frac{1}{D^\frac{2}{3}}$. 
In a similar manner the fractional error to the WKB approximation in the classically 
disallowed region is once again estimated as 
$$E(r) \sim \frac{1}{D \sqrt{1- \frac{4 \omega^2 r^2}{ D^2}}} ,$$
and is once again of order $\frac{1}{D}$ provided we stay away from 
the turning point. In summary, the WKB approximation provides an excellent
approximation to the Greens function at large $D$ except within a distance 
of order $\frac{1}{D^{\frac{2}{3}}}$ of the turning point.  

We end this subsection with a qualitative description of the retarded Green's 
function in the large $D$ limit. There are four qualitatively distinct regions in the Greens function. 
Deep into the classically allowed region, for $r \omega \gg D^2$,  
the Greens function is in the radiation zone. In this regime \eqref{gfaa} 
applies, and the modulus of Greens function is proportional to 
$\frac{e^{i\omega r}}{r^\frac{D-2}{2}} $. It follows that the mod squared 
Greens function is proportional to the inverse volume of the $D-2$ sphere 
of radius $r$ in this region, and so represents radiation whose integrated 
flux is independent of $r$. 

Moving further in we reach the intermediate radiation zone 
$ \frac{D}{2} < \frac{\omega r}{2} \ll D^2$. In this region the Greens 
function represents an oscillating radiation field that has not propagated 
far enough to settle into its large $r$  asymptotic value. 

Moving to smaller $r$ we pass the 
turning point of the potential and enter the classically forbidden region. 
In this intermediate static regime  
$ \sqrt{D} \ll  \frac{\omega r}{2} \ll \frac{D}{2}$, 
$\psi$ no longer oscillates as a function of $r$. Instead the 
Greens function turns into a sum of a term that grows as $r$ 
increases and another that decays as $r$ decreases.  
The decaying and growing pieces are comparable in magnitude near the turning point. However
the decaying term grows towards small $r$ and quickly dominates.

Moving to still smaller $r$ we reach the static zone 
$ \frac{\omega r}{2} \ll \sqrt{D}$. 
In this region the first of \eqref{gfaa} applies, and $G(\omega, r)$ becomes 
independent of $\omega$ (justifying the name static zone). The 
decaying term in \eqref{webbpp} is much larger than the growing term 
in this region; in particular the the ratio of the growing term  to the 
decaying term of order $\frac{1}{D^D}$ when $\omega r$ is of order unity. 

\section{Action of the Greens function on scalars, vectors and tensors in a spherical basis} \label{segf}

\subsubsection{Results for the off centred Green's function}\label{ocgf}
In our analysis of radiation we will will find it useful to have a 
generalization of the exact expression \eqref{Gofs} to a Greens function 
whose source point is displaced away from the origin. In the next subsection
we demonstrate that 
\begin{equation}\label{gfexpa0}
G(\omega,|{\vec r}-{\vec r}'|)= \frac{i\pi}{2} \sum_{l=0}^\infty
\frac{1}{(r'r)^{\frac{D-3}{2}}}
H_\frac{D-3+2 l}{2}^{(1)} (\omega r) 
J_\frac{D-3+2 l}{2} (\omega r') {\mathcal P}_l(\theta, \theta') .
\end{equation}

This result applies provided $|r|>|r'|'$, i.e. provided that the observation point is located 
further from the origin than the source point. The Hankel function  $H^{(1)}(r)$ which appears in 
\eqref{gfexp} is the unique solution to the Bessel function that is  
is purely outgoing at infinity. On the other hand the Bessel function $J(r')$ that also appears in 
this expression is the unique solution to the Bessel equation that is regular at the origin.
 $\theta$ collectively denotes all the angles of the point ${\vec r}$ 
on $S^{D-2}$, $\theta'$ similarly denotes all angles of the point 
${\vec r'}$ and  $\mathcal{P}_l(\theta, \theta')$ is the projector onto the 
angular dependence of the $l^{th}$ spherical harmonic defined in 
\eqref{scspproj}.
 
It is easily verified that \eqref{gfexp} reduces to \eqref{Gofs} in the 
limit $r ' \to 0$. As a consistency check on this formula we have 
explicitly verified that the expansion \eqref{gfexp} is translationally 
invariant, i.e. that 
\begin{equation}\label{dergf}
(\partial_{{\vec r}} + \partial_{{\vec r}'} ) G(\omega, |{\vec r}-{\vec r}'|)=0.
\end{equation}
 
Note also that in the limit $\omega \rightarrow 0$,
\begin{equation}\label{greenstat}
G(0,|\vec{r}-\vec{r}'|) = \frac{1}{r^{D-3}}\sum_{l=0}^{\infty}\left(\frac{r'}{r}\right)^l\frac{1}{2l+D-3}{\mathcal P}_l(\theta, \theta').
\end{equation}

\subsubsection{Derivation}

The expression \eqref{gfexp} may be derived as follows.
 Provided that ${\vec r} \neq {\vec r}'$ (and so in particular when 
$|{\vec r} | > |{\vec r}'|$ ) the Greens function is annihilated by the 
action of $$\left( \omega^2 + {\vec \nabla}^2 \right) , $$
separately on the variables ${\vec r}$ and ${\vec r}'$. 
The most general solution of the equation 
\begin{equation}\label{phe}
\left( \omega^2 + {\vec \nabla}^2 \right) \phi(\omega,\vec r)=0 ,
\end{equation}
is a linear superposition of modes of the form $\phi_l(\omega,r) Y_{lm}(\theta)$ , where $Y_{lm}$ represents 
an arbitrary scalar spherical harmonic \footnote{See Appendix \ref{sphericalharmonics} for a 
discussion of Spherical harmonics and their properties in arbitrary dimensions.} in the representation 
$(l, 0, 0, \ldots, 0)$ of $SO(D-1)$. Using the fact that 
\begin{equation}\label{evssh}
\nabla^2 Y_{lm}=-l (l+D-3) Y_{lm} ,
\end{equation}
where the Laplacian is taken on the unit sphere, see \eqref{nls}) it follows from 
\eqref{phe} that 
\begin{equation}\label{phen}
\left( \omega^2 + \frac{1}{r^{D-2}} \partial_r(r^{D-2} \partial_r)
 - \frac{l(l+D-3)}{r^2} \right) \phi_l(\omega,r)=0 .
\end{equation}
Solving this equation we find that 
\begin{equation}\label{plrs}
\phi_l(\omega,r)= \frac{1}{r^{\frac{D-3}{2}}}
\left( A_{l, \omega} H_\frac{D-3+2 l}{2}^{(1)}(\omega r) + B_{l, \omega} J_\frac{D-3+2 l}{2} (\omega r) \right) .
\end{equation}

The boundary conditions on our Greens function require it to be regular at every finite value of ${\vec r}'$ other than ${\vec r}$; 
and requires the Greens function to be an outgoing function of ${\vec r}$; these considerations force us to use the Hankel function 
with argument $r$ and the Bessel function with argument $r'$. The Greens 
function must also be rotationally invariant under simultaneous 
rotations of $\theta$ and $\theta'$. As we have explained above, the 
unique rotationally invariant function of two angles constructed using 
functions only in in the $l^{th}$ spherical harmonic sector is the 
projector ${\cal P}_l$ defined in \eqref{scspproj}. 
It follows from all these considerations that the Greens function must be given by an expression of the form 
\begin{equation}\label{gfexpa}
G(\omega,|{\vec r}-{\vec r}'|)= \sum_{l=0}^\infty
\frac{a_l}{(r'r)^{\frac{D-3}{2}}}
H_\frac{D-3+2 l}{2}^{(1)} (\omega r) 
J_\frac{D-3+2 l}{2} (\omega r') {\cal P}_l(\theta, \theta') ,
\end{equation}
for some as yet unknown coefficients $a_l$. We will now demonstrate that 
\begin{equation} \label{anff}
a_l = \frac{i\pi}{2} ;~~~ {\rm for~ all ~} l
\end{equation}
using which \eqref{gfexp} follows.

In order to obtain \eqref{anff} we use the large argument expansion of the 
Hankel function  \eqref{hfa}. \eqref{gfexpa} simplifies to 
\begin{equation}
G(\omega,|{\vec r}-{\vec r}'|) \approx i\sqrt{\frac{\pi}2}\left(\frac{-i\omega}{r}\right)^{\frac{D-2}{2}}e^{i\omega r}\sum_l J_{\frac{D-3+2l}{2}}(\omega r')(\omega r')^{-\frac{D-3}{2}} {\cal P}_l(\theta, \theta')
\end{equation}
We also specialize \eqref{gfexpa} to the case in which the source and 
observation points are 
at the same angle. In this special case the LHS of \eqref{gfexp} is simply 
$G(\omega,(r-r'))$ (see \eqref{Gofs}) and \eqref{gfexpa} reduces to 
\begin{equation}\label{gfmod}
G(\omega,(r-r'))= \sum_{l=0}^\infty
\frac{ a_{l} {\mathcal P}_l(0)}{(r'r)^{\frac{D-3}{2}}}
H_\frac{D-3+2 l}{2}^{(1)} (\omega r) 
J_\frac{D-3+2 l}{2} (\omega r') 
\end{equation}
(where $G(\omega, r)$ is defined in \eqref{Gofs} and ${\mathcal P}_l(0)$ 
is presented 
in \eqref{pea}). 
In order to determine the coefficients $a_l$ it is sufficient to further 
specialize \eqref{gfmod} to large $r$ and retain only 
leading order terms on both sides in the $\frac{1}{r}$ expansion. 
\eqref{gfmod} reduces to 
\begin{equation}\label{gffmod} \begin{split}
& \frac{i}{4}\left(\frac{\omega}{2\pi r}\right)^{\frac{D-3}{2}}\left(i^{-\frac{D-2}{2}}\sqrt{\frac{2}{\pi\omega r}}e^{i\omega (r-r')}\right)\\
&=  i^{-\frac{D-2}{2}}\sqrt{\frac{2}{\pi\omega r}}\left(\frac{\omega}{r}\right)^{\frac{D-3}{2}}\frac{e^{i\omega r}}{\omega}\sum_l i^{-l}a_l {\mathcal P}_l(0)
 J_{\frac{D-3+2l}{2}}(\omega r')(\omega r')^{-\frac{D-3}{2}}\\
& {\rm i.e.}~~~ e^{ix} = -4i(2\pi)^{\frac{D-3}2}\sum_l a_l {\mathcal P}_l(0)J_{\frac{D-3+2l}{2}}(x)(x)^{\frac{D-3}{2}} \\
\end{split}
\end{equation}
(where we have used \eqref{gfaa}, and $x= \omega r'$). Taylor expanding 
the LHS and RHS in $x$ about $x=0$ and using the well known series expansion 
for the Bessel function 
\begin{equation}
J_{\frac{D-3+2l}{2}}(x)(x)^{\frac{D-3}{2}} = \sum_{m=0}^{\infty}\frac{(-1)^m}{m!\Gamma\left(l+m+\frac{D-1}{2}\right)}\frac{(x)^{l+2m}}{2^{l+2m+\frac{D-3}{2}}} ,
\end{equation}
we find the following recursion relations
\begin{equation}\label{recur}
\tilde{a}_n = \Gamma\left(n+\frac{D-1}{2}\right)\left(\frac{2^n}{n!} - \sum_{m=1}^{\lfloor\frac{n}{2}\rfloor} \frac{\tilde{a}_{n-2m}}{m!~\Gamma\left(n-m+\frac{D-1}{2}\right)}\right), 
\end{equation}
where
\begin{equation}\label{tildrel}
\tilde{a}_l = -4i\pi^{\frac{D-3}2}{\cal P}_l(0)a_l.
\end{equation}
Using the explicit value of ${\cal P}_l(0)$ listed in \eqref{pea} 
it may be verified that 
\begin{equation}\label{al}
a_l = \frac{i\pi}{2} ,
\end{equation}
solves the recursion relation \eqref{recur}, establishing \eqref{anff}.

\subsection{Action of the retarded Greens function on the arbitrary spherically decomposed vector field}

We will now use the results the previous subsections together with those of 
Appendix \ref{sphericalharmonics} to present the general solution to the 
equation 
\begin{equation} \label{gfovf}
\left( -\nabla^2 - \omega^2 \right){\vec E}= {\vec J}_{\rm eff} , 
\end{equation}
where the field $J_{\rm eff}$ is a general vector field in $R^{D-1}$ that 
admits the expansion \eqref{arbitvector}. We search for the unique solution 
to this problem subject to the restriction that it behaves as
$e^{i \omega r}$ at infinity. 

In Cartesian coordinates the solution to this problem is simply given by 
\begin{equation}\label{sov}
{\vec E}({\vec r} )= \int dr' G(\omega, |{\vec r}-{\vec r}'|) {\vec J}_{\rm eff}(r') ,
\end{equation}
where the Green's function $G(\omega, |{\vec r}-{\vec r}'|)$ was 
defined in \eqref{gfexp}. Using \eqref{projrel} the solution 
\eqref{sov} can be rewritten in terms of a spherical decomposition as 
\begin{equation}\label{sovv}
{\vec E}(\omega,{\vec r} )= {\vec {\mathcal A}}^-[\xi_\alpha] 
+ {\vec {\mathcal A}}^+[\xi_\beta]+\vec{\upsilon}_\gamma ,
\end{equation} 
where
\begin{equation} \label{vrgf} \begin{split}
\xi_\alpha(\omega,{\vec r} ) &= \sum_l\frac{i \pi}{2} \frac{H^{(1)}_{\frac{D-3+2 (l-1)}{2}}(\omega r)}{r^{\frac{D-3}2}}\int dr' J_{\frac{D-3+2 (l-1)}{2}} (\omega r')~
r'^{\frac{D-1}{2}}\alpha_l(\omega,r', \theta) , \\
\xi_\beta(\omega,{\vec r} ) &= \sum_l\frac{i \pi}{2}\frac{H^{(1)}_{\frac{D-3+2 (l+ 1)}{2}}(\omega r)}{r^{\frac{D-3}2}} \int dr' J_{\frac{D-3+2 (l+1)}{2}} (\omega r')~
r'^{\frac{D-1}{2}}\beta_l(\omega,r', \theta) , \\
{\vec \upsilon}_\gamma(\omega,{\vec r} ) &=\sum_l \frac{i \pi}{2}\frac{H^{(1)}_{\frac{D-3+2l}{2}}(\omega r) }{r^{\frac{D-3}2}} \int dr' J_{\frac{D-3+2 l}{2}} (\omega r')~
r'^{\frac{D-1}{2}}{\vec \gamma}_l(\omega,r', \theta) . \\
\end{split}
\end{equation}

\subsection{Action of the retarded Greens function on the arbitrary spherically decomposed tensor field}

In this brief subsection we study the equation 
\begin{equation} \label{gfovft}
\left( -\nabla^2 - \omega^2 \right){{\cal H}_{ij}}= T_{ij}  ,
\end{equation}
where $T_{ij}$ is a given symmetric tensor field. We will find the 
unique solution to \eqref{gfovft} 
subject to the condition that ${\cal H}_{ij}$ is outgoing at infinity.

Let the source function $T_{ij}$ have the spherical decomposition listed in 
\eqref{arbitscalarg}. Proceeding as in the previous subsection, it is 
easy to verify that the unique outgoing solution to \eqref{gfovft} is given by  
\begin{equation}\label{sott}
{\cal H}_{ij}(\omega,{\vec r} )=  {\mathcal C}_{ij}^-[\xi_\alpha] + 
{\mathcal C}_{ij}^+[\xi_\beta] +
 {\mathcal C}_{ij}^0[\xi_\gamma] +
\delta_{ij}\xi_\kappa
+ {\mathcal B}_{ij}^-[\vec{\upsilon}_\phi]+ {\mathcal B}_{ij}^+[\vec{\upsilon}_\psi]
+ \tau^\chi_{ij} ,
\end{equation} 
where
\begin{equation}\label{trgf}\begin{split}
\xi_{\alpha}(\omega,{\vec r} )&=\sum_l\frac{i \pi}{2} \frac{H^{(1)}_{\frac{D-3+2 (l-2)}{2}}(\omega r)}{r^{\frac{D-3}2}}\int dr' J_{\frac{D-3+2 (l-2)}{2}} (\omega r')~
r'^{\frac{D-1}{2}}\alpha_l(\omega,r', \theta) , \\
\xi_{\beta}(\omega,{\vec r} )&=\sum_l\frac{i \pi}{2} \frac{H^{(1)}_{\frac{D-3+2 (l+2)}{2}}(\omega r)}{r^{\frac{D-3}2}}\int dr' J_{\frac{D-3+2 (l+2)}{2}} (\omega r')~
r'^{\frac{D-1}{2}}\beta_l(\omega,r', \theta) , \\
\xi_{\gamma}(\omega,{\vec r} )&=\sum_l\frac{i \pi}{2} \frac{H^{(1)}_{\frac{D-3+2l}{2}}(\omega r)}{r^{\frac{D-3}2}}\int dr' J_{\frac{D-3+2 l}{2}} (\omega r')~
r'^{\frac{D-1}{2}}\gamma_l(\omega,r', \theta) , \\
\xi_{\kappa}(\omega,{\vec r} )&=\sum_l\frac{i \pi}{2} \frac{H^{(1)}_{\frac{D-3+2l}{2}}(\omega r)}{r^{\frac{D-3}2}}\int dr' J_{\frac{D-3+2 l}{2}} (\omega r')~
r'^{\frac{D-1}{2}}\kappa_l(\omega,r', \theta) , \\
\vec{\upsilon}_{\phi}(\omega,{\vec r} )&=\sum_l\frac{i \pi}{2} \frac{H^{(1)}_{\frac{D-3+2 (l-1)}{2}}(\omega r)}{r^{\frac{D-3}2}}\int dr' J_{\frac{D-3+2 (l-1)}{2}} (\omega r')~
r'^{\frac{D-1}{2}}\vec{\phi}_l(\omega,r', \theta) , \\
\vec{\upsilon}_{\psi}(\omega,{\vec r} )&=\sum_l\frac{i \pi}{2} \frac{H^{(1)}_{\frac{D-3+2 (l+1)}{2}}(\omega r)}{r^{\frac{D-3}2}}\int dr' J_{\frac{D-3+2 (l+1)}{2}} (\omega r')~
r'^{\frac{D-1}{2}}\vec{\chi}_l(\omega,r', \theta) , \\
\tau^\chi_{ij}(\omega,{\vec r} )&=\sum_l\frac{i \pi}{2} \frac{H^{(1)}_{\frac{D-3+2l}{2}}(\omega r)}{r^{\frac{D-3}2}}\int dr' J_{\frac{D-3+2 l}{2}} (\omega r')~
r'^{\frac{D-1}{2}}(\chi_l)_{ij}(\omega,r', \theta) .
\end{split}
\end{equation}

\section{Details relating to the general theory of radiation} \label{consist}

\subsection{Static Limit of Electromagnetic Radiation}

It is useful 
to separately consider the scalar part of the electric field 
(first line of \eqref{emad}) and the vector part (second line of \eqref{emad}).
Let us first focus on the scalar part of this field. Using \eqref{rbs} 
and the fact that $H_n(x) \sim x^{-n}$ , we see that the first term in the 
first line of \eqref{emad} is negligible compared to the second term in the 
same line and at small $\omega$ and we find 
\begin{equation}\label{Estatl}
\begin{split}
{\vec E} (\omega,\vec{x}) &= 
\sum_{l=0}^{\infty} \left(\frac{H_{\frac{D+ 2 l-1}{2}} (\omega r)}{r^{\frac{D-3}2}}
 {\vec {\cal A}}^{+} [
S^+_l(\omega,\theta) ] \right)\\
&=\frac{-i}{\pi} \sum_{l=0}^{\infty} 
\left(2^{\frac{2l+D-1}{2}}\frac{\Gamma\left(\frac{2l+D-1}{2}\right)
\left((l+D-3)\hat{r}S^+_l-r\vec{\tilde{\nabla}}S^+_l\right)}{\omega^{\frac{2l+D-1}{2}}r^{l+D-2}}\right)\\
&=-\vec{\nabla}\Phi^E ,
\end{split}
\end{equation}
where
\begin{equation} \label{effphE}
\Phi^E=\frac{-i}{\pi} \sum_{l=0}^{\infty} \left(2^{\frac{2l+D-1}{2}}
\frac{\Gamma\left(\frac{2l+D-1}{2}\right)}{\omega^{\frac{2l+D-1}{2}}}
\frac{S^+_l}{r^{l+D-3}} \right) ,
\end{equation}

\begin{equation} \label{effscE}
\begin{split} 
S^+_l &= \frac{i\pi}{2}\int dr' (r')^{\frac{D-1}{2}}J_{\frac{2l+D-1}{2}}(\omega r')
\mathfrak{b}_l(\omega,r',\theta)\\
&= \frac{i\pi}{2}\frac{\omega^{\frac{2l+D-1}{2}}}{2^{\frac{2l+D-1}{2}}\Gamma\left(\frac{2l+D+1}{2}\right)}\int dr' (r')^{l+D-1}\mathfrak{b}_l(\omega,r',\theta).
\end{split}
\end{equation} 
\footnote{In going from 
the first to the second lines of \eqref{Estatl} we replaced the 
Hankel function by its leading term in a small argument expansion 
(this is appropriate in the small $\omega$ limit). 
The equations \eqref{effphE} and \eqref{effscE} are expressions for the
effective potential. In going from the first to the second line of 
\eqref{effscE} we have used the fact that $\omega$ is small to replace 
the Bessel function by the leading piece in a small argument expansion.}

\eqref{Estatl} is simply the statement that the electric field in the 
stationary limit is the gradient of a scalar potential. There is, of course, 
a simple explanation and interpretation of this fact. Recall that t
he effective source ${\vec{\mathcal J}}_{eff}$ - from which $\mathfrak{b}$ 
is built - is a linear combination of two terms. One of the two terms is 
the time derivative of the spatial current, and is subleading compared to the 
other term (the spatial derivative of the charge current) in the small 
$\omega$ limit. In this limit, consequently, the formula for the electric 
field reduces to 
\begin{equation}\label{efg} \begin{split}
{\vec E}&= 
\int G(\omega \to 0 , |{\vec r} -{\vec r}'| \nabla' {\mathcal J}_0(r')\\
&= \nabla  \left( \int G(\omega \to 0 , |{\vec r} -{\vec r}'| {\mathcal J}_0(r')
\right) \\
&= - \nabla \Phi_E ,\\
&\Phi_E= -\int G(\omega \to 0 , |{\vec r} -{\vec r}'| {\mathcal J}_0(r').
\end{split}
\end{equation} 
This is simply Coulomb's law. Indeed it may directly be verified that 
$\Phi_E$ defined in \eqref{effphE} and \eqref{efg} agree with each other. 
\footnote{In order to perform this verification it is useful to note 
in the limit of small $\omega$ 
$${\vec {\mathcal J}}_{eff} \rightarrow \nabla \mathcal{J}_0$$
\begin{equation} \label{effsct}
\mathfrak{b}_l = \frac{(r')^l}{2l+D-3}\partial_{r'}\left(\frac{\mathcal{J}_{0l}}{(r')^l}\right)
\end{equation}
where ${\mathcal J}_{0l}$ is the $l^{th}$ spherical harmonic piece of the 
charge current ${\mathcal J}_{0}$.}.

In the static limit magnetic field can be written using the Bianchi identity 
as:
\begin{equation}\label{maff}
F_{ij}(0,r,\theta) = 
\lim_{\omega \to 0}\frac{i( \nabla_i {\vec E}_j(\omega,r,\theta) - \nabla_j {\vec E}_i (\omega,r,\theta))}{\omega} .
\end{equation} 
The only term in \eqref{arbitscalarJ} that contributes to the RHS of 
\eqref{maff} is the pure vector piece $\mathfrak{c}$, which only receives 
contributions from the the term in ${\vec {\mathcal J}}_{eff}$ equal to 
$\partial_0 {\vec {\mathcal J}}$. 

In the limit of small $\omega$ ${\tilde {\vec {\mathfrak c}}}$ and 
${\vec{\mathfrak{c}}}$ differ only by a factor of $\omega$ at it is 
easily verified that 
\begin{equation} \label{magstat} \begin{split}
F_{ij}(\omega,r,\theta) &= 
( \nabla_i {\vec A }_j(\omega,r,\theta) - 
\nabla_j {\vec A }_i (\omega,r,\theta)) ,\\
& {\vec  A } = \lim_{\omega \to 0} \frac {i\vec E}{\omega} \\
\therefore & {\vec A}= \sum_{l} \frac{i}{(2l+D-3)r^{D-3}}\int dr' (r')^{D-2}
\left(\frac{r'}r\right)^l({\tilde {\vec { \mathfrak c}}}_l(0,r',\theta)).
\end{split}
\end{equation}
In other words the magnetic field is given by $dA$ where 
$\nabla^2 {\vec A} = {\vec {\mathcal J}}$, and we 
recover the usual formulae of magnetostatics.

\subsection{Constraints from current conservation and $\nabla\cdot E = 0$}
\label{consistem}

As we have explained in the main text the fact that $\nabla. {\vec E}$ 
vanishes in vacuum implies that the scalar functions ${\mathcal S}^\pm$ that 
characterize a general radiation field (see \eqref{emad}) are not 
independent but are related by \eqref{rbs}. In \eqref{avvv}, however, 
we have presented separate formulae for ${\mathcal S}^\pm$ in terms of 
integrals over scalar components of charge currents. The consistency 
of \eqref{avvv} requires that these results for ${\mathcal S}^\pm$ 
automatically obey \eqref{rbs}. We will now demonstrate that this 
is indeed the case. 

At the structural level the way this works is very simple. If we take the 
divergence of \eqref{nablasqv} and use \eqref{fcc} we obtain \eqref{boxe}, 
which guarantees that $\nabla . E$ vanishes in vacuum. In this section 
we will rerun this structural argument on the explicit formulae \eqref{avvv}. 
The fact that we land on our feet serves as a consistency check of the 
algebra that led to \eqref{avvv} and \eqref{rbs}.

In the rest of this subsection we proceed to algebraically demonstrate that 
\begin{itemize}
\item The equation $\nabla\cdot E = 0$ implies that the coefficient 
functions in \eqref{emad} obey \eqref{rbs}
\item That the relations \eqref{avvv} automatically obey \eqref{rbs} once 
we account for the fact that the current is conserved.
\end{itemize} 

\begin{center}{\it  Demonstration that  $\nabla\cdot E = 0$ implies 
\eqref{rbs}} \end{center}

According to \eqref{emad} the vacuum electromagnetic solution is given by
\begin{equation}
\vec{E} = \sum_l\left(\frac{H^{(1)}_{l+\frac{D-5}2}(\omega r)}{r^{\frac{D-3}2}}\vec{\mathcal{A}^-}[S^-_l(\omega,\theta)]+\frac{H^{(1)}_{l+\frac{D-1}2}(\omega r)}{r^{\frac{D-3}2}}\vec{\mathcal{A}^+}[S^+_l(\omega,\theta)]+\frac{H^{(1)}_{l+\frac{D-3}2}(\omega r)}{r^{\frac{D-3}2}}\vec{V}_l(\omega,\theta)\right).
\end{equation}
Using 
\begin{equation}
\begin{split}
\nabla\cdot\vec{\mathcal{A}}^-[S^-_l(\omega,\theta)] &= lr^{l-1}\partial_r\left(\frac{H^{(1)}_{l+\frac{D-5}2}(\omega r)}{r^{l+\frac{D-5}2}}\right)S^-_l(\omega,\theta)\\
&= -lS^-_l(\omega,\theta)\frac{H^{(1)}_{l+\frac{D-3}2}(\omega r)}{r^{\frac{D-3}2}}\\
\nabla\cdot\vec{\mathcal{A}}^+[S^+_l(\omega,\theta)] &= \frac{l+D-3}{r^{l+D-2}}\partial_r\left(H^{(1)}_{l+\frac{D-1}2}(\omega r)r^{l+\frac{D-1}2}\right)S^+_l(\omega,\theta)\\
&= +(l+D-3)S^+_l(\omega,\theta)\frac{H^{(1)}_{l+\frac{D-3}2}(\omega r)}{r^{\frac{D-3}2}}\\
\nabla\cdot\vec{V}_l(\omega,\theta) &= 0 ,
\end{split}
\end{equation}
it follows that $\nabla\cdot E = 0$ provided 
\begin{equation}
-lS^-_l(\omega,\theta)+(l+D-3)S^+_l(\omega,\theta) = 0.
\end{equation}

\begin{center}{{\it Proof that current conservation satisfies this requirement}}
\end{center}

Recall that according to \eqref{arbitscalarJ} the effective current admits 
the following decomposition
\begin{equation}
\vec{\mathcal{J}}_{eff} = \sum_l\left(\mathcal{A}^-[\mathfrak{a}_l(\omega,r',\theta')]+\mathcal{A}^+[\mathfrak{b}_l(\omega,r',\theta')]+\vec{\mathfrak{c}}_l(\omega,r',\theta')\right).
\end{equation}
Using \eqref{divbasis} yields
\begin{equation}
\nabla'\cdot\vec{\mathcal{J}}_{eff} =  \sum_l\left(l{r'}^{l-1}\partial_{r'}\left(\frac{\mathfrak{a}_l(\omega,r',\theta')}{{r'}^{l-1}}\right)+\frac{l+D-3}{{r'}^{l+D-2}}\partial_{r'}\left(\mathfrak{b}_l(\omega,r',\theta'){r'}^{l+D-2}\right)\right).
\end{equation}
From \eqref{fcc} we conclude that
\begin{equation}
l{r'}^{l-1}\partial_{r'}\left(\frac{\mathfrak{a}_l(\omega,r',\theta')}{{r'}^{l-1}}\right)+\frac{l+D-3}{{r'}^{l+D-2}}\partial_{r'}\left(\mathfrak{b}_l(\omega,r',\theta'){r'}^{l+D-2}\right)-\Box'(J_0)_l = 0.
\end{equation}
Multiplying by $\frac{i\pi}2J_{l+\frac{D-3}2}(\omega r'){r'}^{\frac{D-1}2}$ integrating by parts w.r.t. $r'$ and noting also that $\int dr'\Box'\left(J_{l+\frac{D-3}2}(\omega r'){r'}^{\frac{D-1}2}\right)=0$ we get ,
\begin{equation}
\frac{i\pi l}2\int dr'J_{l+\frac{D-5}2}(\omega r'){r'}^{\frac{D-1}2}\mathfrak{a}_l(\omega,r',\theta')-\frac{i\pi (l+D-3)}2\int dr'J_{l+\frac{D-1}2}(\omega r'){r'}^{\frac{D-1}2}\mathfrak{b}_l(\omega,r',\theta') = 0
\end{equation}
Which from \eqref{avvv} translates to
\begin{equation}
lS^-_l(\omega,\theta)-(l+D-3)S^+_l(\omega,\theta) = 0.
\end{equation}

\subsection{Static Limit of Gravitational Radiation}\label{statgrav}

It is useful 
to separately consider the scalar vector and tensor parts of the 
curvature given in \eqref{gmad}. 

Focusing first on the scalar part we see from \eqref{gmad}, \eqref{grbs}
and the fact that $h_n(x) \sim x^{-n}$ , that in the limit $\omega \to 0$ 
the scalar part of \eqref{gmad} reduces to

\begin{equation}\label{Gstatl}
\begin{split}
R_{0i0j}(\omega,\vec{x}) &= 
\lim_{\omega \to 0} \sum_{l=0}^{\infty} \left(\frac{H_{\frac{D+ 2 l+1}{2}} (\omega r)}{r^{\frac{D-3}2}}\mathcal{C}^{+}_{ij} [
S^+_l(\omega,\theta) ] \right)\\
&=
\frac{-i}{\pi}  \lim_{\omega \to 0} \sum_{l=0}^{\infty} 
\left(2^{\frac{2l+D+1}{2}}\frac{\Gamma\left(\frac{2l+D+1}{2}\right)
\left((l+D-2)\hat{r}-r\vec{\tilde{\nabla}}\right)\left((l+D-3)\hat{r}S^+_l-r\vec{\tilde{\nabla}}S^+_l\right)}{\omega^{\frac{2l+D+1}{2}}r^{l+D-1}}\right)\\
&=\nabla_i\nabla_j\Phi^G ,
\end{split}
\end{equation}
where
\begin{equation} \label{Geffph} 
\begin{split}
\Phi^G&=\frac{-i}{\pi} \sum_{l=0}^{\infty} \left(2^{\frac{2l+D+1}{2}}
\frac{\Gamma\left(\frac{2l+D+1}{2}\right)}{r^{l+D-3}}
 \right) \lim_{\omega \to 0}\frac{S^+_l} {\omega^{\frac{2l+D+1}{2}}}\\
&=\sum_{l=0}^{\infty} 
\frac{1}{(2 l+ D + 1)r^{l+D-3}}
\int dr' (r')^{l+D}\mathfrak{b}_l(0,r',\theta) ,
\end{split}
\end{equation}
where we have used
\begin{equation} \label{Geffsc}
 \lim_{\omega \to 0}\frac{S^+_l} {\omega^{\frac{2l+D+1}{2}}}
= \frac{i\pi}{2}\frac{1}{2^{\frac{2l+D+1}{2}}\Gamma\left(\frac{2l+D+3}{2}\right)}\int dr' (r')^{l+D}\mathfrak{b}_l(0,r',\theta).
\end{equation}

\eqref{Geffph} is simply the statement that $R_{0i0j}$ in the 
stationary limit is the double gradient of a suitably scaled version 
of the Newtonian potential 
$\phi^G$. Indeed it is easily verified that $\phi^G$ is given by 
\begin{equation}\label{npe}
\nabla^2 \phi^G= -8\pi\left(\mathcal{T}_{00}+\frac{\mathcal{T}}{D-2}\right)
\end{equation}
\footnote{To see this note that, 
in the strict limit $\omega \to 0$ the effective stress 
tensor reduces to 
$$\mathcal{T}^{eff}_{ij} = 8\pi\nabla_i'\nabla_j'\left(\mathcal{T}_{00}+\frac{\mathcal{T}}{D-2}\right).$$  
It follows that 
\begin{equation}
\mathfrak{b}_l = \frac{8\pi (r')^{l+1}}{(2l+D-1)(2l+D-3)}\partial_{r'}\left(\frac{1}{r'}\partial_{r'}\left(\frac{\mathcal{T}_{00}+\frac{\mathcal{T}}{D-2}}{(r')^l}\right)\right) .
\end{equation}
}

Let us now turn to the vector part of $R_{0i0j}$. Once again 
using  \eqref{gmad}, \eqref{grbs} and the small argument expansion of the 
Hankel function we see that the vector part of $R_{0i0j}$ simplifies to 
\begin{equation}\label{Gstatlv}
\begin{split}
R_{0i0j}(\omega,\vec{x}) &= 
\sum_{l=0}^{\infty} \left(\frac{H_{\frac{D+ 2 l-1}{2}} (\omega r)}{r^{\frac{D-3}2}}
 \mathcal{B}^{+}_{ij} [
V^+_l(\omega,\theta) ] \right)\\
&=\frac{-i}{\pi} \sum_{l=0}^{\infty} 
\left(2^{\frac{2l+D-1}{2}}\frac{\Gamma\left(\frac{2l+D-1}{2}\right)
\left((l+D-3)\hat{r}_i(V^+_l)_j-r\tilde{\nabla}_i(V^+_l)_j\right)}{\omega^{\frac{2l+D-1}{2}}r^{l+D-2}}\right)+\{i\leftrightarrow j\}\\
&=-i\omega\left(\nabla_iA^G_j + \nabla_jA^G_i \right) ,
\end{split}
\end{equation}
so that (using the Bianchi identity) 
\begin{equation} \label{ous}
\begin{split}
R_{0ijk} &= \frac{i}{\omega}\left(\nabla_jR_{0i0k}-\nabla_kR_{0i0j}\right)\\
&= -\nabla_i\left(\nabla_jA^G_k-\nabla_kA^G_j\right) ,
\end{split}
\end{equation}
where
\begin{equation}
A^G_i = \frac{1}{2\pi}\left(\frac{2}{\omega}\right)^{\frac{2l+D+1}2}\Gamma\left(\frac{2l+D-1}2\right)\frac{(V^+_l)_i}{r^{l+D-3}} .
\end{equation}
It is easily verified that 
$$\nabla^2 A^G_i = -8\pi T_{0i}.$$
\footnote{ To see this note that 
the term in $({\mathcal T}_{eff})_{ij}$ (see \eqref{nablasqg}) 
that contributes to the the vector in \eqref{gmad} 
$$i\omega(\partial_i\mathcal{T}_{0j}+\partial_j\mathcal{T}_{0i}).$$
It follows that 
$$\mathfrak{v}_i=-\frac{8\pi i \omega {r'}^l}{2l+D-3}\partial_{r'}\left(\frac{T_{0i}}{{r'}^l}\right) .$$ .}
Note that ${\vec A}^G$ obeys the same equation obeyed by the `vector potential' 
magnetostatics with the role of the current being played by $T_{0i}$. 
Indeed \eqref{ous} asserts that $R_{0ijk}$ is proportional to $\nabla_i F^G_{jk}$ 
where $F^G_{jk}$ is the magnetic field constructed from the effective 
vector potential $A^G_i$.

Finally we turn to the tensor part of $R_{0i0j}$. It follows immediately from 
$$\nabla^2R_{0i0j} = 8\pi\omega^2 T_{ij}$$

In the small $\omega$ limit the contribution of tensor sources to to $R_{ijkm}$ takes the form
\begin{equation}
\begin{split}
R_{0ijk} &= \frac{i}{\omega}\left(\nabla_jR_{0i0k}-\nabla_kR_{0i0j}\right) ,\\
R_{ijkm} &= \frac{i}{\omega}\left(\nabla_iR_{0jkm}-\nabla_jR_{0ikm}\right)\\
&= \nabla_i\nabla_k\mathfrak{T}_{jm}+\nabla_j\nabla_m\mathfrak{T}_{ik}-\nabla_j\nabla_k\mathfrak{T}_{im}-\nabla_i\nabla_m\mathfrak{T}_{jk},\\
\mathfrak{T}_{ij} &= \frac{-R_{0i0j}}{\omega^2} .
\end{split}
\end{equation}

The tensor contribution from the source is
\begin{equation}
\mathfrak{z}_{ij} = -8\pi\omega^2\left(\mathcal{T}_{ij}-\delta_{ij}\frac{\mathcal{T}^k_k}{D-2}\right).
\end{equation}

The scalar sector also contributes to $R_{ijkm}$, but its closed form is a bit ugly unlike the other beautiful results in this section, and that can be obtained from the scalar contribution to $R_{0i0j}$. We don't present it here.

\subsection{Tracelessness and divergenceless of gravitational radiation} \label{consistg}

In this subsection we rerun some of the discussion of section \ref{consistem} but this time in the 
context of gravitational radiation. In particular we will explain how the explicit gravitational 
radiation formulae ensure that gravitational radiation is traceless and divergence free. 
At the formal level these results follow immediately once we use that fact that 
when a box of something  (e.g. $\Box \zeta$) is convoluted with Green's function, the 
resulting integral vanishes. We will now use this fact to demonstrate

{\bf Result 1: Gravitational Radiation is traceless.}\\

\begin{equation}\label{hgrint}
h_{ij}(\omega,\vec{x}) = -\frac{2}{\omega^2}\int G(\omega,|\vec{x}-\vec{x}'|)\hat{\mathcal{T}}_{ij}(\omega,\vec{x'})d^{D-1}x' .
\end{equation}
Hence
\begin{equation}
h_{ij}\eta^{ij}(\omega,\vec{x}) = -\frac{2}{\omega^2}\int G(\omega,|\vec{x}-\vec{x}'|)(\eta^{ij}\hat{\mathcal{T}}_{ij}(\omega,\vec{x'}))d^{D-1}x' .
\end{equation}
Using Conservation of stress tensor, we have
\begin{equation}
\eta^{ij}\hat{\mathcal{T}}_{ij} = -\Box'\left(\mathcal{T}_{00}+\frac{\mathcal{T}}{D-2}\right) ,
\end{equation}
hence the integration vanishes, i.e. $h_{ij}\eta^{ij}=0$

{\bf Result 2: Gravitational Radiation is divergenceless.}\\
Taking spatial divergence of \eqref{hgrint},
\begin{equation}
\begin{split}
\partial_ih_{ij}(\omega,\vec{x}) &= -\frac{2}{\omega^2}\partial_i\int G(\omega,|\vec{x}-\vec{x}'|)\hat{\mathcal{T}}_{ij}(\omega,\vec{x'})d^{D-1}x'\\
&= -\frac{2}{\omega^2}\int G(\omega,|\vec{x}-\vec{x}'|)\left(\partial_i\hat{\mathcal{T}}_{ij}(\omega,\vec{x'})\right)d^{D-1}x' .
\end{split}
\end{equation}
Using Conservation of stress tensor, we have
\begin{equation}
\partial_i'\hat{\mathcal{T}}_{ij} = -\Box'\left(i\omega\mathcal{T}_{0j}+\mathcal{T}_{00}+\frac{\mathcal{T}}{D-2}\right),
\end{equation}
hence the integration vanishes, i.e. $\partial_ih_{ij}=0 .$

\section{Variation of the first order gravitational counterterm action}\label{cact}
 
In this brief Appendix we demonstrate that the variation of \eqref{couteraction} yields the 
stress tensor \eqref{tabin}.
 
Varying the first term inside the bracket in \eqref{couteraction} we find
\begin{equation}\label{vary1}
\begin{split}
\int\delta\sqrt{{\cal R}}=&~\int\frac{\delta {\cal R}}{2\sqrt{{\cal R}}}\\
=&~\int\frac{1}{2\sqrt{{\cal R}}}\left[-{\cal R}_{\mu\nu} {\delta g}^{\mu\nu} + \hat\nabla_\mu\hat\nabla_\nu {\delta g}^{\mu\nu} - \hat\nabla^2 {\delta g}\right]\\
=&~\int\frac{1}{2}\left[-\left(\frac{{\cal R}_{\mu\nu} }{\sqrt{{\cal R}}}\right) +\left(\hat\nabla_\mu\hat\nabla_\nu - g^{(in)}_{AB}\hat\nabla^2\right)\left(\frac{1}{\sqrt{{\cal R}}}\right)\right]{\delta g}^{\mu\nu}\\
=&~\int\frac{1}{2}\left[-\left(\frac{{\cal R}_{\mu\nu} }{\sqrt{{\cal R}}}\right) - g^{(in)}_{\mu\nu}\hat\nabla^2\left(\frac{1}{\sqrt{{\cal R}}}\right) + {\cal O}\left(\frac{1}{D}\right)\right]{\delta g}^{\mu\nu}\\
=&~\int\frac{1}{2}\left[-\left(\frac{{\cal R}_{\mu\nu} }{\sqrt{{\cal R}}}\right)+ g^{(in)}_{\mu\nu}\left(\frac{\hat\nabla^2{\cal R}}{2{\cal R}^\frac{3}{2}}\right) + {\cal O}\left(\frac{1}{D}\right)\right]{\delta g}^{\mu\nu} ,
\end{split}
\end{equation}
where for convenience,  we have used the notation $\delta g^{(ind)}_{\mu\nu} = \delta g_{\mu\nu}$ 
and we have used the formula
\begin{equation}\label{varyr}
\begin{split}
&\delta {\cal R}_{\mu\nu} = \frac{1}{2}\left[\hat\nabla_\alpha\hat\nabla_\mu {\delta g}^\alpha_\nu + \hat\nabla_\alpha\hat\nabla_\nu {\delta g}^\alpha_\mu -\hat\nabla_\mu\hat\nabla_\nu {\delta g} -\hat\nabla^2 {\delta g}_{\mu\nu}\right]\\
\Rightarrow~&\delta {\cal R} = -{\delta g}^{\mu\nu}{\cal R}_{\mu\nu} + \left(\hat\nabla_\mu\hat\nabla_\nu-g^{(ind)}_{\mu\nu}\hat\nabla^2\right) h^{AB} , \\
\text{where}~~& {\delta g}= g_{(ind)}^{\mu\nu}{\delta g}_{\mu\nu},~~\delta g^{\mu\nu}= g_{(ind)}^{\mu\alpha}~\delta g_{\alpha\beta}~g_{(ind)}^{\alpha\nu} .
\end{split}
\end{equation}

Varying the second term inside the bracket in \eqref{couteraction} we find
\begin{equation}\label{vary2}
\begin{split}
&\frac{1}{2}~~\delta\left(\frac{{\cal R}_{\mu\nu}{\cal R}^{\mu\nu}}{{\cal R}^\frac{3}{2}}\right)\\
=&~-\frac{3}{4}{\cal R}^{-\frac{5}{2}}~{\cal R}_{\mu\nu}{\cal R}^{\mu\nu}~\delta {\cal R}-\frac{{\cal R}_{\mu\alpha}{\cal R}^\alpha_\nu~ {\delta g}^{\mu\nu}}{{\cal R}^\frac{3}{2}}+\frac{{\cal R}^{\mu\nu}~\delta {\cal R}_{\mu\nu} }{{\cal R}^\frac{3}{2}} .
\end{split}
\end{equation}
Now from equation \eqref{varyr} it follows that 
\begin{equation}\label{midstep1}
\begin{split}
&\int {\cal R}^{-\frac{5}{2}}~{\cal R}_{\mu\nu}{\cal R}^{\mu\nu}~\delta R\\
=~&\int {\cal R}^{-\frac{5}{2}}~{\cal R}_{\alpha\beta}{\cal R}^{\alpha\beta}\left[ -{\delta g}^{\mu\nu}{\cal R}_{\mu\nu} + \left(\hat\nabla_\mu\hat\nabla_\nu-g^{(ind)}_{\mu\nu}\hat\nabla^2\right) {\delta g}^{\mu\nu}\right]\\
=~&\int {\delta g}^{\mu\nu}\left[-{\cal R}^{-\frac{5}{2}}~{\cal R}_{\alpha\beta}{\cal R}^{\alpha\beta}{\cal R}_{\mu\nu} + \left(\hat\nabla_\mu\hat\nabla_\nu-g_{\mu\nu}\hat\nabla^2\right)\left({\cal R}^{-\frac{5}{2}}~{\cal R}_{\alpha\beta}{\cal R}^{\alpha\beta}\right)\right]\\
=~&\int {\delta g}^{\mu\nu}\left[{\cal O}\left(\frac{1}{D}\right)\right]
\end{split}.
\end{equation}
Similarly the second term in equation \eqref{vary2} is also of order ${\cal O}\left(\frac{1}{D}\right)$. 
In the third term of equation \eqref{vary2} if we substitute the formula equation \eqref{varyr} we get one term which is of order ${\cal O}(1)$.
\begin{equation}\label{midstep2}
\begin{split}
&\int\frac{{\cal R}^{\mu\nu}~\delta{\cal R}_{\mu\nu} }{{\cal R}^\frac{3}{2}}\\
=~&\int\left[- g^{(ind)}_{\mu\nu} \left(\frac{\hat\nabla_\alpha\hat\nabla_\beta {\cal R}^{\alpha\beta}}{R_{(in)}^\frac{3}{2}}\right)+{\cal O}\left(\frac{1}{D}\right)\right]\delta g^{\mu\nu}\\
=~&\int\left[- g^{(ind)}_{\mu\nu} \left(\frac{\hat\hat\nabla^2{\cal R}}{2{\cal R}^\frac{3}{2}}\right)+{\cal O}\left(\frac{1}{D}\right)\right]\delta g^{\mu\nu} .
\end{split}
\end{equation}

Using equation \eqref{vary1}, \eqref{vary2}, \eqref{midstep1} and \eqref{midstep2} we find
the equation \eqref{tabin}

\section{Perturbative solution  for $\rho$}\label{rhoconstruct}

In this section we find the solution of \eqref{rhodef}. In order to do this
we find it convenient  to use the following coordinate system. Choose 
any point on the membrane. We treat this point as the origin of our 
coordinate system. We now erect a Cartesian coordinate system about this 
point, making sure to orient a special coordinate, $z$, so that the normal 
to the membrane at that point is $dz$. Let the remaining Cartesian coordinates 
(which are all orthogonal to each other and to 
$z$ but are otherwise arbitrary) be denoted 
by $x^\mu$).  It follows that, in this coordinate 
system, the equation of the membrane takes the following form 
\begin{equation} \label{memdef}
z(y_\mu)=-\frac{K_{\mu\nu} }{2} y_{\mu} y_{\nu}-\frac{C_{\m\n\s}}{3} y_{\m} y_{\nu} y_{\s}
-\frac{D_{\m\n\s\rho}}{4} y_{\m} y_{\nu} y_{\s} y_{\rho}+\cdots
\end{equation}

Now consider a point outside the membrane whose coordinates are 
$(z, x_\mu)$. At least in a neighbourhood of the membrane any such 
point may uniquely be associated with a point $(z(y_\mu), y_\mu)$ on the 
membrane by the requirement that a straight line drawn normal through 
this membrane point passes through $(z, x_\mu)$. 

Let $y_\mu(z, x_\mu)$ denote the coordinates of the membrane point associated
with an arbitrary bulk point in this manner, and let $s(z, x_\mu)$ denote
the distance along this line from the given bulk point to the membrane. 
We will now determine  $y_\mu(z, x_\mu)$ and  $s(z, x_\mu)$ in a Taylor 
series expansion in $x_\mu$. 

Working in a Taylor expansion in $y_\m$, the 
normal at any point on the membrane is given by 
\begin{equation}
n=\frac{dz+\left({K_{\mu\nu }}y_{\nu}+{C_{\m\n\s}}y_{\nu}y_{\s}+{D_{\m\n\s\rho}}y_{\nu}y_{\s}y_{\rho}\right)dy_\m}{\mathcal{N}} , 
\end{equation}
where the normalization  $\mathcal{N}$ is chosen to ensure that $n.n=1$.
To solve for $y_\m$ in terms of the $x_\m$ and $z$ we note that, by 
definition 
\begin{equation}
\label{normx}
\begin{split}
\frac{x_\mu-y_\m}{z-z_0}&=\frac{n_\m}{n_z} ,\\
\frac{x_\mu-y_\m}{z-z_0}&= \left(K_{\mu\nu }y_{\nu}+{C_{\m\n\s}}y_{\nu}y_{\s}+{D_{\m\n\s\rho}}y_{\nu}y_{\s}y_{\rho}\right).
\end{split}
\end{equation}
These equations are easily solved in a Taylor series expansion in 
$x_\mu$ (but to all orders in $z$). To the cubic order in $x_\mu$ we have 
\begin{equation}
\label{hatxexp}
\begin{split}
y_\m&=(Px)_\m-z (P \cdot C)_{\m\n\s} (Px)_\n (Px)_\s+2z^2 (P.C.C)_{\m\n\s\rho}(Px)_\n (Px)_\s (Px)_\rho \\&-z(P.D))_{\m\n\s\rho}(Px)_\n (Px)_\s (Px)_\rho-z(P.K)_{\s\rho}K_{\m\n}Px)_\n (Px)_\s (Px)_\rho ,
\end{split}
\end{equation}
where we have defined 
$$P_{\m\n}=\left(\frac{1}{1+zK}\right)_{\m\n}.$$
We now turn to the determination of $s(x_\mu, z)$. First note that 
\begin{equation}
\label{hatz}
\begin{split}
s(x_\mu, z)&=\sqrt{(z-z_0)^2+(x-y)^2}\\
&=(z-z_0)\sqrt{(1+\frac{(x-y)^2}{(z-z_0)^2}}.
\end{split}
\end{equation}
Using \eqref{normx} and retaining terms to cubic order in $y$ we obtain
\begin{equation}
\label{hatzexp}
s(x_\mu, z)=z+\frac{1}{2}\left(K_{\m\n}+z(K\cdot K)_{\m\n}\right)y_\m y_\n+\left(\frac{1}{3}C_{\m\n\s}+z(K\cdot C)_{\m\n\s}\right)y_\m y_\n y_\s+\cdots
\end{equation}
Substituting the expansion of $y$ in \eqref{hatzexp} and retaining terms to the cubic order in $x$ 
\begin{equation} \label{sfin}
\begin{split}
s(x_\mu, z)&=z+\frac{1}{2}\left(K_{\m\n}+z(K\cdot K)_{\m\n}\right)\left((Px)_\m (Px)\n-2 z (P \cdot C)_{\m\s\rho}(Px)_\n(Px)_\s(Px)_\rho\right)\\&+\left(\frac{1}{3}C_{\m\n\s}+z(K\cdot C)_{\m\n\s}\right)(Px)_\m(Px)_\n(Px)_\s+\cdots
\end{split}
\end{equation}

We now turn to the determination of the function $\rho$. Using the 
Cartesian coordinate system employed in this Appendix it is not difficult 
to solve for $\rho$ in a Taylor series expansion in $x_\mu$. Once that is 
achieved one can re-express the result in terms of $y_\mu$ and $s$ using 
\eqref{hatxexp} and \eqref{sfin}. The algebra involved is tedious and we 
omit all details. Our final result for $\rho$ is 
\begin{equation}\label{rhodfn} \begin{split}
& \rho(x^\mu) -1=
 s(x^\mu) \frac{K(y^\mu)}{D-2} +  \\
& \left( \frac{2 s(x^\mu)}{K} + s(x^\mu)^2 \right) 
\left(\frac{1}{2 K} {\hat \nabla}^2 
\left( \frac{ K}{D-2} \right)  + \frac{ K^2}{2 (D-2)^2}  + 
\frac{K_{MN} K^{MN}}{ K} \right)\\
&+  {\cal O}
\left( \frac{1}{(D-2)^3} \right) .
\end{split} 
\end{equation}

\section{Evolution of the Einstein Constraint Equations }\label{eec}
In this Appendix we derive the equation \eqref{eecc} assuming that the dynamical Einstein equations 
hold everywhere.\\
 Since all components of the Einstein equation are already linear in the metric fluctuation, in this appendix we would simply replace all covariant derivatives $\nabla$ by partial derivatives $\partial$.

Now the dynamical equations are true everywhere and therefore their divergence also vanishes and we find
\begin{equation}\label{divdyna}
\begin{split}
0
&= \partial_A\left[E^A_B -n^A X_B - n_BX^A -n^An_B Y\right]\\
&= -{K}~ X_B -(n\cdot\partial) X_B -X^AK_{AB}-n_B (\partial\cdot X)\\
&~~~ -n_B \left[{K}~Y+(n\cdot\partial Y\right] -Y (n\cdot\partial) n_B .
\end{split}
\end{equation}
Simplifying \eqref{divdyna} further using the expression for $(\nabla\cdot X)$.
\begin{equation}\label{divX}
\begin{split}
\partial\cdot X&= \partial_A\left[\Pi^{AC} E_{CC'}n^{c'}\right]\\
&=\partial_A\left[E^A_Cn^C -n^A Y \right]\\
&= E^{AC}\partial_A n_C - \left[{K}~Y+(n\cdot\partial)Y\right]\\
&= X^C(n\cdot\partial)n_C - \left[{K}~Y+(n\cdot\partial)Y\right] .\\
\end{split}
\end{equation}
Substituting equation \eqref{divX} into equation \eqref{divdyna} we obtain
\begin{equation}\label{divdyna2}
\begin{split}
0&=\partial_A \hat E^{AB}=\partial_A\left[\Pi^{CA} \Pi^{C'}_B E_{CC'}\right]\\
&=\partial_A\left[\Pi^{CA} \Pi^{C'}_B E_{CC'}\right]\\
&= -{K}~ X_B -(n\cdot\partial) X_B -X^AK_{AB}-n_B\left[ X^C(n\cdot\partial)n_C\right]
  -Y (n\cdot\partial) n_B\\
  &= -{K}~ X_B -(n\cdot\partial) X_B -X^AK_{AB}+n_B\left[ n^C(n\cdot\partial)X_C\right]
  -Y (n\cdot\partial) n_B\\
   &= -{K}~ X_B -\Pi_B^C(n\cdot\partial) X_C -X^AK_{AB}
  -Y (n\cdot\partial) n_B .
\end{split}
\end{equation}
It follows that
\begin{eqnarray}
\partial\cdot X +\left[{K}~Y+(n\cdot\partial)Y\right] -X^C(n\cdot\partial)n_C&=&0\label{eq1} , \\
{K}~ X_B +\Pi_B^C(n\cdot\partial) X_C +X^AK_{AB}
  +Y (n\cdot\partial) n_B&=&0\label{eq2}
\end{eqnarray}
These are the equations \eqref{eecc}.

\section{Derivation of the large D foliation adapted solution to Maxwell's equations and Charge Current}\label{Maxeq}

In this Appendix we present the derivation of some of the results reported in subsections 
\ref{linearout} and \ref{linearin}. 

\subsection{$\rho>1$}
As reported in \eqref{out1}, the Maxwell field in the region $\rho>1$ is assumed to take the form 
$${\cal A}_M = \rho^{-(D-3)}G_M = \rho^{-(D-3)}\sum_{k}(\rho -1)^kG_M^{(k)} .$$
Maxwell's equations take the form\footnote{In this subsection all raising, lowering and contraction of indices have been done using the flat Minkowski metric $\eta_{AB} .$}
\begin{equation}\label{stepgauge1}
\begin{split}
0={\partial}_A {F^A}  _B 
&= 2({\partial}_A \rho^{-(D-3)})({\partial}^A G_B) - ({\partial}_B \rho^{-(D-3)})({\partial}\cdot G) +  \rho^{-(D-3)} {\partial}_A ({\partial}^A G_B- {\partial}_B G^A)  .
\end{split}
\end{equation}
To derive expression for ${\partial}_A {F^A}  _B $ in 
\eqref{stepgauge1} we have used the subsidiary condition 
\eqref{subsidonGa}, the gauge choice
\eqref{outgauge} and the harmonicity condition \eqref{rhof}. Note also that
\begin{equation}\label{stepgauge2}
\begin{split}
n^A G_A=0\Rightarrow (\partial_A\partial_B\rho)G^A =-(\partial_A\rho)(\partial_B G^A).
\end{split}
\end{equation}

It is convenient to rewrite the Maxwell equation \eqref{stepgauge1} in the form 
\begin{equation}\label{mesym}
T^{(1)}_B + T^{(2)}_B + T^{(3)}_B=0 ,
\end{equation}
where

\begin{equation}\label{simpgauge}
\begin{split}
T^{(1)}_B=&2({\partial}^A\rho^{-(D-3)})({\partial}_A G_B) \\
&= - \frac{2(D-3)}{\rho^{D-3}} \left[ \sum_{k=0}^\infty k \frac{(\rho -1)^{k-1}}{\rho} N^2 G_B ^{(k)}  + \sum_{k=0}^\infty \frac{{(\rho -1)}^{k} }{\rho} N(n.{\partial}) G_B ^{(k)} \right] ,\\
\\
T^{(2)}_B=&-({\partial}_B \rho^{-(D-3)}) \left({\partial}_A G^A\right)= \frac{(D-3)}{\rho^{D-2}}( N n_B) \sum_{k=0}^\infty  (\rho -1)^k(\partial^A G_A ^{(k)}) , \\
\\
T^{(3)}_B=&\rho^{-(D-3)}{\partial}_A ({\partial}^A G_B- {\partial}_B G^A) \\
=& {\rho}^{-(D-3)} \sum_{k=0}^\infty \Bigg\{ k {(\rho -1)}^{(k-1)}\bigg\{ \left( \left( n\cdot\partial\right) \left(N G_B ^{(k)} \right)+ {K} N G_B ^{(k)}  +2 N n_B (\partial\cdot G^{(k)})\right) \\
&~~~ - n_B(G^{(k)} \cdot \partial) N -N n_B (\partial \cdot G^{(k)}) \bigg\}  
+   {(\rho -1)}^{k} {\partial ^A}F_{AB} ^{(k)} \Bigg\} , \\
\text{where}&~~
F_{AB} ^{(k)}= \partial_A G_B^{(k)} - \partial_B G_A^{(k)} .
\end{split}
\end{equation}
We now simply plug \eqref{simpgauge} into \eqref{mesym} and equate the coefficients of distinct powers 
of $(\rho-1)$. As explained in the main text, at this stage we are only interested in solving the dynamical 
Maxwell equations \eqref{dyndef}. We find the first nontrivial constraint by equating to zero the coefficient 
of $(\rho-1)^0$ in the projected version (\eqref{dyndef}) of the  Maxwell equation \eqref{mesym}.
This procedure yields the equation
\begin{equation}\label{projmeq}
\begin{split}
0=&~\rho^{(D-3)}\Pi^C_B\partial^AF_{AC}\\
=&-2(D-3) N^2 G_B^{(1)} +\left[{K} N + (n\cdot\partial)N\right] G_B^{(1)}  + \Pi^C_B\partial^A F^{(0)}_{AC} + 2 N^2 G_B^{(2)}\\
&+{\cal O}(\rho-1) .
\end{split}
\end{equation}

Solving equation \eqref{projmeq} at leading order in $\left(\frac{1}{D}\right)$ we get
\begin{equation}\label{10}
\begin{split}
G_B^{(1)} &= \frac{\Pi_B^C~\partial^A F_{AC}^{(0)}}{2(D-3) N^2 -N{K}} +{\cal O}\left(\frac{1}{D}\right)\\
&= \frac{\Pi_B^C~\partial^A F_{AC}^{(0)}}{N{K}} +{\cal O}\left(\frac{1}{D}\right) . \\
\end{split}
\end{equation}
In the second line we have used the fact that ${K} = D N + {\cal O}(1)$.

As we have explained around \eqref{defboor}, the solution \eqref{10} for $G_B^{(1)}$ is only 
valid on the membrane  i.e. at $\rho=1$. $G_B^{(1)}$ can be determined off the membrane using 
\eqref{outgaugec} to evolve the result \eqref{10} off the membrane.

\subsubsection{Exterior current}
The exterior current for the solution determined above is given by 
\begin{eqnarray} \label{extcurform}
J_B = n_A {F^A} _B \bigg{\vert} _{\rho = 1} .
\end{eqnarray}
In order to explicitly evaluate this current we note that 
\begin{eqnarray} 
n_A {F^{A}}  _B &=& -(D-3) \left(\frac{{\rho}^{-(D-3)}}{\rho} \right) N G_B +  {\rho}^{-(D-3)} \left[(n.\partial)G_B - n_A \partial_B G^A \right]  \nonumber\\
&=& -(D-3) \left(\frac{{\rho}^{-(D-3)}}{\rho} \right) N \sum_{k=0}^\infty (\rho -1)^k G_B^{(k)}  + {\rho}^{-(D-3)} \sum_{k=0}^\infty  k (\rho -1)^{(k-1)} N G_B^{(k)} \nonumber \\ \nonumber \\
&+&  {\rho}^{-(D-3)} \sum_{k=0}^\infty  (\rho -1)^{(k)} {{ K}^A}  _B G_A^{(k)}  . \label{nf}
\end{eqnarray}
In the derivation of the last equation we have used
\begin{eqnarray}\label{manip}
-n_A \partial_B G^A &=& -\partial_B ( n_A G^A) + (\partial_B n_A)G^A \nonumber \\
&=&\delta^C_B (\partial_C n_A) G^A \nonumber \\
&=& \left( \Pi ^C_B + n^C n_B  \right) (\partial_C n_A) G^A \nonumber \\
&=& K_{BA} G^A + n_B G^A {(n.\partial)n_A} \nonumber \\
&=& {K^A} _B G_A - n_B n^A {(n.\partial)G_A} .
\end{eqnarray}
Setting $\rho=1$ in  \eqref{nf} we obtain
\begin{equation}\label{outcurrent}
J_B ^{(out)} = -(D-3) N G^{(0)}_B + N G_B^{(1)} + K_B^A G_A^{(0)}
\end{equation}

\subsection{$\rho<1$}
For $\rho<1$ the form of the gauge field is given by 
$$A^{(in)}_M =\tilde G_M = \sum_{k}(\rho -1)^k\tilde G_M^{(k)} .$$
Maxwell equation takes the form
\begin{equation} \label{Fab}
\begin{split}
\partial^A F^{(in)}_{AB} =&  \sum_{k=0}^\infty\bigg[ k {(\rho -1)}^{(k-1)} \bigg\{\tilde G_B ^{(k)} \left( n\cdot\partial\right) N + 2N(n \cdot \partial){\tilde G_B ^{(k)}} + {K} N  \tilde G_B ^{(k)}+N n_B (\partial\cdot \tilde G^{(k)}) \bigg\} \\
&~~~~~+k (k-1) {(\rho -1)}^{(k-2)} N^2 \tilde G_B ^{(k)}  
+   {(\rho -1)}^{k} {\partial ^A}\tilde F_{AB} ^{(k)} \bigg] , \\
\text{where}&~~
\tilde F_{AB} ^{(k)}= \partial_A\tilde  G_B^{(k)} - \partial_B \tilde G_A^{(k)} .
\end{split}
\end{equation}
Here also we could determine $\tilde G_B^{(k)},~~k>0$ in terms of $\tilde G_B^{(0)} = G_B^{(0)}$ using equation \eqref{Fab} projected in the direction perpendicular to $n_B$.
The leading order $\tilde G_B^{(1)}$ could be determined from $\left(\Pi^C_B\partial^A F^{(in)}_{AC}\right)$ by setting the coefficient of $(\rho-1)^0$ to zero :
\begin{equation}\label{projin}
\begin{split}
\left[{K} N + (n\cdot\partial) N\right]\tilde G_B^{(1)} +\Pi^C_B\partial^A F^{(0)}_{AC} + N^2\tilde G_B^{(2)}=0 .
\end{split}
\end{equation}

%
Here all lowering and raising of indices have been done using the flat metric $\eta_{AB}$ .
\begin{equation}\label{insolution}
\begin{split}
\tilde G_B^{(1)} &= -\frac{\Pi_B^C~\partial^A F_{AC}^{(0)}}{N{K}} + {\cal O}\left(\frac{1}{D}\right) .\\
\end{split}
\end{equation}
\subsubsection{Inside current}
The inside current on the $\rho=1$ surface is given as 
\begin{eqnarray}
J^{(in)}_B = n^A F^{(in)} _{AB} \bigg{\vert} _{\rho = 1} , 
\end{eqnarray}
so that
\begin{eqnarray} 
n^AF^{(in)}_{A B} &=&\sum_{k=0}^\infty  k (\rho -1)^{(k-1)} N \tilde G_B^{(k)}  
+  \sum_{k=0}^\infty  (\rho -1)^{(k)} {{ K}^A} _B \tilde G_A^{(k)} . \label{nfin}
\end{eqnarray}
Here also to simplify we have used the equation \eqref{manip}.
Substituting $\rho=1$ in equation \eqref{nfin} we find the inside current 
\begin{equation}\label{incurrent}
J_B^{(in)} =  N\tilde G^{(1)}_B + {K^A}_B {G}_A^{(0)} .
\end{equation}

\section{Derivation of the large D foliation adapted solution to Einstein's equations}\label{Eineq}

\subsection{$\rho>1$}
As explained in subsection \ref{einlinearout}, the metric in the region $\rho>1$ is assumed to take the 
form \eqref{hexp} which we repeat here for convenience
\begin{equation}\label{einout1}
\begin{split}
g_{AB} = \eta_{AB} + \rho^{-(D-3)}{\mathfrak h}_{AB} =\eta_{AB} + \rho^{-(D-3)}\sum_k (\rho-1)^kh^{(k)}_{AB} .
\end{split}
\end{equation}

Einstein's equations (linearized around $\eta_{AB}$) take the form :
\begin{equation}\label{stepein1}
\begin{split}
0=R^{(out)}_{AB}
= t^{(1)}_{AB} + t^{(2)}_{AB} + t^{(3)}_{AB}  , 
\end{split}
\end{equation}
where
\begin{eqnarray}
t^{(1)}_{AB}&=& {\partial _A} \left[\left (\partial _C \rho^{-(D-3)}\right)~ {\mathfrak h}^C_B + \rho^{-(D-3)}{\partial _C}{\mathfrak h}^C_B \right] + (A \leftrightarrow B) \nonumber   \\
&=&\left[{\partial _A}\rho^{-(D-3)}\right]\left[ {\partial _C}{\mathfrak h}^C_B\right] + \rho^{-(D-3)}{\partial _A}{\partial _C}{\mathfrak h}^C_B+ (A \leftrightarrow B)\label{tab1} , \\
t^{(2)}_{AB} &=& -\partial^2 {\left[ \rho^{-(D-3)} {\mathfrak h}_{AB} \right]}   \nonumber \\
&=& -2\left[\partial _C \rho^{-(D-3)}\right][\partial^C {\mathfrak h}_{AB}] - \rho^{-(D-3)}\partial^2 {\mathfrak h}_{AB}\label{tab2} , \\
t^{(3)}_{AB} &=& -{\partial _A}{\partial _B} \left[\rho^{-(D-3)}{\mathfrak h}\right]\nonumber \\
 &=& -[\partial _A \partial_ B \rho^{-(D-3)}]{\mathfrak h} -(\partial _A \rho^{-(D-3)}){\partial _B {\mathfrak h}}\nonumber\\
 &&-(\partial _B \rho^{-(D-3)}){\partial _A {\mathfrak h}} - \rho^{-(D-3)} (\partial _A \partial _B {\mathfrak h}) .\label{tab3}
\end{eqnarray}
In deriving equations \eqref{tab1}, \eqref{tab2} and \eqref{tab3} we have used \eqref{metoutgauge}, 
\eqref{outsubsidiary} and \eqref{rhof}.

We now substitute equation \eqref{einout1} in equation \eqref{stepein1} and expand it in powers of 
$(\rho-1)$. Equating powers of $\rho-1$ in the dynamical equation allows us to solve for the 
unknown coefficients $h_{AB}^{(k)},~~k>0$ in terms of $h_{AB}^{(0)}$, order by order in 
$\left(\frac{1}{D}\right)$. In particular $h_{AB}^{(1)}$ is determined at  leading order in 
$\left(\frac{1}{D}\right)$ by equating terms of order $(\rho-1)^0$ on both sides of the 
projected Einstein equation
\begin{eqnarray}\label{projeinsol}
&&0=\rho^{D-3}\Pi^C_B\Pi^{C'}_AR^{(out)}_{CC'}\nonumber\\
 &=& \left(\frac{D-3}{2}\right) N K_{AB} h^{(0)} + (D-3) N^2 h_{AB} ^{(1)} -\left(\frac{1}{2}\right)\left( (n.\partial)N + N{ K}\right)h_{AB} ^{(1)} \nonumber\\
&&+ \frac{1}{2}\Pi^C_B\Pi^{C'}_A\bigg[{\partial _C}\partial ^M h_{MC'}^{(0)}+ \partial _{C'}{\partial^M}h_{MC}^{(0)}-\partial^2h^{(0)}_{CC'}-\partial_C\partial_{C'}h^{(0)}\bigg] \nonumber\\
&&+{\cal O}(\rho-1) .
\end{eqnarray}
Solving equation \eqref{projeinsol} at leading order in $\left(\frac{1}{D}\right)$ we get ,
\begin{equation}\label{h10d}
\begin{split}
&h_{AB} ^{(1)}\\
 =&-\Pi^C_B\Pi^{C'}_A\left[\frac{{\partial _C}\partial ^M h_{MC'}^{(0)}+ \partial _{C'}{\partial^M}h_{MC}^{(0)}-\partial^2h^{(0)}_{CC'}-\partial_C\partial_{C'}h^{(0)} + (D-3) h^{(0)}K_{CC'} }{2(D-3)N^2 - N{K}}\right] \\
 &+ {\cal O}\left(\frac{1}{D}\right)\\
 =&-\Pi^C_B\Pi^{C'}_A\left[\frac{{\partial _C}\partial ^M h_{MC'}^{(0)}+ \partial _{C'}{\partial^M}h_{MC}^{(0)}-\partial^2h^{(0)}_{CC'}-\partial_C\partial_{C'}h^{(0)} + D h^{(0)}K_{CC'} }{ N{K}}\right]\\
&+ {\cal O}\left(\frac{1}{D}\right) .\\
\end{split}
\end{equation}
In equation \eqref{h10d} naively it seems that the last term is of order ${\cal O}(D)$. 
But we shall see that for our case $h^{(0)}$ is actually of order ${\cal O}\left(\frac{1}{D}\right)$, so the last two terms do not even contribute to the leading solution for $h_{AB}^{(1)}$.

\subsubsection{External stress tensor}\label{outstress}
The stress tensor $T_{AB}^{out}$ is given by 
\begin{equation}\label{outstresabs}
\begin{split}
T_{AB}^{(out)} &= \left[ {\cal K}^{(out)}_{AB} -  {\cal K}^{(out)} ~{\mathfrak p}^{(out)}_{AB}\right] ,
\end{split}
\end{equation}
where $ {\cal K}^{(out)}_{AB}$ is the extrinsic curvature of the $(\rho = 1)$ surface (approached from the 
outside) viewed as a submanifold of the full space-time with bulk metric $g_{AB} = \eta_{AB} + \rho^{-(D-3)} {\mathfrak h}_{AB}$. The trace of $  {\cal K}^{(out)}_{AB}$ is denoted by $ {\cal K}^{(out)}$ and $ {\mathfrak p}^{(out)}_{AB}$ is the projector onto the surface $(\rho=1)$.
Let the normal to the surface is denoted by $ n^{(out)}_A= \frac{\partial_A\rho}{\sqrt{g^{AB} (\partial_A\rho)(\partial_B\rho)}}$. It follows from the gauge condition \eqref{metoutgauge} that the denominator of this 
expression - the norm of the one-form $\partial_A\rho$ in the metric $g_{AB}$ - differs from the norm of the same oneform in the metric $\eta_{AB}$ only at quadratic order in $h_{AB}$. If we work only to linear 
order in $h_{AB}$ it follows that 
$ (n^{(out)}_A = n_A)$
and also since $n_A {\mathfrak h}^{AB} =0$, it implies $g^{AB} n_B^{(out)} =g^{AB} n_B = n^A$.\\
It thus also follows that 
\begin{equation*}
\begin{split}
&{\mathfrak p}^{(out)}_{AB} \equiv g_{AB} -n^{(out)}_A n^{(out)}_B = 
g_{AB} -n_A n_B=  \Pi_{AB} + \rho^{-(D-3)}{\mathfrak h}_{AB} ,\\
&[{\mathfrak p}^{(out)}]^C_A = \delta^C_A -n^Cn_A = \Pi^C_A .
\end{split}
\end{equation*}
Where in the last step we have used the definition $\Pi_{AB}= \eta_{AB}-n_A n_B$ and the 
definition $g_{AB}=\eta_{AB}+ \frac{{\mathfrak h}_{AB}}{\rho^{D-3}}$.

The extrinsic curvature evaluates to  \footnote{ In this section $ \nabla $ means covariant derivative with respect to full linearised space-time from outside. }
\begin{equation}\label{hatkab}
\begin{split}
{\cal K}^{(out)}_{AB}&= [{\mathfrak p}^{(out)}]^C_A [{\mathfrak p}^{(out)}]^{C^\prime}_B{{\nabla}_C  n_{C^\prime}}\vert_{\rho=1} \\
 &=  \Pi^C_A ~\Pi^{C^\prime}_B\left ({{\partial}_C n_{C^\prime} - n_q {\Gamma}^q _{CC^\prime}}  \right)\vert_{\rho=1}  \\
&=K_{AB} -  \Pi^C_A \Pi^{C^\prime}_B~(n_q { {\Gamma}^q _{CC^\prime}} )\vert_{\rho=1}  ,
\end{split}
\end{equation}
where $K_{AB}$ is the extrinsic curvature of $(\rho=1)$ surface as embedded in flat Minkowski space-time $\eta_{AB}$.
The last term in equation \eqref{hatkab} can be evaluated by determining the Christoffel symbol with respect to the metric $g_{AB}$ to linear order in ${\mathfrak h}_{AB}$. We find
\begin{equation}\label{gammaout}
\begin{split}
&-\Pi^C_A \Pi^{C^\prime}_B~ n_q { {\Gamma}^q _{CC^\prime}} {\mid}_{\rho=1} \\
=&-\left(\frac{\Pi^C_A \Pi^{C^\prime}_B}{2}\right) n^q\left[\partial_C\left(\rho^{-(D-3)} {\mathfrak h}_{C'q}\right) +\partial_{C'}\left(\rho^{-(D-3)} {\mathfrak h}_{Cq}\right)-\partial_q \left(\rho^{-(D-3)} {\mathfrak h}_{C'C}\right)\right]_{\rho=1}\\
 =&~\bigg[ \frac{N}{2}h_{AB} ^{(1)} - \frac{N}{2}(D-3)h_{AB} ^{(0)} + \frac{1}{2} \left( h^{(0)} _{Aq}  K^q_B + h^{(0)} _{Bq}  K^q_A  \right)\bigg]_{\rho=1} \\
\end{split}
\end{equation}
In the last step of equation \eqref{gammaout} we have used the following manipulation :
\begin{equation}\label{gammamani}
\begin{split}
\Pi^{AC} \Pi^{B{C^\prime}} n_q {\partial _C}h^q _{C^\prime} &= -\Pi^{AC} \Pi^{B{C^\prime}} {\mathfrak h}^q _{C^\prime}({\partial _C} n_q) \\
&= -{\Pi}^{AC} {\Pi}^{B{C^\prime}} {\mathfrak h}^q _{C^\prime}  K_{Cq}\\
&= -h^{qB} K^A _q .
\end{split}
\end{equation}
Substituting equation \eqref{gammaout} in equation \eqref{hatkab} we finally get
\begin{equation}\label{exfinal}
{\cal K}^{(out)}_{AB}= K_{AB} + \bigg[\frac{N}{2} h_{AB}^{(1)} - \frac{N}{2} (D-3)h_{AB}^{(0)} + \frac{1}{2} \left( h^{(0)} _{Aq}  K^q_B + h^{(0)} _{Bq}  K^q_A  \right)\bigg]_{\rho=1} .
\end{equation}
It follows that the trace of the trace of the external extrinsic curvature is given by 
\begin{equation}\label{trexfinal}
\begin{split}
{\cal K}^{(out)} = \left[\eta^{AB} - h^{AB}_{(0)}\right] {\cal K}^{(out)}_{AB} = {K} +\bigg[\frac{N}{2} h^{(1)} - \frac{N}{2} (D-3)h^{(0)}\bigg]_{\rho=1} ,
\end{split}
\end{equation}
where ${K} = \eta^{AB}K_{AB} =$ Trace of the extrinsic curvature of $(\rho=1)$ surface as embedded in flat space-time and $h^{(k)}$ denotes $\left[\eta^{AB} h_{AB}^{(k)}\right]$.\\
Note, if we assume the membrane to be embedded in an auxiliary space with metric $(\eta_{AB} +h^{(0)}_{AB})$ and denote the extrinsic curvature as $\tilde K_{AB}$, then  ${\cal K}^{(out)}_{AB}$ and ${\cal K}^{(out)}$ could simply be written as
\begin{equation}\label{hiji}
{\cal K}^{(out)}_{AB} = \tilde K_{AB} +\frac{N}{2}\bigg[ h_{AB}^{(1)} -  (D-3)h_{AB}^{(0)}\bigg],~~~{\cal K}^{(out)}=\tilde K +\frac{N}{2} \bigg[h^{(1)} - (D-3)h^{(0)}\bigg]
\end{equation}
Substituting equations \eqref{exfinal}, \eqref{trexfinal} and \eqref{hiji} in equation \eqref{outstresabs} we obtain our final expression for the stress tensor from outside $(\rho=1)$ surface as given in \eqref{stressout}.

\subsection{$\rho<1$}
For $\rho<1$ the metric is assumed to take the form \eqref{ein} which we reproduce for convenience
$$\tilde g_{AB} =\eta_{AB} + \tilde h_{AB} =\eta_{AB} + \sum_{k} (\rho-1)^k\tilde h_{AB}^{(k)} .$$
Einstein equation takes the form
\begin{equation} \label{Rinab}
\begin{split}
R^{(in)}_{AB}&=\left(\frac{1}{2}\right)\left[ \partial_C\partial_A \tilde h^C_B +\partial_C\partial_B \tilde h^C_A-\partial^2\tilde h_{AB} -\partial_A\partial_B\tilde h\right]=0 .
\end{split}
\end{equation}
As in the previous subsection $\tilde h_{AB}^{(k)},~~k>0$ can be determined in terms of 
$\tilde h_{AB}^{(0)} = h_{AB}^{(0)}$ using the dynamical Einstein equations. In particular 
$\tilde h_{AB}^{(1)}$ may be determined from the coefficient of $(\rho-1)^0$ in 
\begin{equation}\label{projRin}
\begin{split}
0=\left(\Pi^C_B\Pi^{C'}_AR^{(in)}_{CC'}\right)=& \left(\frac{\Pi^C_A\Pi^{C'}_B}{2}\right)\left[ \partial^M\partial_{C'} \tilde h^{(0)}_{MC} +\partial^M\partial_C \tilde h^{(0)}_{MC'}-\partial^2\tilde h^{(0)}_{CC'} -\partial_C\partial_{C'}\tilde h^{(0)} \right]\\
&-\left(\frac{1}{2}\right) [N{K}+(n\cdot\partial) N] ~\tilde h^{(1)}_{AB}-K_{AB}~\tilde h^{(1)} + {\cal O}(\rho-1)
\end{split}
\end{equation}
(Here all lowering and raising of indices have been done using the flat metric $\eta_{AB}$).
Solving equation \eqref{projRin} in leading order in ${\cal O}\left(\frac{1}{D}\right)$ we find :
\begin{equation}\label{inmet}
\begin{split}
\tilde h^{(1)}_{AB}&=\left(\frac{\Pi^C_A\Pi^{C'}_B}{N{K}}\right)\left[ \partial^M\partial_{C'} \tilde h^{(0)}_{MC} +\partial^M\partial_C \tilde h^{(0)}_{MC'}-\partial^2\tilde h^{(0)}_{CC'} -\partial_C\partial_{C'}\tilde h^{(0)} \right]+{\cal O}\left(\frac{1}{D}\right) .
\end{split}
\end{equation}
\subsubsection{Interior stress tensor}
The interior stress tensor is given by  
\begin{equation}\label{tindef}
T^{(in)}_{AB} ={\cal K}^{(in)}_{AB} -{\cal K}^{(in)}{\mathfrak p}^{(in)}_{AB} \bigg{\vert} _{\rho = 1} , 
\end{equation}
where ${\cal K}^{(in)}_{AB}$ is the extrinsic curvature of the $\rho=1$ surface (as approached from the interior) 
viewed as a submanifold of the full space-time with bulk metric $\tilde g_{AB} = \eta_{AB} + \tilde h_{AB}$. The trace of $ {\cal K}^{(in)}_{AB}$ is denoted by ${\cal K}^{(in)}$ and ${ \mathfrak p}^{(in)}_{AB}$ is the projector onto the surface $(\rho=1)$. As in the previous subsection, working to linear order in the metric fluctuations 
$$ n^{(in)}_A = n_A; ~~~{\mathfrak p}^{(in)}_{AB} = \Pi_{AB} + \tilde h_{AB} . $$
The extrinsic curvature evaluates to
\begin{equation}\label{hatkabin}
\begin{split}
{\cal K}^{(in)}_{AB}&= [{\mathfrak p}^{(in)}]^C_A [{\mathfrak p}^{(in)}]^{C^\prime}_B{{\nabla}_C \hat n_{C^\prime}}\vert_{\rho=1} \\
 &=  \Pi^C_A \Pi^{C^\prime}_B\left ({{\partial}_C n_{C^\prime} - n_q {\Gamma}^q _{CC^\prime}}  \right)\vert_{\rho=1}  \\
&= K_{AB} -  \Pi^C_A \Pi^{C^\prime}_B n_q { {\Gamma}^q _{CC^\prime}} \vert_{\rho=1}  ,
\end{split}
\end{equation}
where $K_{AB}$ is the extrinsic curvature of $(\rho=1)$ surface as embedded in flat Minkowski space-time $\eta_{AB}$. The last term in equation \eqref{hatkabin} is simplified further by evaluating the Christoffel
symbol as :
\begin{equation}\label{gammain}
\begin{split}
&-\Pi^C_A \Pi^{C^\prime}_B n_q { {\Gamma}^q _{CC^\prime}} {\mid}_{\rho=1} \\
=&-\left(\frac{1}{2}\right)\Pi^C_A \Pi^{C^\prime}_B n^q\left[\partial_C\tilde h_{C'q} +\partial_{C'}\tilde h_{Cq}-\partial_q \tilde h_{C'C}\right]_{\rho=1}\\
 =&~\bigg[ \frac{N}{2}\tilde h_{AB} ^{(1)}+ \frac{1}{2} \left( \tilde h^{(0)} _{Aq}  K^q_B + \tilde h^{(0)} _{Bq}  K^q_A  \right)\bigg]_{\rho=1} \\
  =&~\bigg[ \frac{N}{2}\tilde h_{AB} ^{(1)}+ \frac{1}{2} \left(  h^{(0)} _{Aq}  K^q_B +  h^{(0)} _{Bq}  K^q_A  \right)\bigg]_{\rho=1} .\\
\end{split} 
\end{equation}

%
Substituting equation \eqref{gammain} in equation \eqref{hatkabin} we finally get
\begin{equation}\label{exinfinal}
\begin{split}
{\cal K}^{(in)}_{AB}&= K_{AB} + \bigg[\frac{N}{2} \tilde h_{AB}^{(1)}  + \frac{1}{2} \left( h^{(0)} _{Aq}  K^q_B + h^{(0)} _{Bq}  K^q_A  \right)\bigg]_{\rho=1}\\
&=\tilde K_{AB} +\left(\frac{N}{2}\right) \tilde h_{AB}^{(1)} .
\end{split}
\end{equation}
The trace of the extrinsic curvature is given by
\begin{equation}\label{trinfinal}
\begin{split}
{\cal K}^{(in)} = \left[\eta^{AB} -\tilde  h^{AB}_{(0)}\right]{\cal  K}^{(in)}_{AB} = \bigg[{\tilde K} +\left(\frac{N}{2} \right)\tilde h^{(1)} \bigg]_{\rho=1} ,
\end{split}
\end{equation}
where ${\tilde K} = \left(\eta^{AB}-h^{AB}_{(0)}\right)\tilde K_{AB}$ and $\tilde h^{(k)}$ denotes $\left[\eta^{AB} \tilde h_{AB}^{(k)}\right]$.

Substituting equations \eqref{exinfinal} and \eqref{trinfinal} in equation \eqref{tindef} we get the final expression for the stress tensor from interior of the $(\rho=1)$ surface as given in equation \eqref{stressin2}.

\section{Details Related to the Large D black hole membrane current}\label{memour}

In this Appendix we first perform the consistency check described in subsection \ref{acc}. 
We then go onto supply some of the algebraic details of the derivation of the final form 
of the charge current on the large D black hole membrane \eqref{memcur}.

\subsection{Details of the consistency check described in subsection 6.3}

\subsubsection{Gauge Transformation}

In this subsection we gauge transform the gauge field presented in 
\eqref{out1m} to put it in the gauge employed in subsection \ref{linearout}.

Let us apply a  gauge transformation parametrized by the gauge function $\Lambda$ on the gauge field of \eqref{out1m}, where
\begin{equation}\label{gaugef1}
\begin{split}
\Lambda & = \rho^{-(D-3)}\left[\Lambda^{(0)} + (\rho-1)\Lambda^{(1)} + (\rho-1)^2\Lambda^{(2)} +\cdots\right] ,\\
\tilde M_B &= \partial_B\Lambda + M_B,~~
0=n^B\tilde M_B  =n^B M_B +(n\cdot\partial)\Lambda\\
\Rightarrow& (n\cdot\partial)\Lambda=-n^B M_B .
\end{split}
\end{equation}

Equating different powers of $(\rho-1)$ on both sides of the  last equation in \eqref{gaugef1} we get the following equations for $\Lambda^{(0)}$, $\Lambda^{(1)}$ and $\Lambda^{(2)}$.
\begin{equation}\label{gaugef2}
\begin{split}
&-(D-3) N \Lambda^{(0)} + (n\cdot\partial)\Lambda^{(0)} + N\Lambda^{(1)} = -\sqrt{2} Q\\
&-(D-3) N [\Lambda^{(1)}-\Lambda^{(0)}] + (n\cdot\partial)\Lambda^{(1)} + N\Lambda^{(2)} = -\sqrt{2} \left(\frac{D}{K}\right)\left(\frac{\bar\nabla^2 Q}{K}\right) , \\
\end{split}
\end{equation}
where $\bar\nabla^2 Q = \Pi^{AB}\partial_A\left[\Pi_B^C\partial_C Q\right]. $
Solving equation \eqref{gaugef2}
\begin{equation}\label{gaugef03}
\begin{split}
\Lambda^{(0)} &= \left(\frac{1}{D-3}\right)\frac{\sqrt{2} Q}{ N} + \left(\frac{1}{D}\right)^2\left(\frac{\sqrt{2}}{N}\right)  \left[ Q+(n\cdot\partial) \left(\frac{Q}{N}\right) + \left(\frac{D}{K}\right)\left(\frac{\bar\nabla^2 Q}{K}\right)\right] \\
&+ {\cal O}\left(\frac{1}{D}\right)^3 , \\
\Lambda^{(1)} &= \left(\frac{1}{D-3}\right)\left(\frac{\sqrt{2}}{N}\right)  \left[ Q+ \left(\frac{D}{K}\right)\left(\frac{\bar\nabla^2 Q}{K}\right)\right] 
+ {\cal O}\left(\frac{1}{D}\right)^2 . \\
\end{split}
\end{equation}

Now after applying the gauge transformation 
\begin{equation}\label{gaugef3}
\begin{split}
\tilde M_B &= M_B + \partial_B\Lambda =\rho^{-(D-3)} \left[\tilde M_B^{(0)} + (\rho-1)\tilde M_B^{(1)}+\cdots\right] ;\\
\tilde M_B^{(0)} &= - \sqrt{2} Q ~u_B  + \frac{\sqrt{2}Q^3}{D}\left(\frac{D}{K}\right)\left(\frac{\partial_A{K}}{K} - (u\cdot\partial)u_A\right)P^A_B\\
& + \frac{\sqrt{2}}{D}\Pi^A_B\left[\frac{\partial_A Q}{N} - \frac{Q\partial_A N}{N^2}\right]+ {\cal O}\left(\frac{1}{D}\right)^2 ,\\
\tilde M_B^{(1)} &= -\sqrt{2} \left(\frac{D}{K}\right) \left(\frac{\bar\nabla^2 Q}{K} \right)u_B - \sqrt{2}Q \left(\frac{D}{K}\right) \left(\frac{\bar\nabla^2 u_A}{K} + u^C K_{CA}\right)p^A_B+ {\cal O}\left(\frac{1}{D}\right) ,
\end{split}
\end{equation}
where 
\begin{equation}\label{notderi2}
\begin{split}
&\bar\nabla^2u_A \equiv \Pi^{CB}\partial_C\left[\Pi^{A'}_A\Pi^{B'}_B\partial_{B'} u_{A'}\right] ,~~\bar \nabla^2 Q \equiv \Pi^{AB}\partial_A\left[\Pi^{B'}_B\partial_{B'}Q\right]\\
\end{split}
\end{equation}
Note that $\tilde M_B$ now satisfies the gauge condition of the previous section, i.e., $n^B\tilde M_B=0$. 

\subsubsection{Change in the subsidiary condition} \label{csc}

In this section we re-expand the coefficients of the gauge field 
of the previous subsection so that these coefficients obey the subsidiary 
conditions of subsection \ref{linearout}.

$\tilde M_B$ satisfies the gauge condition imposed in the previous section, and consequently 
can be identified with the field $G_B$ of \eqref{out1}. However the expansion coefficients 
${\tilde M_B^{(k)}}$ cannot yet be identified with the expansion coefficients $G_B^{(k)}$ of 
\eqref{out1} as ${\tilde M_B^{(k)}}$ do not obey \eqref{subsidonGa}, i.e. 
$$P^B_A(n\cdot\partial)\tilde M_B^{(k)} \neq 0 . $$
 
The coefficient functions $G_B^{(k)}$ are easily extracted from the expansion of 
$\sqrt{16 \pi} G_B={\tilde M_B}$ by following a recursive procedure we now outline. On the surface $\rho=1$, the quantity $\sqrt{16 \pi}{G_B}^{(0)}$ simply equals ${\tilde M_B^{(0)}}$. 
Away from $\rho=1$,  $\sqrt{16 \pi}{G_B}^{(0)}$ (which no longer agrees with ${\tilde M_B^{(0)}}$) can be determined 
from knowledge of its value on the $\rho=1$ surface using the equation
$$P^B_A(n\cdot\partial) G_B^{(0)} =0 . $$

Now that $G_B^{(0)}$ is known everywhere consider  
$$ G-G_B^{(0)}$$\\
This expression is a known power series expansion in $(\rho-1)$ which starts at $(\rho-1)^1$. 
On the surface $\rho=1$ the quantity $G_B^{(1)}$ is simply the coefficient of the linear term in 
$(\rho-1)$ in this expansion. We have thus determined  $G_B^{(1)}$ at $\rho=1$. However this 
information together with the subsidiary condition 
$$P^B_A(n\cdot\partial) G_B^{(1)} =0  , $$
determines $G_B^{(1)}$ everywhere.

Now that we know $G_B^{(1)}$ also everywhere consider the quantity 
$$ G-G_B^{(0)}-G_B^{1}(\rho-1) .$$
This quantity is a known power series that starts at order $(\rho-1)^2$. The coefficient of 
$(\rho-1)^2$ is simply $G_B^{(2)}$ evaluated at $\rho=1$ \ldots, and so on. We can thus proceed to 
evaluate $G_B^{(n)}$ for all $n$.

As the black hole membrane solution is known only to a very low order, we need to implement the 
recursive procedure described above only to very low order. This is very easily done. Clearly 
$$\sqrt{16 \pi}G_B^{(0)}={\tilde M_B^{(0)}} - (\rho-1) C_B^{(0)} + {\cal O}(\rho-1)^2 , $$
for some as yet unknown function $C_B^{(0)}$. Now the operator $P^A_B(n\cdot\partial)$ annihilates the LHS so it must also 
kill the RHS. Applying this operator to both sides of this equation, Taylor expanding in 
$\rho-1$ and equating the coefficient of $(\rho-1)^0$ to zero we find 
$$C_A^{(0)} = \frac{1}{N}P^B_A(n\cdot\partial)\tilde M_B^{(0)} .$$
It follows that 
$$\sqrt{16 \pi}G_B=G_B^{(0)}+ \left({\tilde M_B^{(1)}} + C_B^{(0)}\right)(\rho-1) . $$
so that on the surface $\rho=1$ 
\begin{equation}\label{gbmb1}
G_B^{(1)}=\left({\tilde M_B^{(1)}} + C_B^{(0)}\right) .
\end{equation}
From the explicit black hole membrane solution we know ${\tilde M_B^{(1)}}$ only to leading order in the 
$1/D$ expansion (though we know ${\tilde M_B^{(0)}}$ and so $C_B^{(0)}$ to first subleading order). 
It follows that our current knowledge of the black hole membrane solution is detailed enough only 
to allow to determine $G_B^{(1)}$ only at leading order in $1/D$ on the membrane surface. 
\footnote{As explained above, once $G_B^{(1)}$ has been determined on the surface $\rho=1$ it is easily continued away from this surface. We will, however, have no need for 
this continuation.}

We now turn to simplifying the expression for $C_B^{(0)}$. 
Plugging in the actual value of $\tilde M_B^{(0)}$ for the black hole membrane 
we may simplify this expression as follows :
\begin{equation}\label{sub2}
\begin{split}
C_A^{(0)} &= \frac{1}{N}\Pi^B_A(n\cdot\partial)\tilde M_B^{(0)}\\
& = -\frac{\sqrt{2} Q}{N}\Pi^B_A(n\cdot\partial) u_B + {\cal O}\left(\frac{1}{D}\right)\\
&= -\frac{\sqrt{2} Q}{N}P^B_A(n\cdot\partial) u_B+ {\cal O}\left(\frac{1}{D}\right)\\
&=\frac{\sqrt{2} Q}{N}P^B_A(u\cdot\partial) n_B+ {\cal O}\left(\frac{1}{D}\right)\\
&= \frac{\sqrt{2} Q}{N}u^CK_{CB}~ P^B_A+ {\cal O}\left(\frac{1}{D}\right) ,
\end{split}
\end{equation}
where we have plugged in the explicit expressions listed in \eqref{gaugef3}.
In the second and last line of equation \eqref{sub2} we have used the fact that the membrane
charge density and velocity field in \cite{Bhattacharyya:2015fdk} obey the subsidiary conditions
$$(n\cdot\partial) Q=0,~~P^B_A(n\cdot\partial) u_B +p^B_A(u\cdot\partial) n_B=0 . $$

From equation \eqref{gaugef3} and \eqref{sub2} it is not difficult to read off the values of  $\tilde M_B^{(1)}$ and $C_B^{(0)}$ . Using
$${K} = D N + {\cal O}(1)~~~~\text{and}~~~~ \frac{\bar\nabla^2 u_A}{K} = \frac{P^B_A\bar\nabla^2 u_B}{K}+ {\cal O}\left(\frac{1}{D}\right), $$
 we find
\begin{equation}\label{sub4}
\begin{split}
\sqrt{16 \pi} G_B^{(0)}&=- \sqrt{2} Q ~u_B  + \frac{\sqrt{2}Q^3}{D}\left(\frac{D}{K}\right)\left(\frac{\partial_A{K}}{K} - (u\cdot\partial)u_A\right)p^A_B\\
& + \frac{\sqrt{2}}{D}\Pi^A_B\left[\frac{\partial_A Q}{N} - \frac{Q\partial_A N}{N^2}\right]+ {\cal O}\left(\frac{1}{D}\right)^2 ,\\
\sqrt{16 \pi} G_B^{(1)}&=\left[\tilde M_B^{(1)} + C_B^{(0)}\right]=-\sqrt{2} \left(\frac{D}{K}\right) \left(\frac{\bar\nabla^2 Q}{K} \right)u_B - \sqrt{2}Q \left(\frac{D}{K}\right) \left(\frac{\bar\nabla^2 u_A}{K} \right)+ {\cal O}\left(\frac{1}{D}\right) .
\end{split}
\end{equation}

\subsubsection{Consistency}

In the previous subsubsections we have transformed the linearized part of the large D black gauge field into gauge and subsidiary conditions used in subsubsection \ref{linearout}, and have thus 
managed to read off the expressions for the quantities $G_B^{(0)}$ and $G_B^{(1)}$ listed in that subsection. 
However, according to the analysis of subsubsection \ref{linearout} the quantities $G_B^{(0)}$ and $G_B^{(1)}$
are not independent. In fact $G_B^{(1)}$ is given in terms of  $G_B^{(0)}$ by the equations
\eqref{10mt} and \eqref{10mti}. 

In other words the linearized part of the large $D$ black hole metric 
is fits into the general framework of  subsubsection \ref{linearout} if and only if the explicit results 
\eqref{sub4} obey \eqref{10mt} upto corrections of order $ {\cal O}\left(\frac{1}{D}\right)$. We have 
explicitly verified that this is indeed the case. This completes our check of the consistency of the 
large D black hole solutions at linearized order. 

\subsection{Details of the derivation of equation 6.13}

\subsubsection{The Membrane Current from Outside}

Now that we have recast the solution \eqref{out1m} in the form 
of the solutions presented in subsection \ref{linearout} we can 
use any of the formulae of that subsection to evaluate the membrane current. 
The external contribution to the current, $J^{out}$, is most simply obtained from 
the equation \eqref{outcurrentexact} which we quote again here for convenience
\begin{equation*}
J^{out}_B = -(D-3) N G^{(0)}_B + N G_B^{(1)} + K_B^A G_A^{(0)}.
\end{equation*}
Substituting $G_B^{(0)}$ and $G_B^{(1)}$ from equation \eqref{sub4} we find that upto  corrections of order ${\cal O}\left(\frac{1}{D}\right),$
\begin{equation}\label{outcur}
\begin{split}
\sqrt{16 \pi} J^{out}_B=&~\sqrt{2}Q\left[
 (1-Q^2)\left(\frac{\partial_A{K}}{K}\right)+(1+ Q^2) (u\cdot\partial)u_A-\left(\frac{\bar\nabla^2 u_A}{K}\right)-~K^C_A u_C\right]P^A_B\\
&+\sqrt{2}\left[(D-3)N Q +(u\cdot\partial) Q - \left(\frac{\partial^2 Q +Q(u\cdot \partial){K}}{K}\right) + Q(u\cdot K\cdot u)\right] u_B\\
&-\sqrt{2}Q\left[\left(\frac{\partial_A Q}{Q}\right) + (u\cdot\partial)u_A \right]P^A_B
+ {\cal O}\left(\frac{1}{D}\right).\\
\end{split}
\end{equation}
In the next subsection we shall see that the first line in the final expression of $J_B^{out}$ (the third step) vanishes as consequence of the stress tensor conservation equation on the membrane. So the final form of the outside current after removing the first line
\begin{equation}\label{outcurour}
\begin{split}
\sqrt{16 \pi} J^{out}_B  
=&~\sqrt{2}\bigg[Q\left( {K} + \frac{\bar\nabla^2{K}}{{K}^2}- \frac{2K}{D}\right)+(u\cdot\partial) Q \\
&~~~~~~- \left(\frac{\bar\nabla^2 Q +Q(u\cdot \partial){K}}{K}\right) + Q(u\cdot K\cdot u)\bigg] u_B\\
&-\sqrt{2}Q\left[\left(\frac{\partial_A Q}{Q}\right) + (u\cdot\partial)u_A \right]P^A_B
+ {\cal O}\left(\frac{1}{D}\right) . \\
\end{split}
\end{equation}
To simplify in equation \eqref{outcurour}  we have used the identities (see equations \eqref{I10}, \eqref{I11}, \eqref{I12}, \eqref{I9} and \eqref{I13} for derivation) that
$$(D-3)N = {K} + \frac{\bar\nabla^2{K}}{{K}^2}- \frac{2K}{D}+ {\cal O}\left(\frac{1}{D}\right) . $$

\subsubsection{The membrane current from inside}
In order to compute $J^{in}_B$ we use \eqref{incurrentexact} which we quote here again for convenience
\begin{equation}\label{incurrentrep}
J_B^{in} =  N\tilde G^{(1)}_B + K_B^A {{G}}_A^{(0)} .
\end{equation}
By comparing \eqref{insolutionexact} and \eqref{10mt} we see that it is a general feature of the solutions 
obtained in subsections \ref{linearout} and \ref{linearin} that 
$$\tilde G_B^{(1)}= - G_B^{(1)} + {\cal O}\left(\frac{1}{D}\right) .$$
It follows that \eqref{incurrentrep} can be rewritten as 
\begin{equation}\label{incurrentrepfn}
J_B^{in} =  -N G^{(1)}_B + K_B^A {{G}}_A^{(0)} .
\end{equation}
Using \eqref{sub4} it follows that 
\begin{equation}\label{incurour}
\begin{split}
\sqrt{16 \pi} J^{(in)}_B 
&= \sqrt{2}\bigg[ \left(\frac{\bar\nabla^2 Q}{K} \right)u_B + Q  \left(\frac{P^A_B\bar\nabla^2 u_A}{K} \right)-QK_B^A~u_A\bigg]+ {\cal O}\left(\frac{1}{D}\right) .\\
\end{split}
\end{equation}

\section{Details Related to the large D black hole Membrane Stress Tensor}

This Appendix mirrors the previous one except for the fact that it focuses on the membrane stress 
tensor rather than the charge current. In the first part of this Appendix we check that the 
large $D$ black hole metrics - upon linearization - do indeed fit into the general structure of 
linearized solutions to Einstein's equations at large $D$ developed in this paper. In the second 
part of the Appendix we provided details of our computation of the precise form of the large $D$ 
black hole stress tensor. 

\subsection{Consistency}

As we have described above, the large $D$ black hole metric of \cite{Bhattacharyya:2015fdk}
simplifies in the `matching' region to the linearized form \eqref{expmet} with \eqref{metricread}. 
For the convenience of the reader we reproduce those equations here:  
\begin{equation}\label{expmetapp}
G_{AB} =\eta_{AB} +\rho^{-(D-3)} M_{AB}
=\eta_{AB}+ \rho^{-(D-3)}\sum_n (\rho-1)^n M_{AB}^{(n)} ,
\end{equation}
where
\begin{equation}\label{matricread}
\begin{split}
M_{AB} ^{(0)} &= (1+Q^2)O_A O_B 
+ 2Q^4 \left( O_A V_B ^{(2)} + O_B V_A ^{(2)} \right) -Q^2 O_A O_B -2Q^2 {\tau}_{AB} \\
& + {\cal O}\left(\frac{1}{D}\right)^2 , \\
M_{AB}^{(1)} &= 2Q^2 S^{(1)} O_A O_B - (1+Q^2) \left[ V_A ^{(1)}O_B + O_A V_B ^{(1)} \right]    \\
& + {\cal O}\left(\frac{1}{D}\right) ,\\
\end{split}
\end{equation}
with 
\begin{equation}\label{metnot2}
\begin{split}
&V_A ^{(1)} = \left( \frac{D}{{K}}\right)\left[\frac{\bar\nabla^2 u_B} {{K}} +u^C K_{CB}\right]~{ P}^B_A  ,\\
&V_A ^{(2)} =  \left( \frac{D}{K} \right)   \left[ \frac{\partial _C{K} }{K} - (u\cdot\partial)u_C \right]P_A^C  , \\
&S^{(1)} =  \left( \frac{D}{K^2} \right)  \bar\nabla^2 Q  , \\
&\tau_{AB} =  { P}_A^{A'}~\left(\frac{D}{{K}}\right)\left[\frac{\partial_{A'} O_{B'}+\partial_{B'} O_{A'}}{2}-\eta_{A' B'}\left(\frac{\partial\cdot O}{D-2}\right) \right]~{ P}_B^{B'} , \\
&\text{where}\\
&\bar\nabla^2 Q =\Pi^A_B \partial_A\left[\Pi^{BC}\partial_C Q\right],~~\bar\nabla^2u_A= \Pi_{AA'}\Pi^{B}_{C}\partial_{B}\left[\Pi^{CC'}\Pi^{A'A''}(\partial_{C'}u_{A''})\right].
\end{split}
\end{equation}

In this section we will recast the results \eqref{expmetapp} and \eqref{matricread} into the 
general form obtained subsection \ref{einlinearout}. As in the previous Appendix, this requires 
us to perform first a coordinate (gauge) transformation on the solution \eqref{expmetapp},
\eqref{matricread}. We then read off the expansion coefficients of the general solution 
described in subsection \ref{einlinearout} by imposing the subsidiary conditions defined 
in that subsection.

\subsubsection{Gauge transformation} 
Starting with the solution \eqref{expmetapp} and \eqref{matricread} we perform the 
infinitesimal coordinate transformation 
$$x_A\rightarrow x^A + \rho^{-(D-3)}\xi^A , $$
which recasts the solution into the form
\begin{equation}\label{gtm}
\tilde M_{AB} =  M_{AB} + \rho^{(D-3)}\partial _A \left( {\rho}^{-(D-3)} \xi _B  \right) + \partial _B \left( {\rho}^{-(D-3)} \xi _A  \right) . 
 \end{equation}
We wish to choose our coordinate transformation to ensure that 
$ \hat h_{ab} $ satisfies the gauge condition of subsubsection \ref{einlinearout}, namely
\begin{equation}\label{coorrep}
n^A\tilde M_{AB}=0 .
\end{equation}
It follows that the infinitesimal coordinate transformation must be chosen to ensure that 
\begin{equation}\label{gaugecond}
 \begin{split}
 -n^A M_{AB} = (n\cdot\partial) \left[\rho^{-(D-3)} \xi_B\right] + n^A\partial_B\left[\rho^{-(D-3)} \xi_A\right] .
 \end{split}
 \end{equation}
Our general strategy for determining the vector field $\xi_A$ that satisfied \eqref{gaugecond} is to 
assume that like $h_{AB}$, the vector $\xi_A$ generating the coordinate transformation also admits an expansion in the powers of $(\rho-1)$ :
 \begin{equation}\label{xiexp}
\xi _A = \sum_{m=0}^\infty (\rho -1)^m \xi _{A}^{(m)} .  
 \end{equation}
We then substitute the expansion equations \eqref{expmet} and \eqref{xiexp} into \eqref{gaugecond} and determine
the expansion coefficients  $\xi^{(m)}$, order by order in the $\left(\frac{1}{D}\right)$ expansion by equating powers of $(\rho-1)$ on both sides for the equation \eqref{gaugecond}.

For the practical purposes of this paper we only need to implement this programme to the first couple of orders. 
Equating the coefficient of $(\rho-1)^0$ on both sides of equation \eqref{gaugecond} we find
 \begin{equation}\label{xi0}
 \begin{split}
 n^AM_{AB}^{(0)} 
= &~(D-3)N\left[\xi_B^{(0)} + n_B (n\cdot\xi^{(0)})\right] -\left[(n\cdot\partial)\xi_B^{(0)} +n^A\partial_B\xi_A^{(0)} \right] \\
 &- N \left[\xi_B^{(1)} + n_B (n\cdot\xi^{(1)})\right] .
 \end{split}
 \end{equation}
 Similarly equating the coefficient of $(\rho-1)^1$ we find
  \begin{equation}\label{xi1}
 \begin{split}
 n^AM_{AB}^{(1)} 
= &~(D-3)N\left[(\xi_B^{(1)} -\xi_B^{(0)}) + n_B (\xi_A^{(1)}- \xi_A^{(0)})n^A\right]\\ &-\left[(n\cdot\partial)\xi_B^{(1)} +n^A\partial_B\xi_A^{(1)} \right] 
 -2 N \left[\xi_B^{(2)} + n_B (n\cdot\xi^{(2)})\right] .
 \end{split}
 \end{equation}
 
Solving equation \eqref{xi0} and \eqref{xi1} simultaneously we find ,
\begin{equation}\label{xisol}
\begin{split}
\xi_A^{(1)} &=\left[ \frac{1}{(D-3)N}\right] \bigg[n^B [M^{(1)}_{AB} + M^{(0)}_{AB}] -\left(\frac{n_A}{2}\right)\left(n\cdot [M^{(1)}+ M^{(0)}]\cdot n\right)\bigg] \\
&+ {\cal O}\left(\frac{1}{D}\right) , \\
\xi^{(0)}_A &= \xi^{(0,1)}_A +\left(\frac{1}{D}\right)\xi_A^{(0,2)} +\left(\frac{1}{D}\right)^2 , \\
\text{where}~~&\\
\xi^{(0,1)}_A&=\left[ \frac{D}{(D-3)N}\right] \bigg[n^B  M^{(0)}_{AB} -\left(\frac{n_A}{2}\right)\left(n\cdot [M^{(0)}]\cdot n\right)\bigg],\\
\xi_A^{(0,2)}&= \left[ \frac{1}{(D-3)N}\right] \bigg[n^B  \left(\partial_A \xi^{(0,1)}_B + \partial_B \xi^{(0,1)}_A\right)-n_A\left(n^C[\partial_C\xi^{(0,1)}_{C'}]n^{C'}\right)\bigg]\\
&+\left[ \frac{D}{(D-3)^2N^2}\right] \bigg[n^B [M^{(1)}_{AB} + M^{(0)}_{AB}] -\left(\frac{n_A}{2}\right)\left(n\cdot [M^{(1)}+ M^{(0)}]\cdot n\right)\bigg] .\\
\end{split}
\end{equation} 

After substituting equation \eqref{xisol} in equation \eqref{gtm} we find
\begin{equation}\label{gtmf}
\begin{split}
\tilde M_{AB}&=\Pi^C_A\Pi^{C'}_B\left[M_{CC'}^{(0)} + \partial_C\xi^{(0)}_{C'} +\partial_{C'}\xi^{(0)}_{C} + (\rho-1) M^{(1)}_{CC'} +{\cal O}(\rho-1)^2 \right] .
\end{split}
\end{equation} 

\subsubsection{Change in subsidiary condition}\label{submet}

In order to extract the expansion coefficients $h^{(m)}_{MN}$ defined in subsection \ref{einlinearout}
we need to `Taylor' expand the metric \eqref{gtmf} in a power series expansion in $\rho$ while 
ensuring that the Taylor coefficients of this expansion obey the subsidiary conditions 
\eqref{outsubsidiary}. This is easily accomplished using the method outlined in the previous 
Appendix for the case of the gauge field. Let ${\tilde M}_{MN}^{(j)}$ represent the expansion 
coefficients of the metric \eqref{gtmf} where these coefficients don't necessarily obey the 
subsidiary condition \eqref{outsubsidiary}, i.e. 
$$\Pi^C_A\Pi^{C'}_B~(n\cdot\partial)~\tilde M_{CC'}^{(k)} \neq 0. $$
It must be that 
$$h_{AB}^{(0)}=\tilde M_{AB}^{(0)} - (\rho-1) C_{AB}^{(0)} + {\cal O}(\rho-1)^2 , $$
for some as yet unknown function $C_{AB}^{(0)}$. Now the operator $\Pi^{C'}_A\Pi^C_B(n\cdot\partial)$ annihilates the LHS so it must also 
kill the RHS. Applying this operator to both sides of this equation, Taylor expanding in 
$\rho-1$ and equating the coefficient of $(\rho-1)^0$ to zero we find 
$$C_{AB}^{(0)} = \frac{1}{N}\Pi^{C'}_A\Pi^C_B~(n\cdot\partial)\tilde M_{CC'}^{(0)} .$$
It follows that 
$$h_{AB}=h_{AB}^{(0)}+ \left({\tilde M_{AB}^{(1)}} + C_{AB}^{(0)}\right)(\rho-1) , $$
so that on the surface $\rho=1$ 
\begin{equation}\label{gbmb}
 h_{AB}^{(1)}=\left({\tilde M_{AB}^{(1)}} + C_{AB}^{(0)}\right) .
\end{equation}
Using equations \eqref{gtmf} and \eqref{gbmb} it follows that the coefficients $h_{AB}^{(0)}$ and $h_{AB} ^{(1)}$ 
corresponding to metric \eqref{gtmf} are given by : 
\begin{equation}\label{readp}
\begin{split}
h_{AB}^{(0)} &= (1+ Q^2)~ u_A u_B\\
&+\left(\frac{1}{D}\right)\bigg[-2Q^4 \left( u_A V_B^{(2)} + u_B V_A ^{(2)}  \right)  - Q^2 u_A u_B- 2Q^2~ {\tau}_{AB}  \\
& ~~~~~~~~~~~~~~~+  \Pi^C_A  \left[\nabla _C \xi _{C^ \prime} + \nabla _{C^\prime} \xi _C \right] \Pi^{C'}_B \bigg]
+{\cal O}\left(\frac{1}{D}\right)^2 ,\\
h_{AB} ^{(1)} &= \left( \frac{D}{{K}^2} \right) \left[ 2Q\bar \nabla ^2 Q ~u_A u_B + (1+ Q^2)~ \Pi^C_B\Pi^{C'}_A\left(u_{C'}\bar\nabla ^2 u_{C}+u_{C}\nabla ^2 u_{C'} \right) \right]\\ 
&+ {\cal O}\left(\frac{1}{D}\right) ,  \\
\end{split}
\end{equation}
where
\begin{equation}\label{notmet}
\begin{split}
\xi _A &= (1+Q^2)\left( \frac{D}{K} \right)\left( \frac{n_A}{2} -u_A \right)  , \\
V_A ^{(2)} &=  \left( \frac{D}{K} \right)   \left[ \frac{\partial _C{K} }{K} - (u.\partial)u_C \right] P_A^C , \\
{\tau}_{AB} &= \left( \frac{D}{K} \right)P_A^C   \bigg[ K_{CD} -  \left( \frac{{\partial _C}u_D+{\partial _D}u_C }{2}   \right)\\  
&~~~~~~~~~~~~~~~~~~~-\eta_{CD} \left(\frac{{K} -(\partial\cdot u)}{D-3}\right)    \bigg]  P_B^D .\\
\end{split}
\end{equation}
Here  $ \bigg[p_{AB} = \eta_{AB} - n_A n_B + u_A u_B\bigg]$ 
 and $\nabla$ denotes covariant derivative with respect to the intrinsic metric on the membrane as embedded in flat space.\\
 Note that the trace of $h^{(0)}_{AB}$ vanishes till order ${\cal O}(1)$ in our $\left(\frac{1}{D}\right)$ expansion.
\begin{equation}\label{traceh}
\begin{split}
\therefore h^{(0)}= \eta^{AB}h_{AB}^{(0)}  &=-(1+Q^2) + \frac{ \Pi^{AB}\nabla_A\xi_B}{D} + {\cal O}\left(\frac{1}{D}\right)\\
 &=-(1+Q^2) +2\left(\frac{1+Q^2}{D}\right) \left(\frac{D}{K}\right) \nabla_A \left(\frac{n^A}{2}-u^A\right)+ {\cal O}\left(\frac{1}{D}\right)\\
 &= {\cal O}\left(\frac{1}{D}\right). \\
\end{split}
\end{equation}
\subsubsection{Consistency}
As in the previous Appendix, it is not difficult to verify that the second equation in \eqref{readp} is consistent with \eqref{h10out} upto corrections of order $ {\cal O}\left(\frac{1}{D}\right)$.

\subsubsection{Derivation of equation 6.36}\label{fincal}
 
\begin{equation}\label{eman}
\begin{split}
E\equiv&~u^\mu\hat\nabla_\nu [T^{(NT)}]^\nu_\mu\\
=&~\left(\frac{K}{2}\right)(1+Q^2)(\hat\nabla\cdot u)  + \left(\frac{1+Q^2}{2}\right )(u\cdot\hat\nabla){K}+ \left(\frac{K}{2}\right)(u\cdot\hat\nabla)Q^2\\
&-\left(\frac{1-Q^2}{2}\right)u_\mu\hat\nabla_\nu K^{\mu\nu} +u_\nu\hat\nabla_\mu\left(\frac{\hat\nabla^\nu u^\mu +\hat\nabla^\mu u^\nu}{2}\right)-(\hat\nabla\cdot{\cal V}) +{\cal O}(1)\\
\\
=&~\left(\frac{K}{2}\right)(1+Q^2)(\hat\nabla\cdot u)  + Q^2(u\cdot\hat\nabla){K}+ \left(\frac{K}{2}\right)(u\cdot\hat\nabla)Q^2\\
& +{K}(u^\alpha K_{\alpha\beta} u^\beta)-(\hat\nabla\cdot{\cal V}) +{\cal O}(1)\\
\\
=&~\left(\frac{K}{2}\right)(1+Q^2)(\hat\nabla\cdot u) -(1+Q^2)(u\cdot\hat\nabla){K}+ \left(\frac{K}{2}\right)(u\cdot\hat\nabla)Q^2\\
&-Q\hat\nabla^2Q- \left(\frac{2Q^4-Q^2-1}{2}\right)\left(\frac{\hat\nabla^2{K}}{K}\right)
+\left(1+\frac{Q^2 + 2Q^4}{2}\right){K}~ (u^\alpha K_{\alpha\beta} u^\beta)\\
&+ {\cal O}(1). \\
\end{split}
\end{equation}
In the second last line we have used identities \eqref{I3} and \eqref{I4}. In the last line we have used identity \eqref{I7}.\\
Now we could simplify equation \eqref{eman} further by using the current conservation equation equation \eqref{divcur}. For convenience we are quoting the equation here.
\begin{equation}\label{concurrep}
\hat\nabla^2Q = Q{K}(\hat\nabla\cdot u)  + {K}(u\cdot\hat\nabla)Q + Q(u\cdot\hat\nabla) {K} - Q{K}(u^\alpha K_{\alpha\beta} u^\beta)+ {\cal O}(1).
\end{equation}
Substituting equation \eqref{concurrep} in equation \eqref{eman} we find
\begin{equation}\label{emansimp1}
\begin{split}
E= &~-\left(\frac{1+2Q^2}{2}\right)\left[2(u\cdot\hat\nabla{K}) -(1-Q^2)\left(\frac{\hat\nabla^2{K}}{K}\right) - (1+Q^2){K}(u^\alpha K_{\alpha\beta} u^\beta)\right]\\
&+\left(\frac{K}{2}\right)(1-Q^2)(\hat\nabla\cdot u)+ {\cal O}(1)
\end{split} .
\end{equation}

Now we shall show that the term in the first line of equation \eqref{emansimp1} could be re-expressed as $\bigg[-\left(\frac{1+2Q^2}{{K}}\right)(\hat\nabla_\mu E^\mu)\bigg]$, where $E^\mu$ is the projection of stress tensor conservation equation in the direction perpendicular to $u^\mu$.
\begin{equation*}
\begin{split}
E^A &=-\left(\frac{K}{2}\right)\bigg[(1+Q^2) (u\cdot\nabla)u^A + (1-Q^2) p^{AC}\left(\frac{\nabla_C {K}}{K}\right)\\
&~~~~~~~~~~~~ -p^{AC}\left(\frac{\nabla^2 u_C}{K} + K_{CB}u^B\right)\bigg]
+ {\cal O}(1) .\\
\end{split}
\end{equation*}
Taking the divergence of the above equation we find
\begin{equation}\label{divp}
\begin{split}
\hat\nabla_\mu E^\mu &= -\left(\frac{K}{2}\right)\hat\nabla_\mu\bigg[(1+Q^2) (u\cdot\hat\nabla)u^\mu+ (1-Q^2) p^{\mu\nu}\left(\frac{\hat\nabla_\nu {K}}{K}\right)\\
&~~~~~~~~~~~~ -p^{\mu\nu}\left(\frac{\hat\nabla^2 u_\nu}{K} + K_{\nu\alpha}u^\alpha\right)\bigg]
+ {\cal O}(D)\\
&= -\left(\frac{K}{2}\right)\bigg[(1+Q^2){K}(u^\alpha K_{\alpha\beta} u^\beta) +(1-Q^2)\left(\frac{\hat\nabla^2{K}}{K}\right)-2(u\cdot\hat\nabla){K}\bigg]+ {\cal O}(D) .\\
\end{split}
\end{equation}
Here in the last line we have used identities \eqref{I3}, \eqref{I6} and \eqref{I8}.
Substituting equation \eqref{divp} in equation \eqref{emansimp1} we get equation \eqref{emanmntxt}.

\section{Identities}\label{identity}
In this appendix we shall prove several identities and equations that we have used at different steps in our  calculations. 

\subsection{Membrane embedded in flat-spacetime}
In this subsection all identities are derived on $\rho=1$ hypersurface as embedded in flat space-time. Usually all contractions (often denoted by `$\cdot$') are with respect to flat Minkowski metric $\eta_{AB}$. In few cases we have to use contraction and covariant derivative with respect to the induced metric on the membrane. In those cases we have used Greek indices and the covariant derivatives are denoted as $\hat\nabla$. Sometimes we have used $\bar\nabla_A$ to denote $\hat\nabla$ in the language of the embedding space. For example,
$$\hat\nabla_\mu u_\nu\rightarrow \bar\nabla_A u_B \equiv \Pi_A^{A'}\Pi_B^{B'}\nabla_{A'} u_{B'} ,$$
where $\Pi_{AB}$ is the projector on the membrane.\footnote{
Most of the identities that are derived here involve indices, functions and derivatives that are defined entirely along the membrane. Therefore they could be very easily re expressed in the language of the intrinsic geometry of the membrane,
(by simply replacing $\bar\nabla\rightarrow\hat\nabla,~~\{A,B\} \rightarrow\{\mu,\nu\},~~\Pi_{AB}\rightarrow g^{(ind,f)}_{\mu\nu}$). In the main text we have often used these identities with such replacement.}\\

{\bf Identity-1}:
\begin{equation}\label{I1}
\begin{split}
&\hat\nabla_\mu\left[(u^\nu\hat\nabla_\nu)u^\mu\right]\\
=&~\partial_B [\Pi^{AB}(u\cdot\partial)u_A]-n_B (n\cdot\partial) [\Pi^{AB}(u\cdot\partial)u_A]\\
=&~\partial_B\left[ (u\cdot\partial) u^B\right] - \partial_B\left[ n^B n^C(u\cdot\partial) u^c\right] +{\cal O}(1)\\
=&~\partial_B\left[ (u\cdot\partial) u^B\right] + \partial_B\left[ n^B u^C(u\cdot\partial) n^c\right] +{\cal O}(1)\\
=&~\partial_B\left[ (u\cdot\partial) u^B\right] + \partial_B\left[ n^B u^C(u\cdot\partial) n^c\right] +{\cal O}(1)\\
=&~ (u\cdot\partial)\left[ \partial\cdot u\right]+(\partial_A u^B)(\partial_B u^A) + \partial_B\left[ n^B (u^A u^{A'} K_{AA'})\right] +{\cal O}(1)\\
=&~{K} (u^Au^{A'} K_{AA'} ) +{\cal O}(1) .\\
\end{split}
\end{equation}
Here $\hat\nabla_\mu$ denotes covariant derivative with respect to the induced metric on the membrane as embedded in the flat space, $g^{(ind,f)}_{\mu\nu}$.\\
{\bf Identity-2}:
\begin{equation}\label{midI2}
\begin{split}
n^A\partial^2 u_A=&~\partial_C(n^A\partial^C u_A)+{\cal O}(1)\\
=&-\partial_C(u^A\partial^C n_A)+{\cal O}(1)\\
=&-\partial_C(n^C u^k(n\cdot\partial)n_k + K^C_A u^A)+{\cal O}(1)\\
=&- (u\cdot\partial){K} -\partial_A( K^C_A u^A)+{\cal O}(1)\\
=&- (u\cdot\partial){K} -\partial_A( K^C_A )u^A+{\cal O}(1)\\
=&-2 (u\cdot\partial){K} +{\cal O}(1) .\\
\end{split}
\end{equation}

{\bf Identity-3}:
\begin{equation}\label{I3}
\begin{split}
&\Pi^{A'}_A\partial_{A'}\left[K^{AB} - K\Pi^{AB}\right] =0\\
\Rightarrow&~\Pi^{A'}_A\partial_{A'} K^{AB} = \Pi^{AB} \partial_A{K}.
\end{split}
\end{equation}

{\bf Identity-4}:
\begin{equation}\label{I5}
\begin{split}
&u_A\bar\nabla^2 u^A = -\Pi^{BB'}(\partial_B u_A)( \partial_{B'} u^A) = {\cal O}(1) ,\\
&\text{since}~{\Pi^{AB}\partial_A u_B}\sim{\cal O}\left(\frac{1}{D}\right) .
\end{split}
\end{equation}
Here $\bar\nabla^2u_A$ denotes the following.
$$\bar\nabla^2u_A\equiv \Pi_A^{A'}\Pi^{BB'} \partial_{B}\left( \Pi_{B'}^{B''}\Pi_{A'}^C \partial_{B''}u_C\right) . $$

{\bf Identity-5}:
\begin{equation}\label{I8}
\begin{split}
\bar\nabla_A \bar\nabla^2u^A &=\Pi^A_{A'}\partial_{A}\left[\Pi^{A'A''} \bar\nabla^2 u_{A''}\right]\\
&=-{K}n^A\bar\nabla^2 u_A + {\cal O}(D)\\
&=-{K}\left[\partial_B(n^A\partial^Bu_A) -( \partial_B n^A)\partial^Bu_A)\right]+ {\cal O}(D)\\
&=-{K}\left[\partial_B(n^A\partial^Bu_A)\right]+ {\cal O}(D)\\
&={K}\left[\partial_B(u^A\partial^Bn_A)\right]+ {\cal O}(D)\\
&={K}\left[\partial_B(K^B_A u^A)\right]+ {\cal O}(D)\\
&={K}\left[(u\cdot\partial){K}\right]+ {\cal O}(D) . \\
\end{split}
\end{equation}
In the last line we have used identity \eqref{I3}.\\

{\bf Identity-6}:
\begin{equation}\label{I2}
\begin{split}
&\Pi^{B'}_B\partial_{B'} \left[p^{AB}Q\left( \frac{\bar\nabla^2 u_A}{K} -K^C_A u_C\right)\right]\\
=&~Q~\Pi^{B'}_B\partial_{B'} \left[p^{AB}\left( \frac{\bar\nabla^2 u_A}{K} -K^C_A u_C\right)\right]+{\cal O}(1)\\
=&~ {\cal O}(1) . \\
\end{split}
\end{equation}
Here $p_{AB}$ denotes the projector perpendicular to both $n_A$ and $u_A$.
$$p_{AB} =\eta_{AB} -n_A n_B + u_A u_B . $$
In the last step  of equation \eqref{I2} we have used the identities \eqref{midI2}, \eqref{I3}, \eqref{I5} and \eqref{I8}.

{\bf Identity-7}:
\begin{equation}\label{I4}
\begin{split}
\bar\nabla_A\bar\nabla_B u^A &\equiv\Pi^{B'}_B \Pi^A_{A'} \partial_{A}\left[\Pi^{A'A''} \Pi_{B'}^{B''}\left(\partial_{B''}u_{A''}\right)\right]\\
&=-{K}\left[\Pi^{B'}_B  n^A\partial_{B'} u_A\right] +{\cal O}(1)\\
&={K}\left[\Pi^{B'}_B  u^A\partial_{B'} n_A\right] +{\cal O}(1)\\
&={K}(u^AK_{BA}) +{\cal O}(1) . \\
\end{split}
\end{equation}

{\bf Identity-8}:
\begin{equation}\label{I6}
\begin{split}
\bar\nabla_A (u\cdot\bar\nabla) u^A &\equiv\Pi^{A'}_A\partial_{A'}\left[\Pi^{AA''}(u^B\partial_B)u_{A''}\right]\\
&=-{K}~ n^A (u\cdot\partial)u_A + {\cal O}(1)\\
&={K}~ (u\cdot K\cdot u)+ {\cal O}(1) . \\
\end{split}
\end{equation}

{\bf Identity-9}:
\begin{equation}\label{I7}
\begin{split}
{\cal V}_A =&~Q~\Pi_A^B\partial_B Q +Q^2(u^CK_{CA} ) +\left(\frac{2Q^4-Q^2-1}{2}\right)\left(\frac{\Pi_A^B\partial_B{K}}{K}\right)\\
&-\left(\frac{Q^2 + 2Q^4}{2}\right)(u\cdot\partial)u_A+ \left(\frac{1+Q^2}{K}\right)\bar\nabla^2u_A .\\
\\
\therefore \Pi^{AB}\partial_A{\cal V}_B=&~Q\bar\nabla^2Q + (1+Q^2)(u\cdot\partial){K} + \left(\frac{2Q^4-Q^2-1}{2}\right)\left(\frac{\bar\nabla^2{K}}{K}\right)\\
&-\left(\frac{Q^2 + 2Q^4}{2}\right){K}~ (u^A u^BK_{AB})+ {\cal O}(1) .
\end{split}
\end{equation}
Here $\bar\nabla^2 Q$ and $\bar\nabla^2 K$ denote
$$\bar\nabla^2 Q = \Pi^{AB}\partial_A\partial_B Q,~~~~\bar\nabla^2 K = \Pi^{AB}\partial_A\partial_B K .$$
In the last line of \eqref{I7} we have used identities \eqref{I3}, \eqref{I6} and \eqref{I8}.\\

{\bf Identity-10}:
\begin{equation}\label{I10}
\begin{split}
&\partial^2\rho^{-(D-3)}=0\\
\Rightarrow&~\partial_A\left[\rho^{-(D-2)}N n^A\right]=0\\ 
\Rightarrow&~{K} N - \frac{(D-2)N^2}{\rho} + (n\cdot\partial)N =0\\ 
\Rightarrow&~{K} N -(D-2)N^2 + (n\cdot\partial)N =0~~~~\because~ \rho=1\\ 
\Rightarrow&~ (D-3)N={K}-N + \frac{(n\cdot\partial)N}{N}\\
\Rightarrow&~ (D-3)N={K}-\frac{K}{D} + \frac{(n\cdot\partial){K}}{K}+ {\cal O}\left(\frac{1}{D}\right) .
\end{split}
\end{equation}

{\bf Identity-11}
\begin{equation}\label{I11}
\begin{split}
\partial_A N &= \frac{\partial_A\left[(\partial_B\rho)(\partial^B\rho)\right]}{2N}
= \frac{(\partial^B\rho)\partial_A\partial_B\rho}{N}\\
&= \frac{(\partial^B\rho)\partial_B\partial_A\rho}{N}= (n\cdot\partial)(Nn_A)\\
\Rightarrow~(n\cdot\partial)n_A&=   \frac{\Pi^B_A\partial_B N}{N}= \frac{\Pi^B_A\partial_B {K}}{K} + {\cal O}\left(\frac{1}{D}\right) .
\end{split}
\end{equation}

{\bf Identity-12}:
\begin{equation}\label{I12}
\begin{split}
(n\cdot\partial){K}&= n^A\partial_A\partial_B n^B\\
&= n^A\partial_B\partial_A n^B\\
&= \partial_B\left[(n\cdot\partial) n^B\right]-(\partial_B n^A)(\partial_A n^B)\\
&= \partial_B\left(\frac{\Pi^{BA}\partial_A{K}}{K}\right) -K_{AB} K^{AB}\\
&= \frac{\bar\nabla^2{K}}{K} -\frac{{K}^2}{D} +{\cal O}(1) .
\end{split}
\end{equation}
Here in the last line we have used identity \eqref{I11}. Combining \eqref{I10}. \eqref{I11} and \eqref{I12} we find\\

{\bf Identity-13}:
\begin{equation}\label{I9}
\begin{split}
(D-3)N ={K} +  \left(\frac{\bar\nabla^2{K}}{{K}^2} \right)- 2\left(\frac{{K}}{D}\right) + {\cal O}\left(\frac{1}{D}\right) .
\end{split}
\end{equation}

{\bf Identity-14}:
\begin{equation}\label{I13}
\begin{split}
\partial^2Q &= \partial_A \left(\Pi^{AB}\partial_B Q\right) ~~~~\because~(n\cdot\partial)Q=0\\
&=\bar\nabla^2 Q + {\cal {O}}(1) .
\end{split}
\end{equation} 

\subsection{Relating intrinsic and extrinsic curvature of  membrane with  curvature of embedding space-time}
Here we shall relate the intrinsic curvatures of a timelike membrane with the extrinsic curvature of the membrane and the curvatures of the full space-time. For our derivation we shall follow \cite{Wald:1984rg}.\\
Define the coordinates along the full-space time  as $$\{X^A\}\equiv\{\rho,x^\mu\},~~~A=\{1,2,\cdots,D\},~~~\mu=\{2,\cdots,D\}$$
The equation of the membrane is given by $(\rho=1)$. $\{x^\mu\}$ are the coordinates that can vary along the membrane.
The unit normal to the surface is denoted as $n_A$ .\\
Suppose $\omega_A$ is a vector tangent to the membrane. $\hat\nabla_A$ denotes the covariant derivative with respect to the intrinsic metric of the membrane and $\nabla_A$ denotes the  covariant derivative with respect to the full space-time metric.
It follows that 
\begin{equation}\label{Gp1}
\begin{split}
[\hat\nabla_A,\hat\nabla_B]\omega_C &= {{\cal R}^P}_{CBA}~\omega_P\\
[\nabla_A,\nabla_B]\omega_C &= {R^P}_{CBA}~\omega_P ,\\
\end{split}
\end{equation}
where ${{\cal R}^P}_{CBA}$ denotes the intrinsic Riemann tensor of the membrane and ${R^P}_{CBA}$ is the Riemann tensor of the full space-time. We shall use $~{\mathfrak p}_{AB}~$ as the projector on the membrane surface.
\begin{equation}\label{Gp2}
\begin{split}
&~\hat\nabla_A\hat\nabla_B\omega_C\\
=&~ {\mathfrak p}^{A'}_A {\mathfrak p}^{B'}_B{\mathfrak p}^{C'}_C  \nabla_{A'}\left({\mathfrak p}^{B''}_{B'}{\mathfrak p}^{C''}_{C'}\nabla_{B''}\omega_{C''}\right) \\
=&~ {\mathfrak p}^{A'}_A {\mathfrak p}^{B'}_B{\mathfrak p}^{C'}_C  \nabla_{A'}\nabla_{B'}\omega_{C'} +  {\mathfrak p}^{A'}_A {\mathfrak p}^{B'}_B{\mathfrak p}^{C'}_C  \nabla_{A'}\left({\mathfrak p}^{B''}_{B'}{\mathfrak p}^{C''}_{C'}\right)\left(\nabla_{B''}\omega_{C''}\right)\\
=&~ {\mathfrak p}^{A'}_A {\mathfrak p}^{B'}_B{\mathfrak p}^{C'}_C  \nabla_{A'}\nabla_{B'}\omega_{C'} +{\cal K}_{AC}{\cal K}_{BC'}\omega^{C'}-  {\cal K}_{AB}\left[ (n\cdot\nabla)\omega_{C'}\right]{\mathfrak p}^{C'}_C .
\end{split}
\end{equation}
Here in the last line we have used the fact that $n^C\omega_C =0$\\
Using equations \eqref{Gp1} and \eqref{Gp2} we find
\begin{equation}\label{Gp3}
\begin{split}
{\cal R}_{PCBA}~\omega^P=  {\mathfrak p}^{A'}_A {\mathfrak p}^{B'}_B{\mathfrak p}^{C'}_C  R_{PC'B'A'}~\omega^P+\left[{\cal K}_{AC}{\cal K}_{BP}-{\cal K}_{AP}{\cal K}_{BC}\right]\omega^{P} .
\end{split}
\end{equation}
Since equation \eqref{Gp3} is true for any $\omega^P$ we find 
\begin{equation}\label{Gf1}
\begin{split}
{\cal R}_{PCBA}=  {\mathfrak p}^{A'}_A {\mathfrak p}^{B'}_B{\mathfrak p}^{C'}_C  R_{PC'B'A'}+\left[{\cal K}_{AC}{\cal K}_{BP}-{\cal K}_{AP}{\cal K}_{BC}\right] .
\end{split}
\end{equation}
Contracting equation \eqref{Gf1} with ${\mathfrak p}^{AC}$ and ${\mathfrak p}^{AC}{\mathfrak p}^{BP}$ we find 
\begin{equation}\label{gencurv}
\begin{split}
&{\mathfrak p}^C_A~{\mathfrak p}^{C'}_BR_{CC'} = {\cal R}_{AB} -{\cal K} {\cal K}_{AB} +{\cal K}_{AC} {\cal K}^C_B +
R_{AkBk'}~n^k n^{k'}  , \\
&R= {\cal R} +2 R_{CC'}~n^C n^{C'} -{\cal K}^2 + {\cal K}_{AB}{\cal K}^{AB}  .
\end{split}
\end{equation}
Note that the second equation of \eqref{gencurv} could be rewritten as
\begin{equation}\label{G1}
\begin{split}
\left[R_{CC'} - \frac{R}{2}G_{CC'}\right]n^C n^{C'}
\equiv n^C n^{C'} {\cal E}_{CC'}
= -{\cal R} +{\cal K}^2 - {\cal K}_{AB}{\cal K}^{AB}  .
\end{split}
\end{equation}
Note also that for Ricci flat geometries equation \eqref{gencurv} reduces to
\begin{equation}\label{G2}
\begin{split}
&0= {\cal R}_{AB} -{\cal K} {\cal K}_{AB} +{\cal K}_{AC} {\cal K}^C_B +
R_{AkBk'}~n^k n^{k'} \\
& \therefore 0= {\cal R} -{\cal K}^2 + {\cal K}_{AB}{\cal K}^{AB}  .
\end{split}
\end{equation}

\bibliography{lmp}{}
\bibliographystyle{JHEP}

\end{document}